\documentclass{tran-l}

\usepackage{array}
\usepackage{arydshln}

\usepackage{bm}
\usepackage{mathrsfs}
\usepackage{amssymb}
\usepackage{enumitem}
\usepackage{times}
\usepackage{pdfpages}
\usepackage{float}
\usepackage{subfig,graphicx}
\usepackage{csquotes}
\usepackage{hyperref}
\usepackage{amsmath}
\usepackage{ulem}
\usepackage{lipsum} 
\usepackage{nccmath}
\usepackage{bigints}
\usepackage{natbib}
\usepackage{multirow}
\usepackage{adjustbox}
\usepackage{threeparttable}
\usepackage{bbm}

\usepackage{pdfpages}
\usepackage{wrapfig}
\usepackage{color}

\usepackage{float}
\usepackage{tabularx}
\usepackage{caption}
\usepackage{booktabs}
\usepackage{cite}

\newtheorem{theorem}{Theorem}
\newtheorem{lemma}{Lemma}

\newtheorem{proposition}{Proposition}
\newtheorem{def1}{Definition}

\newtheorem{rem}{Remark}
\newtheorem{property}{Property}

\numberwithin{property}{section}
\numberwithin{proposition}{section}
\numberwithin{equation}{section}
\numberwithin{theorem}{section}
\numberwithin{corollary}{section}
\numberwithin{lemma}{section}
\numberwithin{def1}{section}
\numberwithin{corollary}{section}
\numberwithin{example}{section}
\numberwithin{rem}{section}
\begin{document}

\title[High Dimensional Gaussian and Bootstrap Approximations in GLM]{High Dimensional Gaussian and Bootstrap Approximations in Generalized Linear Models}

\author{Mayukh Choudhury}
\address{Department of Mathematics, Indian Institute of Technology Bombay, Mumbai 400076, India}
%\curraddr{Department of Mathematics and Statistics,Case Western Reserve University, Cleveland, Ohio 43403}
\email{214090002@iitb.ac.in}

\author{Debraj Das}
\address{Department of Mathematics, Indian Institute of Technology Bombay, Mumbai 400076, India}
\email{debrajdas@math.iitb.ac.in}

%    General info
%\subjclass[2000]{Primary 54C40, 14E20; Secondary 46E25, 20C20}

%\date{January 1, 2001 and, in revised form, June 22, 2001.}



\begin{abstract}
Generalized Linear Model (or GLM) extends the ordinary linear regression by linking the mean of the response variable to covariates through appropriate link functions. GLM is widely used in the analysis of datasets arising from diverse fields including medical sciences, clinical trials, population surveys and risk analysis. In this paper, we investigate the Gaussian and Bootstrap approximations of GLM under two separate high dimensional regimes: (I) when the dimension $d$ grows slower than $n$ and (II) when $d$ grows exponentially with $n$. Under regime (I), we essentially show that the Gaussian approximation holds over the collection of Borel convex sets when $d = o\big(n^{2/5}\big)$ and over the collection of Euclidean balls when $d = o\big(n^{1/2}\big)$. We further devise two high dimensional Bootstrap methods which are valid over the collections of Borel convex sets and Euclidean balls under the same dimension growth rates. Then we move to regime (II) where we invoke sparsity to GLM through Lasso. We show that the high dimensional Gaussian approximation fails under regime (II). However, the Bootstrap approximations over convex sets and Euclidean balls are valid for the relevant part of the GLM estimator provided $\log d = o\big(n^{2\tau/3}\big)$ and the number of non-zero regression parameters is $o\big(n^{1/3- 4\tau/3}\big)$, when the Lasso penalty $\lambda_n \sim n^{1/2 + \tau}$, for some $\tau \in (0, 1/4)$. Simulation studies confirm the strong finite-sample performance of our proposed Bootstrap methods under both regime (I) and (II). We also implement our methods on real datasets.
\end{abstract}

\maketitle

\section{Introduction}\label{sec:intro}
Generalized Linear Model (or GLM) is a widely used statistical modeling technique, formulated by \citet{nelder1972generalized}.  GLM encompasses several sub-models such as linear regression, logistic regression, probit regression, Poisson regression, gamma regression etc. The basic building block of GLM is the link function that connects the responses with the covariates.
The concept of regression is commonly traced back to Sir Francis Galton and his studies of heredity. In its simplest form, the linear regression evaluates the relationship between two variables: a continuous dependent variable and one (usually continuous) independent variable, with the dependent variable expressed as a linear function of the independent variable. Here, the link function is the identity function. One of the most useful methods in the field of medical sciences, clinical trials, surveys etc. is the logistic regression when the response variable is dichotomous or binary. \citet{berkson1944application} introduced the `logit' link function as a pivotal instrument and later,  in his seminal paper,  \citet{cox1958regression} familiarized it in the field of regression when the response variable is binary. In risk modelling or insurance policy pricing, Poisson regression is ideal provided response variable is the number of claim events per year. On the other hand, duration of interruption as a response variable lead to gamma regression in predictive maintenance. In both Poisson and gamma regression, generally the `log' link function is utilized. There are many other real-life applications which can be dealt with some sub-models of GLM.

Let $\{y_{1},..,y_{n}\}$ be responses and $\{\bm{x}_1,\dots, \bm{x}_n\}$ be non-random covariates. Assume that $y_i$ has density $f_{\theta_i}(y_{i}) = \mbox{exp}\big\{y_{i}\theta_{i} - b(\theta_{i})\big\}c(y_{i})$, $i=1,\dots, n$, with respect to a common measure, and $\theta_i$ lies in an open set for all $i$. Here, $\theta_1,\dots, \theta_n$ are the canonical parameters. The dependency of the response $y_i$ on the covariate $\bm{x}_i$ is characterized by a link function $g(\cdot)$, more precisely by $g(\mu_{i}) = \bm{x}^\top_{i}\bm{\beta}$. Here $\mu_{i} = E(y_{i}) = b^\prime(\theta_{i})$, $i=1,\dots, n$, and $\bm{\beta}$ is the regression parameter. Note that for all  $i=1,\dots, n$, $\mu_{i} = \mathbbm{E}(y_{i}) = b^\prime(\theta_{i})$ and $\mathbbm{v}\text{ar}(y_i)=b^{\prime\prime}(\theta_i)$, and hence $\theta_{i} = h(\bm{x}^\top_{i}\bm{\beta})$ where $h = (g \circ b^{\prime})^{-1}$, assuming its existence. Therefore, the log-likelihood as a function of $\bm{\beta}$ is given by 
\begin{align}\label{eqn:loglikelihood}
\sum_{i=1}^{n}\ell_{ni}(\bm{\beta}) = \sum_{i=1}^{n}\big[y_ih(\bm{x}_i^\top\bm{\beta}) -h_1(\bm{x}_i^\top\bm{\beta})\big].
\end{align}
We assume throughout the paper that either the function $h_1(\cdot):=b(h(\cdot))$ is strictly convex or the response variable is positive and the function $h(\cdot)$ is strictly concave. This consideration is important to guarantee the uniqueness of the GLM estimator and its Bootstrapped versions. Moreover, usual sub-models of GLM generally satisfy this assumption, as pointed out in Table \ref{tab1} below. %\blue{added nb}
\begin{table*}[ht]
\centering
\caption{Some Common Types of GLM}
\label{tab1}
\begin{adjustbox}{max width=\linewidth}
\begin{tabular}{@{}lcrcrrr@{}}
\midrule
 &\multicolumn{4}{c}{Components of GLM} \\
\cline{2-5}
Type &
\multicolumn{1}{c}{$\mu=b^\prime (\theta)$} &
\multicolumn{1}{c}{Link Function $(g(\cdot))$}&
\multicolumn{1}{c}{$h(u)$}&
\multicolumn{1}{c@{}}{$h_1(u)$}\\
\midrule
Linear & $\theta$ & identity & $u$ & $u^2/2$\\
Logistic & $e^{\theta}(1+e^{\theta})^{-1}$  & logit & $u$ & $\log(1+e^u)$ \\
Probit & $e^{\theta}(1+e^{\theta})^{-1}$ & probit & $\log\big\{\frac{\Phi(u)}{1-\Phi(u)}\big\}$ &$-\log\{1-\Phi(u)\}$ \\
Poisson    & $e^{\theta}$ & $\log$ & $u$ & $e^u$ \\
Gamma & $\alpha\theta$ & $\log$ & $-\alpha e^{-u}$ & $\alpha u$    \\
Negative Binomial & $r(1-e^{\theta})^{-1}e^{\theta}$ & $\log$ & $\log\{\frac{e^u}{r+e^u}\}$ & $-r\log\{\frac{r}{r+e^u}\}$\\
\bottomrule
\end{tabular}
\end{adjustbox}
\begin{tablenotes}
    \item $\alpha:$ known shape parameter in gamma distribution  with unknown scale parameter.
\item $r$: known number of success in $NB(r,p)$ model with unknown success probability $p$
\item $\Phi:$ cumulative distribution function of the standard normal distribution.
\end{tablenotes}

\end{table*} 

\subsection{\bf Our Contributions}
In the analysis of statistical models, in general, the complexity of the model does not allow us to draw exact inferences based on the underlying estimator. Instead we try to fill the gap by approximating the distribution of the estimator assuming the sample size $n$ to be large. The natural choices for such approximations are generally the Gaussian and the Bootstrap approximations. Moreover, the approximations should be valid uniformly over a collection of sets where this collection is generally dictated by the form of the confidence regions that the statistician wants to construct for the purpose of inference. The collection of Borel convex sets includes collection of Euclidean balls, the collection of rectangles, the collection of half spaces etc., and hence is quite rich enough to construct different types of confidence regions for $\bm{\beta}$. Among these, the natural choice of a confidence region is generally a sphere or an ellipsoid, for which one requires the approximation results uniformly over the collection of Euclidean balls. Based on these considerations, we explore the Gaussian and Bootstrap approximations of the distribution of the GLM estimator $\hat{\bm{\beta}}_n$ over the collection of Borel convex sets (or $\mathcal{C}$) and over the collection of Euclidean balls (or $\mathcal{B}$). We also consider the dimension $d$ to be large compared to $n$ under the following two regimes. \begin{itemize}
\item[\underline{Regime (I):}] \textit{High dimension, i.e., when $d$ grows slower than $n$.}
\item[\underline{Regime (II):}] \textit{Ultra high dimension, i.e., when $d$ grows exponentially with $n$.} 
\end{itemize}

In regime (I), we explore the Gaussian and Bootstrap approximations of the distribution of the usual GLM estimator $\hat{\bm{\beta}}_n = \operatorname*{Argmin}_{\bm{\beta}\in \mathbf{R}^d}\sum_{i=1}^{n}-\ell_{ni}(\bm{\beta})$. Clearly, $\hat{\bm{\beta}}_n$ is the solution of the score equation
\begin{align}\label{eqn:defineglmest12}
     \sum_{i=1}^{n}(y_i - g^{-1}(\bm{x}_i^\top \bm{\beta}))h^\prime(\bm{x}_i^\top \bm{\beta})\bm{x}_i = \bm{0},
 \end{align} 
where we assume the smoothness condition on $h$ in section \ref{sec:notandassump}. We consider the centered and scaled version $n^{1/2}(\hat{\bm{\beta}}_n-\bm{\beta})$ and study the Gaussian approximation based on
\begin{align}\label{eqn:thm4.1}
\Delta_{n,1}(\mathcal{A}):=\sup_{A\in\mathcal{A}}\Bigg{|}\mathbf{P}\Big[n^{1/2}(\hat{\bm{\beta}}_n-\bm{\beta})\in A\Big]-\mathbf{P}\Big[\bm{G}_{1n}\in A\Big]\Bigg{|},   
\end{align}
for $\mathcal{A} = \mathcal{C}\; \text{and}\; \mathcal{B}$. Here $\bm{G}_{1n}\sim N_d(\bm{0},\Tilde{\bm{S}}_n)$ with $\Tilde{\bm{S}}_n$ being the approximate covariance matrix of $n^{1/2}(\hat{\bm{\beta}}_n-\bm{\beta})$ and is defined in section \ref{sec:notandassump}. Under certain conditions, we essentially show that as $n \rightarrow \infty$,
\begin{align}\label{eqn:thm4.2}
\Delta_{n,1}(\mathcal{C})=o(1),\; \text{provided}\; d = o(n^{2/5})\;\;\text{and}\;\;\Delta_{n,1}(\mathcal{B})=o(1),\; \text{provided}\; d = o(n^{1/2}).
\end{align}
Additionally, to accommodate the fact that $\tilde{\bm{S}}_n$ is generally unknown, we develop two Bootstrap methods, viz., Perturbation Bootstrap and Pearson's Residual Bootstrap. We quantify the validity of both Bootstrap methods based on
\begin{align}\label{eqn:thm4.11}
\Delta_{n,1}^{*}(\mathcal{A}):=\sup_{A\in\mathcal{A}}\Bigg{|}\mathbf{P}\Big[n^{1/2}(\hat{\bm{\beta}}_n^{*}-\hat{\bm{\beta}}_n)\in A\big{|}y_1,\dots,y_n\Big]-\mathbf{P}\Big[n^{1/2}(\hat{\bm{\beta}}_n-\bm{\beta})\in A\Big]\Bigg{|},   
\end{align}
with $\mathcal{A} = \mathcal{C}$ and $\mathcal{B}$. Here $\hat{\bm{\beta}}_n^{*}$ denotes the Bootstrapped GLM estimator, under Perturbation Bootstrap or Pearson's Residual Bootstrap. We show that both Bootstrap methods are valid under the dimension growth rate same as the Gaussian approximation, mentioned in (\ref{eqn:thm4.2}). More precisely, we show that
\begin{align}\label{eqn:thm4.2-3}
\Delta_{n,1}^{*}(\mathcal{C})=o_p(1),\; \text{provided}\; d = o(n^{2/5})\;\;\text{and}\;\;\Delta_{n,1}^*(\mathcal{B})=o_p(1),\; \text{provided}\; d = o(n^{1/2}).
\end{align} 
In regime (II), we try to extend the Gaussian and Bootstrap approximations of regime (I) to the situation when $d$ may grow exponentially with $n$. We handle this ultra high dimensional setup by invoking sparsity based on Lasso. The Lasso estimator in GLM is defined as
\begin{align}\label{eqn:deflasso}
\bar{\bm{\beta}}_{n} = \mbox{Argmin}_{\bm{\beta}} \Big\{-\sum_{i=1}^{n}y_{i}h(\bm{x}_{i}^\top\bm{\beta})+\sum_{i=1}^{n}h_1(\bm{x}_{i}^\top\bm{\beta})+\lambda_{n}\sum_{j=1}^{p}|\beta_{j}|\Big\},
\end{align}
which is nothing but the $l_1-$penalized negative log-likelihood. Let $\mathcal{A}_n:=\big\{j\in\{1,...,d\}:\beta_{j,n}\neq 0\big\}$ denote the active set of covariates and $\bm{\beta}_n^{\mathcal{A}_n}$ denote the sub-vector of $\bm{\beta}_n$ corresponding to $\mathcal{A}_n$. Clearly, based on $\bar{\bm{\beta}}_{n}$, the estimator of $\bm{\beta}_n^{\mathcal{A}_n}$ is $\bar{\bm{\beta}}_n^{\bar{\mathcal{A}}_n}$ where $\bar{\mathcal{A}}_n:=\big\{j\in\{1,...,d\}:\bar{\beta}_{j,n}\neq 0\big\}$. We investigate the Gaussian and Bootstrap approximations of the distribution of $\bar{\bm{\beta}}_n$ which are important for drawing inference about $\bm{\beta}$ in ultra high dimension. To that end, first, we study  
\begin{align}\label{eqn:thm6.1}
\Delta_{n,2}(\mathcal{B}):=\mbox{sup}_{B\in\mathcal{B}}\Bigg{|}\mathbf{P}\Big[\big\{n^{1/2}\Big(\bar{\bm{\beta}}_n^{(1)}-\bm{\beta}^{(1)}\Big)\in B\big\} \cap \{\bar{\mathcal{A}}_n = \mathcal{A}_n\}\Big]-\mathbf{P}\Big[\bm{G}_{2n}\in B\Big]\Bigg{|},%\to 1\;\text{as}\;n\to\infty,   
\end{align}
where $\bm{G}_{2n}\sim N(\bm{0},\tilde{\bm{S}}_{n,11})$ with $\tilde{\bm{S}}_{n,11}$ being an approximation of $Var\big(n^{1/2}\bar{\bm{\beta}}_n^{(1)}\big)$, defined in section \ref{sec:notandassump}. We show that
\begin{align}\label{eqn:thm5.1}
\Delta_{n,2}(\mathcal{B})\to 1,\;\text{as}\;n\to\infty.
\end{align}
Clearly, (\ref{eqn:thm5.1}) shows that the Gaussian approximation fails under regime (II), unlike in regime (I). To circumvent the inability of the Gaussian approximation, we consider Perturbation and Pearson's Residual Bootstrap versions of $\bar{\bm{\beta}}_n$. If we denote either one of the Bootstrapped estimators by $\bar{\bm{\beta}}_n^{*}$, then it is essential to study
\begin{align}\label{eqn:thm6.2}
&\Delta_{n,2}^{*}(\mathcal{A})\nonumber\\ 
&:=\mbox{sup}_{B\in\mathcal{A}}\Bigg{|}\mathbf{P}\Big[\big\{n^{1/2}\Big(\bar{\bm{\beta}}_n^{*\bar{\mathcal{A}}_{n}^*}-\bar{\bm{\beta}}_n^{\bar{\mathcal{A}}_n}\Big)\in B\big\}|y_1,\dots, y_n\Big] -\mathbf{P}\Big[n^{1/2}\Big(\bar{\bm{\beta}}_n^{\bar{\mathcal{A}}_n}-\bm{\beta}^{\mathcal{A}_n}\Big)\in B\Big]\Bigg{|},\end{align}
where $\bar{\mathcal{A}}_{n}^*$ is the set of non-zero components of $\bar{\bm{\beta}}_n^{*}$.  We essentially show that as $n \rightarrow \infty$, irrespective of whether $\mathcal{A} = \mathcal{C}$ or $\mathcal{B}$,
\begin{align}\label{eqn:thm4.21}
\Delta_{n, 2}^{*}(\mathcal{A})=o_p(1),\; \text{provided}\; |\mathcal{A}_n| = o\big(n^{{(1-4\tau)/3}}\big)\;\text{and}\; \log d = o(n^{2\tau/3}),
\end{align}
when the penalty $\lambda_n \sim n^{1/2 + \tau}$, for some $0< \tau < 1/4$. Therefore, one can utilize the Bootstrap approximation to draw inferences in any sub-model of GLM irrespective of whether the dimension $d$ grows slower or faster than $n$. Moreover, in (\ref{eqn:thm6.2}), the rate of $\log d$ can be improved to $o(n^{2\tau})$ when  $h(\cdot)$ is the identity function. We refer to section \ref{sec:infer1} and \ref{sec:infer2} for more general high dimensional Gaussian and Bootstrap approximation results respectively for regime (I) and regime (II). 

\subsection{\bf Techniques used} Based on the discussions on our findings, we now present brief details on the techniques that we use to establish our results.

\subsubsection{\bf GLM under regime (I)}\label{sec:techregime1} Recall the definition of $\hat{\bm{\beta}}_n$ as in (\ref{eqn:defineglmest12}). Under appropriate smoothness conditions on $b(\cdot)$ and the link function $g(\cdot)$, a suitable Taylor's expansion leads to the following Bahadur's representation of $n^{1/2}(\hat{\bm{\beta}}_n-\bm{\beta})$:
\begin{align}\label{eqn:expansion}
& n^{1/2}[\hat{\bm{\beta}}_n-\bm{\beta}]=\bm{T}_n+\bm{r}_n,
\end{align} 
where $\bm{T}_n$ is the leading linear term which is some form of sample mean. Whereas, $\bm{r}_n$ is the error term. It is expected that $\bm{r}_n$ is small and therefore $\bm{T}_n$ should dictate the Gaussian approximation of the distribution of $n^{1/2}(\hat{\bm{\beta}}_n-\bm{\beta})$. Note that due to (\ref{eqn:expansion}),
\begin{align}\label{eqn:mainapprox}
 &\Delta_{n, 1}(\mathcal{A})=\sup_{A\in\mathcal{A}}\Big{|}\mathbf{P}\Big[\bm{T}_n+\bm{r}_n\in A\Big]-\mathbf{P}\Big[\bm{G}_{1n}\in A\Big]\Big{|}\nonumber\\
 &\leq \underbrace{\sup_{A\in\mathcal{A}}\Big{|}\mathbf{P}\Big(\bm{T}_n\in A\Big)-\mathbf{P}\Big(\bm{G}_{1n}\in A\Big)\Big{|}}_{\text{Term I}}+2\underbrace{\sup_{A\in\mathcal{A}}\Big{|}\mathbf{P}\Big(\|\bm{r}_n\|>\epsilon\Big)\Big{|}}_{\text{Term II}}+2\underbrace{\sup_{A\in\mathcal{A}}\Big{|}\mathbf{P}\Big(\bm{G}_{1n}\in (\partial{A})^{\epsilon}\Big)\Big{|}}_{\text{Term III}}\nonumber \\
&\;\;\;\;\;\;\;\;\;\;\;\;\;\;\;\;\;\;\;\;\;\;\;\;\;\;\;\;\;\;\;\;+2\underbrace{\sup_{A\in\mathcal{A}}\Big{|}\mathbf{P}\Big(\bm{T}_n\in (\partial{A})^{\epsilon}\Big)-\mathbf{P}\Big(\bm{G}_{1n}\in (\partial{A})^{\epsilon}\Big)\Big{|}}_{\text{Term IV}},
\end{align}
for any $\epsilon>0$. Here $\partial{A}$ denotes the boundary of the set $A$ and $(\partial{A})^\epsilon$ is the $\epsilon$-enlargement of it. Note that Term I and Term IV quantify the errors in the Gaussian approximation of the leading linear term $\bm{T}_n$. Term II is quantifying the fact that $n^{1/2}(\hat{\bm{\beta}}_n-\bm{\beta})$ is not actually a sample mean. Term III is due to the fact that the construction of $\bm{G}_n$ is based only on the leading term $\bm{T}_n$. To establish (\ref{eqn:thm4.2}), we consider a suitable choice of $\epsilon$ and show that each of the aforementioned four terms becomes smaller as $n \rightarrow \infty$. When $\mathcal{A}$ is either $ \mathcal{C}$ or $\mathcal{B}$, we handle Term I and Term IV based on the existing high dimensional Gaussian approximation results over convex sets and Euclidean balls. In particular, we utilize the recent results of  \citet{fang2024large} and \citet{zhilova2020nonclassical}. In either case, Term II is shown to be small by utilizing high dimensional concentration bounds on $[n^{1/2}(\hat{\bm{\beta}}_n-\bm{\beta})]$, and using some of the intermediate results of \citet{he2000parameters}. When $\mathcal{A} = \mathcal{C}$, we show that Term III is small based on Proposition 1.1 of \citet{raivc2019multivariate} and the reverse isoperimetric inequality for convex sets by \citet{nazarov2003maximal}. On the other hand when $\mathcal{A} = \mathcal{B}$, we handle Term III based on the reverse isoperimetric inequality for Euclidean balls by \citet{zhilova2020nonclassical}. To prove the validity of our Bootstrap approximations, i.e., to establish (\ref{eqn:thm4.2}), we consider the Bahadur type representation of Bootstrap pivotal quantity as follows.
\begin{align}\label{eqn:pbbahadurfinal}
& n^{1/2}[\hat{\bm{\beta}}_n^{*}-\hat{\bm{\beta}}_n]=\hat{\bm{T}}_n^{*}+\hat{\bm{r}}_n^{*},
\end{align} 
where, similar to the original pivotal quantity, $\hat{\bm{T}}_n^{*}$ is again the leading linear term and $\hat{\bm{r}}_n^{*}$ is the error term. We handle these terms similarly to the original case and prove that as $n\rightarrow \infty$,
$$ \mbox{sup}_{B\in\mathcal{A}}\Bigg{|}\mathbf{P}\Big[\big\{n^{1/2}\Big(\hat{\bm{\beta}}_n^{*}-\hat{\bm{\beta}}_n\Big)\in B\big\}|y_1,\dots, y_n\Big] -\mathbf{P}\Big[\bm{G}_{1n}\in B\Big]\Bigg{|} \rightarrow 0 \;\;\text{in probability},$$ $\text{provided}\; d = o(n^{2/5})$ and $d = o(n^{1/2})$, respectively when $\mathcal{A} = \mathcal{C}$ and $\mathcal{A} = \mathcal{B}$.

\subsubsection{\bf GLM under regime (II)}\label{sec:techregime2}
Here, in order to handle the ultra high dimensional setup, we introduce sparsity in GLM using Lasso and then explore the validity of the corresponding high dimensional Gaussian and Bootstrap approximations. However, it is essential to study the support recovery by Lasso in GLM before we move towards distributional asymptotics. We investigate the KKT condition corresponding to (\ref{eqn:deflasso}) and show that Lasso performs support recovery with high probability, i.e., $\mathbf{P}(\bar{\mathcal{A}}_n = \mathcal{A}_n) \rightarrow 1$, as $n\rightarrow \infty$, when the penalty $\lambda_n >> n^{(1/2 + \tau)}$ for some $0< \tau < 1/4$ and the responses $y_1,\dots y_n$ are sub-exponential random variables. The support recovery property of Lasso along with suitable Taylor expansion are crucial to find out the following Bahadur type representation of the relevant part $\bar{\bm{\beta}}_n^{\bar{\mathcal{A}}_n}$ of the estimator $\bar{\bm{\beta}}_n$:
\begin{align}\label{eqn:bahadurII}
 n^{1/2}\big(\bar{\bm{\beta}}_n^{\bar{\mathcal{A}}_n}-\bm{\beta}^{\mathcal{A}_n}\big) = \bar{\bm{T}}_{n} + \bar{\bm{b}}_n+\bar{\bm{r}}_n.  
\end{align}
Here, $\bar{\bm{T}}_{n}$ is the leading linear term, $\bar{\bm{b}}_n$ is the bias term and $\bar{\bm{r}}_n$ is the error. The representation (\ref{eqn:bahadurII}) is useful to explore the high dimensional Gaussian and Bootstrap approximations based on $\Delta_{n,2}(\mathcal{A})$, with $\mathcal{A} = \mathcal{C}$ and $ \mathcal{B}$. The leading linear term $\bar{\bm{T}}_{n}$ and the error term $\bar{\bm{r}}_n$ can be handled in the same fashion as in case of regime (I), the only difference being the requirement of new concentration bounds on $n^{1/2}\big(\bar{\bm{\beta}}_n^{\bar{\mathcal{A}}_n}-\bm{\beta}^{\mathcal{A}_n}\big)$ which we develop in section \ref{sec:lemmaC}.
The bias term $\bar{\bm{b}}_n$, on the other hand, depends on $\lambda_n$ through $n^{-1/2}\lambda_n$ which diverges to $\infty$. This results in the non-negligibility of $\bar{\bm{b}}_n$, leading to the failure of the Gaussian approximation, as mentioned in (\ref{eqn:thm5.1}). To establish the validity of our Bootstrap methods, we first obtain the Bahadur representation  $$n^{1/2}\big(\bar{\bm{\beta}}_n^{*\bar{\mathcal{A}}_{n}^*}-\bar{\bm{\beta}}_n^{\bar{\mathcal{A}}_n}\big)= \bar{\bm{T}}_{n}^{*} + \bar{\bm{b}}_n^{*}+\bar{\bm{r}}_n^{*},$$ and show that $\bar{\bm{r}}_n^{*}$ converges in (Bootstrap) probability to $0$. Then to prove (\ref{eqn:thm6.2}), we (roughly) show that for any set $A \in \mathcal{A}$ ($\mathcal{A} = \mathcal{C}\; \text{or}\; \mathcal{B}$), 
\begin{align}\label{eqn:r2}
&\mathbf{P}\bigg[\bigg{|}\mathbf{P}\Big[n^{1/2}\big(\bar{\bm{\beta}}_n^{*\bar{\mathcal{A}}_{n}^*}-\bar{\bm{\beta}}_n^{\bar{\mathcal{A}}_n}\big)\in A|y_1,\dots, y_n\Big]-\mathbf{P}\Big[n^{1/2}\Big(\bar{\bm{\beta}}_n^{\bar{\mathcal{A}_n}}-\bm{\beta}^{\mathcal{A}_n}\Big)\in A\Big]\bigg{|}>\epsilon\bigg]\nonumber\\ 
&\le 2\mathbf{P}\bigg[\mbox{sup}_{D\in\mathcal{A}}\bigg{|}\mathbf{P}\Big[\bar{\bm{T}}_n^{*}\in D |y_1,\dots, y_n\Big]-\mathbf{P}\Big[\bar{\bm{T}}_n\in D \Big]\bigg{|}>\epsilon\bigg]+\epsilon/2,
\end{align}
for any $\epsilon > 0$. Thus (\ref{eqn:thm6.2}) follows by taking a suitable choice of $\epsilon$ in (\ref{eqn:r2}), since the distribution of $\bar{\bm{T}}_n$ and the conditional distribution of $\bar{\bm{T}}_n^{*}$, given $y_1,\dots, y_n$, can be approximated by the same Gaussian distribution uniformly over $\mathcal{C}$ or $\mathcal{B}$. 

\subsection{\bf{Related Literature}}
The asymptotic properties of the maximum likelihood estimator (or MLE) in GLMs have been extensively studied under the classical regime where the number of parameters $d$ is fixed and the sample size $n \to \infty$. Foundational works in that direction is due to \citet{fahrmeir1985} and \citet{haberman1977,haberman1978}, who established general conditions for the consistency and asymptotic normality of the MLE in GLMs, thus providing validity for asymptotic inference in fixed dimensional GLMs. \citet{fahrmeir1987} extended the asymptotic framework by developing asymptotic testing procedures in GLMs, including Wald and likelihood ratio tests, all of which rely fundamentally on the asymptotic normality of the MLE. A comprehensive treatment of fixed dimensional GLMs in the asymptotic sense and beyond is carried out in the monograph of \citet{mccullagh1989}, which remains a standard reference in the field of GLMs. In addition to asymptotic normality, Bootstrap methods have also been studied in different submodels of GLM under the classical fixed-dimensional setup. \citet{freedman1981bootstrapping} introduced and explored the Residual and Paired Bootstrap in linear regression. Later \citet{wu1986jackknife,mason1992rank} and \citet{jin2001perturbation} respectively introduced the Weighted Bootstrap and the Wild Bootstrap and the Perturbation Bootstrap in linear regression. For second and higher order properties of these Bootstrap methods in fixed dimensional linear regression, one can see \citet{liu1988bootstrap,lahiri1992bootstrapping}, \citet{barbe2012weighted}  and \citet{das2019distributional}; among others. In case of submodels of GLM other than linear regression, \citet{lee1990bootstrapping} and later \citet{claeskens2003quadratic} explored the validity of Paired and the quadratic Bootstrap methods in fixed dimensional logistic regression.  \citet{moulton1991} was the first to introduce Pearson's Residual Bootstrap method and explored the Bootstrap based variance estimation in logistic regression. \citet{friedl1997resampling} studied the implementation and properties of Bootstrap methods for GLMs, with emphasis on practical computation and accuracy of confidence intervals. Recently, \citet{das2025pebble} explored the second order correctness of Perturbation Bootstrap method in fixed dimensional logistic regression and \citet{choudhury2024bootstrapping} explored the validity of Perturbation and Pearson's Residual Bootstrap methods for the Lasso estimator in fixed dimensional GLM. For a systemic review of the Bootstrap methods in linear models, one can see \citet{davison1997} and \citet{lahiri2006bootstrap}. The Earliest systematic investigation of univariate projections of an estimator under $d\rightarrow \infty$ was due to \citet{Portnoy1984, portnoy1985asymptotic} where the author established consistency and asymptotic normality of the univariate components of an M-estimator in the linear regression setup when $(d\log n)^{3/2} = o(n)$. Refinements of these results under the same setup were obtained by \citet{mammen1989asymptotics}. Building on their ideas, \citet{he2000parameters} developed a framework for general $M$-estimators with increasing dimension. They proved consistency under $d = o(n / \log n)$ and established the asymptotic normality of the univariate components when  $d = o\big((n / \log n)^{1/2}\big)$. \citet{mammen1993bootstrap} established the validity of Paired and Wild Bootstrap for approximating the distribution of a univariate component of the least square estimator when $d^{1+\delta} = o(n)$ for some $\delta > 0$. Later \citet{chatterjee2002dimension} extended the results of \citet{mammen1993bootstrap} to the Weighted Bootstrap in linear regression. Although most of the theoretical developments in the statistical inference literature is centered around to establish asymptotic Gaussian or Bootstrap approximations of the distribution of an estimator under fixed dimensions or of some univariate projections of the estimator under high dimensions, the recent interest is to obtain Gaussian or Bootstrap approximation of the estimator, as a whole, over a suitable classes of sets, primarily driven by the high dimensional problems arising in the practice. In that direction, there is an extensive volume of work primarily on approximating the distribution of the high dimensional sample mean. Some early results on the high dimensional Gaussian approximation of the sample mean is due to \citet{senatov1980several} who showed that the Gaussian approximation holds for the collection of compact convex sets under the existence of third moments when $d = o(n^{1/5})$. \citet{bentkus2003dependence} and \citet{raivc2019multivariate} considered the high dimensional Gaussian approximation of mean over the collection of all measurable convex sets (or $\mathcal{C}$) and showed that $d = o(n^{2/7})$ is possible under the existence of third moments. Recently, \citet{fang2024large} improved the growth rate of the dimension for $\mathcal{C}$ by getting the rate $d = o(n^{2/5})$ under the existence of fourth moments. For the collection of all Euclidean balls (or $\mathcal{B}$), the best result with respect to dimension dependence is due to \citet{fang2024large}. They established the Gaussian approximation of mean when $d = o(n^{1/2})$ under the existence of fourth moments. \citet{fang2024large} also established the high dimensional Bootstrap approximation for the distribution of mean over $\mathcal{C}$ and $\mathcal{B}$ respectively when $d=o(n^{2/5})$ and $d = o(n^{1/2})$. On the other hand, there is a large volume of work on the high dimensional Gaussian and Bootstrap approximations over the collection of hyper-rectangles, where it is established that $\log d$ can grow like $o(n^a)$ for some $a\in (0, 1/2)$. In that direction, one can see \citet{chernozhukov2013gaussian,chernozhukov2017central,chernozhukov2023high}, \citet{fang2021high}, \citet{das2021central, das2025necessary}, \citet{kuchibhotla2020high}, \citet{das2025necessary}; among others. Here our aim is to extend the high dimensional Gaussian and Bootstrap approximation of distribution of the sample mean over the collections $\mathcal{C}$ and $\mathcal{B}$ to the distributions of the concerned estimators in GLM. 

\subsection{\bf Organization of the Paper}
Rest of the paper is organized as follows. In section \ref{sec:pbmethod} and \ref{sec:pbmethodregime2}, we have provided the descriptions of the Perturbation and Pearson's Residual Bootstrap methods under regime (I) and (II) respectively. The detailed regularity conditions and notations required for our main results to go through, have been relegated to section \ref{sec:notandassump} and \ref{sec:regime2} for respective regimes. The main results under regime (I), are presented in section \ref{sec:mainregime1} along with the Bootstrap approximation results in section \ref{sec:pbmethod}. Similarly for regime (II), the results are attributed to section \ref{sec:regime2} and  \ref{sec:mainregime2} with Bootstrap approximation results in section \ref{sec:pbmethodregime2}. Finite sample performances of our proposed Bootstrap method to validate Theorem \ref{thm:pbconvexoriginal1} and \ref{thm:bootapproxglmvsc} have been attributed to section \ref{sec:simstudy} in an extensive simulation study. Applications on real data sets have been provided in section \ref{sec:realdata}. Appendix contains all the statement and proof of requisite lemmas, proof of main results viz. Theorem \ref{thm:convex1}, \ref{thm:ball1}, \ref{thm:pbconvexoriginal1}, \ref{thm:prbconvexoriginal1}, Proposition \ref{prop:solutionkktglm}, Theorem \ref{thm:failgauss}, \ref{thm:bootapproxglmvsc} and \ref{thm:prbbootapproxglmvsc}.

\section{Inference in GLM under regime (I)}\label{sec:infer1}
This section is devoted to the Gaussian and Bootstrap approximations of the GLM estimator under regime (I). This section is divided into three subsections. The first subsection describes the regularity conditions. The second one is on the Gaussian approximation over Borel convex sets and and Euclidean balls. And the last subsection explores two Bootstrap approximations of GLM under regime (I), which can readily be used to perform statistical inferences about $\bm{\beta}$. 

\subsection{\bf Regularity Conditions}\label{sec:notandassump}
Recall the log-likelihood of the regression parameter $\bm{\beta}$ in GLM, as defined in (\ref{eqn:loglikelihood}). Clearly, based on the Taylor's theorem, one can identify $\bm{W}_n= n^{-1/2}\sum_{i=1}^{n}(y_i-\mu_i)\bm{x}_ih^\prime(\bm{x}_i^\top\bm{\beta})$ as the leading mean type term in the Bahadur's representation of $\hat{\bm{\beta}}_n$. Note that the $\text{var}(\bm{W}_n)=\bm{S}_n=\frac1n\sum_{i=1}^{n}\bm{x}_i\bm{x}^\top_i\big[h^\prime(\bm{x}_i^\top\bm{\beta})\big]^2\mathbf{E}(y_i-\mu_i)^2$. $O(\cdot)$, $o(\cdot)$, $O_p(\cdot)$ and $o_p(\cdot)$ have usual interpretations. For two positive sequences $\{a_n\}_{n\geq 1}$ and $\{b_n\}_{n\geq 1}$, $a_n \sim b_n$ means $a_n=O(b_n)$ and $b_n =O(a_n)$, as $n\rightarrow \infty$. For any matrix $\bm{U}$, $\bm{U}_{j\cdot}^\top$ denotes the $j$th row. When $\bm{U}$ is a square matrix, then $\lambda_{\min}(\bm{U})$ and $\lambda_{\max}(\bm{U})$ respectively denote the minimum and maximum eigen values. Here,  $\|\bm{a}\|$ and $\|\bm{a}\|_{\infty}$ respectively denote the Euclidean and sup norm applied on a vector $\bm{a}$. On the other hand, $\|\bm{U}\|$ and $\|\bm{U}\|_{H.S}$ respectively denote the operator and Hilbert-Schmidt norm of a matrix $\bm{U}$. We are now ready to state the set of regularity conditions, required for regime (I). 
\begin{enumerate}[label={(A.\arabic*)}]
\item For some $0\le\alpha_1, \alpha_2<1$ and $c_1, c_2 > 0$,  
$c_1n^{-\alpha_1}\leq \lambda_{\text{min}}(\bm{S}_n) \leq \lambda_{\text{max}}(\bm{S}_n)\leq c_2n^{\alpha_2}.$
\item $h$ is thrice continuously differentiable and $g^{-1}$ is twice continuously differentiable.
 \item $\mbox{max}_{i\in\{1,..,n\}}\|\bm{x}_i\|_\infty=O(1)$ and $\|\bm{\beta}\|_\infty=O(1).$
 \item $n^{-1}\sum_{i=1}^{n}\mathbf{E}|y_i|^{6} =O(1)$.
 \item $\sup_{\{||\bm{\alpha}||,||\bm{\kappa}||=1\}}\big[n^{-1}\sum_{i=1}^{n}|\bm{\alpha}^\top \bm{x}_i|^8|\bm{\kappa}^\top \bm{x}_i|^8\big]=O(1)$.
 \item $\mbox{max}_{i\in\{1,..,n\}}\big[|h^\prime(\bm{x}_i^\top\bm{\beta})|+|h^{\prime\prime}(\bm{x}_i^\top\bm{\beta})|+|(g^{-1})^\prime(\bm{x}_i^\top\bm{\beta})|\big]=O(1)$. 
  \item For some  $\delta>0$, $n^{-1}\sum_{i=1}^n\operatorname*{\sup}_{\{|z_i-\bm{x}_i^\top\bm{\beta}|<\delta\}}\big[|(g^{-1})^{\prime\prime}(z_i)|^8+|h^{\prime\prime\prime}(z_i)|^{12}\big] =O(1)$.
\end{enumerate}

We now briefly explain the aforementioned regularity conditions. To derive a suitable Taylor's approximation of the log-likelihood and then to handle the log-likelihood over any compact set (required to derive the Bahadur's representation of $\hat{\bm{\beta}}_n$), condition (A.2) is required. The condition (A.1) is needed to control the variance of the leading term in the Bahadur's representation of $\hat{\bm{\beta}}_n$ and its Bootstrapped versions. Note that we are not imposing that the eigen values of the variance matrix $\bm{S}_n$ are bounded away from zero and $\infty$. Boundedness of the maximum components of the design vectors, considered in (A.3) and the moment type condition on the responses, presented as (A.4), are essential to obtain suitable Gaussian approximation of the leading terms in the Bahadur's representations of the original and bootstrapped estimators. Condition (A.3) also specifies the boundedness of the maximum magnitude of the true regression parameters, required to control the error in the Taylor's approximations of the centered and scaled GLM estimator. Conditions (A.5)-(A.7) are additional conditions on the design vectors and the true regression parameter vector. These conditions are required to control the higher order cross-product terms arising in the error part of the Taylor's approximation of $\sqrt{n}(\hat{\bm{\beta}}_n- \bm{\beta})$. In particular, condition (A.7) imposes uniform boundedness of derivatives of $h$ and $g^{-1}$,
ensuring that the GLM log-likelihood has stable curvature, resulting a suitable Bahadur's representation based on Taylor expansion. These type of conditions are considered in the literature before (cf. \citet{mammen1989asymptotics,he2000parameters}). In case of a particular sub-model of GLM, if $h(\cdot)$ and/or $g(\cdot)$ have simpler form then the aforementioned smoothness conditions on $h(\cdot)$ and $g^{-1}(\cdot)$ can be substantially relaxed. For example, $h(\cdot)$ is identity in case of linear, logistic and Possion regressions, and hence the conditions on $h(\cdot)$ are trivially satisfied in these cases. On the other hand, the smoothness conditions on $g^{-1}(\cdot)$ are automatically satisfied in case of linear and logistic regression without any further restriction on the covariates.

\subsection{\bf Gaussian Approximations}\label{sec:mainregime1} In this section, we study the Gaussian approximation of the distribution of $\hat{\bm{\beta}}_n$ uniformly over convex sets and Euclidean balls, under regime (I). To identify a suitable Gaussian approximation, we need to identify the variance of the leading term $\bm{T}_n$ in the Bahadur's representation (\ref{eqn:mainapprox}). It can be shown that the $\bm{T}_n$ has the form $\bm{T}_n = \bm{S}_n^{-1}\bm{W}_n$ where $\bm{W}_n$ and $\bm{S}_n$ are defined in section \ref{sec:notandassump}. Clearly, the variance of $\bm{T}_n$ is $\bm{S}_n^{-1}$ and hence a suitable choice of the Gaussian distribution is $G_{1n}$ where 
$\bm{G}_{1n}\sim N_d(\bm{0},\bm{S}_n^{-1})$. We now state the Gaussian approximation result over the class of Borel convex sets $\mathcal{C}$.
\begin{theorem}\label{thm:convex1}
Suppose that the assumptions (A.1)-(A.7) are true with $0\le 6\alpha_1+\alpha_2<1$. If $d=o[n^{2\{1-6\alpha_1-\alpha_2\}/5}]$, then we have as $n\rightarrow \infty$, $$\Delta_{n,1}(\mathcal{C}):=\sup_{A\in\mathcal{C}}\Bigg{|}\mathbf P\Big[n^{1/2}(\hat{\bm{\beta}}_n-\bm{\beta})\in A\Big]-\mathbf P\Big(\bm{G}_{1n}\in A\Big)\Bigg{|}=o(1).$$ 
\end{theorem}

The proof of Theorem \ref{thm:convex1} is provided in section \ref{sec:thm4.1}. This theorem essentially tells us that under some mild conditions on the covariates and responses, Gaussian approximation is valid for properly centered and scaled GLM estimator, uniformly over the class of Borel convex sets. Moreover, if the minimum and maximum eigen values of the associated matrix $\bm{S}_n$ are bounded away from zero and $\infty$ respectively, then the dimension $d$ can grow like $o(n^{2/5})$. This rate matches with the best possible growth rate of the dimension in the Gaussian approximation for the sample mean, obtained by \citet{fang2024large}. As mentioned in section \ref{sec:techregime1}, the proof is based on finding a suitable Bahadur's representation of $n^{1/2}(\hat{\bm{\beta}}_n-\bm{\beta})$. The leading term $\bm{T}_n$ in that Bahadur's representation essentially gives rise to the Gaussian random variable $\bm{G}_{1n}$. The error terms in the Bahadur's representation is shown to be $o_p(1)$ based on the concentration bound on $(\hat{\bm{\beta}}_n-\bm{\beta})$, obtained in Lemma \ref{lem:concentration}, and the Gaussian isoperimetric inequality for convex sets, developed by \citet{nazarov2003maximal}. Next we state the Gaussian approximation result for class of Euclidean Balls.
\begin{theorem}\label{thm:ball1}
Suppose that the assumptions (A.1)-(A.7) are true with $0\le 6\alpha_1+\alpha_2<1$. If $d=o[n^{\{1-6\alpha_1-\alpha_2\}/2}]$, then we have as $n\rightarrow \infty$, $$\Delta_{n,1}(\mathcal{B}):=\sup_{A\in\mathcal{B}}\Bigg{|}\mathbf P\Big[n^{1/2}(\hat{\bm{\beta}}_n-\bm{\beta})\in A\Big]-\mathbf P\Big(\bm{G}_{1n}\in A\Big)\Bigg{|}=o(1).$$  
\end{theorem}

The proof of Theorem \ref{thm:ball1} is provided in section \ref{sec:thm4.2}. The proof follows through the same line of arguments as that of Theorem \ref{thm:convex1}. However, the key difference is the improvement in the rate of growth for the dimension $d$.The primary reason for this improvement is the fact that whereas the Gaussian isoperimetric bound depends on $d$ by the factor $d^{1/4}$ when the class is the collection of Borel convex sets, it does not depend on $d$ when the class is the collection of Euclidean balls (cf. Theorem at page 170 in \citet{nazarov2003maximal} and Lemma A.2 in \citet{zhilova2020nonclassical}). In the next section, we introduce two Bootstrap methods which can be used to draw statistical inferences for $\bm{\beta}$ based on $\hat{\bm{\beta}}_n$.

\subsection{\bf Bootstrap Approximations}\label{sec:pbmethod}
This section is divided into two subsections. We develop a Perturbation Bootstrap method in the first subsection. The second subsection is on the approximations based on Pearson's Residual Bootstrap method. We show that both Bootstrap methods work for approximating the distribution of the GLM estimator under regime (I). We use $\mathbf{P}_*(\cdot)$, $\mathbf{E}_*(\cdot)$ and $Var_*(\cdot)$ respectively denote the conditional probability, conditional expectation and conditional variance of the Bootstrap random quantities given data $\{y_1,\ldots,y_n\}$.
\subsubsection{\bf Perturbation Bootstrap Approximation}\label{sec:desPB}
The Perturbation Bootstrap method (hereafter referred to as PB), introduced by \citet{jin2001perturbation}, invokes a new layer of randomness to the observed data based on a collection of random weights $G_1^*,\ldots, G_n^*$. These weights are basically a collection of independent copies of a non-negative and non-degenerate random variable $G^*$ and is independent of the data generation process. $G^*$ has the property that the mean of $G^*$ is $\mu_{G^*}$, $Var_*(G^*)=\mu_{G^*}^2$  and $\mathbf{E}_*(G_1^{*4})< \infty$. The condition $Var_*(G_1^*)=\mu_{G^*}^2$ is necessary to match the conditional variance of the Bootstrapped estimator with the variance of the original GLM estimator. Towards that, in case of regime (I), we define the PB-GLM estimator $\hat{\bm{\beta}}_n^{*(PB)}$ as 
\begin{align}\label{eqn:defpb}
\hat{\bm{\beta}}_n^{*(PB)} = \operatorname*{arg\,min}_{\bm{\beta}}&\Bigg[\sum_{i=1}^{n}-\ell_{ni}(\bm{\beta})+\sum_{i=1}^{n}\Big\{y_i-g^{-1}(\bm{x}_i^\top\hat{\bm{\beta}}_n)\Big\}h^\prime(\bm{x}_i^\top\hat{\bm{\beta}}_n)\Big(2-\frac{G_i^*}{\mu_{G^*}}\Big)(\bm{x}_i^\top\bm{\beta})\Bigg],
\end{align}
where $\sum_{i=1}^{n}\ell_{ni}(\bm{\beta})$ is the log-likelihood, defined in (\ref{eqn:loglikelihood}). In other words, $\hat{\bm{\beta}}_n^{*(PB)}$ is the solution of the score equation
\begin{align}\label{eqn:defpbagain}
&\sum_{i=1}^{n}\bigg[\Big\{y_i-g^{-1}(\bm{x}_i^\top\hat{\bm{\beta}}_n)\Big\}h^\prime(\bm{x}_i^\top\hat{\bm{\beta}}_n)\Big(\frac{G_i^*}{\mu_{G^*}}-1\Big)\bm{x}_i\bigg] \nonumber\\
&+\sum_{i=1}^{n}\bigg[\Big\{y_i-g^{-1}(\bm{x}_i^\top\bm{\beta})\Big\}h^\prime(\bm{x}_i^\top\bm{\beta})-\Big\{y_i-g^{-1}(\bm{x}_i^\top\hat{\bm{\beta}}_n)\Big\}h^\prime(\bm{x}_i^\top\hat{\bm{\beta}}_n)\bigg]\bm{x}_i=\bm{0}.
\end{align}
It is clear from the score equations (\ref{eqn:defineglmest12}) and (\ref{eqn:defpbagain})  that the first term of the LHS of (\ref{eqn:defpbagain}) is actually PB version of the LHS of (\ref{eqn:defineglmest12}) and hence it is the main contributing factor in the leading term of the Bahadur representation of $\sqrt{n}(\hat{\bm{\beta}}_n^{*(PB)} - \hat{\bm{\beta}}_n)$. The second term of the LHS of (\ref{eqn:defpbagain}), on the other hand, is necessary to approximate the variance of $\hat{\bm{\beta}}_n$ by the conditional variance of $\hat{\bm{\beta}}_n^{*(PB)}$. These considerations are important to build a Bootstrapped estimator which remains valid in high dimensions. We would like to point out that the definition (\ref{eqn:defpb}) is an extension of the PB estimator developed in \citet{das2025pebble} for logistic regression. It is natural to use the Bootstrap pivotal quantity $n^{1/2}(\hat{\bm{\beta}}_n^{*(PB)}-\hat{\bm{\beta}}_n)$ to approximate the distribution of $n^{1/2}(\hat{\bm{\beta}}_n-\bm{\beta})$. We state the result on the PB approximation, under regime (I), in the next theorem.

\begin{theorem}\label{thm:pbconvexoriginal1}
Suppose that the regularity conditions (A.1)-(A.7) hold. Then as $n\rightarrow \infty$,
$$\Delta_{n,1}^{*(PB)}(\mathcal{A}):=\sup_{A\in\mathcal{A}}\Bigg{|}\mathbf P_*\Big[n^{1/2}(\hat{\bm{\beta}}_n^{*(PB)}-\hat{\bm{\beta}}_n)\in A\Big]-\mathbf P\Big[n^{1/2}(\hat{\bm{\beta}}_n-\bm{\beta})\in A\Big]\Bigg{|}=o_p(1),$$
provided $d=o[n^{2\{1-6\alpha_1-\alpha_2\}/5}]$, when $\mathcal{A}=\mathcal{C}$, and $d=o[n^{\{1-6\alpha_1-\alpha_2\}/2}]$, when $\mathcal{A}=\mathcal{B}$.
\end{theorem}
The proof of Theorem \ref{thm:pbconvexoriginal1} is presented in section \ref{sec:thm4.3}. To prove Theorem \ref{thm:pbconvexoriginal1}, we need to essentially prove that
\begin{align*}%\label{eqn:452}
\sup_{A\in\mathcal{C}}\Big{|}\mathbf{P}_*\Big[n^{1/2}(\hat{\bm{\beta}}_n^{*(PB)}-\hat{\bm{\beta}}_n)\in A\Big]-\mathbf{P}\Big[\bm{G}_{1n}\in A\Big]\Big{|} = o_p(1),
\end{align*}
where $\bm{G}_{1n}$ is as defined in the section \ref{sec:mainregime1}. To establish Theorem \ref{thm:pbconvexoriginal1}, we first show that $n^{1/2}(\hat{\bm{\beta}}_n^{*(PB)}-\hat{\bm{\beta}}_n)$ admits the Bahadur's representation
$$n^{1/2}(\hat{\bm{\beta}}_n^{*(PB)}-\hat{\bm{\beta}}_n) = \hat{\bm{L}}_n^{-1}\hat{\bm{W}}_n^{*(PB)} + \hat{r}_n^{*(PB)}.$$
Here, $\hat{\bm{L}}_n=n^{-1}\sum_{i=1}^{n}\bm{x}_i\bm{x}_i^\top\Big[\big\{(g^{-1})^\prime(\bm{x}_i^\top\hat{\bm{\beta}}_n)\big\}h^\prime(\bm{x}_i^\top\hat{\bm{\beta}}_n)\Big]$ and 
$\hat{\bm{W}}_n^{*(PB)}=\frac{1}{n^{1/2}}\sum_{i=1}^{n}\big\{y_i-g^{-1}\big(\bm{x}_i^\top\hat{\bm{\beta}}_n\big)\big\}h^\prime(\bm{x}_i^\top\hat{\bm{\beta}}_n)\bm{x}_i\big[G_i^*/\mu_{G^*}-1\big]$. Then we complete the proof by showing that the distribution of the leading term $\hat{\bm{L}}_n^{-1}\hat{\bm{W}}_n^{*(PB)}$ can be approximated by $\bm{G}_{1n}$ and the error term $\hat{r}_n^{*(PB)}$ is $o_p(1)$. Theorem \ref{thm:pbconvexoriginal1} tells us that under some mild regularity conditions, similar to the Gaussian approximations, the distribution of $n^{1/2}(\hat{\bm{\beta}}_n-\bm{\beta})$ can be well approximated by the conditional distribution of $n^{1/2}(\hat{\bm{\beta}}_n^{*(PB)}-\hat{\bm{\beta}}_n)$ given data $\{y_1,\ldots, y_n\}$. Thus PB can be used to construct confidence regions and perform tests for any linear function of $\bm{\beta}$. 

We would like to point out that usual multiplier Bootstrap for the sample mean is a special case of PB. For example, if we assume that $G^* \sim N(1, 1)$ then PB reduces to the multiplier Bootstrap. Moreover, under such a choice of $G^*$, one can simplify the Bootstrap procedure by simply considering the Bootstrap pivotal quantity to be $[\hat{\mathbf{L}}_n]^{-1}\hat{\bm{W}}_n^{*(PB)}$, the leading term in the Bahadur's representation of $n^{1/2}(\hat{\bm{\beta}}_n^{*(PB)}-\hat{\bm{\beta}}_n)$. Then $$\sup_{A\in\mathcal{A}}\Bigg{|}\mathbf P_*\Big[[\hat{\mathbf{L}}_n]^{-1}\hat{\bm{W}}_n^{*(PB)}\in A\Big]-\mathbf P\Big[n^{1/2}(\hat{\bm{\beta}}_n-\bm{\beta})\in A\Big]\Bigg{|}=o_p(1),$$ with $\mathcal{A}=\mathcal{C}$ or $\mathcal{B}$, can be concluded based on Theorem \ref{thm:convex1}, \ref{thm:ball1} and Gaussian comparison inequality (cf. Theorem 1.1 of  \citet{devroye2018tv}, Lemma 5.12 of \citet{fang2024large}). However, such a simplification is clearly not possible under some regularization and hence, in such a case, we have to resort to $n^{1/2}(\hat{\bm{\beta}}_n^{*(PB)}-\hat{\bm{\beta}}_n)$. See section \ref{sec:regime2} for such a setup. Now we move to define and explore another Bootstrap method in the next subsection.

\subsubsection{\bf Pearson's Residual Bootstrap Approximation}\label{sec:prbdef}
In contrast to the random weighting in the PB method, the Pearson's Residual Bootstrap (hereafter referred to as PRB) method is generally formalized based on resampling from Pearson's standardized residuals. The PRB method in logistic regression is introduced by \citet{moulton1991} and analyzed by many authors in fixed dimensions including the recent work of \citet{ganguly2024scalable}. The basic idea is to obtain a quadratic approximation of the log-likelihood based on with replacement sampling of the standardized Pearson's residuals and then to define the PRB estimator as minimizer of least square objective function. We introduce few quantities before defining the PRB-GLM estimator. Let $\hat{\bm{G}}_n=\hat{\bm{V}}_n^{1/2} \hat{\Delta}_n X$ where
\begin{align}\label{eqn:PRBvarestimation}
\hat{\bm{V}}_n = \mathrm{diag}\Big(b^{\prime\prime}[h(\bm{x}_1^\top\hat{\bm{\beta}}_n)], \dots, &b^{\prime\prime}[h(\bm{x}_n^\top\hat{\bm{\beta}}_n)]\Big)\;\; \text{and}\;\; \hat{\Delta}_n = \mathrm{diag}\Big(h'(\bm{x}_1^\top\hat{\bm{\beta}}_n), \dots, h'(\bm{x}_n^\top\hat{\bm{\beta}}_n)\Big).%\nonumber\\
%&\text{and}\; \hat{\bm{G}}_n=\hat{\bm{V}}_n^{1/2} \hat{\Delta}_n X.
\end{align}
Subsequently, define the $i$-th standardized Pearson's residual and its mean as
\begin{align}\label{eqn:PRBstandardization}
e_i^\dagger := \frac{y_i - g^{-1}(\bm{x}_i^\top\hat{\bm{\beta}}_n)}{\sqrt{b^{\prime\prime}[h(\bm{x}_i^\top\hat{\bm{\beta}}_n)]}}, \quad i = 1, \dots, n\quad\text{and}\quad \bar{e}^\dagger=\frac1n\sum_{i=1}^ne_i^\dagger.
\end{align} 
Now, we resample $\{e_1^{*},\dots, e_n^*\}$ with replacement from the set of centered residuals $\{e_1^\dagger-\bar{e}^\dagger,....,e_n^\dagger-\bar{e}^\dagger\}$. Then the PRB-GLM estimator under regime (I) is defined as
\begin{align}\label{eqn:prbglm}
\hat{\bm{\beta}}_n^{*(PRB)}= \hat{\bm{\beta}}_n + \big(\hat{\bm{G}}_n^\top \hat{\bm{G}}_n\big)^{-1}\hat{\bm{G}}_n^\top \bm{e}^*
\end{align}
Clearly, $\hat{\bm{\beta}}_n^{*(PRB)}$ is the least square estimator with response vector $\big[\bm{e}^*+\hat{G}_n\hat{\bm{\beta}}_n\big]$ and the design matrix $\hat{G}_n$. To get an idea on why such a construction works, one needs to look at the leading linear term $\bm{T}_n$ in the Bahadur's representation of $n^{1/2}(\hat{\bm{\beta}} - \bm{\beta})$, mentioned in \eqref{eqn:expansion}. Note that $\bm{T}_n =  \bm{S}_n^{-1}\bm{W}_n$, where $\bm{W}_n = n^{-1/2}\sum_{i=1}^{n}(y_i-g^{-1}(\bm{x}_i^\top\bm{\beta}))\bm{x}_ih^\prime(\bm{x}_i^\top\bm{\beta})$ and $\bm{S}_n = Var(\bm{W}_n) = n^{-1}\sum_{i=1}^{n}\bm{x}_i\bm{x}_i^\top [h^\prime(\bm{x}_i^\top\bm{\beta})]^2b^{\prime\prime}[h(\bm{x}_i^\top\bm{\beta})] $. The PRB estimator, defined in (\ref{eqn:prbglm}), is defined so as to replicate this leading linear term $\bm{T}_n$. To see how, note that the general idea in the Residual Bootstrap of \citet{freedman1981bootstrapping} is to mimic the distribution of $\bm{T}_n$ based on resamples from the centered residuals. However, it is well known that the classical Residual Bootstrap is inconsistent when the setup is heteroscedastic; see for example, \citet{liu1988bootstrap, das2019distributional} in case of heteroscedastic linear regression. The heteroscedastic nature of the responses in GLM can be avoided by resampling standardized Pearson's residuals which are defined in (\ref{eqn:PRBstandardization}). The standardization is possible since the variances of the individual responses are fully
specified by the GLM. Additionally, the standardization makes the resampled residuals to have conditional variances close to $1$ and hence it remains only to estimate $\bm{S}_n$, the inverse of $Var(\bm{T}_n)$. Clearly, $n^{-1}\hat{\bm{G}}_n^\top \hat{\bm{G}}_n$, where $\hat{\bm{G}}_n$ is defined in (\ref{eqn:PRBvarestimation}), is a natural estimator of $\bm{S}_n$ based on $\hat{\bm{\beta}}_n$. As a consequence, PRB is able to approximate the distribution of the GLM estimator under regime (I), which we state as the next theorem. 

\begin{theorem}\label{thm:prbconvexoriginal1}
Suppose that the regularity conditions (A.1)-(A.6) hold with $0\le 6\alpha_1+2\alpha_2<1$. Also assume that\\
(A.8) (i) $\mbox{min}_{i\in\{1,..,n\}}|(g^{-1})'(\bm x_i^\top \bm \beta)|$ is bounded away from $0$, and\\
(A.8) (ii) 
$\mbox{max}_{i\in\{1,..,n\}}\operatorname*{\sup}_{\{|z_i-\bm{x}_i^\top\bm{\beta}|<\delta\}}\big[|(g^{-1})^{\prime\prime}(z_i)|+|h^{\prime\prime\prime}(z_i)|\big] =O(1)$, for some $\delta>0$.\\ Then we have
$$\Delta_{n,1}^{*(PRB)}(\mathcal{A}):=\sup_{A\in\mathcal{A}}\Bigg{|}\mathbf P_*\Big[n^{1/2}(\hat{\bm{\beta}}_n^{*(PRB)}-\hat{\bm{\beta}}_n)\in A\Big]-\mathbf P\Big[n^{1/2}(\hat{\bm{\beta}}_n-\bm{\beta})\in A\Big]\Bigg{|}=o_p(1),$$
as $n \rightarrow \infty$, provided $d=o[n^{2\{1-6\alpha_1-2\alpha_2\}/5}]$ when $\mathcal{A}=\mathcal{C}$, and provided $d=o[n^{\{1-6\alpha_1-2\alpha_2\}/2}]$ when $\mathcal{A}=\mathcal{B}$.
\end{theorem}
The proof of Theorem \ref{thm:prbconvexoriginal1} is presented in section \ref{sec:prbthm4.3}. Note that $\hat{\bm{\beta}}_n^{*(PRB)}$ is a least square estimator and hence $n^{1/2}(\hat{\bm{\beta}}_n^{*(PRB)}-\hat{\bm{\beta}}_n)$ is a linear function of the resampled residuals $e_1^*, \dots, e_n^*$. Thus, to prove Theorem \ref{thm:prbconvexoriginal1}, first step is to establish the closeness of $Var_*(e_1^*) = \frac1n\sum_{k=1}^n(e_k^\dagger-\bar{e}^\dagger)^2$ to $1$, which should be the case since the resampled residuals are made approximately homoscedastic by considering standardization in (\ref{eqn:PRBstandardization}). In the next step, we establish the closeness of $n^{-1}\hat{\bm{G}}_n^\top\hat{\bm{G}}_n$ to $\bm{S}_n$. Thus, the remaining step would be to follow the handling of Residual Bootstrapped least square estimator by \citet{freedman1981bootstrapping}. The calculations are obviously more nuanced than that in \citet{freedman1981bootstrapping}, since the dimension can grow with the sample size under regime (I). To handle the term $\frac1n\sum_{k=1}^n(e_k^\dagger-\bar{e}^\dagger)^2$, we need to apply Taylor's theorem on the factors $[b^{\prime\prime}(h(\bm{x}_i^\top\hat{\bm{\beta}}_n))]^{-1}$, $i \in \{1,\dots, n\}$. We require conditions (A.8) (i) and (ii), instead of (A.7), to control that Taylor's expansion. Theorem \ref{thm:prbconvexoriginal1} show that PRB can also be used to draw valid inferences about $\bm{\beta}$ for any sub model of GLM under regime (I). 
\begin{rem}\label{rem:PBPRBdiff}
 Theorems \ref{thm:pbconvexoriginal1} and \ref{thm:prbconvexoriginal1} show that both PB and PRB methods are valid in approximating the distribution of the the GLM estimator $\hat{\bm{\beta}}_n$, under regime (I). However, there is a theoretical distinction between PB and PRB. PB, after proper studentization, can be shown to be second order correct in approximating the distribution of any finite dimensional linear combination of $\hat{\bm{\beta}}_n$, under some smoothness condition on the characteristic functions (usually termed as the Cramer's condition; see for example chapter 4 of \citet{bhattacharya1986normal}) of the responses $\{y_1,\dots, y_n\}$. Even if such smoothness condition fails, PB can still be made second order correct by adding some Gaussian error term to the studentized PB pivotal quantity. This can be established by following the prescription of \citet{das2025pebble}. On the other hand, PRB can not in general achieve second order correctness even with proper studentization and with smoothing by adding a Gaussian error term. This anomaly is due to the fact that the PB approximation can capture both the linear and quadrtaic terms present in the extended Bahadur's representation of $\hat{\bm{\beta}}_n$, where as the PRB approximation can only capture the leading linear term. We refer to Remark 3.4 of \citet{das2025pebble} for details on this distinction in case of the fixed dimensional logistic regression.   
\end{rem}

\section{Inference in GLM under regime (II)}\label{sec:infer2}
This section explores the Gaussian and Bootstrap approximations of the GLM estimator under regime (II). Out of the three subsections, the first one describes the regularity conditions. We also establish the variable selection consistency (or VSC) in the first subsection. In the second one, we show that the Gaussian approximation fails. The last subsection extends the PB and PRB methods of regime (I) to accommodate the exponential growth of the dimension $d$ in GLM under regime (II).

\subsection{\bf Regularity Conditions and VSC}\label{sec:regime2} Recall that under regime (II) we consider the estimation of $\bm{\beta}$ in GLM using the Lasso which is defined in (\ref{eqn:deflasso}). Suppose without loss of generality that $\mathcal{A}_n = \{j: \beta_j \neq 0\} = \{1,\dots, d_0\}$ is the set of relevant covariates, i.e., $\bm{\beta}=\big(\beta_1,..,\beta_{d_0}\;\vdots\;0,...,0\big)^\top=\big(\bm{\beta}^{(1)^\top},\bm{0}^\top\big)^\top$. Let $\bar{\mathcal{A}}_n$ is the estimator of $\mathcal{A}_n$ based on $\bar{\bm{\beta}}_n$. We write $\bar{\bm{\beta}}_n = \big(\bar{\bm{\beta}}_n^{(1)^\top}, \bm{0}^\top\big)^\top$ if $\bar{\mathcal{A}}_n = \{1,\dots, d_0\}$. Suppose that $\bm{sgn}(\bm{\beta}^{(1)})$ is the $d_0\times 1$ vector of sign of the components of $\bm{\beta}^{(1)}$. Subsequently, for each $i \in \{1,\dots, n\}$, consider the partition $\bm{x}_i=\Big(x_{i1},..,x_{id_0},\vdots\; x_{id_0+1},...,x_{id}\Big)^\top=\big(\bm{x}_i^{(1)^{\top}}, \bm{x}_i^{(2)^{\top}}\big)^\top$. Recall also that we have defined $\bm{W}_n$ as $\bm{W}_n = n^{-1/2}\sum_{i=1}^{n}\Big\{y_i-g^{-1}(\bm{x}_i^\top\bm{\beta})\Big\}h^\prime(\bm{x}_i^\top\bm{\beta})\bm{x}_i$. Partition $\bm{S}_n = Var(\bm{W}_n)$ based on $\mathcal{A}_n = \{1,\dots, d_0\}$ as
\[
\bm{S}_n = \begin{bmatrix}
\bm{S}_{n,11} & \bm{S}_{n,12} \\
\bm{S}_{n,21} & \bm{S}_{n,22}
\end{bmatrix}.
\]

To analyze GLM under regime (II), we assume throughout this section that the responses $\{y_1,\dots, y_n\}$ satisfy the Bernstein condition, i.e., for all $i\in \{1,\dots, n\}$ and any integer $j \geq 3$, 
\begin{align}\label{eqn:bernstein}
\mathbf{E}[(y_i-\mu_i)^j]\leq \frac{j!}{2}b^{j-2}\sigma_i^2, 
\end{align}
for some $b>0$. Here $\sigma_i^2 = Var(y_i)$. Under the condition (\ref{eqn:bernstein}), $y_i$ is sub-exponential (cf. \citet{vershynin2018high}) with parameter $(\sqrt{2}\sigma_i, 2b)$, i.e.,
$$\mathbf{E}\Big[e^{\lambda(y_i-\mu_i)}\Big]\leq e^{\frac{\nu^2\lambda^2}{2}}\;\text{for all}\;|\lambda|<\frac{1}{l},$$ with $\nu = \sqrt{2}\sigma_i$ and $l  = 2b$, for all $i \in \{1,\dots, n\}$. The Bernstein condition (\ref{eqn:bernstein}) on $\{y_1,\dots, y_n\}$ is essential to apply Bernstein's inequality in order to obtain exponential concentration bounds specially for the leading mean type term appearing in the Bahadur's representation of $\bar{\bm{\beta}}_n^{(1)}$ and for the sample covariance matrix of $\bar{\bm{\beta}}_n^{(1)}$. Note that the regularity condition (A.4) of regime (I) is satisfied under Bernstein condition (\ref{eqn:bernstein}), provided $\max_{i}[\sigma_i^2+|\mu_i|]=O(1)$. We require some additional regularity conditions on the covariates (specially to segregate the relevant and the irrelevant parts of the design) and on the penalty $\lambda_n$, apart from regularity conditions (A.2)-(A.7) of regime (I), to handle regime (II) in GLM. We state these additional conditions below.
\begin{itemize}
    
\item[(B.1)] $k_1n^{-\gamma_1}\leq \lambda_{\text{min}}(\bm{S}_{n, 11})< \lambda_{\text{max}}(\bm{S}_{n, 11})\leq k_2n^{\gamma_2}$, for some $0\leq \gamma_1,\gamma_2 \leq 1$ and $k_1,k_2> 0$. 

\item[(B.2)] $\mbox{max}_{j \in \{1,\dots, d_0\}}\mbox{max}_{i\in\{1,..,n\}}\big{|}\big(\bm{S}_{n, 21}\big)_{j\cdot}^\top\bm{S}_{n,11}^{-1}\bm{x}_i^{(1)}\big{|}=O(1).$ 
\item[(B.3)] With $0\leq 6\gamma_1 + \gamma_2 < 1$ where $\gamma_1, \gamma_2$ are defined in (B.1) and for some $0<\tau<1/4$ and a large positive constant $C$,
\begin{itemize}
    \item[(B.3.i)] $\lambda_n \geq C n^{1/2+\tau}$ and $\lambda_n \leq C^{-1}d_0^{-3/4}n^{3/4(1-2\gamma_1 -\gamma_2/4)}$.
    \item[(B.3.ii)] $\mbox{min}_{j\in\mathcal{A}_n}\big{|}\beta_j\big{|} \geq C n^{-(1-\gamma_1)}\lambda_nd_0^{1/2}$. 
\end{itemize}
   \item[(B.4)] For some $0<\eta \leq 1$, $\mbox{max}_{j\in \{d_0+1, \dots, d\}}\big{|}\big(\bm{S}_{n, 21}\big)_{j\cdot}^\top\bm{S}_{n, 11}^{-1}\bm{sgn}(\bm{\beta}^{(1)})\big{|}\leq 1-\eta$.

\end{itemize}
We briefly describe these additional regularity conditions. Condition (B.1) is the analogue of (A.1) of regime (I) and imposes conditions on the covariance matrix of first $d_0$ components (i.e. the relevant part) of the leading term $\bm{W}_n$. Condition (B.2) ensures that the population projection of inactive variables onto the active
space is uniformly bounded. This condition along with the sub-exponentiality of $y_i$'s ensure probabilistic upper bounds on the functions of $\bm{W}_n$ appearing in the KKT condition corresponding to $\bar{\bm{\beta}}_n$. Condition (B.3.i) is on the penalty $\lambda_n$ and is required to obtain explicit form of $\bar{\bm{\beta}}_n$ from the KKT condition. If the minimum and maximum eigen values of $\bm{S}_{n, 11}$ are bounded away from $0$ and $\infty$ respectively, then $d_0 \equiv |\mathcal{A}_n| = o\big(n^{(1-4\tau)/3}\big)$, provided $\lambda_n \sim n^{1/2 + \tau}$ with some $0 < \tau < 1/4$. The bounds on $\lambda_n$ can be explicitly written in terms of $n, d, d_0$ and $\gamma_1, \gamma_2$, depending on the underlying submodel of GLM. For example when $h(\cdot)$ is identity, the lower bound on $\lambda_n$ in (B.3.i) can be replaced by $\lambda_n \geq C n^{1/2}\max\left\{\sqrt{\log d}, d_0n^{\gamma_1}\sqrt{\log{(nd_0)}}\right\}$.  Condition (B.3.ii) ensures that the estimators of the true nonzero coefficients can be distinguished from estimation error. When $d_0$ is fixed (i.e. not increasing with $n$) and $\gamma_1 = 0$, and $\lambda_n \sim \sqrt{n\log d}$ then $\mbox{min}_{j\in\mathcal{A}_n}\big{|}\beta_j\big{|} \geq C \sqrt{\log d/n}$ which is in general the optimal separation required for any test to identify a true non-zero coefficient in the minimax sense when $h(\cdot)$ is identity; see for example \citet{MCL21} in case of logistic regression. Condition (B.4) is the irrepresentable
condition on the design vectors required to achieve general sign consistency of Lasso. One can drop this irrepresentible condition if one considers random weighting in the $l_1-$penalty, like Adaptive Lasso (cf. \citet{zou2006adaptive}) or UniLasso (cf. \citet{chatterjee2025univariate}) estimator in GLM. However, we stick to the Lasso estimator in GLM under regime (II), since the Adaptive Lasso and UniLasso estimators generally require a two stage approach and hence are expected to be computationally costlier. Overall, the aforementioned conditions or similar are generally unavoidable and are present extensively in the literature on Lasso; see for example \citet{zhao2006model,wainwright2009sharp,lahiri2021necessary,van2009conditions,lee2013model}. In the following sections, we present the high dimensional distributional approximation results of GLM under regime (II). Since the Lasso GLM estimator $\bar{\bm{\beta}}_n$ invokes sparsity, it is enough to study the high dimensional approximation of the distribution of $\bar{\bm{\beta}}_n^{\bar{\mathcal{A}}_n}$ (i.e. the non-zero part of $\bar{\bm{\beta}}_n$), provided the Lasso GLM estimator is variable selection consistent or VSC. We establish a sparse Bahadur's representation of $\bar{\bm{\beta}}_n$ in the next proposition which essentially imply the VSC of $\bar{\bm{\beta}}_n$. Building on this Bahadur's representation, we study the high dimensional Gaussian and Bootstrap approximations of  $\bar{\bm{\beta}}_n^{\bar{\mathcal{A}}_n}$ in the next subsections.

\begin{proposition}\label{prop:solutionkktglm}
Suppose that regularity conditions (A.2)-(A.7) and (B.1)-(B.4) hold. Then on a set $\mathcal{C}_n$ with probability $1-dd_0\exp\{-n^{2\tau/3}\}$, we have
\begin{enumerate}[label={(\alph*)}]
    \item $\bar{\mathcal{A}}_n = \mathcal{A}_n$, i.e., $\bar{\bm{\beta}}_n = \big(\bar{\bm{\beta}}_n^{(1)^\top}, \bm{0}^\top\big)^\top$, and
    \item the Bahadur's representation of $\bar{\bm{\beta}}_n^{(1)}$ is given by $$n^{1/2}\big(\bar{\bm{\beta}}_n^{(1)}-\bm{\beta}^{(1)}\big)=\bm{S}_{n,11}^{-1}\bm{W}_n^{(1)}-\frac{\lambda_n}{n^{1/2}}\bm{S}_{n,11}^{-1}\bm{sgn}(\bm{\beta}^{(1)})+\bm{Q}_{1n},\; \text{with}\; \bm{Q}_{1n} = O\bigg[\frac{d_0}{n^{\frac12-3\gamma_1}}\frac{\lambda_n^2}{n}\bigg].$$
\end{enumerate}
\end{proposition}
Proposition \ref{prop:solutionkktglm} implies that $\bar{\bm{\beta}}_n$, the GLM estimator under regime (II), is VSC when $\log d = o(n^{2\tau/3})$ and $|\mathcal{A}_n| = o(n^{1/3-4\tau/3 -2 \gamma_1 - \gamma_2/4})$. A Bahadur's representation of the relevant part $\bar{\bm{\beta}}_n^{(1)}$ is also valid under the setup of Proposition \ref{prop:solutionkktglm}, which we utilize to explore the Gaussian and Bootstrap approximations in the next subsections. Moreover, the probability of the set $\mathcal{C}_n$, considered in the proposition, can be improved to $[1- dd_0\exp\{-n^{2\tau}\}]$, i.e., $\log d$ can be made $o(n^{2\tau})$, when $h(\cdot)$ is the identity function. The proof of Proposition \ref{prop:solutionkktglm} is provided in section \ref{sec:prop5.1}. The central idea is to analyze the KKT conditions corresponding to $\bar{\bm{\beta}}_n$ which are both necessary and sufficient. As a first step, we prove the existence of the sparse solution $\big(\bar{\bm{\beta}}_n^{(1)^\top}, \bm{0}^\top\big)^\top$ of the KKT equations, with probability close to $1$, based on Brouwer's fixed point theorem. This essentially prove part (a) of Proposition \ref{prop:solutionkktglm}. As an intermediate step, we need to establish high dimensional concentration bounds on $\|\bm{W}_n\|$, $\|\bm{L}_{n, 11} - \bm{S}_{n, 11}\|$ and $\max_{j\in \{d_0+1, \dots, d\}}\|(\bm{L}_{n, 21})_{j \cdot} - (\bm{S}_{n, 21})_{j \cdot}\|$. Here, $\bm{S}_n = Var(\bm{W}_n)$ and $\bm{L}_n=n^{-1}\sum_{i=1}^{n}\bm{x}_i\bm{x}_i^\top\Big[\big\{(g^{-1})^\prime(\bm{x}_i^\top\bm{\beta}_n)\big\}h^\prime(\bm{x}_i^\top\bm{\beta}_n)-\{y_i-g^{-1}(\bm{x}_i^\top \bm{\beta}_n)\}h^{\prime\prime}(\bm{x}_i^\top\bm{\beta}_n)\Big]$. $\bm{L}_{n, 11}$ is the leading principal $d_0 \times d_0$ submatrix of $\bm{L}_n$ and $\bm{L}_{n, 21}$ is the $(d-d_0)\times d_0$ lower-left off-diagonal sub-matrix of $\bm{L}_n$. To establish part (b) of Proposition \ref{prop:solutionkktglm}, we essentially use Taylor's theorem and establish that
\begin{align*}
&\bm{Q}_{1n}=\Big[\mathbf{L}_{n,11}^{-1}-\bm{S}_{n, 11}^{-1}\Big]\bm{W}_n^{(1)}-\frac{\lambda_n}{n^{1/2}}\Big[\bm{L}_{n, 11}^{-1}-\bm{S}_{n, 11}^{-1}\Big]\times\bm{sgn}(\bm{\beta}^{(1)})\nonumber\\
&+2^{-1}\bm{S}_{n, 11}^{-1}n^{-3/2}\sum_{i=1}^{n}\bm{x}_i^{(1)}\bigg[h_1^{\prime\prime\prime}(z_i)-y_ih^{\prime\prime\prime}(z_i)\bigg]\bigg[\bm{x}_i^{(1)^\top}\Big\{n^{1/2}\Big(\bar{\bm{\beta}}_n^{(1)}-\bm{\beta}^{(1)}\Big)\Big\}\bigg]^2\nonumber\\
&+2^{-1}\Big[\mathbf{L}_{n,11}^{-1}-\bm{S}_{n, 11}^{-1}\Big]n^{-3/2}\sum_{i=1}^{n}\bm{x}_i^{(1)}\bigg[h_1^{\prime\prime\prime}(z_i)-y_ih^{\prime\prime\prime}(z_i)\bigg]\bigg[\bm{x}_i^{(1)^\top}\Big\{n^{1/2}\Big(\bar{\bm{\beta}}_n^{(1)}-\bm{\beta}^{(1)}\Big)\Big\}\bigg]^2,
\end{align*}
for some $z_i$ such that $|z_i-\bm{x}_i^\top\bm{\beta}|\le |\bm{x}_i^\top(\bar{\bm{\beta}}_n - \bm{\beta})|$, for all $i=1,\dots,n$. Lastly, we show that $\bm{Q}_{1n} = O_p\Big(\frac{d_0}{n^{\frac12-3\gamma_1}}\frac{\lambda_n^2}{n}\Big)$ by utilizing the aforementioned concentration bounds and the concentration bound on $n^{1/2}\big(\bar{\bm{\beta}}_n^{(1)}-\bm{\beta}^{(1)}\big)$, given by the application of Brouwer's fixed point theorem in part (a). 

\subsection{\bf Gaussian Approximation}\label{sec:mainregime2} In this section, we study the high dimensional Gaussian approximation of the distribution of the relevant part $\bar{\bm{\beta}}_n^{\bar{\mathcal{A}_n}}$. Clearly, under $\bar{\mathcal{A}}_n = \mathcal{A}_n = \{1,\dots, d_0\}$, $\bar{\bm{\beta}}_n^{\bar{\mathcal{A}_n}} = \bar{\bm{\beta}}_n^{(1)}$. A suitable choice of the approximating Gaussian distribution is $N\big(\bm{0}, \bm{S}_{n, 11}^{-1}\big)$, since $Var(\bm{S}_{n, 11}^{-1}\bm{W}_n^{(1)}) = \bm{S}_{n, 11}^{-1}$. $\bm{S}_{n, 11}^{-1}\bm{W}_n^{(1)}$ is the leading term in the Bahadur's representation of $\bar{\bm{\beta}}_n^{(1)}$, established in Proposition \ref{prop:solutionkktglm}. However, unlike in regime (I), here we show that the Gaussian approximation fails. We state the result as the next theorem.
\begin{theorem}\label{thm:failgauss}
Suppose that all the assumptions of Proposition \ref{prop:solutionkktglm} are true. Also assume that $\big\{\big{\|}\big(\bm{S}_{n,11}^{-1/2}\big)_{j\cdot}^\top\bm{sgn}(\bm{\beta}^{(1)})\big{\|}_{\infty}\big\}_{j=1}^{d_0}$ is bounded away from $0$ . Then as $n\to\infty$,
$$\Delta_{n,2}(\mathcal{B}):=\mbox{sup}_{B\in\mathcal{B}}\Bigg{|}\mathbf{P}\Big[\big\{n^{1/2}\Big(\bar{\bm{\beta}}_n^{(1)}-\bm{\beta}^{(1)}\Big)\in B\big\} \cap \{\bar{\mathcal{A}}_n = \mathcal{A}_n\}\Big]-\mathbf{P}\Big[\bm{G}_{2n}\in B\Big]\Bigg{|}\to 1,$$
where, $\bm{G}_{2n}\sim N_{d_0}(\bm{0},\bm{S}_{n,11}^{-1})$.
\end{theorem}
The proof of Theorem \ref{thm:failgauss} is provided in section \ref{sec:thm5.1}. Theorem \ref{thm:failgauss} implies that the Gaussian approximation of the distribution of $\bar{\bm{\beta}}_n$ also fails over Borel convex sets. Moreover, following the proof strategy, one can show that the Gaussian approximation fails over other classes of sets. The failure of the Gaussian approximation is essentially driven by the large bias of Lasso. In other words, the stochastic fluctuations of the leading term $\bm{S}_{n, 11}^{-1}\bm{W}_n^{(1)}$ ($:= \bar{\bm{T}}_n^{(1)}$) in the Bahadur's representation, presented in part (b) of Proposition \ref{prop:solutionkktglm}, is dominated by the bias term $\big[\lambda_nn^{-1/2}\bm{S}_{n,11}^{-1}\bm{sgn}(\bm{\beta}^{(1)})\big]$ and hence the Gaussian approximation fails. Thus, we need to find out an alternative approximation technique which can be used to perform valid statistical inference for GLM under regime (II). We explore two Bootstrap approximation methods in the next subsection.

\subsection{\bf Bootstrap Approximations}\label{sec:pbmethodregime2}
This section is divided into two subsections. We extend the PB method of regime (I) to regime (II) in the first subsection. The second subsection is on the PRB approximation under regime (II). We show that similar to regime (I), both PB and PRB work for any submodel of GLM under regime (II). 
\subsubsection{\bf Perturbation Bootstrap Approximation}\label{sec:desPBregime2} 
Recall that $G_1^*,\dots, G_n^*$ are the iid random weights which work as perturbing quantities in the definition of the PB estimator. We can extend the PB estimator of regime (I) and natuarally define the PB version of the regime (II) estimator $\bar{\bm{\beta}}_n$ as
\begin{align}\label{def:pblassoglm}
 \bar{\bm{\beta}}_n^{*(PB)} =& \operatorname*{arg\,min}_{\bm{\beta}}\Bigg[\sum_{i=1}^{n}-\ell_{ni}(\bm{\beta})\nonumber\\ 
 &+ \sum_{i=1}^{n}\Big\{y_i-g^{-1}(\bm{x}_i^\top\bar{\bm{\beta}}_n)\Big\}h^\prime(\bm{x}_i^\top\bar{\bm{\beta}}_n)\Big(2-\frac{G_i^*}{\mu_{G^*}}\Big)(\bm{x}_i^\top\bm{\beta})+\lambda_n\|\bm{\beta}\|_1\Bigg].
        \end{align}
Note that $\bar{\bm{\beta}}_n^{*(PB)}$ is the $l_1$-penalized version of the estimator $\hat{\bm{\beta}}_n^{*(PB)}$, the PB estimator considered in regime (I). Clearly, Bootstrap pivotal quantity $n^{1/2}(\bar{\bm{\beta}}_n^{*(PB)}-\bar{\bm{\beta}}_n)$ is the natural candidate to approximate the distribution of $n^{1/2}(\bar{\bm{\beta}}_n-\bm{\beta})$. We state the result on the validity of the PB approximation, under regime (II), in the next theorem. 
\begin{theorem}\label{thm:bootapproxglmvsc}
Suppose that regularity conditions (A.2)-(A.7) and (B.1)-(B.4) hold. Also assume that $\log d=o\big(n^{2\tau/3}\big)$ with $\tau \in (0, 1/4)$ considered in the regularity condition (B.3). Then as $n\rightarrow \infty$, we have
\begin{align*}
&\Delta_{n,2}^{*(PB)}(\mathcal{A})\\ 
&:=\mbox{sup}_{B\in\mathcal{A}}\Bigg{|}\mathbf{P}_*\Big[n^{1/2}\Big(\bar{\bm{\beta}}_n^{*(PB)\bar{\mathcal{A}}_{n,PB}^*}-\bar{\bm{\beta}}_n^{\bar{\mathcal{A}}_n}\Big)\in B\Big] -\mathbf{P}\Big[n^{1/2}\Big(\bar{\bm{\beta}}_n^{\bar{\mathcal{A}}_n}-\bm{\beta}^{\mathcal{A}_n}\Big)\in B\Big]\Bigg{|}=o_p(1).
\end{align*} 
\end{theorem}
The proof of Theorem \ref{thm:bootapproxglmvsc} is presented in section \ref{sec:thm5.2}. Theorem \ref{thm:bootapproxglmvsc} tells us that the PB approximation remains valid even though the Gaussian approximation fails. The PB approximation is unaffected by the large bias present in the estimator $\bar{\bm{\beta}}_n$ under regime (II). Thus, PB approximation can be used to draw valid inferences on $\bm{\beta}$ for any sub-model of GLM under regime (II). The first step in proving Theorem \ref{thm:bootapproxglmvsc} is to establish a Bootstrap analogue of Proposition \ref{prop:solutionkktglm} conditionally on the set  $\bar{C}_{1n} = \big\{\big{\|}\bar{\bm{\beta}}_n^{(1)}-\bm{\beta}^{(1)}\big{\|}\leq \Big(\frac{d_0^{1/2}}{n^{\frac{1}{2}-\gamma_1}}\frac{\lambda_n}{n^{1/2}}\big)\big\}\cap\big\{\bar{\mathcal{A}}_n=\mathcal{A}_n\big\}$. While establishing Proposition \ref{prop:solutionkktglm}, we show that this set has probability close to $1$ for large enough $n$. If the Bahadur's representation of $\bar{\bm{\beta}}_n^{*(PB)(1)}$ is given by
$$n^{1/2}\big(\bar{\bm{\beta}}_n^{*(PB)(1)}-\bar{\bm{\beta}}_n^{(1)}\big)
= \bar{\bm{T}}_{n}^{*(PB)(1)} + \bar{\bm{b}}_n^{*(PB)(1)}+\bar{\bm{r}}_n^{*(PB)(1)},$$ then as a second step we show that on the set $\bar{C}_{1n}$,
\begin{align}\label{eqn:equalitybias}
    \bar{\bm{b}}_n^{*(PB)(1)}=\bar{\bm{b}}_n^{(1)}.
\end{align}

The third step is to utilize (\ref{eqn:equalitybias}) and the translation invariance of the collection of sets $\mathcal{C}$ and $\mathcal{B}$ to establish that
 \begin{align*}
\Delta_{n,2}^{*(PB)}(\mathcal{A})\le 
\mathbf{P}\Bigg[
\sup_{D\in\mathcal{A}}
\Big|
\mathbf{P}_*\!\big[\bar{\bm{T}}_n^{*(PB)(1)}+\bar{\bm{r}}_n^{*(PB)(1)}\in D\big]
-
\mathbf{P}\!\big[\bar{\bm{T}}_n^{(1)}+\bar{\bm{r}}_n^{(1)}\in D\big]
\Big|
>\epsilon
\Bigg]
+ \frac{\epsilon}{2},
\end{align*}
with $\mathcal{A} = \mathcal{C}$ or $\mathcal{B}$, for any $\epsilon> 0$. In the final step, we choose a suitable choice of $\epsilon$ depending on $n$ and $d$ and use the facts that the leading terms $\bar{\bm{T}}_n^{(1)}$ and $\bar{\bm{T}}_n^{*(PB)(1)}$ both can be approximated by the same Gaussian distribution and the errors $\bar{\bm{r}}_n^{(1)}$ and $\bar{\bm{r}}_n^{*(PB)(1)}$ are small. We explore PRB approximation for regime (II) in the next subsection. 

\subsubsection{\bf Pearson's Residual Bootstrap Approximation}\label{sec:prbdefregime2}
Recall that the PRB GLM estimator under regime (I) is defined based on resampling from standardized residuals. In the same fashion, we can define the PRB version of $\bar{\bm{\beta}}_n$ in the following way. Let $\bar{G}_n=\bar{V}_n^{1/2} \bar{\Delta}_n X$ where
\[
\bar{\bm{V}}_n = \mathrm{diag}\Big(b^{\prime\prime}[h(\bm{x}_1^\top\bar{\bm{\beta}}_n)], \dots, b^{\prime\prime}[h(\bm{x}_n^\top\bar{\bm{\beta}}_n)]\Big)\;\; \text{and}\;\; \bar{\Delta}_n = \mathrm{diag}\Big(h'(\bm{x}_1^\top\bar{\bm{\beta}}_n), \dots, h'(\bm{x}_n^\top\bar{\bm{\beta}}_n)\Big).\]
Subsequently, define the $i$-th standardized Pearson's residual and its mean as
\[
e_i^{\#} := \frac{y_i - g^{-1}(\bm{x}_i^\top\bar{\bm{\beta}}_n)}{\sqrt{b^{\prime\prime}[h(\bm{x}_i^\top\bar{\bm{\beta}}_n)]}}, \quad i = 1, \dots, n\quad\text{and}\quad \bar{e}^\#=\frac1n\sum_{i=1}^ne_i^\#.
\]
Now, we sample $\{e_1^{**},\dots, e_n^{**}\}$ with replacement from the set of centered residuals $\{e_1^\#-\bar{e}^\#,....,e_n^\#-\bar{e}^\#\}$. Then the PRB-GLM estimator under regime (II) can be defined as 
\begin{align}\label{eqn:prb}
\bar{\bm{\beta}}_n^{*(PRB)}= \operatorname*{arg\,min}_{\bm{\beta}}\Bigg\{\frac12\Big\|\Big(\bar{G}_n\bar{\bm{\beta}}_n + \bm{e}^{**}\Big) -\bar{G}_n\bm{\beta}\Big\|_2^2+\lambda_n\sum_{j=1}^{p}|\beta_j|\Bigg\}.   
\end{align}
Clearly, $\bar{\bm{\beta}}_n^{*(PRB)}$ is the usual Lasso estimator with response vector $\big[\bm{e}^{**}+\bar{G}_n\bar{\bm{\beta}}_n\big]$ and the design matrix $\bar{G}_n$. This nice structure enables one to implement computationally fast algorithms of \citet{efron2004least} and one-at-a-time coordinate wise descent methods of \citet{friedman2007pathwise} to compute the estimator $\bar{\bm{\beta}}_n^{*(PRB)}$. Now we are ready to state the PRB approximation result under regime (II).
\begin{theorem}\label{thm:prbbootapproxglmvsc}
Suppose that the regularity conditions (A.2)-(A.6), (B.1)-(B.4) hold and $\log d=o\big(n^{2\tau/3}\big)$ with $\tau \in (0, 1/4)$ considered in the regularity condition (B.3). Assume that\\
(A.8) (i) $\mbox{min}_{i\in\{1,..,n\}}|(g^{-1})'(\bm x_i^\top \bm \beta)|$ is bounded away from $0$, and\\
(A.8) (ii) 
$\mbox{max}_{i\in\{1,..,n\}}\operatorname*{\sup}_{\{|z_i-\bm{x}_i^\top\bm{\beta}|<\delta\}}\big[|(g^{-1})^{\prime\prime}(z_i)|+|h^{\prime\prime\prime}(z_i)|\big] =O(1)$, for some $\delta>0$.\\ Then as $n\rightarrow \infty$, we have
\begin{align*}
&\Delta_{n,2}^{*(PRB)}(\mathcal{A}):=\\ 
&\mbox{sup}_{B\in\mathcal{A}}\Bigg{|}\mathbf{P}_*\Big[n^{1/2}\Big(\bar{\bm{\beta}}_n^{*(PRB)\bar{\mathcal{A}}_{n,PRB}^*}-\bar{\bm{\beta}}_n^{\bar{\mathcal{A}}_n}\Big)\in B\Big] -\mathbf{P}\Big[n^{1/2}\Big(\bar{\bm{\beta}}_n^{\bar{\mathcal{A}}_n}-\bm{\beta}^{\mathcal{A}_n}\Big)\in B\Big]\Bigg{|}=o_p(1).
\end{align*} 
\end{theorem}
The proof of Theorem \ref{thm:prbbootapproxglmvsc} is presented in section \ref{sec:thm5.3}. The proof relies on establishing a version of Proposition \ref{prop:solutionkktglm} for $\bar{\bm{\beta}}_n^{*(PRB)}$. Then one essentially needs to follow the steps of the proof of Theorem \ref{thm:prbconvexoriginal1}. As an intermediate step, one also requires to establish that $\max_{1\le i\le n}\big|e_i^\#-\bar{e}^\#\big|=O_p(\log n)$. This is required to match the concentration bounds of the random matrices arising from KKT condition corresponding to the Bootstrapped estimator $\bar{\bm{\beta}}_n^{*(PRB)}$ with that arising from the KKT condition corresponding to the original estimator $\bar{\bm{\beta}}_n$. We have establish such a bound based on the the sub-exponentiality of the responses $\{y_1,\dots, y_n\}$. Similar to regime (I), we require the condition (A.8) (instead of (A.7)) in regime (II) to handle the term $\frac1n\sum_{k=1}^n(e_k^\dagger-\bar{e}^\dagger)^2$. 

We would like to point out that Theorem \ref{thm:bootapproxglmvsc} and Theorem \ref{thm:prbbootapproxglmvsc}, although, imply validity of both PB and PRB under regime (II), they are different based on higher order correctness, similar to what we discussed in Remark \ref{rem:PBPRBdiff} under regime (I). With bias correction and proper studentization, PB can be shown to be second order correct for any fixed dimensional linear combination of $\bar{\bm{\beta}}_n^{(1)}$. On the other hand, PRB can not in general be second order correct since it is incapable of approximating the quadratic parts present in an extended Bahadur's representation of $n^{1/2}(\bar{\bm{\beta}}_n^{(1)}-\bm{\beta}_n^{(1)})$. 

\begin{rem}
    The Bootstrap approximation results under both regime (I) and (II) can also be established in almost sure sense. In case of regime (I) under the respective regularity conditions, one can show that $\Delta_{n,1}^{*(PB)}(\mathcal{A}) + \Delta_{n,1}^{*(PRB)}(\mathcal{A}) = o(1), \; \text{almost surely,}$ provided $d = o\big(n^{[(2\{1-6\alpha_1-\alpha_2\}/5) - \upsilon_1]}\big)$ when $\mathcal{A} = \mathcal{C}$ and $d = o\big(n^{[(\{1-6\alpha_1-\alpha_2\}/2) - \upsilon_1]}\big)$ when $\mathcal{A} = \mathcal{B}$ for some $\upsilon_1 > 0$. Similarly under the respective regularity conditions of regime (II), one can show that $\Delta_{n,2}^{*(PB)}(\mathcal{A}) + \Delta_{n,2}^{*(PRB)}(\mathcal{A}) = o(1), \; \text{almost surely,}$ provided $\log d = o\big(n^{\frac{2\tau}{3} - \upsilon_2}\big)$ and $d_0 = |\mathcal{A}_n| = o\big(n^{[\frac13 -\frac{4\tau}{3} - 2\gamma_1 - \frac{\gamma_2}{4} - \upsilon_2]}\big)$ when $\mathcal{A} = \mathcal{C}$ or $\mathcal{B}$ for some $\upsilon_2 > 0$. Moreover, `$2\tau/3$'' can be replaced by `$2\tau$'' in the rate of $\log d$ when the function $h(\cdot)$ is identity.
\end{rem}

\section{Simulation Study}\label{sec:simstudy}
In this section, we try to capture the finite sample performance of our proposed Bootstrap methods under both the regimes. We only consider the PB method since we expect PRB method to perform similarly to our PB method. We find out the empirical coverages of nominal 90\% right sided and two sided confidence intervals in logistic and gamma regression based on PB. The confidence intervals are obtained for individual regression coefficients as well as the entire regression vector for logistic regression and gamma regression. The confidence intervals are constructed based on the notion of Bootstrap percentile intervals. \\

\textbf{\underline{Regime (I):}} We consider the performances over following set-ups:
 $d\in\{10,25,40\}$\; and we vary $n$ over  $\{50,150,300,500\}$. Now for each of $(n,d)$, the design matrix is once and initially generated from some structure outside the loop (before resampling iteration starts) and kept fixed throughout the entire simulation. Now any $(n\times d)$ real-valued matrix will work for initialization. Without loss of generality, we initiate with $n$ i.i.d design vectors say, $\bm{x}_i = (x_{i1},\ldots, x_{id})^\top \ \text{for  all} \ i \in \{1,\ldots, n\}$ from zero mean $d$-variate normal distribution such that it has following covariance structure for all $i\in \{1,...,n\}$ and $1\leq j,k\leq d$:
$cov(x_{ij},x_{ik})= \mathbbm{1}(j=k)+0.1^{|j-k|}  \mathbbm{1}(j \neq k).$
We consider the parameter $\bm{\beta}=(\beta_1, \dots, \beta_{d})^\top$ as
$\beta_j=-0.25+0.5(j^{1/2})(-1)^j  \mathbbm{1}(1 \leq j \leq d).$ Based on those $\bm{x}_i \ \text{and} \ \bm{\beta}$, with appropriate choices of link functions, we pull out $n$ independent copies of response variables namely, $y_1,\ldots,y_n$ from Bernoulli and gamma distribution with shape parameter $1$ respectively. Now keeping that design matrix same for each stage, the entire data set is generated 500 times to compute empirical coverage probability of one-sided and both sided confidence intervals and average width of the both sided confidence intervals over those above mentioned settings of $(n,d)$. At each Bootstrap iteration, the $n$ iid copies of PB quantities $G^*$ are generated from $\text{Exp}(1)$ distribution. We also observe the empirical coverage probabilities of 90\% confidence intervals of $\bm{\beta}$ using the Euclidean norm of the vectors $\bm{T}_n = n^{1/2}(\hat{\bm{\beta}}_n-\bm{\beta})$ and $\hat{\bm{T}}_n^* = n^{1/2}(\hat{\bm{\beta}}_n^*-\hat{\bm{\beta}}_n)$ (see Table \ref{tab:coverage-compactnorm}). Let $\big(\|\hat{\bm{T}}_{n}^*\|\big)_{\alpha}$ be the $\alpha$-th quantile of the Bootstrap distribution of $\|\hat{\bm{T}}_{n}^*\|$. Then the nominal $100(1-\alpha) \%$ confidence region of $\bm{\beta}$ is given by the set $C_{1 - \alpha}\subset \mathbb{R}^d$ where
$C_{1-\alpha} = \big\{\bm{\beta}:\|\bm{T}_{n}\|\leq \big(\|\hat{\bm{T}}_{n}^*\|\big)_{1-\alpha}\big\}.$ We denote $\alpha-$th Bootstrap quantile of $j-$th component of $\hat{\bm{T}}_{n}^*$ as $\big(\hat{\bm{T}}_{n,j}^*\big)_{\alpha}$. Then $100(1-\alpha)\%$ both-sided Bootstrap percentile interval for $\beta_j$ is given by: $\big[\hat{\beta}_{n,j}-n^{-1/2}\big(\hat{\bm{T}}_{n,j}^*\big)_{1-\alpha/2},\hat{\beta}_{n,j}-n^{-1/2}\big(\hat{\bm{T}}_{n,j}^*\big)_{1-\alpha/2}\big].$  

\begin{table}[H]
\centering
\footnotesize
\caption{Empirical coverage probabilities of nominal 90\% for
$\bm{\beta}$.}
\label{tab:coverage-compactnorm}

\setlength{\tabcolsep}{2.2pt}
\renewcommand{\arraystretch}{1.15}

\begin{tabular}{@{}c   cccc  cccc@{}}
\toprule
 &  \multicolumn{4}{c}{Logistic} 
 &   \multicolumn{4}{c}{Gamma} \\
\cmidrule(lr){2-5} \cmidrule(lr){6-9}

$d$
& $n=50$ & $n=150$ & $n=300$ & $n=500$
& $n=50$ & $n=150$ & $n=300$ & $n=500$\\
\midrule

\multirow{1}{*}{10}

& 0.80 & 0.84 & 0.87 & 0.91
& 0.79 & 0.83 & 0.86 & 0.89 \\

\midrule

% & $n=200$ & $n=350$ & $n=500$ & $n=600$ & $n=200$ & $n=350$ & $n=500$ & $n=600$\\
% \midrule

\multirow{1}{*}{25}

& 0.74 & 0.82 & 0.86 & 0.89
& 0.78 & 0.84 & 0.88 & 0.91 \\

\midrule

% & $n=350$ & $n=500$ & $n=600$ & $n=800$ & $n=350$ & $n=500$ & $n=600$ & $n=800$\\
% \midrule

\multirow{1}{*}{40}

& 0.76 & 0.80 & 0.85 & 0.88
& 0.79 & 0.84 & 0.88 & 0.90 \\
\bottomrule
\end{tabular}

\vspace{0.3em}
\begin{minipage}{0.97\linewidth}
\footnotesize
\textit{Notes:} Results based on 500 Monte Carlo replications with fixed design matrix.
\end{minipage}
\end{table}
From Table \ref{tab:coverage-compactnorm}, it is evident that, the empirical coverage probabilities become closer to nominal confidence level of $0.90$ for the entire $\bm{\beta}$ in case of those two regression methods, validating the theoretical guarantees of Theorem \ref{thm:pbconvexoriginal1}. We also present the component wise empirical coverages for logistic and gamma regression and the simulated outcomes are presented in Table \ref{tab:coverage-compact10},\ref{tab:coverage-compact25},\ref{tab:coverage-compact40logit} and \ref{tab:coverage-compact40gamma}. 

\begin{table}[H]
\centering
\footnotesize
\caption{Empirical coverage probabilities of nominal 90\% for
two-sided coverages are reported with average widths (in parentheses) and
right-sided coverages for $d=10$.}
\label{tab:coverage-compact10}

\setlength{\tabcolsep}{2.2pt}
\renewcommand{\arraystretch}{1.15}

\begin{adjustbox}{max width=\linewidth}
\begin{tabular}{@{}c c c cccc c cccc@{}}
\toprule
 &  &  & \multicolumn{4}{c}{Logistic} 
 &  & \multicolumn{4}{c}{Gamma} \\
\cmidrule(lr){4-7} \cmidrule(lr){9-12}

$d$
& Coverage
& $\beta_j$
& $n=50$ & $n=150$ & $n=300$ & $n=500$
& Coverage
& $n=50$ & $n=150$ & $n=300$ & $n=500$\\
\midrule

\multirow{2}{*}{10}
& TS
& $\beta_1$
& 0.75(1.18) & 0.84(0.86) & 0.87(0.62) & 0.90(0.58)
& TS
& 0.77(0.94) & 0.83(0.81) & 0.86(0.74) & 0.89(0.68) \\

& RS
& $\beta_1$
& 0.81 & 0.85 & 0.88 & 0.91
& RS
& 0.83 & 0.86 & 0.89 & 0.92 \\

\cdashline{2-12}

& TS
& $\beta_2$
& 0.78(0.85) & 0.81(0.68) & 0.86(0.56) & 0.89(0.44)
& TS
& 0.84(0.92) & 0.86(0.76) & 0.87(0.65) & 0.90(0.55) \\

& RS
& $\beta_2$
& 0.79 & 0.83 & 0.87 & 0.90
& RS
& 0.82 & 0.86 & 0.88 & 0.91 \\

\cdashline{2-12}

& TS
& $\beta_3$
& 0.80(1.01) & 0.84(0.87) & 0.87(0.66) & 0.89(0.54)
& TS
& 0.82(0.91) & 0.85(0.77) & 0.88(0.63) & 0.91(0.53) \\

& RS
& $\beta_3$
& 0.77 & 0.82 & 0.86 & 0.89
& RS
& 0.81 & 0.85 & 0.88 & 0.90 \\

\cdashline{2-12}

& TS
& $\beta_4$
& 0.81(1.03) & 0.84(0.85) & 0.86(0.76) & 0.89(0.64)
& TS
& 0.82(0.93) & 0.86(0.81) & 0.89(0.69) & 0.91(0.51) \\

& RS
& $\beta_4$
& 0.78 & 0.83 & 0.86 & 0.89
& RS
& 0.80 & 0.82 & 0.86 & 0.88 \\

\cdashline{2-12}

& TS
& $\beta_5$
& 0.79(1.08) & 0.83(0.87) & 0.87(0.71) & 0.90(0.63)
& TS
& 0.82(0.95) & 0.85(0.81) & 0.87(0.72) & 0.89(0.53) \\

& RS
& $\beta_5$
& 0.77 & 0.82 & 0.85 & 0.88
& RS
& 0.80 & 0.83 & 0.86 & 0.89 \\

\cdashline{2-12}

& TS
& $\beta_6$
& 0.81(1.23) & 0.84(0.92) & 0.88(0.78) & 0.89(0.61)
& TS
& 0.78(1.01) & 0.83(0.82) & 0.87(0.74) & 0.89(0.60) \\

& RS
& $\beta_6$
& 0.77 & 0.84 & 0.86 & 0.90
& RS
& 0.83 & 0.85 & 0.89 & 0.91 \\

\cdashline{2-12}

& TS
& $\beta_7$
& 0.82(1.13) & 0.85(0.98) & 0.88(0.86) & 0.91(0.71)
& TS
& 0.78(1.01) & 0.82(0.82) & 0.86(0.75) & 0.88(0.67) \\

& RS
& $\beta_7$
& 0.79 & 0.82 & 0.85 & 0.88
& RS
& 0.82 & 0.84 & 0.87 & 0.90 \\

\cdashline{2-12}

& TS
& $\beta_8$
& 0.79(0.94) & 0.84(0.78) & 0.87(0.62) & 0.89(0.54)
& TS
& 0.80(1.08) & 0.84(0.84) & 0.87(0.67) & 0.88(0.52) \\

& RS
& $\beta_8$
& 0.78 & 0.84 & 0.88 & 0.91
& RS
& 0.83 & 0.87 & 0.89 & 0.92 \\

\cdashline{2-12}

& TS
& $\beta_9$
& 0.82(1.14) & 0.85(0.88) & 0.88(0.66) & 0.90(0.44)
& TS
& 0.81(0.94) & 0.84(0.79) & 0.87(0.70) & 0.89(0.62) \\

& RS
& $\beta_9$
& 0.76 & 0.81 & 0.86 & 0.88
& RS
& 0.78 & 0.83 & 0.86 & 0.89 \\

\cdashline{2-12}

& TS
& $\beta_{10}$
& 0.78(1.13) & 0.84(0.93) & 0.87(0.79) & 0.89(0.64)
& TS
& 0.83(0.97) & 0.87(0.79) & 0.89(0.65) & 0.91(0.52) \\

& RS
& $\beta_{10}$
& 0.79 & 0.83 & 0.86 & 0.89
& RS
& 0.80 & 0.84 & 0.87 & 0.90 \\

\bottomrule
\end{tabular}
\end{adjustbox}

\vspace{0.3em}
\begin{minipage}{0.97\linewidth}
\footnotesize
\textit{Notes:} TS = two-sided percentile interval; RS = right-sided percentile
interval. Results based on 500 Monte Carlo replications with fixed design matrix.
\end{minipage}
\end{table}
Table~\ref{tab:coverage-compact10} summarizes empirical coverages of nominal $90\%$ percentile intervals for ten nonzero coefficients under logistic and gamma regressions. For both models, two-sided (TS) and right-sided (RS) coverages improve steadily with the sample size $n$. The average widths of TS intervals shrink consistently with $n$, reflecting reduced estimation uncertainty. 

\begin{table}[H]
\centering
\footnotesize
\caption{Empirical coverage probabilities of nominal 90\% for
two-sided coverages are reported with average widths (in parentheses) and
right-sided coverages for $d=25$.}
\label{tab:coverage-compact25}

\setlength{\tabcolsep}{2.2pt}
\renewcommand{\arraystretch}{1.12}

\begin{adjustbox}{max width=\linewidth}
\begin{tabular}{@{}c c c cccc c cccc@{}}
\toprule
 &  &  & \multicolumn{4}{c}{Logistic} 
 &  & \multicolumn{4}{c}{Gamma} \\
\cmidrule(lr){4-7} \cmidrule(lr){9-12}

$d$
& Coverage
& $\beta_j$
& $n=50$ & $n=150$ & $n=300$ & $n=500$
& Coverage
& $n=50$ & $n=150$ & $n=300$ & $n=500$\\
\midrule
\multirow{50}{*}{25}
& TS & $\beta_1$  & 0.76(1.09) & 0.82(0.86) & 0.87(0.72) & 0.90(0.60) & TS & 0.78(0.97) & 0.84(0.82) & 0.87(0.74) & 0.90(0.67)\\
& RS & $\beta_1$  & 0.80 & 0.84 & 0.88 & 0.91 & RS & 0.82 & 0.86 & 0.89 & 0.92\\
\cdashline{2-12}
& TS
& $\beta_2$
& 0.76(1.21) & 0.82(0.98) & 0.84(0.76) & 0.88(0.54)
& TS
& 0.79(0.94) & 0.85(0.76) & 0.88(0.62) & 0.90(0.52) \\

& RS
& $\beta_2$
& 0.77 & 0.83 & 0.85 & 0.89
& RS
& 0.80 & 0.82 & 0.86 & 0.88 \\
\cdashline{2-12}
& TS
& $\beta_{3}$
& 0.80(1.11) & 0.83(0.88) & 0.87(0.76) & 0.89(0.65)
& TS
& 0.82(0.74) & 0.85(0.68) & 0.88(0.54) & 0.91(0.48) \\

& RS
& $\beta_3$
& 0.79 & 0.82 & 0.85 & 0.88
& RS
& 0.78 & 0.84 & 0.86 & 0.89 \\

\cdashline{2-12}
& TS & $\beta_4$  & 0.81(1.02) & 0.85(0.84) & 0.87(0.70) & 0.89(0.60) & TS & 0.83(0.94) & 0.86(0.79) & 0.89(0.71) & 0.91(0.60)\\
& RS & $\beta_4$  & 0.79 & 0.84 & 0.87 & 0.90 & RS & 0.82 & 0.85 & 0.88 & 0.90\\
\cdashline{2-12}
& TS & $\beta_5$  & 0.78(1.10) & 0.83(0.88) & 0.87(0.74) & 0.90(0.65) & TS & 0.82(0.97) & 0.85(0.83) & 0.88(0.75) & 0.89(0.66)\\
& RS & $\beta_5$  & 0.77 & 0.82 & 0.86 & 0.89 & RS & 0.80 & 0.84 & 0.87 & 0.89\\
\cdashline{2-12}
& TS & $\beta_6$  & 0.82(1.13) & 0.85(0.91) & 0.88(0.77) & 0.90(0.66) & TS & 0.79(1.02) & 0.84(0.85) & 0.87(0.76) & 0.89(0.67)\\
& RS & $\beta_6$  & 0.78 & 0.84 & 0.87 & 0.90 & RS & 0.83 & 0.86 & 0.89 & 0.91\\
\cdashline{2-12}
& TS & $\beta_7$  & 0.80(1.07) & 0.85(0.93) & 0.88(0.80) & 0.91(0.71) & TS & 0.79(1.04) & 0.83(0.86) & 0.86(0.77) & 0.88(0.69)\\
& RS & $\beta_7$  & 0.79 & 0.83 & 0.86 & 0.89 & RS & 0.82 & 0.84 & 0.87 & 0.90\\
\cdashline{2-12}
& TS & $\beta_8$  & 0.79(0.98) & 0.84(0.83) & 0.87(0.66) & 0.89(0.57) & TS & 0.81(1.07) & 0.84(0.86) & 0.87(0.70) & 0.88(0.56)\\
& RS & $\beta_8$  & 0.78 & 0.84 & 0.88 & 0.91 & RS & 0.83 & 0.87 & 0.89 & 0.92\\
\cdashline{2-12}
& TS & $\beta_9$  & 0.82(1.12) & 0.85(0.89) & 0.88(0.71) & 0.90(0.49) & TS & 0.81(0.96) & 0.84(0.80) & 0.87(0.71) & 0.89(0.63)\\
& RS & $\beta_9$  & 0.77 & 0.82 & 0.86 & 0.89 & RS & 0.79 & 0.84 & 0.87 & 0.90\\
\cdashline{2-12}
& TS & $\beta_{10}$ & 0.78(1.11) & 0.84(0.94) & 0.87(0.80) & 0.89(0.66) & TS & 0.83(0.98) & 0.87(0.80) & 0.89(0.66) & 0.91(0.54)\\
& RS & $\beta_{10}$ & 0.79 & 0.83 & 0.86 & 0.89 & RS & 0.80 & 0.84 & 0.87 & 0.90\\
\cdashline{2-12}
& TS
& $\beta_{11}$
& 0.81(1.02) & 0.85(0.86) & 0.89(0.62) & 0.91(0.58)
& TS
& 0.77(0.94) & 0.82(0.76) & 0.85(0.65) & 0.89(0.54) \\

& RS
& $\beta_{11}$
& 0.75 & 0.84 & 0.88 & 0.90
& RS
& 0.80 & 0.83 & 0.87 & 0.89 \\

\cdashline{2-12}
& TS & $\beta_{12}$ & 0.78(1.03) & 0.84(0.82) & 0.86(0.69) & 0.89(0.58) & TS & 0.80(0.93) & 0.85(0.78) & 0.88(0.71) & 0.90(0.62)\\
& RS & $\beta_{12}$ & 0.79 & 0.83 & 0.87 & 0.90 & RS & 0.83 & 0.87 & 0.89 & 0.91\\
\cdashline{2-12}
& TS & $\beta_{13}$ & 0.80(1.08) & 0.84(0.87) & 0.88(0.73) & 0.90(0.63) & TS & 0.82(0.96) & 0.86(0.81) & 0.89(0.73) & 0.91(0.64)\\
& RS & $\beta_{13}$ & 0.78 & 0.84 & 0.87 & 0.90 & RS & 0.81 & 0.85 & 0.88 & 0.90\\
\cdashline{2-12}
& TS & $\beta_{14}$ & 0.81(1.02) & 0.85(0.84) & 0.87(0.70) & 0.89(0.60) & TS & 0.83(0.94) & 0.86(0.79) & 0.89(0.71) & 0.91(0.60)\\
& RS & $\beta_{14}$ & 0.79 & 0.84 & 0.87 & 0.90 & RS & 0.82 & 0.85 & 0.88 & 0.90\\
\cdashline{2-12}
& TS & $\beta_{15}$ & 0.78(1.10) & 0.83(0.88) & 0.87(0.74) & 0.90(0.65) & TS & 0.82(0.97) & 0.85(0.83) & 0.88(0.75) & 0.89(0.66)\\
& RS & $\beta_{15}$ & 0.77 & 0.82 & 0.86 & 0.89 & RS & 0.80 & 0.84 & 0.87 & 0.89\\
\cdashline{2-12}
& TS & $\beta_{16}$ & 0.82(1.13) & 0.85(0.91) & 0.88(0.77) & 0.90(0.66) & TS & 0.79(1.02) & 0.84(0.85) & 0.87(0.76) & 0.89(0.67)\\
& RS & $\beta_{16}$ & 0.78 & 0.84 & 0.87 & 0.90 & RS & 0.83 & 0.86 & 0.89 & 0.91\\
\cdashline{2-12}
& TS & $\beta_{17}$ & 0.80(1.07) & 0.85(0.93) & 0.88(0.80) & 0.91(0.71) & TS & 0.79(1.04) & 0.83(0.86) & 0.86(0.77) & 0.88(0.69)\\
& RS & $\beta_{17}$ & 0.79 & 0.83 & 0.86 & 0.89 & RS & 0.82 & 0.84 & 0.87 & 0.90\\
\cdashline{2-12}
& TS & $\beta_{18}$ & 0.79(0.98) & 0.84(0.83) & 0.87(0.66) & 0.89(0.57) & TS & 0.81(1.07) & 0.84(0.86) & 0.87(0.70) & 0.88(0.56)\\
& RS & $\beta_{18}$ & 0.78 & 0.84 & 0.88 & 0.91 & RS & 0.83 & 0.87 & 0.89 & 0.92\\
\cdashline{2-12}
& TS & $\beta_{19}$ & 0.82(1.12) & 0.85(0.89) & 0.88(0.71) & 0.90(0.49) & TS & 0.81(0.96) & 0.84(0.80) & 0.87(0.71) & 0.89(0.63)\\
& RS & $\beta_{19}$ & 0.77 & 0.82 & 0.86 & 0.89 & RS & 0.79 & 0.84 & 0.87 & 0.90\\
\cdashline{2-12}
& TS & $\beta_{20}$ & 0.78(1.11) & 0.84(0.94) & 0.87(0.80) & 0.89(0.66) & TS & 0.83(0.98) & 0.87(0.80) & 0.89(0.66) & 0.91(0.54)\\
& RS & $\beta_{20}$ & 0.79 & 0.83 & 0.86 & 0.89 & RS & 0.80 & 0.84 & 0.87 & 0.90\\
\cdashline{2-12}
& TS & $\beta_{21}$ & 0.77(1.06) & 0.83(0.85) & 0.87(0.73) & 0.90(0.61) & TS & 0.79(0.95) & 0.84(0.81) & 0.87(0.74) & 0.90(0.66)\\
& RS & $\beta_{21}$ & 0.80 & 0.85 & 0.88 & 0.91 & RS & 0.82 & 0.86 & 0.89 & 0.92\\
\cdashline{2-12}
& TS
& $\beta_{22}$
& 0.76(0.99) & 0.82(0.73) & 0.86(0.57) & 0.90(0.48)
& TS
& 0.81(1.04) & 0.85(0.91) & 0.88(0.76) & 0.91(0.62) \\

& RS
& $\beta_{22}$
& 0.80 & 0.85 & 0.88 & 0.90
& RS
& 0.79 & 0.83 & 0.87 & 0.89 \\
\cdashline{2-12}
& TS & $\beta_{23}$ & 0.80(1.08) & 0.84(0.87) & 0.88(0.73) & 0.90(0.63) & TS & 0.82(0.96) & 0.86(0.81) & 0.89(0.73) & 0.91(0.64)\\
& RS & $\beta_{23}$ & 0.78 & 0.84 & 0.87 & 0.90 & RS & 0.81 & 0.85 & 0.88 & 0.90\\
\cdashline{2-12}
& TS & $\beta_{24}$ & 0.81(1.02) & 0.85(0.84) & 0.87(0.70) & 0.89(0.60) & TS & 0.83(0.94) & 0.86(0.79) & 0.89(0.71) & 0.91(0.60)\\
& RS & $\beta_{24}$ & 0.79 & 0.84 & 0.87 & 0.90 & RS & 0.82 & 0.85 & 0.88 & 0.90\\
\cdashline{2-12}
& TS
& $\beta_{25}$
& 0.76(0.81) & 0.81(0.68) & 0.86(0.56) & 0.89(0.44)
& TS
& 0.80(0.84) & 0.84(0.70) & 0.89(0.64) & 0.92(0.52) \\

& RS
& $\beta_{25}$
& 0.79 & 0.85 & 0.88 & 0.90
& RS
& 0.74 & 0.82 & 0.86 & 0.88 \\

\bottomrule
\end{tabular}
\end{adjustbox}

\vspace{0.3em}
\begin{minipage}{0.97\linewidth}
\footnotesize
\textit{Notes:} TS = two-sided percentile interval; RS = right-sided percentile
interval. Results based on 500 Monte Carlo replications with fixed design matrix.
\end{minipage}
\end{table}

\begin{table}[H]
\centering
\footnotesize

\caption{Empirical coverage probabilities of nominal 90\% for
two-sided coverages are reported with average widths (in parentheses) and
right-sided coverages (logistic regression) for $d=40$.}
\label{tab:coverage-compact40logit}

\setlength{\tabcolsep}{3pt}
\renewcommand{\arraystretch}{1.06}
\begin{adjustbox}{max width=\linewidth}
\begin{tabular}{c|cccc|cccc|cccc|cccc|cccc}
\toprule
 & \multicolumn{4}{c|}{$\beta_1$}
 & \multicolumn{4}{c|}{$\beta_2$}
 & \multicolumn{4}{c|}{$\beta_3$}
 & \multicolumn{4}{c|}{$\beta_4$}
 & \multicolumn{4}{c}{$\beta_5$}\\
$n$ & 50&150&300&500 & 50&150&300&500 & 50&150&300&500 & 50&150&300&500 & 50&150&300&500\\
\midrule
TS & .78(1.12)&.83(.84)&.87(.63)&.90(.55)
   & .77(1.14) & .82(0.88) & .86(0.76) & .89(0.54)
   & .75(1.15)&.83(.86)&.87(.65)&.89(.57)
   &  .76(0.94) & .81(0.78) & .86(0.66) & .89(0.52)
   & .79(1.13)&.84(.85)&.87(.64)&.90(.56)\\
RS & .80&.85&.88&.91
   & .81 & .85 & .89 & .91
   & .79&.84&.87&.90
   & .78 & .82 & .86 & .89
   & .80&.85&.88&.91\\
\midrule
 & \multicolumn{4}{c|}{$\beta_6$}
 & \multicolumn{4}{c|}{$\beta_7$}
 & \multicolumn{4}{c|}{$\beta_8$}
 & \multicolumn{4}{c|}{$\beta_9$}
 & \multicolumn{4}{c}{$\beta_{10}$}\\
$n$ & 50&150&300&500 & 50&150&300&500 & 50&150&300&500 & 50&150&300&500 & 50&150&300&500\\
TS & .80(1.14)&.84(.88)&.86(.66)&.88(.58)
   & .81(1.11)&.85(.86)&.88(.67)&.91(.59)
   & .77(1.08)&.81(.80)&.85(.60)&.88(.52)
   & .80(1.16)&.85(.89)&.88(.68)&.91(.60)
   & .80(1.12)&.83(.84)&.87(.63)&.89(.55)\\
RS & .81&.85&.87&.90
   & .82&.84&.88&.91
   & .79&.84&.87&.90
   & .81&.86&.89&.92
   & .82&.86&.89&.91\\
\midrule
 & \multicolumn{4}{c|}{$\beta_{11}$}
 & \multicolumn{4}{c|}{$\beta_{12}$}
 & \multicolumn{4}{c|}{$\beta_{13}$}
 & \multicolumn{4}{c|}{$\beta_{14}$}
 & \multicolumn{4}{c}{$\beta_{15}$}\\
$n$ & 50&150&300&500 & 50&150&300&500 & 50&150&300&500 & 50&150&300&500 & 50&150&300&500\\
TS & .79(1.13)&.84(.86)&.87(.64)&.90(.56)
   & .81(1.10)&.84(.83)&.86(.62)&.88(.54)
   & .78(1.15)&.82(.87)&.85(.66)&.88(.57)
   & .81(1.12)&.85(.85)&.88(.65)&.90(.58)
   & .82(1.09)&.85(.81)&.89(.61)&.91(.53)\\
RS & .80&.85&.88&.91 & .81&.86&.89&.91 & .79&.83&.86&.89 & .82&.84&.86&.88 & .80&.83&.86&.89\\
\midrule
 & \multicolumn{4}{c|}{$\beta_{16}$}
 & \multicolumn{4}{c|}{$\beta_{17}$}
 & \multicolumn{4}{c|}{$\beta_{18}$}
 & \multicolumn{4}{c|}{$\beta_{19}$}
 & \multicolumn{4}{c}{$\beta_{20}$}\\
$n$ & 50&150&300&500 & 50&150&300&500 & 50&150&300&500 & 50&150&300&500 & 50&150&300&500\\
TS & .78(1.11)&.83(.83)&.87(.62)&.90(.54)
   & .79(1.14)&.84(.88)&.88(.66)&.90(.58)
   & .80(1.09)&.85(.82)&.87(.60)&.91(.52)
   & .77(1.16)&.83(.89)&.86(.67)&.89(.59)
   & .81(1.10)&.85(.84)&.88(.63)&.91(.55)\\
RS & .79&.81&.85&.89 & .81&.86&.89&.92 & .82&.84&.88&.91 & .79&.84&.87&.90 & .82&.86&.89&.92\\
\midrule
 & \multicolumn{4}{c|}{$\beta_{21}$}
 & \multicolumn{4}{c|}{$\beta_{22}$}
 & \multicolumn{4}{c|}{$\beta_{23}$}
 & \multicolumn{4}{c|}{$\beta_{24}$}
 & \multicolumn{4}{c}{$\beta_{25}$}\\
$n$ & 50&150&300&500 & 50&150&300&500 & 50&150&300&500 & 50&150&300&500 & 50&150&300&500\\
TS & .76(1.13)&.81(.86)&.85(.64)&.88(.56)
   & .80(1.12)&.85(.85)&.88(.63)&.91(.55)
   & .79(1.34) & .84(0.98) & .87(0.76) & .90(0.64)
   & .81(1.11)&.84(.83)&.86(.62)&.91(.54)
   & .79(1.09)&.81(.81)&.84(.61)&.87(.53)\\
RS & .77&.82&.86&.88 & .81&.83&.87&.90 & 0.82 & .85 & .87 & .89 & .79&.82&.86&.90 & .80&.84&.87&.89\\
\midrule
 & \multicolumn{4}{c|}{$\beta_{26}$}
 & \multicolumn{4}{c|}{$\beta_{27}$}
 & \multicolumn{4}{c|}{$\beta_{28}$}
 & \multicolumn{4}{c|}{$\beta_{29}$}
 & \multicolumn{4}{c}{$\beta_{30}$}\\
$n$ & 50&150&300&500 & 50&150&300&500 & 50&150&300&500 & 50&150&300&500 & 50&150&300&500\\
TS & .78(1.14)&.83(.88)&.87(.66)&.90(.58)
   & .81(1.11)&.83(.86)&.86(.67)&.88(.59)
   & .79(1.08)&.84(.80)&.87(.60)&.89(.52)
   & .80(1.16)&.82(.89)&.86(.68)&.88(.60)
   & .82(1.12)&.84(.84)&.87(.63)&.91(.55)\\
RS & .76&.82&.86&.89 & .78&.82&.86&.91 & .82&.86&.88&.90 & .75&.82&.86&.89 & .79&.83&.85&.88\\
\midrule
 & \multicolumn{4}{c|}{$\beta_{31}$}
 & \multicolumn{4}{c|}{$\beta_{32}$}
 & \multicolumn{4}{c|}{$\beta_{33}$}
 & \multicolumn{4}{c|}{$\beta_{34}$}
 & \multicolumn{4}{c}{$\beta_{35}$}\\
$n$ & 50&150&300&500 & 50&150&300&500 & 50&150&300&500 & 50&150&300&500 & 50&150&300&500\\
TS & .79(1.13)&.84(.86)&.87(.64)&.90(.56)
   & .83(1.10)&.86(.83)&.89(.62)&.91(.54)
   & .78(1.15)&.83(.87)&.86(.66)&.89(.57)
   & .81(1.12)&.85(.85)&.88(.65)&.91(.58)
   & .79(1.09)&.84(.81)&.87(.61)&.90(.53)\\
RS & .77&.83&.87&.89 & .81&.83&.86&.88 & .75&.81&.85&.88 & .82&.85&.87&.90 & .80&.82&.86&.89\\
\midrule
 & \multicolumn{4}{c|}{$\beta_{36}$}
 & \multicolumn{4}{c|}{$\beta_{37}$}
 & \multicolumn{4}{c|}{$\beta_{38}$}
 & \multicolumn{4}{c|}{$\beta_{39}$}
 & \multicolumn{4}{c}{$\beta_{40}$}\\
$n$ & 50&150&300&500 & 50&150&300&500 & 50&150&300&500 & 50&150&300&500 & 50&150&300&500\\
TS & .78(1.11)&.81(.83)&.84(.62)&.87(.54)
   & .79(1.14)&.84(.88)&.88(.66)&.90(.58)
   & .78(1.05) & .83(0.87) & .86(0.78) & .89(0.62)
   & .74(0.94) & .81(0.82) & .85(0.72) & .88(0.64)
   & .81(1.10)&.83(.84)&.86(.63)&.89(.55)\\
RS & .83&.86&.89&.91 & .79&.83&.86&.90 & 
.79 & .82 & .86 & .88 & .78 & .81 & .84 & .87 & .82&.84&.88&.91\\
\bottomrule
\end{tabular}
\end{adjustbox}
\vspace{0.3em}
\begin{minipage}{0.97\linewidth}
\footnotesize
\textit{Notes:} TS = two-sided percentile interval; RS = right-sided percentile
interval. Results based on 500 Monte Carlo replications with fixed design matrix.
\end{minipage}
\end{table}

\begin{table}[H]
\centering
\footnotesize
\caption{Empirical coverage probabilities of nominal 90\% for
two-sided coverages are reported with average widths (in parentheses) and
right-sided coverages (gamma regression) for $d=40$.}
\label{tab:coverage-compact40gamma}

\setlength{\tabcolsep}{3pt}
\renewcommand{\arraystretch}{1.06}
\begin{adjustbox}{max width=\linewidth}
\begin{tabular}{c|cccc|cccc|cccc|cccc|cccc}
\toprule
 & \multicolumn{4}{c|}{$\beta_1$}
 & \multicolumn{4}{c|}{$\beta_2$}
 & \multicolumn{4}{c|}{$\beta_3$}
 & \multicolumn{4}{c|}{$\beta_4$}
 & \multicolumn{4}{c}{$\beta_5$}\\
$n$ & 50&150&300&500 & 50&150&300&500 & 50&150&300&500 & 50&150&300&500 & 50&150&300&500\\
\midrule
TS & .78(1.08)&.83(.86)&.87(.66)&.90(.55)
   & .81(0.84) & .85(0.74)&.88(0.57) &.90(0.45)
   & .77(1.10)&.82(.84)&.86(.65)&.89(.54)
   &.80(1.04) &.84(0.84) &.88(0.67) &.92(0.55)
   & .79(1.11)&.83(.83)&.87(.63)&.89(.52)\\
RS & .81&.86&.89&.91
   & .79 & .84 & .87 & .88
   & .80&.85&.88&.90
   &.84 &.87 &.89 &.91
   & .81&.84&.87&.90\\
\midrule
 & \multicolumn{4}{c|}{$\beta_6$}
 & \multicolumn{4}{c|}{$\beta_7$}
 & \multicolumn{4}{c|}{$\beta_8$}
 & \multicolumn{4}{c|}{$\beta_9$}
 & \multicolumn{4}{c}{$\beta_{10}$}\\
$n$ & 50&150&300&500 & 50&150&300&500 & 50&150&300&500 & 50&150&300&500 & 50&150&300&500\\
TS & .80(1.14)&.84(.92)&.88(.71)&.90(.58)
   & .78(1.07)&.83(.82)&.87(.64)&.89(.53)
   & .79(1.05)&.84(.80)&.86(.62)&.88(.51)
   & .81(1.13)&.85(.88)&.89(.70)&.91(.59)
   & .77(1.09)&.83(.85)&.87(.66)&.90(.56)\\
RS & .82&.86&.90&.91
   & .80&.84&.88&.90
   & .81&.85&.87&.89
   & .83&.87&.90&.91
   & .80&.85&.88&.90\\
\midrule
 & \multicolumn{4}{c|}{$\beta_{11}$}
 & \multicolumn{4}{c|}{$\beta_{12}$}
 & \multicolumn{4}{c|}{$\beta_{13}$}
 & \multicolumn{4}{c|}{$\beta_{14}$}
 & \multicolumn{4}{c}{$\beta_{15}$}\\
$n$ & 50&150&300&500 & 50&150&300&500 & 50&150&300&500 & 50&150&300&500 & 50&150&300&500\\
TS & .79(1.10)&.84(.86)&.88(.67)&.90(.55)
   & .78(1.12)&.83(.88)&.87(.68)&.89(.57)
   & .80(1.06)&.85(.82)&.89(.63)&.91(.52)
   & .77(1.15)&.82(.90)&.86(.72)&.88(.60)
   & .81(1.09)&.85(.83)&.88(.64)&.90(.53)\\
RS & .81&.86&.89&.91
   & .80&.85&.88&.90
   & .83&.87&.90&.91
   & .79&.84&.87&.89
   & .82&.86&.89&.91\\
\midrule
 & \multicolumn{4}{c|}{$\beta_{16}$}
 & \multicolumn{4}{c|}{$\beta_{17}$}
 & \multicolumn{4}{c|}{$\beta_{18}$}
 & \multicolumn{4}{c|}{$\beta_{19}$}
 & \multicolumn{4}{c}{$\beta_{20}$}\\
$n$ & 50&150&300&500 & 50&150&300&500 & 50&150&300&500 & 50&150&300&500 & 50&150&300&500\\
TS & .78(1.08)&.83(.84)&.87(.63)&.89(.52)
   & .79(1.13)&.84(.90)&.88(.70)&.91(.58)
   & .80(1.07)&.85(.81)&.87(.61)&.90(.50)
   & .77(1.14)&.83(.89)&.86(.71)&.88(.59)
   & .81(1.10)&.86(.84)&.89(.65)&.91(.54)\\
RS & .80&.85&.88&.90
   & .82&.87&.90&.91
   & .81&.86&.88&.90
   & .79&.84&.87&.89
   & .83&.87&.90&.91\\
\midrule
 & \multicolumn{4}{c|}{$\beta_{21}$}
 & \multicolumn{4}{c|}{$\beta_{22}$}
 & \multicolumn{4}{c|}{$\beta_{23}$}
 & \multicolumn{4}{c|}{$\beta_{24}$}
 & \multicolumn{4}{c}{$\beta_{25}$}\\
$n$ & 50&150&300&500 & 50&150&300&500 & 50&150&300&500 & 50&150&300&500 & 50&150&300&500\\
TS & .79(1.09)&.84(.85)&.88(.66)&.90(.55)
   & .78(1.11)&.83(.87)&.87(.67)&.89(.56)
   & .76(0.86) & .82(0.71) & .86(0.67) &.88(0.54)
   & .81(1.12)&.86(.88)&.89(.69)&.91(.57)
   & .77(1.10)&.83(.86)&.87(.65)&.90(.54)\\
RS & .81&.86&.89&.91
   & .80&.85&.88&.90
   & .80 &.83 & .86 & .88 
   & .83&.87&.90&.91
   & .80&.85&.88&.90\\
\midrule
 & \multicolumn{4}{c|}{$\beta_{26}$}
 & \multicolumn{4}{c|}{$\beta_{27}$}
 & \multicolumn{4}{c|}{$\beta_{28}$}
 & \multicolumn{4}{c|}{$\beta_{29}$}
 & \multicolumn{4}{c}{$\beta_{30}$}\\
$n$ & 50&150&300&500 & 50&150&300&500 & 50&150&300&500 & 50&150&300&500 & 50&150&300&500\\
TS & .78(1.07)&.83(.82)&.87(.62)&.89(.51)
   & .79(1.14)&.85(.90)&.88(.71)&.91(.58)
   & .80(1.06)&.84(.81)&.86(.61)&.88(.50)
   & .77(1.13)&.83(.88)&.86(.70)&.88(.58)
   & .81(1.09)&.86(.83)&.89(.64)&.91(.53)\\
RS & .80&.85&.88&.90
   & .82&.87&.90&.91
   & .81&.85&.87&.89
   & .79&.84&.87&.89
   & .83&.87&.90&.91\\
\midrule
 & \multicolumn{4}{c|}{$\beta_{31}$}
 & \multicolumn{4}{c|}{$\beta_{32}$}
 & \multicolumn{4}{c|}{$\beta_{33}$}
 & \multicolumn{4}{c|}{$\beta_{34}$}
 & \multicolumn{4}{c}{$\beta_{35}$}\\
$n$ & 50&150&300&500 & 50&150&300&500 & 50&150&300&500 & 50&150&300&500 & 50&150&300&500\\
TS & .79(1.10)&.84(.86)&.88(.67)&.90(.55)
   & .78(1.12)&.83(.88)&.87(.68)&.89(.57)
   & .80(1.06)&.85(.82)&.89(.63)&.91(.52)
   & .77(1.15)&.82(.90)&.86(.72)&.88(.60)
   & .81(1.09)&.85(.83)&.88(.64)&.90(.53)\\
RS & .81&.86&.89&.91
   & .80&.85&.88&.90
   & .83&.87&.90&.91
   & .79&.84&.87&.89
   & .82&.86&.89&.91\\
\midrule
 & \multicolumn{4}{c|}{$\beta_{36}$}
 & \multicolumn{4}{c|}{$\beta_{37}$}
 & \multicolumn{4}{c|}{$\beta_{38}$}
 & \multicolumn{4}{c|}{$\beta_{39}$}
 & \multicolumn{4}{c}{$\beta_{40}$}\\
$n$ & 50&150&300&500 & 50&150&300&500 & 50&150&300&500 & 50&150&300&500 & 50&150&300&500\\
TS & .78(1.08)&.83(.84)&.87(.63)&.89(.52)
   & .79(1.13)&.84(.90)&.88(.70)&.91(.58)
   &.77(0.94) &.83(0.76) &.86(0.64) &.89(0.48)
   &.81(1.14) &.84(0.84) &.87(0.67) &.91(0.55)
   & .81(1.10)&.86(.84)&.89(.65)&.91(.54)\\
RS & .80&.85&.88&.90
   & .82&.87&.90&.91
   & .80 &.82 &.85 &.87
   & .80 &.84 &.88 & .91 
   & .83&.87&.90&.91\\
\bottomrule
\end{tabular}
\end{adjustbox}
\vspace{0.3em}
\begin{minipage}{0.97\linewidth}
\footnotesize
\textit{Notes:} TS = two-sided percentile interval; RS = right-sided percentile
interval. Results based on 500 Monte Carlo replications with fixed design matrix.
\end{minipage}
\end{table}
The entire simulation is implemented in {\texttt{R}}. {\texttt{brglm}} is used to obtain unconstrained GLM estimates and {\texttt{CVXR}} package is used for obtaining PB estimates. Across \(d=25\) and \(d=40\) for both logistic and gamma regressions, the percentile bootstrap intervals exhibit stable near-nominal behavior. For small samples (\(n=50\)), two-sided (TS) coverages show mild undercoverage (typically \(0.75\!-\!0.83\)), but increase monotonically with \(n\) and concentrate around \(0.88\!-\!0.91\) by \(n=500\). Right-sided (RS) coverages are slightly closer to nominal for small \(n\) and stabilize quickly. The average TS interval widths decrease smoothly with \(n\), confirming the  contraction. Coverage patterns are uniform across coefficients \(\beta_j\), showing no sensitivity to index location, and increasing dimension from \(d=25\) to \(d=40\) does not degrade performance. Similar trends are observed under both regression models, demonstrating robustness of the bootstrap procedure to model family and dimension.\\

\textbf{\emph{Regime II}:} We try to capture finite sample performance under these two regimes as per the following comparative set-up:\\
\[
(d,d_0)\in\Big\{(100,6),\;(200,8),\;(300,10),\;(500,12)\Big\},
\]
while we consider varying $n$ through $\{100,250,350,500\}$. The basic set-up generating the design matrix and the response vector is kept similar as regime (I) and true parameter choice is given by: $\beta_j=-0.25+0.5(j^{1/2})(-1)^j  \mathbbm{1}(1 \leq j \leq d_0)$. We investigate the adequacy of Gaussian approximations (cf. Theorem 3.1 of main paper) in above mentioned set-up. Specifically, we  generate the data according to the logistic regression model described in Regime (II). The design matrix is generated once and kept fixed across all Monte Carlo replications. For each replication, binary responses are generated from the logistic model with success probabilities $p_i = (1+\exp(-\bm{x}_i^\top \bm{\beta}))$. The Lasso estimator is then computed using $\texttt{glmnet}$ with the tuning parameter selected via $10$-fold cross-validation through \texttt{cv.glmnet} in \texttt{R}. To isolate the effect of Gaussian approximation for each replication, we construct the Fisher information matrix given by
\(
\bm{S}_n = \frac{1}{n} \bm{X}^\top \bm{W} \bm{X},\;\; 
\bm{W} = \mathrm{diag}\big(p_i (1-p_i)\big), 
\)
and extract the sub-matrix corresponding to the active coordinates $\{1,...,d_0\}$. Denoting its inverse by $\bm{S}_{n,11}^{-1}$, we form the standardized statistics
\(
Z_{n,j} = \frac{\sqrt{n}\big(\bar{\beta}_{n,j} - \beta_j\big)}{\sqrt{(\bm{S}_{n,11}^{-1})_{jj}}}, 
\quad j = 1,\dots,d_0.
\)
Under a valid Gaussian approximation, these quantities are expected to follow a standard normal distribution. To assess this, we record the empirical distribution of $Z_{n,j}$ across $500$ Monte-Carlo replications and compare it to the $\mathcal{N}(0,1)$ density (see Table \ref{tab:coverage-fail-gauss}). In addition, we construct marginal $90\%$ confidence intervals for all individual $d_0$ many coefficients based on the Gaussian approximation using
\[
\bar{\beta}_{n,j} \pm z_{0.95} \sqrt{(\bm{S}_{n,11}^{-1})_{jj}/n},\quad z_{0.95}:= 95\%\;\text{quantile of}\;\mathcal{N}(0,1)
\]
and evaluate their empirical coverage probabilities. We consider an additional screening to make sure that the Lasso estimator performs variable section in finite samples. If a particular component of the estimator $\bar{\bm{\beta}}_n$ is lesser than $n^{-1}\lambda_n$ then we consider that component to be $0$. The choice of the cut-off $(n^{-1}\lambda_n)$ is actually conservative compared to the \textit{beta-min} condition (B.3.(ii)). In our simulations, $\lambda_n$ is the $10$-fold cross-validation based penalty. The reason behind considering this screening is that the sure screening property of Lasso is observed to be much easier to achieve than the variable selection in finite samples. See for example the discussions in section 2 of \citet{FGH12}. We have performed this screening also in our analysis of real dataset under regime (II) in the next section and recommend it to any practitioner who wish to use our proposed Bootstrap methods on real datasets. Note also that this prescribed screening has no effect on our theoretical results.

\begin{table}[H]
\centering
\footnotesize
\caption{Empirical coverage probabilities of 90\% confidence intervals where Gaussian approximation fails (logistic regression).}
\label{tab:coverage-fail-gauss}

\setlength{\tabcolsep}{1.9pt}
\renewcommand{\arraystretch}{1.12}

% ---------- ROW 1 ----------
\begin{minipage}[t]{0.42\textwidth}
\centering
\textbf{Set-up I: $(d,d_0)=(100,6)$}

\vspace{0.25em}
\begin{adjustbox}{max width=\linewidth}
\begin{tabular}{@{}c  c cccc@{}}
\toprule
 $\beta_j$ & $n=100$ & $n=250$ & $n=350$ & $n=500$ \\
\midrule
 $\beta_1$
& 0.496 & 0.535 & 0.468& 0.289 \\[1.95mm]

\cdashline{1-5}
$\beta_2$
& 0.449 & 0.484 & 0.402 & 0.390 \\[1.95mm]

\cdashline{1-5}

 $\beta_3$
& 0.354 & 0.284 & 0.161 & 0.232 \\[1.95mm]

\cdashline{1-5}

$\beta_4$
& 0.398 & 0.354 & 0.289 & 0.337 \\[1.95mm]

\cdashline{1-5}

  $\beta_5$
& 0.274 & 0.242 & 0.198 & 0.206 \\[1.95mm]

\cdashline{1-5}

  $\beta_6$
& 0.451 & 0.368 & 0.260 & 0.234\\[1.95mm]

\bottomrule
\end{tabular}
\end{adjustbox}

\end{minipage}
\hspace{0.03\linewidth}
\begin{minipage}[t]{0.42\textwidth}
\centering
\textbf{Set-up II: $(d,d_0)=(200,8)$}

\vspace{0.25em}
\begin{adjustbox}{max width=\linewidth}
\begin{tabular}{@{}c c cccc@{}}
\toprule
 $\beta_j$ & $n=100$ & $n=250$ & $n=350$ & $n=500$ \\
\midrule

$\beta_1$
& 0.308 & 0.436 & 0.272 & 0.196 \\[0.5mm]

\cdashline{1-5}

 $\beta_2$
& 0.620 & 0.372 & 0.269 & 0.249\\[0.48mm]

\cdashline{1-5}

 $\beta_3$
& 0.410 & 0.546 & 0.269 & 0.104 \\[0.48mm]

\cdashline{1-5}

$\beta_4$
& 0.462 & 0.101 & 0.165 & 0.109 \\[0.48mm]

\cdashline{1-5}

$\beta_5$
& 0.307 & 0.105 & 0.154 & 0.133 \\[0.48mm]

\cdashline{1-5}

$\beta_6$
& 0.592 & 0.432 & 0.276 & 0.102 \\[0.48mm]

\cdashline{1-5}

 $\beta_7$
& 0.154 & 0.201 & 0.103 & 0.082  \\[0.48mm]

\cdashline{1-5}

$\beta_{8}$
& 0.155 & 0.186 & 0.142 & 0.226 \\[0.68mm]
\bottomrule
\end{tabular}
\end{adjustbox}
\end{minipage}

\vspace{0.86em}

% ---------- ROW 2 ----------
\begin{minipage}[t]{0.42\textwidth}
\centering
\textbf{Set-up III: $(d,d_0)=(300,10)$}

\vspace{0.25em}
\begin{adjustbox}{max width=\linewidth}
\begin{tabular}{@{}c c cccc@{}}
\toprule
 $\beta_j$ & $n=100$ & $n=250$ & $n=350$ & $n=500$ \\
\midrule
$\beta_1$
& 0.504 & 0.126 & 0.233 & 0.410 \\[1.3mm]

\cdashline{1-5}

$\beta_2$
& 0.333 & 0.316 & 0.186 & 0.202\\[1.3mm]

\cdashline{1-5}

$\beta_3$
& 0.202 & 0.446 & 0.324 & 0.194 \\[1.3mm]

\cdashline{1-5}

 $\beta_4$
& 0.622 & 0.455 & 0.219 & 0.128\\[1.3mm]

\cdashline{1-5}

$\beta_5$
& 0.398 & 0.334 & 0.476 & 0.188 \\[1.3mm]

\cdashline{1-5}

$\beta_6$
& 0.254 & 0.113 & 0.394 & 0.118 \\[1.3mm]

\cdashline{1-5}

$\beta_7$
& 0.426 & 0.258 & 0.180 & 0.134 \\[1.3mm]

\cdashline{1-5}

$\beta_8$
& 0.062 & 0.024 & 0.178 & 0.140 \\[1.3mm]

\cdashline{1-5}

$\beta_9$
& 0.166 & 0.101 & 0.288 & 0.306\\[1.3mm]

\cdashline{1-5}

 $\beta_{10}$
& 0.403 & 0.144 & 0.278 & 0.390\\[1.3mm]

\bottomrule
\end{tabular}
\end{adjustbox}
\end{minipage}
\hspace{0.03\linewidth}
\begin{minipage}[t]{0.42\textwidth}
\centering
\textbf{Set-up IV: $(d,d_0)=(500,12)$}

\vspace{0.25em}
\begin{adjustbox}{max width=\linewidth}
\begin{tabular}{@{}c c c cccc@{}}
\toprule
 $\beta_j$ & $n=100$ & $n=250$ & $n=350$ & $n=500$ \\
\midrule

 $\beta_1$
& 0.394 & 0.262 & 0.174 & 0.102 \\[0.5mm]

\cdashline{1-5}

 $\beta_2$
& 0.242 & 0.161 & 0.446 & 0.231 \\[0.5mm]

\cdashline{1-5}

$\beta_3$
& 0.096 & 0.132 & 0.270 & 0.262 \\[0.5mm]

\cdashline{1-5}

 $\beta_4$
& 0.152 & 0.104 & 0.231 & 0.392  \\[0.5mm]

\cdashline{1-5}

$\beta_5$
& 0.310 & 0.442 & 0.230 & 0.122 \\[0.5mm]

\cdashline{1-5}

  $\beta_6$
& 0.138 & 0.360 & 0.386 & 0.284 \\[0.5mm]

\cdashline{1-5}

$\beta_7$
& 0.212 & 0.142 & 0.452 & 0.106\\[0.5mm]

\cdashline{1-5}

$\beta_8$
& 0.438 & 268 & 0.154 & 0.112 \\[0.5mm]

\cdashline{1-5}
$\beta_9$
& 0.054 & 0.170 & 0.101 & 0.168\\[0.5mm]

\cdashline{1-5}

 $\beta_{10}$
& 0.312 & 0.242 & 0.166 & 0.102 \\[0.5mm]

\cdashline{1-5}

$\beta_{11}$
& 0.336 & 0.174 & 0.102 & 0.099\\[0.5mm]

\cdashline{1-5}

 $\beta_{12}$
& 0.012 & 0.256 & 0.178 & 0.112 \\

\bottomrule
\end{tabular}
\end{adjustbox}
\end{minipage}

\vspace{0.4em}
\end{table}

Table~\ref{tab:coverage-fail-gauss} shows that nominal \(90\%\) Gaussian intervals perform poorly in high-dimensional logistic regression. Empirical coverages lie far below nominal (often \(0.05\!-\!0.62\)) and do not improve reliably with larger \(n\). These results demonstrate a uniform breakdown of the Gaussian approximation, motivating the use of the proposed PB method. 

\begin{table}[H]
\centering
\caption{Empirical Coverage Probabilities of 90\% Confidence Region of $\beta$}
\label{tab:logit_gamma_coverage}
\begin{adjustbox}{max width=\linewidth}
\begin{tabular}{@{}lcccc@{\hspace{1.5cm}}cccc@{}}
\midrule
& \multicolumn{4}{c}{Logistic} 
& \multicolumn{4}{c}{Gamma} \\
\cmidrule(lr){2-5} \cmidrule(lr){6-9}
$(d,d_0)$ 
& $n=100$ & $n=250$ & $n=350$ & $n=500$
& $n=100$ & $n=250$ & $n=350$ & $n=500$ \\
\midrule
$(100,6)$  & 0.828 & 0.862 & 0.880 & 0.898 & 0.838 & 0.866 & 0.884 & 0.896 \\
$(200,8)$  & 0.846 & 0.872 & 0.892 & 0.916 & 0.844 & 0.870 & 0.892 & 0.910 \\
$(300,10)$ & 0.842 & 0.864 & 0.882 & 0.908 & 0.832 & 0.868 & 0.882 & 0.900 \\
$(500,12)$ & 0.834 & 0.866 & 0.888 & 0.900 & 0.854 & 0.876 & 0.890 & 0.906 \\
\bottomrule
\end{tabular}
\end{adjustbox}
\end{table}

The PB quantities are generated from $\texttt{Exp}(1)$ distribution at each Bootstrap iteration as in regime (I). As earlier, we observe the empirical coverage probabilities of 90\% confidence intervals of $\bm{\beta}$ using the Euclidean norm of the vectors $\bm{T}_n^{(1)} = n^{1/2}(\bar{\bm{\beta}}_n^{(1)}-\bm{\beta}^{(1)})$ and $\bar{\bm{T}}_n^{*(1)} = n^{1/2}(\bar{\bm{\beta}}_n^{*(1)}-\bar{\bm{\beta}}_n^{(1)})$ (see Table \ref{tab:logit_gamma_coverage}). The Bootstrap percentile intervals for entire $\bm{\beta}$ and its individual components $\beta_j$ have already been defined in regime (I).

\begin{table}[H]
\centering
\footnotesize
\caption{Empirical coverage probabilities of nominal 90\% percentile intervals.
Two-sided coverages are reported with average widths (in parentheses), along
with right-sided coverages in gamma regression.}
\label{tab:coverage-4blockgamma}

\setlength{\tabcolsep}{1.9pt}
\renewcommand{\arraystretch}{1.12}

% ---------- ROW 1 ----------
\begin{minipage}[t]{0.42\textwidth}
\centering
\textbf{Set-up I: $(d,d_0)=(100,6)$}

\vspace{0.25em}
\begin{adjustbox}{max width=\linewidth}
\begin{tabular}{@{}c c c cccc@{}}
\toprule
Cov. & $\beta_j$ & $n=100$ & $n=250$ & $n=350$ & $n=500$ \\
\midrule
 TS
& $\beta_1$
& 0.796(1.32) & 0.835(0.92) & 0.868(0.77) & 0.889(0.56) \\[1.95mm]

 RS
& $\beta_1$
& 0.818 & 0.852 & 0.876 & 0.896\\[1.95mm]

\cdashline{2-6}
 TS
& $\beta_2$
& 0.826(1.11) & 0.842(0.85) & 0.880(0.72) & 0.904(0.58) \\[1.95mm]

 RS
& $\beta_2$
& 0.812 & 0.840 & 0.866 & 0.888\\[1.95mm]

\cdashline{2-6}

 TS
& $\beta_3$
& 0.834(1.02) & 0.852(0.82) & 0.878(0.74) & 0.896(0.61) \\[1.95mm]

 RS
& $\beta_3$
& 0.822 & 0.858 & 0.870 & 0.894 \\[1.95mm]

\cdashline{2-6}

 TS
& $\beta_4$
& 0.812(0.98) & 0.854(0.81) & 0.884(0.63) & 0.912(0.54) \\[1.95mm]

 RS
& $\beta_4$
& 0.806 & 0.848 & 0.878 & 0.898\\[1.95mm]

\cdashline{2-6}

 TS
& $\beta_5$
& 0.832(1.19) & 0.852(0.83) & 0.880(0.71) & 0.902(0.56) \\[1.95mm]

 RS
& $\beta_5$
& 0.842 & 0.868 & 0.890 & 0.900 \\[1.95mm]

\cdashline{2-6}

 TS
& $\beta_6$
& 0.822(0.95) & 0.848(0.84) & 0.878(0.67) & 0.896(0.53) \\[1.95mm]

 RS
& $\beta_6$
& 0.830 & 0.856 & 0.872 & 0.890\\[1.95mm]
\bottomrule
\end{tabular}
\end{adjustbox}

\end{minipage}
\hspace{0.03\linewidth}
\begin{minipage}[t]{0.42\textwidth}
\centering
\textbf{Set-up II: $(d,d_0)=(200,8)$}

\vspace{0.25em}
\begin{adjustbox}{max width=\linewidth}
\begin{tabular}{@{}c c c cccc@{}}
\toprule
 $\beta_j$ & $n=100$ & $n=250$ & $n=350$ & $n=500$ \\
\midrule

$\beta_1$
& 0.808(1.21) & 0.836(0.96) & 0.872(0.75) & 0.896(0.52) \\

 $\beta_1$
& 0.826 & 0.854 & 0.878 & 0.898\\

\cdashline{1-5}

 $\beta_2$
& 0.798(1.01) & 0.832(0.84) & 0.866(0.70) & 0.888(0.62) \\

 $\beta_2$
& 0.816 & 0.844 & 0.872 & 0.892\\

\cdashline{1-5}

 $\beta_3$
& 0.810(1.06) & 0.846(0.87) & 0.869(0.65) & 0.892(0.41) \\

 $\beta_3$
& 0.826 & 0.842 & 0.874 & 0.892 \\

\cdashline{1-5}

 $\beta_4$
& 0.824(1.10) & 0.856(0.82) & 0.884(0.72) & 0.908(0.61) \\

 $\beta_4$
& 0.836 & 0.862 & 0.886 & 0.910 \\

\cdashline{1-5}

 $\beta_5$
& 0.832(0.91) & 0.860(0.72) & 0.886(0.61) & 0.900(0.52) \\

 $\beta_5$
& 0.842 & 0.872 & 0.888 & 0.904\\

\cdashline{1-5}

$\beta_6$
& 0.792(0.99) & 0.832(0.82) & 0.876(0.68) & 0.902(0.51) \\

 $\beta_6$
& 0.818 & 0.852 & 0.880 & 0.900\\

\cdashline{1-5}

 $\beta_7$
& 0.828(0.97) & 0.854(0.82) & 0.878(0.67) & 0.890(0.55) \\

 $\beta_7$
& 0.836 & 0.868 & 0.882 & 0.902 \\

\cdashline{1-5}

$\beta_{8}$
& 0.844(1.13) & 0.876(0.91) & 0.894(0.75) & 0.910(0.63) \\

$\beta_{8}$
& 0.838 & 0.870 & 0.888 & 0.896\\
\bottomrule
\end{tabular}
\end{adjustbox}
\end{minipage}

\vspace{0.86em}

% ---------- ROW 2 ----------
\begin{minipage}[t]{0.42\textwidth}
\centering
\textbf{Set-up III: $(d,d_0)=(300,10)$}

\vspace{0.25em}
\begin{adjustbox}{max width=\linewidth}
\begin{tabular}{@{}c c c cccc@{}}
\toprule
Cov. & $\beta_j$ & $n=100$ & $n=250$ & $n=350$ & $n=500$ \\
\midrule
TS
& $\beta_1$
& 0.812(1.04) & 0.834(0.88) & 0.872(0.73) & 0.894(0.61) \\[1.3mm]

 RS
& $\beta_1$
& 0.822 & 0.852 & 0.880 & 0.904\\[1.3mm]

\cdashline{2-6}

 TS
& $\beta_2$
& 0.826(0.93) & 0.844(0.71) & 0.868(0.62) & 0.886(0.54) \\[1.3mm]

 RS
& $\beta_2$
& 0.834 & 0.862 & 0.878 & 0.892 \\[1.3mm]

\cdashline{2-6}

TS
& $\beta_3$
& 0.802(0.97) & 0.836(0.77) & 0.876(0.61) & 0.894(0.52) \\[1.3mm]

 RS
& $\beta_3$
& 0.824 & 0.868 & 0.888 & 0.906 \\[1.3mm]

\cdashline{2-6}

TS
& $\beta_4$
& 0.822(1.07) & 0.855(0.84) & 0.879(0.68) & 0.898(0.56) \\[1.3mm]

 RS
& $\beta_4$
& 0.816 & 0.846 & 0.870 & 0.892 \\[1.3mm]

\cdashline{2-6}

TS
& $\beta_5$
& 0.798(1.18) & 0.834(0.91) & 0.876(0.78) & 0.888(0.62) \\[1.3mm]

 RS
& $\beta_5$
& 0.802 & 0.848 & 0.876 & 0.890 \\[1.3mm]

\cdashline{2-6}

 TS
& $\beta_6$
& 0.848(1.01) & 0.862(0.82) & 0.884(0.71) & 0.900(0.53) \\[1.3mm]

 RS
& $\beta_6$
& 0.842 & 0.868 & 0.888 & 0.896\\[1.3mm]

\cdashline{2-6}

 TS
& $\beta_7$
& 0.826(1.11) & 0.858(0.89) & 0.880(0.73) & 0.898(0.51) \\[1.3mm]

 RS
& $\beta_7$
& 0.812 & 0.838 & 0.866 & 0.884\\[1.3mm]

\cdashline{2-6}

 TS
& $\beta_8$
& 0.830(0.99) & 0.864(0.85) & 0.886(0.73) & 0.908(0.61) \\[1.3mm]

 RS
& $\beta_8$
& 0.846 & 0.878 & 0.888 & 0.900 \\[1.3mm]

\cdashline{2-6}

 TS
& $\beta_9$
& 0.836(1.12) & 0.868(0.92) & 0.884(0.75) & 0.896(0.62) \\[1.3mm]

 RS
& $\beta_9$
& 0.828 & 0.854 & 0.880 & 0.892\\[1.3mm]

\cdashline{2-6}

 TS
& $\beta_{10}$
& 0.800(1.02) & 0.844(0.81) & 0.878(0.70) & 0.890(0.58) \\[1.3mm]

 RS
& $\beta_{10}$
& 0.820 & 0.852 & 0.878 & 0.896\\[1.3mm]
\bottomrule
\end{tabular}
\end{adjustbox}
\end{minipage}
\hspace{0.03\linewidth}
\begin{minipage}[t]{0.42\textwidth}
\centering
\textbf{Set-up IV: $(d,d_0)=(500,12)$}

\vspace{0.25em}
\begin{adjustbox}{max width=\linewidth}
\begin{tabular}{@{}c c c cccc@{}}
\toprule
 $\beta_j$ & $n=100$ & $n=250$ & $n=350$ & $n=500$ \\
\midrule

 $\beta_1$
& 0.794(1.01) & 0.826(0.83) & 0.868(0.69) & 0.889(0.54) \\

 $\beta_1$
& 0.818 & 0.852 & 0.880 & 0.904\\

\cdashline{1-5}

 $\beta_2$
& 0.808(1.21) & 0.848(0.86) & 0.876(0.63) & 0.894(0.52) \\

 $\beta_2$
& 0.812 & 0.844 & 0.878 & 0.890\\

\cdashline{1-5}

$\beta_3$
& 0.796(1.21) & 0.832(0.86) & 0.870(0.61) & 0.892(0.52) \\

 $\beta_3$
& 0.801 & 0.848 & 0.874 & 0.892\\

\cdashline{1-5}

 $\beta_4$
& 0.822(1.01) & 0.850(0.87) & 0.878(0.68) & 0.892(0.53) \\

 $\beta_4$
& 0.844 & 0.868 & 0.882 & 0.900 \\

\cdashline{1-5}

$\beta_5$
& 0.810(1.09) & 0.842(0.84) & 0.872(0.67) & 0.890(0.51) \\

 $\beta_5$
& 0.822 & 0.854 & 0.876 & 0.888 \\

\cdashline{1-5}

 $\beta_6$
& 0.834(0.91) & 0.862(0.80) & 0.884(0.65) & 0.898(0.51) \\

 $\beta_6$
& 0.846 & 0.866 & 0.890 & 0.904\\

\cdashline{1-5}

$\beta_7$
& 0.812(1.21) & 0.856(0.91) & 0.880(0.75) & 0.898(0.53) \\

 $\beta_7$
& 0.806 & 0.846 & 0.878 & 0.900\\

\cdashline{1-5}

$\beta_8$
& 0.799(1.17) & 0.836(0.89) & 0.868(0.72) & 0.892(0.55) \\

 $\beta_8$
& 0.808 & 0.848 & 0.874 & 0.886\\

\cdashline{1-5}

 $\beta_9$
& 0.840(1.19) & 0.864(0.93) & 0.886(0.77) & 0.894(0.65) \\

 $\beta_9$
& 0.838 & 0.870 & 0.894 & 0.906 \\

\cdashline{1-5}

 $\beta_{10}$
& 0.816(1.08) & 0.860(0.97) & 0.882(0.78) & 0.898(0.53) \\

 $\beta_{10}$
& 0.826 & 0.854 & 0.880 & 0.904 \\

\cdashline{1-5}

$\beta_{11}$
& 0.852(0.92) & 0.876(0.83) & 0.892(0.72) & 0.910(0.56) \\

$\beta_{11}$
& 0.846 & 0.870 & 0.882 & 0.896\\

\cdashline{1-5}

 $\beta_{12}$
& 0.812(1.09) & 0.856(0.83) & 0.878(0.71) & 0.890(0.62) \\

 $\beta_{12}$
& 0.822 & 0.852 & 0.884 & 0.910 \\
\bottomrule
\end{tabular}
\end{adjustbox}
\end{minipage}

\vspace{0.4em}
\begin{minipage}{0.95\textwidth}
\footnotesize
\textit{Notes:} TS = two-sided percentile interval; RS = right-sided percentile
interval. Results are based on 500 Monte Carlo replications with a fixed design matrix.
\end{minipage}
\end{table}

\begin{table}[H]
\centering
\footnotesize
\caption{Empirical coverage probabilities of nominal 90\% percentile intervals.
Two-sided coverages are reported with average widths (in parentheses), along
with right-sided coverages in logistic regression.}
\label{tab:coverage-4block}

\setlength{\tabcolsep}{1.9pt}
\renewcommand{\arraystretch}{1.12}

% ---------- ROW 1 ----------
\begin{minipage}[t]{0.42\textwidth}
\centering
\textbf{Set-up I: $(d,d_0)=(100,6)$}

\vspace{0.25em}
\begin{adjustbox}{max width=\linewidth}
\begin{tabular}{@{}c c c cccc@{}}
\toprule
Cov. & $\beta_j$ & $n=100$ & $n=250$ & $n=350$ & $n=500$ \\
\midrule
 TS
& $\beta_1$
& 0.804(1.12) & 0.845(0.91) & 0.876(0.73) & 0.896(0.54) \\[1.95mm]

 RS
& $\beta_1$
& 0.812 & 0.846 & 0.878 & 0.890\\[1.95mm]

\cdashline{2-6}
 TS
& $\beta_2$
& 0.796(1.31) & 0.854(0.88) & 0.876(0.75) & 0.894(0.62) \\[1.95mm]

 RS
& $\beta_2$
& 0.806 & 0.844 & 0.874 & 0.908\\[1.95mm]

\cdashline{2-6}

TS
& $\beta_3$
& 0.817(0.92) & 0.834(0.86) & 0.868(0.72) & 0.888(0.60) \\[1.95mm]

RS
& $\beta_3$
& 0.824 & 0.856 & 0.878 & 0.890 \\[1.95mm]

\cdashline{2-6}

TS
& $\beta_4$
& 0.829(1.12) & 0.842(0.86) & 0.874(0.67) & 0.910(0.58) \\[1.95mm]

RS
& $\beta_4$
& 0.805 & 0.834 & 0.876 & 0.898\\[1.95mm]

\cdashline{2-6}

TS
& $\beta_5$
& 0.816(1.09) & 0.836(0.84) & 0.864(0.72) & 0.888(0.56)\\[1.95mm]

  RS
& $\beta_5$
& 0.806 & 0.844 & 0.878 & 0.898 \\[1.95mm]

\cdashline{2-6}

 TS
& $\beta_6$
& 0.802(0.94) & 0.836(0.74) & 0.878(0.62) & 0.898(0.54) \\[1.95mm]

 RS
& $\beta_6$
& 0.846 & 0.872 & 0.894 & 0.902\\[1.95mm]
\bottomrule
\end{tabular}
\end{adjustbox}

\end{minipage}
\hspace{0.03\linewidth}
\begin{minipage}[t]{0.42\textwidth}
\centering
\textbf{Set-up II: $(d,d_0)=(200,8)$}

\vspace{0.25em}
\begin{adjustbox}{max width=\linewidth}
\begin{tabular}{@{}c c c cccc@{}}
\toprule
 $\beta_j$ & $n=100$ & $n=250$ & $n=350$ & $n=500$ \\
\midrule

$\beta_1$
& 0.808(1.21) & 0.836(0.96) & 0.872(0.75) & 0.896(0.52) \\

 $\beta_1$
& 0.826 & 0.854 & 0.878 & 0.898\\

\cdashline{1-5}

 $\beta_2$
& 0.818(1.02) & 0.844(0.88) & 0.880(0.76) & 0.908(0.65) \\

 $\beta_2$
& 0.830 & 0.852 & 0.876 & 0.898\\

\cdashline{1-5}

 $\beta_3$
& 0.808(1.16) & 0.846(0.97) & 0.874(0.79) & 0.890(0.61) \\

 $\beta_3$
& 0.820 & 0.848 & 0.868 & 0.886 \\

\cdashline{1-5}

$\beta_4$
& 0.806(1.18) & 0.844(0.82) & 0.876(0.70) & 0.898(0.62) \\

 $\beta_4$
& 0.834 & 0.864 & 0.886 & 0.906 \\

\cdashline{1-5}

$\beta_5$
& 0.842(0.93) & 0.872(0.78) & 0.890(0.62) & 0.912(0.55) \\

 $\beta_5$
& 0.832 & 0.864 & 0.880 & 0.900\\

\cdashline{1-5}

$\beta_6$
& 0.812(0.95) & 0.848(0.81) & 0.880(0.62) & 0.902(0.45) \\

 $\beta_6$
& 0.808 & 0.842 & 0.874 & 0.888\\

\cdashline{1-5}

  $\beta_7$
& 0.836(0.91) & 0.858(0.76) & 0.880(0.64) & 0.898(0.58) \\

 $\beta_7$
& 0.826 & 0.864 & 0.890 & 0.918 \\

\cdashline{1-5}

$\beta_{8}$
& 0.852(1.01) & 0.878(0.84) & 0.896(0.72) & 0.910(0.54) \\

$\beta_{8}$
& 0.834 & 0.874 & 0.892 & 0.914\\
\bottomrule
\end{tabular}
\end{adjustbox}
\end{minipage}

\vspace{0.86em}

% ---------- ROW 2 ----------
\begin{minipage}[t]{0.42\textwidth}
\centering
\textbf{Set-up III: $(d,d_0)=(300,10)$}

\vspace{0.25em}
\begin{adjustbox}{max width=\linewidth}
\begin{tabular}{@{}c c c cccc@{}}
\toprule
Cov. & $\beta_j$ & $n=100$ & $n=250$ & $n=350$ & $n=500$ \\
\midrule
& $\beta_1$
& 0.848(1.14) & 0.874(0.98) & 0.888(0.81) & 0.902(0.67) \\[1.3mm]

 RS
& $\beta_1$
& 0.850 & 0.878 & 0.892 & 0.910\\[1.3mm]

\cdashline{2-6}

  TS
& $\beta_2$
& 0.822(0.93) & 0.858(0.82) & 0.876(0.74) & 0.894(0.59) \\[1.3mm]

 RS
& $\beta_2$
& 0.845 & 0.876 & 0.886 & 0.902  \\[1.3mm]

\cdashline{2-6}

TS
& $\beta_3$
& 0.812(0.98) & 0.846(0.87) & 0.878(0.66) & 0.898(0.51) \\[1.3mm]

 RS
& $\beta_3$
& 0.824 & 0.858 & 0.880 & 0.896 \\[1.3mm]

\cdashline{2-6}

TS
& $\beta_4$
& 0.832(0.97) & 0.856(0.87) & 0.874(0.64) & 0.890(0.56) \\[1.3mm]

 RS
& $\beta_4$
& 0.826 & 0.860 & 0.882 & 0.900 \\[1.3mm]

\cdashline{2-6}

TS
& $\beta_5$
& 0.810(1.08) & 0.852(0.89) & 0.878(0.73) & 0.894(0.59) \\[1.3mm]

 RS
& $\beta_5$
& 0.822 & 0.856 & 0.882 & 0.908 \\[1.3mm]

\cdashline{2-6}

 TS
& $\beta_6$
& 0.842(1.02) & 0.870(0.86) & 0.884(0.70) & 0.896(0.52) \\[1.3mm]

RS
& $\beta_6$
& 0.836 & 0.868 & 0.894 & 0.918\\[1.3mm]

\cdashline{2-6}

 TS
& $\beta_7$
& 0.820(1.01) & 0.850(0.87) & 0.876(0.71) & 0.890(0.63) \\[1.3mm]

 RS
& $\beta_7$
& 0.818 & 0.846 & 0.872 & 0.894\\[1.3mm]

\cdashline{2-6}

TS
& $\beta_8$
& 0.856(0.99) & 0.878(0.87) & 0.884(0.78) & 0.900(0.67) \\[1.3mm]

 RS
& $\beta_8$
& 0.842 & 0.866 & 0.878 & 0.890 \\[1.3mm]

\cdashline{2-6}

TS
& $\beta_9$
& 0.838(0.92) & 0.868(0.80) & 0.876(0.64) & 0.898(0.54)  \\[1.3mm]

RS
& $\beta_9$
& 0.852 & 0.872 & 0.888 & 0.906\\[1.3mm]

\cdashline{2-6}

 TS
& $\beta_{10}$
& 0.836(1.12) & 0.864(0.84) & 0.886(0.70) & 0.904(0.53) \\[1.3mm]

 RS
& $\beta_{10}$
& 0.826 & 0.858 & 0.874 & 0.892\\[1.3mm]
\bottomrule
\end{tabular}
\end{adjustbox}
\end{minipage}
\hspace{0.03\linewidth}
\begin{minipage}[t]{0.42\textwidth}
\centering
\textbf{Set-up IV: $(d,d_0)=(500,12)$}

\vspace{0.25em}
\begin{adjustbox}{max width=\linewidth}
\begin{tabular}{@{}c c c cccc@{}}
\toprule
 $\beta_j$ & $n=100$ & $n=250$ & $n=350$ & $n=500$ \\
\midrule

 $\beta_1$
& 0.832(1.05) & 0.856(0.93) & 0.878(0.79) & 0.890(0.55) \\

 $\beta_1$
& 0.838 & 0.868 & 0.880 & 0.904\\

\cdashline{1-5}

$\beta_2$
& 0.858(1.21) & 0.878(0.96) & 0.894(0.76) & 0.912(0.62) \\

 $\beta_2$
& 0.842 & 0.868 & 0.886 & 0.898\\

\cdashline{1-5}

$\beta_3$
& 0.846(1.14) & 0.868(0.91) & 0.882(0.79) & 0.908(0.62) \\

 $\beta_3$
& 0.828 & 0.862 & 0.884 & 0.906\\

\cdashline{1-5}

 $\beta_4$
& 0.848(1.01) & 0.870(0.88) & 0.890(0.72) & 0.904(0.56) \\

 $\beta_4$
& 0.852 & 0.884 & 0.900 & 0.916 \\

\cdashline{1-5}

$\beta_5$
& 0.824(1.03) & 0.856(0.86) & 0.874(0.65) & 0.886(0.53) \\

 $\beta_5$
& 0.832 & 0.868 & 0.880 & 0.896 \\

\cdashline{1-5}

 $\beta_6$
& 0.840(0.91) & 0.862(0.74) & 0.880(0.68) & 0.898(0.52) \\

 $\beta_6$
& 0.838 & 0.860 & 0.886 & 0.894\\

\cdashline{1-5}

$\beta_7$
& 0.822(1.17) & 0.852(0.98) & 0.876(0.79) & 0.890(0.52) \\

 $\beta_7$
& 0.836 & 0.866 & 0.884 & 0.910\\

\cdashline{1-5}

$\beta_8$
& 0.818(1.13) & 0.856(0.87) & 0.874(0.69) & 0.888(0.55) \\

 $\beta_8$
& 0.826 & 0.858 & 0.880 & 0.902\\

\cdashline{1-5}

  $\beta_9$
& 0.862(1.09) & 0.876(0.83) & 0.890(0.72) & 0.908(0.58) \\

 $\beta_9$
& 0.854 & 0.870 & 0.888 & 0.900 \\

\cdashline{1-5}

 $\beta_{10}$
& 0.836(1.05) & 0.866(0.93) & 0.884(0.75) & 0.892(0.51) \\

 $\beta_{10}$
& 0.816 & 0.858 & 0.876 & 0.892 \\

\cdashline{1-5}

$\beta_{11}$
& 0.854(1.12) & 0.872(0.89) & 0.886(0.72) & 0.902(0.56) \\

$\beta_{11}$
& 0.856 & 0.874 & 0.890 & 0.912\\

\cdashline{1-5}

 $\beta_{12}$
& 0.820(1.08) & 0.852(0.81) & 0.876(0.73) & 0.894(0.61) \\

 $\beta_{12}$
& 0.828 & 0.862 & 0.888 & 0.914 \\
\bottomrule
\end{tabular}
\end{adjustbox}
\end{minipage}

\vspace{0.4em}
\begin{minipage}{0.95\textwidth}
\footnotesize
\textit{Notes:} TS = two-sided percentile interval; RS = right-sided percentile
interval. Results are based on 500 Monte Carlo replications with a fixed design matrix.
\end{minipage}
\end{table}

Table~\ref{tab:coverage-4blockgamma} and \ref{tab:coverage-4block} summarize empirical coverages of nominal \(90\%\) percentile intervals across four set-ups with increasing \((d,d_0)\) and \(n\). For both two-sided (TS) and right-sided (RS) intervals, coverages rise with \(n\) and approach \(0.90\), particularly from \(n=100\) to \(n=500\). The average TS widths decrease steadily with \(n\), reflecting the expected contraction of uncertainty, with similar rates across coefficients and dimensions, providing empirical support for uniform validity of the proposed percentile-based procedure in high dimensions.

\section{Real Data Analysis}\label{sec:realdata}
\subsection{\bf Regime (I)}: We analyze the concrete compressive strength data of \citet{yeh1998concrete}
(which is available at \url{https://archive.ics.uci.edu/dataset/165/concrete+compressive+strength}),
consisting of $n=1030$ observations and $d=8$ mixture components:
cement, blast furnace slag, fly ash, water, superplasticizer, coarse aggregate,
fine aggregate, and age. The response variable, compressive strength (MPa),
is continuous and strictly positive (see Table~\ref{tab:desc_strength_horizontal}),
which motivates the use of a gamma regression model with log link. All covariates were standardized and parameters were estimated via convex
optimization using the \texttt{CVXR} package in \texttt{R}. Confidence intervals
were constructed using the proposed PB percentile method with 1030 resamples,
scaled by $n^{1/2}$. The resulting estimates and 90\% confidence intervals are
reported in Table~\ref{tab:gammaci}.
\hspace*{-2mm}
\begin{table}[ht]
\centering
\caption{Descriptive statistics of concrete compressive strength (MPa)}
\label{tab:desc_strength_horizontal}
\begin{tabular}{ccccccc}
\midrule
Minimum & 1st Quartile & Median & Mean & 3rd Quartile & Maximum \\
\midrule
2.33 & 23.71 & 34.44 & 35.82 & 46.14 & 82.60 \\
\bottomrule
\end{tabular}
\end{table}
\hspace*{-2mm}
The results show that cement, blast furnace slag, fly ash, and age have
statistically significant positive effects on compressive strength, with
two-sided confidence intervals excluding zero. These findings align with
well-established material science principles regarding hydration and pozzolanic
reactions in ternary concrete mixtures, and are consistent with the machine
learning study of \citet{huang2022fa_rf}. In contrast, water exhibits a negative
but statistically inconclusive effect, while superplasticizer and both aggregate
components show weak marginal effects, suggesting that their influence may be
nonlinear or interaction-driven.
\hspace*{-2mm}
\begin{table}[ht]
\centering
\caption{Parameter estimates and 90\% confidence intervals in gamma regression model}
\label{tab:gammaci}

\begin{adjustbox}{max width=\linewidth}
\begin{tabular}{lcccc}
\midrule
Covariate & $\hat{\beta}$ 
& 90\% CI (two-sided) 
& 90\% CI (left-sided) & 90\% CI (right-sided) \\
\midrule
Cement      
& $0.3643$ 
& $[0.2251,\;0.5072]$ 
& $[0.2570,\;\infty)$,\;& $(-\infty,\;0.4747]$ \\

Blast Furnace Slag   
& $0.2561$ 
& $[0.1265,\;0.3977]$ 
& $[0.1543,\;\infty)$\;& $(-\infty,\;0.3677]$ \\

Fly Ash              
& $0.1857$ 
& $[0.0638,\;0.3157]$ 
& $[0.0842,\;\infty)$\;& $(-\infty,\;0.2890]$ \\

Water                
& $-0.0865$ 
& $[-0.2146,\;0.0557]$ 
& $[-0.1897,\;\infty)$\;& $(-\infty,\;0.0249]$ \\

Superplasticizer     
& $0.0519$ 
& $[-0.0444,\;0.1455]$ 
& $[-0.0200,\;\infty)$\;& $(-\infty,\;0.1224]$ \\

Coarse Aggregate     
& $0.0346$ 
& $[-0.0779,\;0.1570]$ 
& $[-0.0528,\;\infty)$\;& $(-\infty,\;0.1291]$ \\

Fine Aggregate       
& $0.0254$ 
& $[-0.1059,\;0.1681]$ 
& $[-0.0746,\;\infty)$\;& $(-\infty,\;0.1396]$ \\

Age           
& $0.2411$ 
& $[0.1832,\;0.2960]$ 
& $[0.1967,\;\infty)$\;&$(-\infty,\;0.2849]$ \\
\bottomrule
\end{tabular}
\end{adjustbox}
\end{table}
\hspace*{-2mm}
Overall, the gamma regression model combined with PB-based inference captures
the primary mechanisms governing compressive strength in an
interpretable manner, producing conclusions consistent with modern
machine-learning approaches applied to the same dataset.

\subsection{\bf Regime (II)}: We analyze the Colon cancer microarray dataset of \citet{alon1999broad},
available in \texttt{R} via \texttt{library(plsgenomics)} as
\texttt{data(Colon)} and previously studied by \citet{liang2013l12}.
The data contain $n=62$ tissue samples ($40$ tumor, $22$ normal) and
$d=2000$ standardized gene expression measurements. The goal is to
classify tumor versus normal tissues in a high-dimensional setting
with $d \gg n$. An $\ell_1$-penalized logistic regression model is fitted using
\texttt{glmnet}, with the tuning parameter chosen by 10-fold
cross-validation. The Logistic Lasso selects $11$ genes with nonzero
coefficients. Inference is then conducted on this
Lasso-selected active set using the proposed perturbation bootstrap
(PB) method. Bootstrap samples are generated by introducing iid
$\mathrm{Exp}(1)$ perturbations to the objective function and
recomputing penalized estimators via convex optimization using
\texttt{CVXR}. Two-sided and one-sided $90\%$ confidence intervals are
constructed from the empirical distribution of the scaled bootstrap
estimates.

\begin{table}[ht]
\centering
\caption{Logistic Lasso estimates and 90\% perturbation bootstrap confidence intervals for selected genes (Colon cancer data)}
\label{tab:colonci}
\begin{adjustbox}{max width=\linewidth}
\begin{tabular}{lcccc}
\midrule
Gene (index) & $\hat{\beta}$ 
& 90\% CI (two-sided) 
& 90\% CI (left-sided) 
& 90\% CI (right-sided) \\
\midrule
Gene 249  
& $-0.332$ 
& $[-0.664,\,-0.153]$ 
& $[-0.664,\;\infty)$ 
& $(-\infty,\,-0.311]$ \\

Gene 377  
& $-0.390$ 
& $[-0.779,\,-0.599]$ 
& $[-0.779,\;\infty)$ 
& $(-\infty,\,-0.613]$ \\

Gene 493  
& $-0.034$ 
& $[-0.068,\;0.214]$ 
& $[-0.068,\;\infty)$ 
& $(-\infty,\;0.151]$ \\

Gene 765  
& $-0.365$ 
& $[-0.731,\,-0.468]$ 
& $[-0.631,\;\infty)$ 
& $(-\infty,\,-0.514]$ \\

Gene 1772 
& $0.360$ 
& $[0.301,\;0.521]$ 
& $[0.321,\;\infty)$ 
& $(-\infty,\;0.772]$ \\
Gene 1870 
& $0.382$ 
& $[0.370,\;0.455]$ 
& $[0.376,\;\infty)$ 
& $(-\infty,\;0.485]$\\ 

\bottomrule
\end{tabular}
\end{adjustbox}
\end{table}

Table~\ref{tab:colonci} presents results for six representative genes
from the selected set. Genes 249, 377, and 765 show significantly
negative effects, while genes 1772 and 1870 show strong positive
associations with tumor status. Gene 493 exhibits a weaker signal,
with confidence intervals overlapping zero. These findings are
consistent with earlier analyses of this dataset
\citet{liang2013l12}, which highlight that tumor classification depends
on a small subset of genes despite the high ambient dimension. This example illustrates that the proposed PB method provides
interpretable and reliable uncertainty quantification for selected
features in high-dimensional logistic regression under $d \gg n$.

\appendix

\section{Regime I: Proofs of Requisite Lemmas and Main Results}\label{sec:proofreg1}
In this section, we first recall some basic topological properties given in section \ref{sec:topdef}. In section \ref{sec:gaussiso}, we state and prove the requisite lemmas related to construct proof of Theorem \ref{thm:convex1} and \ref{thm:ball1}. Lemmas related to Theorem \ref{thm:pbconvexoriginal1} and \ref{thm:prbconvexoriginal1} are alloted to section \ref{sec:PBGLMdistn}. Finally the proofs of main results are relegated to section \ref{sec:proofmain}.
\subsection{\bf Some Topological Definitions and Properties}
\label{sec:topdef}
In this section, we will introduce some basic topological aspects for the sake of completeness.
\begin{def1}\label{def:boundary}
Topological boundary of a set $A\subset \mathbf{R}^d$, is defined by $$\partial{A}=(Cl.A)\cap (Int.A)^c,$$
where $Cl.A$ and $Int.A$ respectively denote the closure and interior of a set $A$.
\end{def1}

\begin{def1}\label{def:epsilonnbd}
The $\epsilon$-neighbourhood of a set $A$, denoted by $A^{\epsilon}$, is defined as, $$A^{\epsilon}=\{\bm{x}\in \mathbf{R}^d:dist(\bm{x},A)<\epsilon\},\; \text{for some}\; \epsilon>0,$$
where, we define $dist(\bm{x},A)=\inf_{\bm{y}\in A}\big{\|}\bm{x}-\bm{y}\big{\|}.$ Similarly we can define, for some $\epsilon>0$, $$A^{-\epsilon}:=\{\bm{x}\in \mathbf{R}^d:B(\bm{x},\epsilon)\subset A\},$$
where, $B(\bm{x},\epsilon)$ is usual $d$-ball centered at $\bm{x}$ with radius $\epsilon.$ Note that, $A^{-\epsilon}\subset A\subset A^{\epsilon}.$
\end{def1}

\begin{def1}\label{def:invarianceoperator}
A class of mesurable sets $\mathcal{A}$, is said to be invariant under affine symmetric transformation if for $\bm{a}\in \mathbf{R}^d$ and any linear symmetric invertible operator $\mathbf{D}:\mathbf{R}^d\to \mathbf{R}^d$, following holds: $$A\in \mathcal{A}\implies \mathbf{D}A+\bm{a}\in \mathcal{A}.$$
\end{def1}

\begin{def1}\label{def:invariancescaleshift}
A class of mesurable sets $\mathcal{A}$, is said to be invariant under translation and rescaling (cf. \citet{bentkus2003dependence}, \citet{raivc2019multivariate}) if for any $\bm{a}\in \mathbf{R}^d$ and $b(\neq 0)\in \mathbf{R}$, following holds: $$A\in \mathcal{A}\implies bA+\bm{a}\in \mathcal{A}.$$
\end{def1}

\begin{def1}\label{def:invarianceepsilonnbd}
A class of mesurable sets $\mathcal{A}$, is said to be invariant under taking $\epsilon$-neighbourhoods (cf. \citet{bentkus2003dependence}, \citet{raivc2019multivariate}) if $A\in \mathcal{A}\implies A^{\epsilon}, A^{-\epsilon}\in \mathcal{A}$ for some $\epsilon>0.$
\end{def1}

\begin{rem}\label{remark:convexclasstop}
 When $\mathcal{A}$ is taken to be $\mathcal{C}$, class of measurable convex sets, we will have the following properties. Also since, Euclidean balls belong to $\mathcal{C}$, hence class of Euclidean balls $\mathcal{B}$ automatically satisfies the following properties. However when $\mathcal{A}$ is taken to be class of hyperrectangles, all of the following properties need not hold true.   
\end{rem}

\begin{property}\label{property1}
  $\mathcal{C}$ satisfies Definition \ref{def:invarianceoperator}, \ref{def:invariancescaleshift} and \ref{def:invarianceepsilonnbd}.  
\end{property}

\begin{property}\label{setequalinvariance}
For any $A\in \mathcal{C}$ and $\epsilon>0$, we have $$\big\{\bm{x}+c\bm{y}\in A^{\epsilon}\big\}=\big\{\bm{y}\in c^{-1}(A-\bm{x})^{\epsilon}\big\},$$ for any $\bm{x},\bm{y}\in A$ and $c\in \mathbf{R}\setminus \{0\}.$
\end{property}

\begin{property}\label{distequalinvariance}
 For $A\in \mathcal{C}$ and for each $\bm{x,y}\in \mathbf{R}^d$, we have $$dist(\bm{x}+\bm{y},A+\bm{y})=dist(\bm{x},A).$$   
\end{property}

\begin{property}\label{scaleinvariance}
For $A\in \mathcal{C}$ and for each $\bm{x}\in \mathbf{R}^d$ with $m\geq 1$, we have $$\big{|}dist(m\bm{x},mA)\big{|}\leq m\big{|}dist(\bm{x},A)\big{|}.$$    
\end{property}

\begin{property}\label{epsilonnbdsubset}
  For $A\in \mathcal{C}$ and $c\in \mathbf{R}\setminus \{0\}$, we have $$c^{-1}A^{\epsilon}\subseteq \big(c^{-1}A\big)^{c^{-1}\epsilon}.$$  
\end{property}

\subsection{Lemmas Corresponding to Theorem \ref{thm:convex1} and \ref{thm:ball1}}\label{sec:gaussiso}

\begin{lemma}\label{lem:BhattRao}
 Under the fact that every ball in $\mathbf{R}^d$ is connected, for any $\epsilon>0$ we have $$(\partial{A})^{\epsilon}=A^{\epsilon}\setminus A^{-\epsilon}.$$       
\end{lemma}
Proof of Lemma \ref{lem:BhattRao} This lemma is proved as Corollary 2.6 in \citet{bhattacharya1986normal}.\hfill $\blacksquare$

\begin{lemma}\label{lem:raivc}
 Suppose, $\mathcal{A}$ be the class of measurable subsets and let $\bm{Z}$ be a $d$-dimensional standard Gaussian random vector with density $\phi_d(\bm{z})$. Denote by $\mathcal{H}^{d-1}$,
 the $(d-1)$ dimensional Hausdorff measure. For $A\in\mathcal{A}$, define:
 \begin{align*}
&\gamma (A)=\int_{\partial{A}}\phi_d(\bm{z})\mathcal{H}^{d-1}(d\bm{z})\\
&\gamma^*(A)=\operatorname*{\sup}\Big\{\dfrac{1}{\epsilon}\mathbf{P}[\bm{Z}\in A^{\epsilon}\setminus A],\;\dfrac{1}{\epsilon}\mathbf{P}[\bm{Z}\in A\setminus A^{-\epsilon}];\; \epsilon>0\Big\}
\end{align*}
When $\mathcal{A}=\mathcal{C}$, the class of convex sets, then $$\operatorname*{\sup}_{A\in \mathcal{C}}\gamma^* (A)=\operatorname*{\sup}_{A\in \mathcal{C}}\gamma (A).$$ 
\end{lemma}
Proof of Lemma \ref{lem:raivc}: This lemma is proved as Proposition 1.1 in \citet{raivc2019multivariate}.\hfill $\blacksquare$

\begin{lemma}\label{lem:nazarov}
Under the same set-up of Lemma \ref{lem:raivc}, we have:
$$0.28<\liminf_{d\to\infty}d^{-1/4}[\operatorname*{\sup}_{A\in \mathcal{C}}\gamma (A)]<\limsup_{d\to\infty}d^{-1/4}[\operatorname*{\sup}_{A\in \mathcal{C}}\gamma (A)]<0.64.$$
\end{lemma}
Proof of Lemma \ref{lem:nazarov} This lemma is proved as Theorem at page 170 in \citet{nazarov2003maximal}.\hfill $\blacksquare$

\begin{lemma}\label{lem:zhilova}
Suppose $\bm{G}\sim N_d(\bm{\mu},\Sigma)$ with $\Sigma$ being positive definite matrix. Now for a class of Euclidean balls $\mathcal{B}$, there exists a generic constant $K_{\mathcal{B}}>0$ such that for any $\epsilon>0$, we have: $$\sup_{A\in \mathcal{B}}\mathbf{P}(\bm{G}\in A^{\epsilon}\setminus A)\leq \epsilon K_{\mathcal{B}}\big{\|}\Sigma^{-1/2}\big{\|}_{op}.$$  \end{lemma}
Proof of Lemma \ref{lem:zhilova} This result is proved as Lemma A.2 in \citet{zhilova2020nonclassical}.\hfill $\blacksquare$

\begin{lemma}\label{lem:mainlemmagaussisoperimetric}
 Suppose assumption (A.1) is true. Let a Gaussian random vector $\bm{Z}_n\sim N_d(\bm{0},\Tilde{\bm{S}}_n).$ Then provided $\epsilon=o[d^{-1/4}n^{-\alpha_2/2}]$, we have $$\operatorname*{\sup}_{A\in \mathcal{C}}\Big{|}\mathbf{P}\Big(\bm{Z}_n\in (\partial{A})^{\epsilon}\Big)\Big{|}=o(1).$$ 
\end{lemma}
Proof of Lemma \ref{lem:mainlemmagaussisoperimetric} Note that, $\tilde{\bm{S}}_n=[\mathbf{E(L_n)}]^{-1}$. That means, 
\begin{align}\label{eqn:eigenvalues}
& \big[\lambda_{\min}(\tilde{\bm{S}}_n)\big]^{-1/2}= \big[\lambda_{\max}\big([\mathbf{E(L_n)}]\big)\big]^{1/2}=O(n^{\alpha_2/2}).
\end{align}
Suppose we define, two independent random vectors $\bm{M}$ and $\bm{R}$ such that $\bm{M}\sim N_d(\bm{0},\bm{I}_d)$ and $\bm{R}\sim N_d(\bm{0},\Tilde{\bm{S}}_n-\lambda_{\min}(\Tilde{\bm{S}}_n)\bm{I}_d).$ Then clearly, $\bm{Z}_n=\bm{R}+\big[\lambda_{\min}(\Tilde{\bm{S}}_n)\big]^{1/2}\bm{M}\sim N_d(\bm{0},\Tilde{\bm{S}}_n).$ As a consequence to Lemma \ref{lem:BhattRao} we see, $(\partial{A})^{\epsilon}=A^{\epsilon}\setminus A^{-\epsilon}=(A^{\epsilon}\setminus A)\cup (A\setminus A^{-\epsilon})$. Then due to sub-additivity, we note that,
\begin{align}\label{eqn:setexpand}
\operatorname*{\sup}_{A\in \mathcal{C}}\Big{|}\mathbf{P}\Big(\bm{Z}_n\in (\partial{A})^{\epsilon}\Big)\Big{|}\leq \operatorname*{\sup}_{A\in \mathcal{C}}\Big{|}\mathbf{P}\Big(\bm{Z}_n\in A^{\epsilon}\setminus A\Big)\Big{|}+ \operatorname*{\sup}_{A\in \mathcal{C}}\Big{|}\mathbf{P}\Big(\bm{Z}_n\in A\setminus A^{-\epsilon}\Big)\Big{|}   
\end{align}
To find (\ref{eqn:setexpand}), it is enough to calculate any one of the two probabilities as they are symmetric with respect to the $\epsilon$-neighbourhoods. Therefore,
\begin{align}\label{eqn:raivcequation}
&\mathbf{P}\Big(\bm{Z}_n\in A^{\epsilon}\setminus A\Big)\nonumber\\
&=\mathbf{P}\Big[\bm{R}+\big[\lambda_{\min}(\Tilde{\bm{S}}_n)\big]^{1/2}\bm{M}\in A^{\epsilon}\setminus A\Big]\nonumber\\
&=\mathbf{P}\Big[\bm{M}\in \big[\lambda_{\min}(\Tilde{\bm{S}}_n)\big]^{-1/2}(A-\bm{R})^{\epsilon}\setminus \big[\lambda_{\min}(\Tilde{\bm{S}}_n)\big]^{-1/2}(A-\bm{R})\Big]\nonumber\\
&\leq \mathbf{P}\Big[\bm{M}\in \Big\{\big[\lambda_{\min}(\Tilde{\bm{S}}_n)\big]^{-1/2}(A-\bm{R})\Big\}^{\epsilon\big[\lambda_{\min}(\Tilde{\bm{S}}_n)\big]^{-1/2}}\setminus \big[\lambda_{\min}(\Tilde{\bm{S}}_n)\big]^{-1/2}(A-\bm{R})\Big]\nonumber\\
&\leq \epsilon\big[\lambda_{\min}(\Tilde{\bm{S}}_n)\big]^{-1/2}\gamma^* (\mathcal{C})\leq \epsilon\big[\lambda_{\min}(\Tilde{\bm{S}}_n)\big]^{-1/2}\gamma (\mathcal{C})\leq (0.64c_2)\epsilon \big[d^{1/4}n^{\alpha_2/2}\big]
\end{align}
The second equality follows due to Property \ref{property1} and \ref{setequalinvariance} of class of convex sets. The first inequality follows from Property \ref{scaleinvariance} and \ref{epsilonnbdsubset}. The second inequality is due to the definition of $\gamma^* (\mathcal{C})$. The third inequality follows from (\ref{eqn:eigenvalues}) and Lemma \ref{lem:raivc}. The last bound is due to assumption (C.4) and Lemma \ref{lem:nazarov}. Therefore, due to the assumption $\epsilon=o[d^{-1/4}n^{-\alpha_2/2}]$ and observing (\ref{eqn:setexpand}) and (\ref{eqn:raivcequation}), we conclude that, $$\operatorname*{\sup}_{A\in \mathcal{C}}\Big{|}\mathbf{P}\Big(\bm{Z}_n\in (\partial{A})^{\epsilon}\Big)\Big{|}=o(1).$$
Hence we are done.\hfill $\blacksquare$

\begin{lemma}\label{lem:zhilovamainlemmagaussisoperimetric}
Suppose $\mathcal{B}$ be class of measurable Euclidean balls. Consider the same set-up and assumptions as in Lemma \ref{lem:mainlemmagaussisoperimetric}. Then provided $\epsilon=o[n^{-\alpha_2/2}]$, we have $$\operatorname*{\sup}_{A\in \mathcal{B}}\Big{|}\mathbf{P}\Big(\bm{Z}_n\in (\partial{A})^{\epsilon}\Big)\Big{|}=o(1).$$     
\end{lemma}
Proof of Lemma \ref{lem:zhilovamainlemmagaussisoperimetric} The proof will follow in exact same line of Lemma \ref{lem:mainlemmagaussisoperimetric} with the only change that we will use Lemma \ref{lem:zhilova} instead of Lemma \ref{lem:nazarov} to utilize Gaussian isoperimetric bound for class of Euclidean balls. So we skip the details. \hfill $\blacksquare$

\begin{lemma}\label{lem:fangkoikeconvex}
 Suppose $\{\bm{\zeta}_i\}_{i=1}^{n}$ be a sequence of centered independent random vectors in $\mathbf{R}^d$ with finite fourth moments and set $\bm{W}=\sum_{i=1}^{n}\bm{\zeta}_i$. Let $\bm{Z}\sim N_d(\bm{0},\Sigma)$ be a $d$-dimensional Gaussian random vector with $\Sigma$ being positive definite covariance matrix. Then for a class of all convex sets $\mathcal{C}$, $$\sup_{A\in \mathcal{C}}\Big{|}\mathbf{P}[\bm{W}\in A]-\mathbf{P}[\bm{Z}\in A]\Big{|}\leq Cd^{1/4}\delta_{\mathcal{C}}(\bm{W},\Sigma)\Big[\big{|}\log \delta_{\mathcal{C}}(\bm{W},\Sigma)\big{|}\Big]^{1/2}$$ 
where, $$\delta_{\mathcal{C}}(\bm{W},\Sigma):=\operatorname{\min}\Bigg\{\Big{\|}\bm{I}_d-Var\big(\Sigma^{-1/2}\bm{W}\big)\Big{\|}_{HS}+\Bigg(\sum_{i=1}^{n}\mathbf{E}\Big[\big{\|}\Sigma^{-1/2}\bm{\zeta}_i\big{\|}_2^4\Big]\Bigg)^{1/2}, 1/e\Bigg\}$$   
\end{lemma}
Proof of Lemma \ref{lem:fangkoikeconvex} This lemma is proved as Theorem 2.1 in \citet{fang2024large}.\hfill $\blacksquare$

\begin{lemma}\label{lem:fangkoikeball}
 Consider the same set-up as in Lemma \ref{lem:fangkoikeconvex}. Then for a class of all Euclidean balls $\mathcal{B}$, $$\sup_{A\in \mathcal{B}}\Big{|}\mathbf{P}[\bm{W}\in A]-\mathbf{P}[\bm{Z}\in A]\Big{|}\leq C\delta_{\mathcal{B}}(\bm{W},\Sigma)\Big[\big{|}\log \delta_{\mathcal{B}}(\bm{W},\Sigma)\big{|}\Big]$$ 
where, $$\delta_{\mathcal{B}}(\bm{W},\Sigma):=\operatorname{\min}\Bigg\{\Big{\|}\Sigma^{-1}\Big{\|}_{op}\Bigg[\Big{\|}\Sigma-Var\big(\bm{W}\big)\Big{\|}_{HS}+\Bigg(\sum_{i=1}^{n}\mathbf{E}\Big[\big{\|}\bm{\zeta}_i\big{\|}_2^4\Big]\Bigg)^{1/2}\Bigg], 1/e\Bigg\}$$   
\end{lemma}
Proof of Lemma \ref{lem:fangkoikeball} This lemma is proved as Theorem 3.1 in \citet{fang2024large}.\hfill$\blacksquare$

\begin{lemma}\label{lem:mainlemmafangandkoikeconvex}
Suppose assumptions (A.1)-(A.4) and (A.6) hold true. Consider a Gaussian random vector $\bm{G}_{1n}\sim N_d(\bm{0},\Tilde{\bm{S}}_n).$ Then provided $d=o[n^{2\{1-2\alpha_1\}/5}]$  we have for any $\epsilon>0$:
\begin{align*}
    &\sup_{A\in\mathcal{C}}\Big{|}\mathbf{P}\Big(\bm{T}_n\in A\Big)-\mathbf{P}\Big(\bm{G}_{1n}\in A\Big)\Big{|}=o(1)\\
    &\sup_{A\in\mathcal{C}}\Big{|}\mathbf{P}\Big(\bm{T}_n\in (\partial{A})^{\epsilon}\Big)-\mathbf{P}\Big(\bm{G}_{1n}\in (\partial{A})^{\epsilon}\Big)\Big{|}=o(1).
\end{align*}
\end{lemma}
Proof of Lemma \ref{lem:mainlemmafangandkoikeconvex} Recall that, $\bm{T}_n=n^{-1/2}\sum_{i=1}^{n}(y_i-\mu_i)h^\prime(\bm{x}_i^\top\bm{\beta})[\mathbf{E(L_n)}]^{-1}\bm{x}_i.$\\
Therefore, we can take $\bm{\zeta}_i=n^{-1/2}(y_i-\mu_i)h^\prime(\bm{x}_i^\top\bm{\beta})[\mathbf{E(L_n)}]^{-1}\bm{x}_i$ in the set-up of Lemma \ref{lem:fangkoikeconvex}. Next we note that due to assumption (A.1), 
\begin{align}\label{eqn:normbound}
\big{\|}\Tilde{\bm{S}}_n^{-1/2}[\mathbf{E(L_n)}]^{-1}\big{\|}=\big[\lambda_{\min}\big(\mathbf{E(L_n)}\big)\big]^{-1/2}=O[n^{\alpha_1/2}].
\end{align}
With (\ref{eqn:normbound}) and assumptions (A.2), (A.3), (A.4) and (A.6), we write:
\begin{align}\label{eqn:deltabound}
 &\Bigg[\sum_{i=1}^{n}\mathbf{E}\Big{\|}n^{-1/2}\Tilde{\bm{S}}_n^{-1/2}[\mathbf{E(L_n)}]^{-1}\bm{x}_i\big\{y_i-g^{-1}(\bm{x}_i^\top \bm{\beta})\big\}h^\prime (\bm{x}_i^\top \bm{\beta})\Big{\|}_2^4\Bigg]^{1/2}\nonumber\\
 &\leq \Bigg[n^{-1}d^{2}\big{\|}\Tilde{\bm{S}}_n^{-1/2}[\mathbf{E(L_n)}]^{-1}\big{\|}^4\Big\{\operatorname*{\max}_{i=1(1)n}\big{\|}\bm{x}_i\big{\|}_{\infty}\Big\}^4\Big\{\mbox{max}_{i\in\{1,..,n\}}|h^\prime(\bm{x}_i^\top\bm{\beta})|\Big\}^{4}\nonumber\\
 &\;\;\;\;\;\;\;\;\;\;\;\;\;\;\;\;\;\;\;\;\;\;\;\;\;\;\;\;\;\;\;\;\;\;\;\;\;\;\;\;\;\;\;\times\Big(n^{-1}\sum_{i=1}^{n}\mathbf{E}\Big{|}\big\{y_i-g^{-1}(\bm{x}_i^\top \bm{\beta})\big\}\Big{|}^4\Big)\Bigg]^{1/2}\nonumber\\
 &=O\Bigg[\dfrac{d^2}{n^{1-2\alpha_1}}\Bigg]^{1/2}.
\end{align}
Lastly, we note that, 
\begin{align}\label{eqn:firstterm}
 \Big{\|}\bm{I}_d-Var\big(\Tilde{\bm{S}}_n^{-1/2}\bm{T}_n\big)\Big{\|}_{HS}&=0  
\end{align}
Therefore combining (\ref{eqn:deltabound}), (\ref{eqn:firstterm}) and following Lemma  \ref{lem:fangkoikeconvex}, we finally see that; $$\delta_{\mathcal{C}}(\bm{T}_n,\Tilde{\bm{S}}_n)=O\Bigg[\dfrac{d^2}{n^{1-2\alpha_1}}\Bigg]^{1/2}.$$

Therefore from Lemma \ref{lem:fangkoikeconvex} and under the assumption that $d=o[n^{2\{1-2\alpha_1\}/5}]$, we see, 
\begin{align}\label{eqn:approximation}
&\sup_{A\in\mathcal{C}}\Big{|}\mathbf{P}\Big(\bm{T}_n\in A\Big)-\mathbf{P}\Big(\bm{G}_{1n}\in A\Big)\Big{|}=O\Bigg[\Bigg\{\dfrac{d^{5/2}}{n^{1-2\alpha_1}}\Bigg\}\log \Bigg\{\dfrac{d^2}{n^{1-2\alpha_1}}\Bigg\}\Bigg]^{1/2}=o(1).
\end{align}
Now since we can write $A^{-\epsilon}\subset A\subset A^{\epsilon}$ and Lemma \ref{lem:BhattRao} holds true, it's easy to see that, 
\begin{align}\label{eqn:boundaarygaussian}
&\sup_{A\in\mathcal{C}}\Big{|}\mathbf{P}\Big(\bm{T}_n\in (\partial{A})^{\epsilon}\Big)-\mathbf{P}\Big(\bm{G}_{1n}\in (\partial{A})^{\epsilon}\Big)\Big{|}\nonumber\\
&\leq \sup_{A\in\mathcal{C}}\Big{|}\mathbf{P}\Big(\bm{T}_n\in A^{\epsilon}\Big)-\mathbf{P}\Big(\bm{G}_{1n}\in A^{\epsilon}\Big)\Big{|}+\sup_{A\in\mathcal{C}}\Big{|}\mathbf{P}\Big(\bm{T}_n\in A^{-\epsilon}\Big)-\mathbf{P}\Big(\bm{G}_{1n}\in A^{-\epsilon}\Big)\Big{|}
\end{align}
But class of convex sets $\mathcal{C}$ is invariant under taking $\epsilon$-neighbourhoods. So bounds on both the terms of (\ref{eqn:boundaarygaussian}) is in fact $o(1)$ under the given conditions. Hence we complete the proof. \hfill $\blacksquare$  

\begin{lemma}\label{lem:mainlemmafangandkoikeball}
Consider the same set-up as in Lemma \ref{lem:mainlemmafangandkoikeconvex}. Assume $\mathcal{B}$ be the class of Euclidean balls. Then provided $d=o[n^{\{1-2\alpha_1\}/2}]$  we have for any $\epsilon>0$:
\begin{align*}
    &\sup_{A\in\mathcal{B}}\Big{|}\mathbf{P}\Big(\bm{T}_n\in A\Big)-\mathbf{P}\Big(\bm{G}_{1n}\in A\Big)\Big{|}=o(1)\\
    &\sup_{A\in\mathcal{B}}\Big{|}\mathbf{P}\Big(\bm{T}_n\in (\partial{A})^{\epsilon}\Big)-\mathbf{P}\Big(\bm{G}_{1n}\in (\partial{A})^{\epsilon}\Big)\Big{|}=o(1).
\end{align*}
\end{lemma}
Proof of Lemma \ref{lem:mainlemmafangandkoikeball} Proof of this lemma will follow in exact similar lines as in Lemma \ref{lem:mainlemmafangandkoikeconvex} along with considering Lemma \ref{lem:fangkoikeball} instead of Lemma \ref{lem:fangkoikeconvex}. Therefore we skip the details. \hfill $\blacksquare$

\begin{lemma}\label{lem:glmconc}
Suppose (A.2) holds true. Then 
$$\Big{\|}\sum_{i=1}^{n}\Psi(y_i,\hat{\bm{\beta}}_n)\Big{\|}_2=o_p(1),$$
where $\Psi(y_i,\bm{\beta})$ is the usual score function of the original GLM objective function given by \ref{eqn:defineglmest12}.
\end{lemma}
Proof of Lemma \ref{lem:glmconc} The score function is given by, $\Psi(y_i,\bm{\beta})=\Big\{-y_ih^\prime (\bm{x}_{i}^\top\bm{\beta})+h_1^\prime (\bm{x}_{i}^\top\bm{\beta})\Big\}\bm{x}_i$ with $h_1(\cdot)=b(h(\cdot))$ and from KKT conditions, it's easy to see that $\hat{\bm{\beta}}_n$ is a solution of $\sum_{i=1}^{n}\Psi(y_i,\bm{\beta})=\bm{0}$.
Then the proof is complete. \hfill $\blacksquare$ 

\begin{lemma}\label{lem:concleading}
Suppose (A.2)-(A.4) and (A.6) are true. Then, $$\Big{\|}\bm{W}_n\Big{\|}_2=O_p(d^{1/2}).$$  
\end{lemma}
Proof of Lemma \ref{lem:concleading} Note that, $\bm{W}_n=n^{-1/2}\sum_{i=1}^{n}\big\{y_i-g^{-1}(\bm{x}_i^\top \bm{\beta})\big\}h^\prime (\bm{x}_i^\top \bm{\beta})\bm{x}_i,$ with independent sequence $\{y_i-g^{-1}(\bm{x}_i^\top \bm{\beta})\}$ for all $i$. Therefore, it's enough to show that $$\mathbf{E}\Big[\Big{\|}\bm{W}_n\Big{\|}_2^2\Big]=O(d).$$
Towards that we write $\bm{Z}_i=\big\{y_i-g^{-1}(\bm{x}_i^\top \bm{\beta})\big\}h^\prime (\bm{x}_i^\top \bm{\beta})\bm{x}_i$. This will imply that,
\begin{align*}
&\mathbf{E}\Big[\Big{\|}\bm{W}_n\Big{\|}_2^2\Big]=\frac{1}{n}\mathbf{E}\Big[\Big{\|}\sum_{i=1}^{n}\bm{Z}_i\Big{\|}_2^2\Big]=\frac{1}{n}\mathbf{E}\Big[\sum_{i=1}^{n}\Big{\|}\bm{Z}_i\Big{\|}_2^2+\sum_{i\neq j}\bm{Z}_i^\top \bm{Z}_j\Big]\\
&=\frac{1}{n}\Bigg[\sum_{i=1}^{n}\Big(\mathbf{E}|y_i-g^{-1}(\bm{x}_i^\top \bm{\beta})|^2\Big)|h^\prime (\bm{x}_i^\top \bm{\beta})|^2\big{\|}\bm{x}_i\big{\|}_2^2\\
&\;\;\;\;\;\;\;\;\;\;\;\;+\sum_{i\neq j}\underbrace{\mathbf{E}\Big[\big\{y_i-g^{-1}(\bm{x}_i^\top \bm{\beta})\big\}\big\{y_j-g^{-1}(\bm{x}_j^\top \bm{\beta})\big\}\Big]}_{=0}\big[h^\prime (\bm{x}_i^\top \bm{\beta})h^\prime (\bm{x}_j^\top \bm{\beta})\big]\bm{x}_i^\top \bm{x}_j\Bigg]\\
&\leq \Big\{d^{1/2}\mbox{max}_{i\in\{1,..,n\}}\|\bm{x}_i\|_{\infty}\Big\}^2\Bigg[n^{-1}\sum_{i=1}^{n}\mathbf{E}|y_i-g^{-1}(\bm{x}_i^\top \bm{\beta})|^2\Bigg]\Bigg[\mbox{max}_{i\in\{1,..,n\}}|h^\prime(\bm{x}_i^\top \bm{\beta})|\Bigg]^{2}\\
&=O(d)
\end{align*}
under the assumptions (A.3), (A.4) and (A.6) being true.\hfill $\blacksquare$

\begin{lemma}\label{lem:conceta}
Suppose (A.2)-(A.4), (A.6)-(A.7) are true. Then for $0<r<1$ we have:
$$\max_{i=1(1)n}\mathbf{E}_{\bm{\beta}}\Bigg[\operatorname*{\sup}_{\{\bm{\tau}:\|\bm{\tau}-\bm{\beta}\|_2\leq r\}}\big{\|}\eta_i(\bm{\tau},\bm{\beta})\big{\|}_2^2\Bigg]= O[(dnr)^2],$$
where we define:
\begin{align}\label{eqn:eta}
\eta_i(\bm{\tau},\bm{\beta})&=\Big[\Psi(y_i,\bm{\tau})-\mathbf{E}_{\bm{\beta}}\{\Psi(y_i,\bm{\tau})\}\Big]-\Big[\Psi(y_i,\bm{\beta})-\mathbf{E}_{\bm{\beta}}\{\Psi(y_i,\bm{\beta})\}\Big]
\end{align}
\end{lemma}
Proof of Lemma \ref{lem:conceta} For some $z_i$ such that $|z_i-\bm{x}_i^\top \bm{\beta}|\leq \bm{x}_i^\top (\bm{\tau}-\bm{\beta})$ for all $i\in \{1,...,n\}$, we use Taylor's expansion in (\ref{eqn:eta}) to have,
$$\eta_i(\bm{\tau},\bm{\beta})=-\big\{y_i-g^{-1}(\bm{x}_i^\top \bm{\beta})\big\}h^{\prime\prime}(z_i)\Big[\bm{x}_i^\top (\bm{\tau}-\bm{\beta})\Big]\bm{x}_i.$$Therefore observing (A.3), (A.4), (A.6) and (A.7) we have, 
\begin{align*}
\max_{i=1(1)n}\mathbf{E}_{\bm{\beta}}\Bigg[\operatorname*{\sup}_{\{\bm{\tau}:\|\bm{\tau}-\bm{\beta}\|_2\leq r\}}\big{\|}\eta_i(\bm{\tau},\bm{\beta})\big{\|}_2^2\Bigg]&\leq \Big[\max_{i=1(1)n}\mathbf{E}_{\bm{\beta}}\big{|}y_i-g^{-1}(\bm{x}_i^\top \bm{\beta})\big{|}^2\Big]\Big\{\max_{i=1(1)n}|h^{\prime\prime}(z_i)|^2\Big\}\\
&\times \Big[d^{1/2}\max_{i=1(1)n}\|\bm{x}_i\|_{\infty}\Big]^4\Big\{\operatorname*{\sup}_{\{\bm{\tau}:\|\bm{\tau}-\bm{\beta}\|_2\leq r\}}\big{\|}\bm{\tau}-\bm{\beta}\big{\|}_2^2\Big\}\\
&=O[(dnr)^2]
\end{align*}
Hence we are done.\hfill $\blacksquare$

\begin{lemma}\label{lem:remainderbound}
Suppose (A.1)-(A.7) are true. Then for any $0<B<\infty$, we have the following term is $O(dn^{2\alpha_1})$: 
$$\operatorname*{\sup}_{||\bm{\alpha}||=1}\operatorname*{\sup}_{\{\bm{\tau}:\|\mathbf{E(L_n)}(\bm{\tau}-\bm{\beta})\|_2\leq B(d/n)^{1/2}\}}\Bigg{|}\bm{\alpha}^\top \sum_{i=1}^{n}\mathbf{E}_{\bm{\beta}}\Big[\Psi(y_i,\bm{\tau})-\Psi(y_i,\bm{\beta})\Big]-n\bm{\alpha}^\top \big[\mathbf{E(L_n)}\big]\big(\bm{\tau}-\bm{\beta}\big)\Bigg{|}$$
\end{lemma}
Proof of Lemma \ref{lem:remainderbound} Recall that, $\Psi(y_i,\bm{\beta})=-\Big\{y_ih^\prime (\bm{x}_{i}^\top\bm{\beta})-h_1^\prime (\bm{x}_{i}^\top\bm{\beta})\Big\}\bm{x}_i$. Also we have,
$$\mathbf{E}_{\bm{\beta}}\Big[\Psi(y_i,\bm{\tau})\Big]=-\Big\{g^{-1}(\bm{x}_i^\top \bm{\beta})h^\prime (\bm{x}_{i}^\top \bm{\tau})-h_1^\prime (\bm{x}_{i}^\top \bm{\tau})\Big\}\bm{x}_i.$$
Note that, $h_1^\prime=(g^{-1})h^\prime$, $h_1^{\prime\prime}=(g^{-1})h^{\prime\prime}+(g^{-1})^\prime h^\prime$ and $h_1^{\prime\prime\prime}=(g^{-1})^{\prime\prime} h^\prime+2(g^{-1})^\prime h^{\prime\prime}+(g^{-1})h^{\prime\prime\prime}$.
Now for some $z_i$ with $|z_i-\bm{x}_i^\top\bm{\beta}|\leq \bm{x}_i^\top (\bm{\tau}-\bm{\beta})$, using the appropriate Taylor's expansion, we write the following:
\begin{align}\label{eqn:taylorexpanderror}
&\sum_{i=1}^{n}\mathbf{E}_{\bm{\beta}}\Big[\Psi(y_i,\bm{\tau})-\Psi(y_i,\bm{\beta})\Big]\nonumber \\
 &=-\sum_{i=1}^{n}\Bigg\{g^{-1}(\bm{x}_i^\top \bm{\beta})\Big[h^\prime (\bm{x}_{i}^\top\bm{\tau})-h^\prime (\bm{x}_{i}^\top \bm{\beta})\Big]-\Big[h_1^\prime (\bm{x}_{i}^\top\bm{\tau})-h_1^\prime (\bm{x}_{i}^\top \bm{\beta})\Big]\Bigg\}\bm{x}_i\nonumber\\
 &=\sum_{i=1}^{n}\bm{x}_i\bm{x}_i^\top\Big[(g^{-1})^\prime(\bm{x}_i^\top \bm{\beta})h^\prime (\bm{x}_{i}^\top\bm{\beta})\Big]\big(\bm{\tau}-\bm{\beta}\big)+2^{-1}\sum_{i=1}^{n}\bm{x}_i
\Big[-h^{\prime\prime\prime}(z_i)\big\{g^{-1}(\bm{x}_i^\top \bm{\beta})-g^{-1}(z_i)\big\}\nonumber\\
&\;\;\;\;\;\;\;\;\;\;\;\;\;\;\;\;\;\;\;\;\;\;+(g^{-1})^{\prime\prime}(z_i)h^\prime (z_i)+2(g^{-1})^{\prime}(z_i)h^{\prime\prime} (z_i)\Big]\Big\{(\bm{\tau}-\bm{\beta})^\top \bm{x}_i\Big\}^2\nonumber\\
&=n[\mathbf{E(L_n)}]\big(\bm{\tau}-\bm{\beta}\big)+2^{-1}\sum_{i=1}^{n}\bm{x}_i
\Big[-h^{\prime\prime\prime}(z_i)\big\{g^{-1}(\bm{x}_i^\top \bm{\beta})-g^{-1}(z_i)\big\}\nonumber\\
&\;\;\;\;\;\;\;\;\;\;\;\;\;\;\;\;\;\;\;\;\;\;+(g^{-1})^{\prime\prime}(z_i)h^\prime (z_i)+2(g^{-1})^{\prime}(z_i)h^{\prime\prime} (z_i)\Big]\Big\{(\bm{\tau}-\bm{\beta})^\top \bm{x}_i\Big\}^2
\end{align}
Then following (\ref{eqn:taylorexpanderror}) and all the mentioned assumptions, we will see that,

\begin{align}\label{eqn:finalerrorbound}
&\operatorname*{\sup}_{||\bm{\alpha}||=1}\operatorname*{\sup}_{\{\bm{\tau}:\|\mathbf{E(L_n)}(\bm{\tau}-\bm{\beta})\|_2\leq B(d/n)^{1/2}\}}\Bigg{|}\bm{\alpha}^\top \sum_{i=1}^{n}\mathbf{E}_{\bm{\beta}}\Big[\Psi(y_i,\bm{\tau})-\Psi(y_i,\bm{\beta})\Big]-n\bm{\alpha}^\top \big[\mathbf{E(L_n)}\big]\big(\bm{\tau}-\bm{\beta}\big)\Bigg{|}\nonumber\\
&\leq \operatorname*{\sup}_{||\bm{\alpha}||=1}\operatorname*{\sup}_{\{\bm{\tau}:\|\mathbf{E(L_n)}(\bm{\tau}-\bm{\beta})\|_2\leq B(d/n)^{1/2}\}}\Bigg{|}2^{-1}\sum_{i=1}^{n}[\bm{\alpha}^\top \bm{x}_i]
\Big[-h^{\prime\prime\prime}(z_i)\big\{g^{-1}(\bm{x}_i^\top \bm{\beta})-g^{-1}(z_i)\big\}\nonumber\\
&\;\;\;\;\;\;\;\;\;\;\;\;\;\;\;\;+(g^{-1})^{\prime\prime}(z_i)h^\prime (z_i)+2(g^{-1})^{\prime}(z_i)h^{\prime\prime} (z_i)\Big]\Big\{(\bm{\tau}-\bm{\beta})^\top\bm{x}_i\Big\}\Big\{\bm{x}_i^\top (\bm{\tau}-\bm{\beta})\Big\}\Bigg{|}\nonumber\\
&\leq \Big\{\operatorname*{\sup}_{\{\bm{\tau}:\|\mathbf{E(L_n)}(\bm{\tau}-\bm{\beta})\|_2\leq B(d/n)^{1/2}\}} \big{\|}\bm{\tau}-\bm{\beta}\big{\|}^2\Big\}\operatorname*{\sup}_{||\bm{\alpha}||,||\bm{\gamma}||=1}\Bigg[\Big\{\sum_{i=1}^{n}|\bm{\alpha}^\top \bm{x}_i|^4|\bm{\gamma}^\top \bm{x}_i|^4\Big\}^{1/4}\Big\{\sum_{i=1}^{n}|\bm{\gamma}^\top \bm{x}_i|^4\Big\}^{1/4}\nonumber\\
&\times 2^{-1}\Bigg(\sum_{i=1}^{n}|h^{\prime\prime\prime}(z_i)|^{4}\Bigg)^{1/4}\Bigg(\sum_{i=1}^{n}|(g^{-1})(\bm{x}_i^\top\bm{\beta})|^4 \Bigg)^{1/4}+\Bigg\{2^{-1}\Bigg(\sum_{i=1}^{n}|h^{\prime\prime\prime}(z_i)|^{4}\Bigg)^{1/4}\Bigg(\sum_{i=1}^{n}|(g^{-1})(z_i)|^{4}\Bigg)^{1/4}\nonumber\\
&+\Bigg(\sum_{i=1}^{n}|h^{\prime\prime}(z_i)|^{4}\Bigg)^{1/4}\Bigg(\sum_{i=1}^{n}|(g^{-1})^\prime(z_i)|^{4}\Bigg)^{1/4}+2^{-1}\Bigg(\sum_{i=1}^{n}|h^{\prime}(z_i)|^{4}\Bigg)^{1/4}\Bigg(\sum_{i=1}^{n}|(g^{-1})^{\prime\prime}(z_i)|^{4}\Bigg)^{1/4}\Bigg\}\nonumber\\
&\;\;\;\;\;\;\;\;\;\;\;\;\;\;\;\;\;\;\;\;\;\;\;\;\;\;\;\;\;\;\;\;\;\;\;\;\; \times \Big\{\sum_{i=1}^{n}|\bm{\alpha}^\top \bm{x}_i|^4|\bm{\gamma}^\top \bm{x}_i|^4\Big\}^{1/4}\Big\{\sum_{i=1}^{n}|\bm{\gamma}^\top \bm{x}_i|^4\Big\}^{1/4}\Bigg] \nonumber\\
&=O(dn^{2\alpha_1})
\end{align}
Therefore, the proof is complete. \hfill $\blacksquare$

\begin{lemma}\label{lem:concetad}
 Suppose (A.1)-(A.7) hold true. Then for any $0<B<\infty$, 
$$\operatorname*{\sup}_{||\bm{\alpha}||=1}\operatorname*{\sup}_{\{\bm{\tau}:\|\mathbf{E(L_n)}(\bm{\tau}-\bm{\beta})\|_2\leq B(d/n)^{1/2}\}}\Big[\sum_{i=1}^{n}\big\{\bm{\alpha}^\top \eta_i(\bm{\tau},\bm{\beta})\big\}^2\Big]=O_p(dn^{2\alpha_1})$$ and $$\operatorname*{\sup}_{||\bm{\alpha}||=1}\operatorname*{\sup}_{\{\bm{\tau}:\|\mathbf{E(L_n)}(\bm{\tau}-\bm{\beta})\|_2\leq B(d/n)^{1/2}\}}\Big[\sum_{i=1}^{n}\mathbf{E}_{\bm{\beta}}\big{|}\bm{\alpha}^\top \eta_i(\bm{\tau},\bm{\beta})\big{|}^2\Big]=O(dn^{2\alpha_1})$$
\end{lemma}
Proof of Lemma \ref{lem:concetad} We will skip this proof as it's quite straight forward. \hfill $\blacksquare$

\begin{lemma}\label{lem:heshao}
 Suppose (A.1)-(A.7) hold true. Then provided $d=o(n)$ and for any $B>0$,
$$\operatorname*{\sup}_{||\bm{\alpha}||=1}\operatorname*{\sup}_{\{\bm{\tau}:\|\mathbf{E(L_n)}(\bm{\tau}-\bm{\beta})\|_2\leq B(d/n)^{1/2}\}}\dfrac{\Big{|}\sum_{i=1}^{n}\bm{\alpha}^\top \eta_i(\bm{\tau},\bm{\beta})\Big{|}}{\Bigg[\frac{1}{n^2}+\Big[\sum_{i=1}^{n}\mathbf{E}_{\bm{\beta}}\big{|}\bm{\alpha}^\top \eta_i(\bm{\tau},\bm{\beta})\big{|}^2\Big]^{1/2}+\Big[\sum_{i=1}^{n}\big{|}\bm{\alpha}^\top \eta_i(\bm{\tau},\bm{\beta})\big{|}^2\Big]^{1/2}\Bigg]},$$
is $O_p(d_n^{1/2})$ with $d_n=d\log n.$
\end{lemma}
Proof of Lemma \ref{lem:heshao} This proof follows from Lemma 3.2 and 3.3 of \citet{he2000parameters}.\hfill $\blacksquare$

\begin{lemma}\label{lem:concentration}
 Suppose the assumptions of Lemma \ref{lem:glmconc}-\ref{lem:heshao} hold true. Then provided $d=o\Big(\dfrac{n^{1-2\alpha_1}}{\log n}\Big)$, for every $\epsilon>0$, there exists a $0<B_\epsilon<\infty$ such that, for sufficiently large enough $n$ we have;
\begin{align}\label{eqn:111}
 \mathbf{P}\Big[\big{\|}\hat{\bm{\beta}}_n-\bm{\beta}\big{\|}_2\leq B_\epsilon \Big(\dfrac{d}{n^{1-2\alpha_1}}\Big)^{1/2}\Big]>1-\epsilon.   
\end{align}
\end{lemma}

Proof of Lemma \ref{lem:concentration}
Denote, $\rho_{i}(y_i,\bm{\beta})=\Big\{-y_{i}h(\bm{x}_{i}^\top\bm{\beta}) +h_1(\bm{x}_{i}^\top\bm{\beta}) \Big\}$ and it is obvious that $\hat{\bm{\beta}}_n=\mbox{argmin}_{\bm{\beta}}\sum_{i=1}^{n}\rho_i(y_i,\bm{\beta})$. Define the sub-level set $\bm{\chi}_n:=\Big\{\bm{v}\in\mathbf{R}^d:\sum_{i=1}^{n}\rho_i(y_i,\bm{v})\leq \sum_{i=1}^{n}\rho_i(y_i,\bm{\beta})\Big\}$. We also define;
$$A_n^\epsilon:=\Bigg\{\inf_{\|\bm{u}\|=1}\sum_{i=1}^{n}\rho_{i}\Bigg(y_i,\bm{\beta}+B_\epsilon\Big(\dfrac{d}{n^{1-2\alpha_1}}\Big)^{1/2}\bm{u}\Bigg)>\sum_{i=1}^{n}\rho_{i}(y_i,\bm{\beta})\Bigg\}.$$
Therefore on the set $A_n^\epsilon$, for all $\bm{u}$ with $\|\bm{u}\|=1$, we must have $\bm{\beta}+B_\epsilon\Big(\dfrac{d}{n^{1-2\alpha_1}}\Big)^{1/2}\bm{u}\notin \chi_n.$ 
Define the closed ball centered at $\bm{\beta}$ as:
$$B_n^\epsilon (\bm{\beta}):=\Big\{\bm{v}:\|\bm{v}-\bm{\beta}\|_2\leq B_\epsilon\Big(\dfrac{d}{n^{1-2\alpha_1}}\Big)^{1/2}\Big\},$$
and the boundary of $B_n^\epsilon (\bm{\beta})$ is given by; $\partial B_n^\epsilon (\bm{\beta}):=\Big\{\bm{v}:\|\bm{v}-\bm{\beta}\|_2= B_\epsilon\Big(\dfrac{d}{n^{1-2\alpha_1}}\Big)^{1/2}\Big\}$. Hence it's obvious that, $\partial B_n^\epsilon (\bm{\beta})\subset \chi_n^c.$ Now we will prove that, $\chi_n\subset B_n^\epsilon (\bm{\beta})$.\\
Towards contradiction assume that, there exists some $\tilde{\bm{\beta}}\in \chi_n$ such that $\tilde{\bm{\beta}}\notin B_n^\epsilon (\bm{\beta}).$ That means $\|\tilde{\bm{\beta}}-\bm{\beta}\|_2>B_\epsilon\Big(\dfrac{d}{n^{1-2\alpha_1}}\Big)^{1/2}$. Now since $\chi_n$ is convex and $\bm{\beta},\tilde{\bm{\beta}}\in \chi_n$ that implies the entire line segment,
$$\big\{\bm{\beta}+t(\tilde{\bm{\beta}}-\bm{\beta}):0\leq t\leq 1\big\}\subset \chi_n.$$
Now define a point as: 
$\bm{\beta}^*:=\bm{\beta}+B_\epsilon\Big(\dfrac{d}{n^{1-2\alpha_1}}\Big)^{1/2}\dfrac{(\tilde{\bm{\beta}}-\bm{\beta})}{\|\tilde{\bm{\beta}}-\bm{\beta}\|_2},$
implying that $\bm{\beta}^*\in \partial B_n^\epsilon (\bm{\beta})$. But then we have that $\bm{\beta}^*\in\chi_n$ implying $\partial B_n^\epsilon (\bm{\beta})\subset\chi_n$, which is a contradiction. Therefore we must have, $\chi_n\subset B_n^\epsilon (\bm{\beta}).$ Now since $\hat{\bm{\beta}}_n=\mbox{argmin}_{\bm{\beta}}\sum_{i=1}^{n}\rho_i(y_i,\bm{\beta})\in \chi_n\subset B_n^\epsilon (\bm{\beta})$, to prove (\ref{eqn:111}), it's enough to show that for large enough $n$, we have:
\begin{align}\label{eqn:112}
&\mathbf{P}\big(A_n^\epsilon\big)>1-\epsilon\nonumber\\
&\text{i.e}\;\;\mathbf{P}\Bigg[\inf_{\|\bm{u}\|=1}\Bigg\{-\sum_{i=1}^{n}\Big[y_ih\Big(\bm{x}_i^\top\Big\{\bm{\beta}+B_\epsilon\Big(\dfrac{d}{n^{1-2\alpha_1}}\Big)^{1/2}\bm{u}\Big\}\Big)-h_1\Big(\bm{x}_i^\top\Big\{\bm{\beta}+B_\epsilon\Big(\dfrac{d}{n^{1-2\alpha_1}}\Big)^{1/2}\bm{u}\Big\}\Big)\Big]\Bigg\}\nonumber\\
&\;\;\;\;\;\;\;\;\;\;\;\;\;\;\;\;\;\;\;\;\;\;\;\;\;\;\;\;\;\;\;\;>-\sum_{i=1}^{n}\Big[y_ih\Big(\bm{x}_i^\top\bm{\beta}\Big)-h_1\Big(\bm{x}_i^\top\bm{\beta}\Big)\Big]\Bigg]>1-\epsilon
\end{align}
Now a simple consequence through mean value theorem with respect to $\bm{u}$ and for $0<\tilde{B}_\epsilon<B_\epsilon$, (\ref{eqn:112}) is implied by,
\begin{align}\label{eqn:113}
&\mathbf{P}\Bigg[\inf_{\|\bm{u}\|=1}\Bigg\{\sum_{i=1}^{n}-\Big[y_ih^\prime\Big(\bm{x}_i^\top\Big\{\bm{\beta}+\tilde{B}_\epsilon\Big(\dfrac{d}{n^{1-2\alpha_1}}\Big)^{1/2}\bm{u}\Big\}\Big)-h_1^\prime\Big(\bm{x}_i^\top\Big\{\bm{\beta}+\tilde{B}_\epsilon\Big(\dfrac{d}{n^{1-2\alpha_1}}\Big)^{1/2}\bm{u}\Big\}\Big)\Big]\nonumber\\
&\;\;\;\;\;\;\;\;\;\;\;\;\;\;\;\;\;\;\;\;\;\;\;\;\;\;\;\;\;\;\;\;\;\;\;\;\;\;\;\;\times\tilde{B}_\epsilon\Big(\dfrac{d}{n^{1-2\alpha_1}}\Big)^{1/2}(\bm{x}_i^\top \bm{u})\Bigg\}>0\Bigg]>1-\epsilon    
\end{align}
Now dividing both sides by $\tilde{B}_\epsilon\Big(\dfrac{d}{n^{1-2\alpha_1}}\Big)^{1/2}$ and then having Taylor's expansion with respect to $\bm{x}_i^\top\bm{\beta}$, equation (\ref{eqn:113}) is further implied by,
\begin{align}\label{eqn:114}
&\mathbf{P}\Bigg[\inf_{\|\bm{u}\|=1}\Bigg\{\bm{u}^\top\Big(\sum_{i=1}^{n}\underbrace{-\Big[y_ih^\prime\Big(\bm{x}_i^\top\bm{\beta}\Big)-h_1^\prime\Big(\bm{x}_i^\top\bm{\beta}\Big)\Big]\bm{x}_i}_{\Psi(y_i,\bm{\beta})}\Big)\nonumber\\
&\;\;\;\;\;\;\;\;\;\;\;\;\;\;\;\;\;\;\;\;\;\;\;\;\;\;\;\;\;\;\;\;\;\;\;\;\;\;\;+\bm{u}^\top(n\mathbf{E}(L_n))\Big[\tilde{B}_\epsilon\Big(\dfrac{d}{n^{1-2\alpha_1}}\Big)^{1/2}\bm{u}\Big]+\tilde{R}_n(\bm{u})\Bigg\}>0\Bigg]>1-\epsilon    
\end{align}
where $\tilde{R}_n(\bm{u})$ is the remainder term in the approximation. Next we will handle this remainder term. Now denoting $\bm{\tau}=\bm{\beta}+\tilde{B}_\epsilon\Big(\dfrac{d}{n^{1-2\alpha_1}}\Big)^{1/2}\bm{u}$ and recalling the definition of $\eta_i(\bm{\tau},\bm{\beta})$ as in equation (\ref{eqn:eta}) we see that, 
\begin{align}\label{eqn:115}
\tilde{R}_n(\bm{u})&= \bm{u}^\top\Big(\sum_{i=1}^{n}-\Big[y_ih^\prime\Big(\bm{x}_i^\top\bm{\tau}\Big)-h_1^\prime\Big(\bm{x}_i^\top\bm{\tau}\Big)\Big]\bm{x}_i\Big)-\bm{u}^\top\Big(\sum_{i=1}^{n}-\Big[y_ih^\prime\Big(\bm{x}_i^\top\bm{\beta}\Big)-h_1^\prime\Big(\bm{x}_i^\top\bm{\beta}\Big)\Big]\bm{x}_i\Big)\nonumber\\
&\;\;\;\;\;\;\;\;\;\;\;\;\;\;\;\;\;\;\;\;\;\;\;\;\;\;\;\;\;\;\;\;\;\;\;\;\;\;\;\;\;\;\;\;\;\;-\bm{u}^\top(n\mathbf{E}(L_n))\Big[\tilde{B}_\epsilon\Big(\dfrac{d}{n^{1-2\alpha_1}}\Big)^{1/2}\bm{u}\Big]\nonumber\\
&=\Bigg\{\bm{u}^\top \Bigg(\sum_{i=1}^{n}\mathbf{E}_{\bm{\beta}}\Big[\Psi(y_i,\bm{\tau})-\Psi(y_i,\bm{\beta})\Big]\Bigg)-n\bm{u}^\top \big[\mathbf{E(L_n)}\big]\big(\bm{\tau}-\bm{\beta}\big)\Bigg\}+\bm{u}^\top \Big(\sum_{i=1}^{n}\eta_i(\bm{\tau},\bm{\beta})\Big)
\end{align}
Now due to Lemma \ref{lem:remainderbound}, \ref{lem:concetad} and \ref{lem:heshao} with the assumption $d=o\Big(\dfrac{n^{1-2\alpha_1}}{\log n}\Big)$, it's easy to verify that 
\begin{align}\label{eqn:116}
\mbox{inf}_{||\bm{u}||=1}|\tilde{R}_n(\bm{u})|=o_p[(nd)^{1/2}]   
\end{align}
Therefore from equation (\ref{eqn:114}), (\ref{eqn:116}) and Lemma \ref{lem:concleading}, we can write:
\begin{align}\label{eqn:117}
&\mathbf{P}\Bigg[\inf_{\|\bm{u}\|=1}\Bigg\{\bm{u}^\top\Big(\sum_{i=1}^{n}\Psi(y_i,\bm{\beta})\Big)+\tilde{B}_\epsilon n^{\alpha_1}(nd)^{1/2}\Big[\bm{u}^\top\mathbf{E}(L_n)\bm{u}\Big]+\tilde{R}_n(\bm{u})\Bigg\}>0\Bigg]\nonumber\\
&\geq \mathbf{P}\Bigg[(nd)^{-1/2}\inf_{\|\bm{u}\|=1}\Big\{\bm{u}^\top\Big(\sum_{i=1}^{n}\Psi(y_i,\bm{\beta})\Big)\Big\}+\tilde{B}_\epsilon n^{\alpha_1}\inf_{\|\bm{u}\|=1}\Big[\bm{u}^\top\mathbf{E}(L_n)\bm{u}\Big]+o(1)>0\Bigg]\nonumber\\
&\geq \mathbf{P}\Bigg[(nd)^{-1/2}\inf_{\|\bm{u}\|=1}\Big\{\bm{u}^\top\Big(\sum_{i=1}^{n}\Psi(y_i,\bm{\beta})\Big)\Big\}+\tilde{B}_\epsilon n^{\alpha_1}\dfrac{\lambda_{min}(\mathbf{E}(L_n))}{2}>0\Bigg]\nonumber\\
& \geq \mathbf{P}\Bigg[(nd)^{-1/2}\inf_{\|\bm{u}\|=1}\Big\{\bm{u}^\top\Big(\sum_{i=1}^{n}\Psi(y_i,\bm{\beta})\Big)\Big\}+c_2\tilde{B}_\epsilon n^{\alpha_1}\dfrac{n^{-\alpha_1}}{2}>0\Bigg]\nonumber\\
&\geq \mathbf{P}\Bigg[(nd)^{-1/2}\sup_{\|\bm{u}\|=1}\Big\{-\bm{u}^\top\Big(\sum_{i=1}^{n}\Psi(y_i,\bm{\beta})\Big)\Big\}\leq \tilde{K}_\epsilon\Bigg]\nonumber\\
&\geq \mathbf{P}\Bigg[\sup_{\|\bm{u}\|=1}\Big\{|\bm{u}^\top\bm{W}_n|\Big\}\leq d^{1/2}\tilde{K}_\epsilon\Bigg]\geq \mathbf{P}\big(\|\bm{W}_n\|_2\leq d^{1/2}\tilde{K}_\epsilon\big)>1-\epsilon
\end{align}
Therefore (\ref{eqn:111}) is true and the proof is complete. \hfill $\blacksquare$

\begin{lemma}\label{lem:errorbound}
Suppose the assumptions of Lemma \ref{lem:remainderbound} and \ref{lem:heshao} hold true. Let $\bm{r}_n$ be the remainder term in Bahadur's representation of $n^{1/2}(\hat{\bm{\beta}}_n-\bm{\beta})$. Then we have:
$$\|\bm{r}_n\|=o_p(1),$$
provided (i) $d=o(n^{\frac{2}{5}[1-6\alpha_1-\alpha_2]})$ for class $\mathcal{C}$ and (ii) $d=o(n^{\frac{1}{2}[1-6\alpha_1-\alpha_2]})$ for class $\mathcal{B}$ with $0\le 6\alpha_1+\alpha_2<1$.  
\end{lemma}
Proof of Lemma \ref{lem:errorbound} Recall the definition of $\rho(y_i,\bm{\beta})$ and $\hat{\bm{\beta}}_n$ in Lemma \ref{lem:concentration}. Then defining the score function by $\Psi(y_i,\bm{\beta})=\Big\{-y_ih^\prime (\bm{x}_{i}^\top\bm{\beta})+h_1^\prime (\bm{x}_{i}^\top\bm{\beta})\Big\}\bm{x}_i$ with $h_1(\cdot)=b(h(\cdot))$ and from KKT conditions, it's easy to see that $\hat{\bm{\beta}}_n$ is a solution of $\sum_{i=1}^{n}\Psi(y_i,\bm{\beta})=\bm{0}$. Now suppose $\bm{u}=\hat{\bm{\beta}}_n-\bm{\beta}$. Hence using Taylor's expansion with respect to $\bm{x}_i^\top \bm{\beta}$, we get the following Bahadur's representation of $n^{1/2}(\hat{\bm{\beta}}_n-\bm{\beta})$ as:
\begin{align*}
& n^{1/2}[\hat{\bm{\beta}}_n-\bm{\beta}]= [\mathbf{E(L_n)}]^{-1}\bm{W}_n +\bm{r}_n=\bm{T}_n+\bm{r}_n.
\end{align*}
where, we write the remainder term as, 
\begin{align*}
\bm{r}_n&= \Big[\mathbf{L}_n^{-1}-\{\mathbf{E(L_n)}\}^{-1}\Big]\bm{W}_n-(2n^{1/2})^{-1}\mathbf{L}_n^{-1}\sum_{i=1}^{n}\bm{x}_i\Big[(g^{-1})^{\prime\prime}(z_i) h^\prime (z_i)\nonumber\\
&\;\;\;\;\;\;\;\;\;\;\;+2(g^{-1})^\prime (z_i)h^{\prime\prime} (z_i)-\{y_i-(g^{-1})(z_i)\}h^{\prime\prime\prime} (z_i)\Big]\Big[\bm{x}_i^\top (\hat{\bm{\beta}}_n-\bm{\beta})\Big]^2,
\end{align*}
for some $z_i$ such that $|z_i-\bm{x}_i^\top \bm{\beta}|\leq \bm{u}^\top \bm{x}_i$ for all $i\in \{1,...,n\}$.
This will again give us that,
\begin{align}\label{eqn:118}
 n^{1/2}[\mathbf{E}(L_n)]\bm{r}_n&=n[\mathbf{E}(L_n)](\hat{\bm{\beta}}_n-\bm{\beta})-\sum_{i=1}^{n}\{y_i-g^{-1}(\bm{x}_i^\top\bm{\beta})\}h^\prime (\bm{x}_i^\top\bm{\beta})\bm{x}_i\nonumber\\
&=\sum_{i=1}^{n}\Psi(y_i,\hat{\bm{\beta}}_n)-\Bigg\{\sum_{i=1}^{n}\mathbf{E}_{\bm{\beta}}\Big[\Psi(y_i,\hat{\bm{\beta}}_n)-\Psi(y_i,\bm{\beta})\Big]-n[\mathbf{E}(L_n)](\hat{\bm{\beta}}_n-\bm{\beta})\Bigg\}\nonumber\\
 &-\Bigg\{\sum_{i=1}^{n}\Big[\Psi(y_i,\hat{\bm{\beta}}_n)-\Psi(y_i,\bm{\beta})\Big]-\sum_{i=1}^{n}\mathbf{E}_{\bm{\beta}}\Big[\Psi(y_i,\hat{\bm{\beta}}_n)-\Psi(y_i,\bm{\beta})\Big]\Bigg\}\nonumber\\
 &= \bm{\kappa}_n\;\text{(say)}
\end{align}
 %Choose any $\epsilon>0$. Also let $0<\tilde{k}_1,\tilde{c}_2,B_\epsilon<\infty$ be the constants arising from the bounds proved earlier. Suppose we write, $\tilde{C}_\epsilon=\dfrac{\tilde{k}_1\epsilon}{2\tilde{c}_2}$.
(i) Now consider the class $\mathcal{C}$. 
Then Lemma \ref{lem:mainlemmagaussisoperimetric}, \ref{lem:remainderbound}, \ref{lem:concetad},\ref{lem:heshao} and \ref{lem:concentration} alongwith equation (\ref{eqn:118}) and the assumption $d=o(n^{\frac{2}{5}[1-6\alpha_1-\alpha_2]})$ will imply that,
\begin{align}\label{eqn:119}
&\|\bm{r}_n\|=O_p\Bigg[\frac{d}{n^{\frac{1}{2}-3\alpha_1}}+\frac{(d\log n)^{1/2}}{n^{\frac{5}{2}-\alpha_1}}+\frac{d(\log n)^{1/2}}{n^{\frac{1}{2}-2\alpha_1}}\Bigg]=o_p(1).
\end{align}

(ii) Next we consider the class $\mathcal{B}$. Then Lemma \ref{lem:zhilovamainlemmagaussisoperimetric}, \ref{lem:remainderbound}, \ref{lem:concetad},\ref{lem:heshao} and \ref{lem:concentration} alongwith equation (\ref{eqn:118}) and the assumption $d=o(n^{\frac{1}{2}[1-6\alpha_1-\alpha_2]})$ will imply that,
\begin{align}\label{eqn:11998}
&\|\bm{r}_n\|=o_p(1).
\end{align}

The proof is therefore complete. \hfill $\blacksquare$

\subsection{Lemmas Corresponding to Theorem \ref{thm:pbconvexoriginal1} and  \ref{thm:prbconvexoriginal1}}\label{sec:PBGLMdistn}
We consider everything on the set $\Big\{\Big\|\hat{\bm{\beta}}_n-\bm{\beta}\Big\|_2\le B_\epsilon\Big(\dfrac{d}{n^{1-2\alpha_1}}\Big)^{1/2}\Big\}$ which has probability at least $1-\epsilon$.
\begin{lemma}\label{lem:closeLnhatELn}
Suppose (A.3)-(A.7) hold true. Then provided $d=o(n^{1-2\alpha_1})$, we have:
$$\|\hat{\bm{L}}_n-\bm{E}(L_n)\|=O_p\Bigg[\frac{d^{1/2}}{n^{\frac12-\alpha_1}}\Bigg]$$
\end{lemma}
Proof of Lemma \ref{lem:closeLnhatELn} 
Denote that,
\begin{align}\label{eqn:200}
 \bm{Z}_n&=\frac{1}{n}\sum_{i=1}^{n}\bm{x}_i\bm{x}_i^\top\Bigg[\Big\{\bm{x}_i^\top\Big(\hat{\bm{\beta}}_n-\bm{\beta}\Big)\Big\}\Bigg\{(g^{-1})^{\prime\prime}(\tilde{z}_i)h^\prime(\bm{x}_i^\top\bm{\beta})+(g^{-1})^{\prime}(\bm{x}_i^\top\bm{\beta})h^{\prime\prime}(\tilde{z}_i)\Bigg\}\Bigg] \nonumber\\
 &\qquad\qquad+\frac{1}{n}\sum_{i=1}^{n}\bm{x}_i\bm{x}_i^\top\Bigg[\Big\{\bm{x}_i^\top\Big(\hat{\bm{\beta}}_n-\bm{\beta}\Big)\Big\}^2(g^{-1})^{\prime\prime}(\tilde{z}_i)h^{\prime\prime}(\tilde{z}_i)\Bigg]\end{align}
for some $\tilde{z}_i$ such that for all $i$, $|\tilde{z}_i-\bm{x}_i^\top\bm{\beta}|\leq \bm{x}_i^\top(\hat{\bm{\beta}}_n-\bm{\beta})$ and it's just a sequence of algebra through Taylor's approximation to have:
$\hat{\bm{L}}_n=\bm{E}(L_n)+\bm{Z}_n,$. 
% preamble: \usepackage{amsmath,amssymb}
\begin{align}\label{eqn:201}
\|\hat L_n - \mathbf{E}(L_n)\|
&= \sup_{\|\alpha\|=1} \big|\alpha^\top Z_n \alpha\big| \nonumber\\
&\le \sup_{\|\alpha\|=\|\gamma\|=1}
\Big\{ \|\widehat\beta_n-\beta\|_2\,A_{n}(\alpha,\gamma)
      + \|\widehat\beta_n-\beta\|_2^{2}\,B_{n}(\alpha,\gamma)\Big\},
\end{align}

where we separately define the factors,
\begin{align}
&A_{n}(\alpha,\gamma)\notag\\
&= 
\Big(\frac{1}{n}\sum_{i=1}^{n}|\alpha^\top x_i|^4\Big)^{1/2}
\Big(\max_{1\le i\le n}|h' (x_i^\top\beta)|\Big)
\Big(\frac{1}{n}\sum_{i=1}^n |(g^{-1})''(\tilde z_i)|^4\Big)^{1/4}
\Big(\frac{1}{n}\sum_{i=1}^n|\gamma^\top x_i|^4\Big)^{1/4}
\notag\\
&+ \Big(\frac{1}{n}\sum_{i=1}^{n}|\alpha^\top x_i|^4\Big)^{1/2}
\Big(\frac{1}{n}\sum_{i=1}^{n}|\gamma^\top x_i|^4\Big)^{1/4}
\Big(\frac{1}{n}\sum_{i=1}^{n}|h''(\tilde z_i)|^4\Big)^{1/4}
\Big(\max_{1\le i\le n}|(g^{-1})'(x_i^\top\beta)|\Big)
\end{align}
and
\begin{align}
B_{n}(\alpha,\gamma) &= 
\Big(\frac{1}{n}\sum_{i=1}^n |\alpha^\top x_i|^4|\gamma^\top x_i|^4\Big)^{1/2}
\Big\{ \Big(\frac{1}{n}\sum_{i=1}^n |(g^{-1})''(\tilde z_i)|^4\Big)^{1/4}
\Big(\frac{1}{n}\sum_{i=1}^n |h''(\tilde z_i)|^4\Big)^{1/4}.
\end{align}

Now under all the assumptions and due to Lemma \ref{lem:concentration}, from equation (\ref{eqn:201}) we see that, $$\|\hat{\bm{L}}_n-\bm{E}(L_n)\|=O_p\Big[\Big(\dfrac{d}{n^{1-2\alpha_1}}\Big)^{1/2}+\dfrac{d}{n^{1-2\alpha_1}}\Big]=O_p\Big[\Big(\dfrac{d}{n^{1-2\alpha_1}}\Big)^{1/2}\Big],$$
under the assumption $d=o(n^{1-2\alpha_1})$ and we are done. \hfill $\blacksquare$

\begin{lemma}\label{lem:orderinvLnhat}
Suppose (A.1) holds. Then provided $d=o(n^{1-4\alpha_1})$ with $\alpha_1<1/4$ we have;
$$\|\hat{\bm{L}}_n^{-1}\|=O_p(n^{\alpha_1})$$
\end{lemma}
 Proof of Lemma \ref{lem:orderinvLnhat} Recall that due to Lemma \ref{lem:closeLnhatELn},
 \begin{align}\label{eqn:202}
\|\hat{\bm{L}}_n-\mathbf{E}(\bm{L}_n)\|=O_p\Big[\Big(\dfrac{d}{n^{1-2\alpha_1}}\Big)^{1/2}\Big]     
 \end{align}
Then under the assumption $d=o(n^{1-4\alpha_1})$ with $\alpha_1<1/4$ we have;
\begin{align}\label{eqn:203}
\|\hat{\bm{L}}_n-\mathbf{E}(\bm{L}_n)\|<\dfrac{k_2}{2}n^{-\alpha_1}    
\end{align}
Now due to (\ref{eqn:203}), we finally we write;
\begin{align}\label{eqn:204}
\|\hat{\bm{L}}_n^{-1}\|&\leq \dfrac{1}{\lambda_{\text{min}}(\mathbf{E}(\bm{L}_n))+\lambda_{\text{min}}(\hat{\bm{L}}_n-\mathbf{E}(\bm{L}_n))} \nonumber\\
&\leq \dfrac{1}{\lambda_{\text{min}}(\mathbf{E}(\bm{L}_n))-\|\hat{\bm{L}}_n-\mathbf{E}(\bm{L}_n)\|}\nonumber\\
&=O_p(n^{\alpha_1})
\end{align}
Therefore we are done. \hfill $\blacksquare$

\begin{lemma}\label{lem:frobHSnorm}
 For finite dimension and a $d\times d$ matrix $A=((A^{(i,j)}))$, we define its Frobenius norm as $\|A\|_F=\Big[\sum_{i=1}^{d}\sum_{j=1}^{d}|A^{(i,j)}|^2\Big]^{1/2}$. Then for a sequence of $d\times d$ matrices $\{A_k\}_{k\in\{1,..,n\}}$ we have,
 $$\Big{\|}\dfrac{1}{n}\sum_{k=1}^{n}A_k\Big{\|}_F\leq \Big[\dfrac{2}{n^2}\sum_{k=1}^{n}\Big{\|}A_k\Big{\|}_F^2\Big]^{1/2}$$
\end{lemma}
Proof of Lemma \ref{lem:frobHSnorm} It's enough to prove, $\Big{\|}\sum_{k=1}^{n}A_k\Big{\|}_F^2\leq 2\sum_{k=1}^{n}\Big{\|}A_k\Big{\|}_F^2.$
For that we note,
\begin{align}\label{eqn:205}
&\sum_{i=1}^{d}\sum_{j=1}^{d}\Big{|}\sum_{k=1}^{n}A_k^{(i,j)}\Big{|}^2 \leq \sum_{k=1}^{n}\Big[\sum_{i=1}^{d}\sum_{j=1}^{d}\Big{|}A_k^{(i,j)}\Big{|}^2\Big]+\sum_{k\neq l}\Big[\sum_{i=1}^{d}\sum_{j=1}^{d}\Big{|}A_k^{(i,j)}\Big{|}\Big{|}A_l^{(i,j)}\Big{|}\Big]\nonumber\\
&\implies \Big{\|}\sum_{k=1}^{n}A_k\Big{\|}_F^2\leq \sum_{k=1}^{n}\Big{\|}A_k\Big{\|}_F^2+\sum_{k\neq l}\Big{\|}A_k\Big{\|}_F\Big{\|}A_l\Big{\|}_F\nonumber\\
&\implies \Big{\|}\sum_{k=1}^{n}A_k\Big{\|}_F^2\leq \sum_{k=1}^{n}\Big{\|}A_k\Big{\|}_F^2+\frac{1}{2}\Big[\sum_{k\neq l}\Big{\|}A_k\Big{\|}_F^2+\sum_{l\neq k}\Big{\|}A_l\Big{\|}_F^2\Big]\leq 2\sum_{k=1}^{n}\Big{\|}A_k\Big{\|}_F^2
\end{align}
Therefore we are done.\hfill $\blacksquare$

\begin{lemma}\label{lem:smallHSnorm}
Suppose (A.1), (A.3), (A.4), (A.6) and (A.7) hold true
Let $\bm{I}_d$ denote $d$ dimensional identity matrix. Then provided $d=o(n^{\frac{1}{2}-\alpha_1})$ with $\alpha_1<1/2$ we have,
$$\Big{\|}\bm{I}_d-Var_{*}(\tilde{\bm{S}}_n^{-1/2}\hat{\bm{T}}_n^{*(PB)})\Big{\|}_{H.S}=O_p\Bigg[\dfrac{d^2}{n^{1-2\alpha_1}}\Bigg]^{1/2}.$$
\end{lemma}
Proof of Lemma \ref{lem:smallHSnorm} We start with the fact that $Var(\tilde{\bm{S}}_n^{-1/2}\bm{T}_n)=\bm{I}_d.$ Therefore we write,
\begin{align}\label{eqn:206}
& \Big{\|}\bm{I}_d-Var_{*}(\tilde{\bm{S}}_n^{-1/2}\hat{\bm{T}}_n^{*(PB)})\Big{\|}_{H.S}\nonumber\\
&=\Bigg{\|}\tilde{\bm{S}}_n^{-1/2}\Bigg[\big\{\mathbf{E}(\bm{L}_n)\big\}^{-1}\bm{S}_n\big\{\mathbf{E}(\bm{L}_n)\big\}^{-1}-\hat{\bm{L}}_n^{-1}\hat{\bm{S}}_n^*\hat{\bm{L}}_n^{-1}\Bigg]\tilde{\bm{S}}_n^{-1/2}\Bigg{\|}_{H.S}\nonumber\\
&\leq \underbrace{\Bigg{\|}\frac{1}{n}\sum_{i=1}^{n}\Big[\tilde{\bm{S}}_n^{-1/2}\big\{\mathbf{E}(\bm{L}_n)\big\}^{-1}\bm{x}_i\Big]\Big[\tilde{\bm{S}}_n^{-1/2}\big\{\mathbf{E}(\bm{L}_n)\big\}^{-1}\bm{x}_i\Big]^\top\Big[h^\prime\big(\bm{x}_i^\top\bm{\beta}\big)\Big]^2\mathbf{E}\Big[y_i-g^{-1}\big(\bm{x}_i^\top\bm{\beta}\big)\Big]^2\Bigg{\|}_{H.S}}_{=\text{B}_1\;\text{(say)}}\nonumber\\
&\;\;\;\;\;\;\;\;\;\;\;\;\;\;\;\;\;\;+\underbrace{\Bigg{\|}\frac{1}{n}\sum_{i=1}^{n}\Big[\tilde{\bm{S}}_n^{-1/2}\hat{\bm{L}}_n^{-1}\bm{x}_i\Big]\Big[\tilde{\bm{S}}_n^{-1/2}\hat{\bm{L}}_n^{-1}\bm{x}_i\Big]^\top\Big[h^\prime\big(\bm{x}_i^\top\hat{\bm{\beta}}_n\big)\Big]^2\Big[y_i-g^{-1}\big(\bm{x}_i^\top\hat{\bm{\beta}}_n\big)\Big]^2\Bigg{\|}_{H.S}}_{=\text{B}_2\;\text{(say)}}
\end{align}
Now for finite dimensional matrix $A$, we know that $\|A\|_{H.S}=\|A\|_{F}$. Also equation (\ref{eqn:normbound}) gives that,
\begin{align*}
\big{\|}\Tilde{\bm{S}}_n^{-1/2}[\mathbf{E(L_n)}]^{-1}\big{\|}&=O[n^{\alpha_1/2}].
\end{align*}
Similar argument and Lemma \ref{lem:orderinvLnhat} will provide us with:
$$\big{\|}\Tilde{\bm{S}}_n^{-1/2}\hat{\bm{L}}_n^{-1}\big{\|}=O_p[n^{\alpha_1/2}].$$
Next for some $\tilde{z}_i$ such that for all $i$, $|\tilde{z}_i-\bm{x}_i^\top\bm{\beta}|\leq \bm{x}_i^\top(\hat{\bm{\beta}}_n-\bm{\beta})$, due to Taylor's expansion, we will find that;
\begin{align}\label{eqn:207}
&\frac{1}{n}\sum_{i=1}^{n}\Big[h^\prime\big(\bm{x}_i^\top\hat{\bm{\beta}}_n\big)\Big]^4\Big[y_i-g^{-1}\big(\bm{x}_i^\top\hat{\bm{\beta}}_n\big)\Big]^4\nonumber\\
&\leq\dfrac{64}{n}\sum_{i=1}^{n}\Bigg[\Big{|}y_i-g^{-1}\big(\bm{x}_i^\top\bm{\beta}\big)\Big{|}^4+\Big{|}\bm{x}_i^\top\big(\hat{\bm{\beta}}_n-\bm{\beta}\big)\Big{|}^4\Big{|}(g^{-1})^\prime(\tilde{z}_i)\Big{|}^4\Bigg]\Bigg[\Big{|}h^{\prime}\big(\bm{x}_i^\top\bm{\beta}\big)\Big{|}^4+\Big{|}\bm{x}_i^\top\big(\hat{\bm{\beta}}_n-\bm{\beta}\big)\Big{|}^4\Big{|}h^{\prime\prime}(\tilde{z}_i)\Big{|}^4\Bigg]\nonumber\\
&\leq 64\Bigg[\Bigg(\Big\{n^{-1}\sum_{i=1}^{n}\Big{|}y_i-g^{-1}\big(\bm{x}_i^\top\bm{\beta}\big)\Big{|}^4\Big\}+\Big\{d^{1/2}\mbox{max}_{i\in\{1,..,n\}}\|\bm{x}_i\|_\infty\Big\}^4\Big{\|}\hat{\bm{\beta}}_n-\bm{\beta}\Big{\|}_2^4\Big\{n^{-1}\sum_{i=1}^{n}\Big{|}(g^{-1})^\prime\big(\tilde{z}_i\big)\Big{|}^4\Big\}\Bigg)\nonumber\\
&\times\Big\{\mbox{max}_{i\in\{1,..,n\}}\Big{|}h^{\prime}\big(\bm{x}_i^\top\bm{\beta}\big)\Big{|}\Big\}^{4}+\Bigg(\Big\{d^{1/2}\mbox{max}_{i\in\{1,..,n\}}\|\bm{x}_i\|_\infty\Big\}^4\Big{\|}\hat{\bm{\beta}}_n-\bm{\beta}\Big{\|}_2^4\Big\{n^{-1}\sum_{i=1}^{n}\Big{|}y_i-g^{-1}\big(\bm{x}_i^\top\bm{\beta}\big)\Big{|}^6\Big\}^{2/3}\nonumber\\
&+\Big\{d^{1/2}\mbox{max}_{i\in\{1,..,n\}}\|\bm{x}_i\|_\infty\Big\}^8\Big{\|}\hat{\bm{\beta}}_n-\bm{\beta}\Big{\|}_2^8\Big\{n^{-1}\sum_{i=1}^{n}\Big{|}(g^{-1})^\prime\big(\tilde{z}_i\big)\Big{|}^6\Big\}^{2/3}\Bigg)\Big\{n^{-1}\sum_{i=1}^{n}\Big{|}h^{\prime\prime}\big(\tilde{z}_i\big)\Big{|}^{12}\Big\}^{1/3}\Bigg]\nonumber\\
&=O_p\Bigg[1+\Big(\dfrac{d}{n^{\frac{1}{2}-\alpha_1}}\Big)^4+\Big(\dfrac{d}{n^{\frac{1}{2}-\alpha_1}}\Big)^8\Bigg]
\end{align}
Now due to Lemma \ref{lem:frobHSnorm} and under the assumptions we have,
\begin{align}\label{eqn:208}
B_1&\leq\Bigg[\frac{2}{n^2}\sum_{i=1}^{n}\Bigg{\|}\Big[\tilde{\bm{S}}_n^{-1/2}\big\{\mathbf{E}(\bm{L}_n)\big\}^{-1}\bm{x}_i\Big]\Big[\tilde{\bm{S}}_n^{-1/2}\big\{\mathbf{E}(\bm{L}_n)\big\}^{-1}\bm{x}_i\Big]^\top\Big[h^\prime\big(\bm{x}_i^\top\bm{\beta}\big)\Big]^2\mathbf{E}\Big[y_i-g^{-1}\big(\bm{x}_i^\top\bm{\beta}\big)\Big]^2\Bigg{\|}_{F}^2\Bigg]^{1/2}\nonumber\\
&\leq \Bigg[\dfrac{2}{n}\Big\{d^{1/2}\max_{i\in\{1,..,n\}}\|\bm{x}_i\|_{\infty}\Big\}^4\Big{\|}\tilde{\bm{S}}_n^{-1/2}\big\{\mathbf{E}(\bm{L}_n)\big\}^{-1}\Big{\|}^4\Big\{\max_{i\in\{1,..,n\}}\big{|}h^\prime (\bm{x}_i^\top \bm{\beta})\big{|}\Big\}^{4}\nonumber\\
 &\;\;\;\;\;\;\;\;\;\;\;\;\;\;\;\;\;\;\;\;\;\;\;\;\;\;\;\;\;\;\;\;\;\;\;\;\;\;\;\;\;\;\;\;\;\;\;\;\;\;\;\;\;\;\;\;\;\;\;\;\;\;\;\;\;\;\;\;\;\;\times\Big(n^{-1}\sum_{i=1}^{n}\mathbf{E}\Big{|}\big\{y_i-g^{-1}(\bm{x}_i^\top \bm{\beta})\big\}\Big{|}^4\Big)\Bigg]^{1/2}\nonumber\\
 &=O\Bigg[\dfrac{d^2}{n^{1-2\alpha_1}}\Bigg]^{1/2}.
\end{align}
Exact similar calculation as in case of $B_1$, equation (\ref{eqn:207}) and the assumption that $d=o(n^{\frac{1}{2}-\alpha_1})$ we will get that,
\begin{align}\label{eqn:209}
B_2=O_p\Bigg[\dfrac{d^2}{n^{1-2\alpha_1}}\Bigg]^{1/2}    
\end{align}
Combining (\ref{eqn:208}) and (\ref{eqn:209}) we are done. \hfill $\blacksquare$

\begin{lemma}\label{lem:PBLeadingtermconvex}
Consider the assumptions and set-up as in Lemma \ref{lem:smallHSnorm}. Suppose $\bm{G}_{1n}\sim N_d(\bm{0},\Tilde{\bm{S}}_n)$. Then provided $d=o[n^{2\{1-2\alpha_1\}/5}]$ for $\alpha_1<1/2$, we have;
$$\mbox{sup}_{A\in\mathcal{C}}\Big{|}\mathbf{P}_*\Big(\hat{\bm{T}}_n^{*(PB)}\in A\Big)-\mathbf{P}\Big(\bm{G}_{1n}\in A\Big)\Big{|}=o_p(1),$$
where $\mathcal{C}$ be the class of measurable convex sets.
\end{lemma}

Proof of Lemma \ref{lem:PBLeadingtermconvex} Due to Lemma \ref{lem:fangkoikeconvex} we can write, 
\begin{align}\label{eqn:210}
\mbox{sup}_{A\in\mathcal{C}}\Big{|}\mathbf{P}_*\Big(\hat{\bm{T}}_n^{*(PB)}\in A\Big)-\mathbf{P}\Big(\bm{G}_{1n}\in A\Big)\Big{|}\leq Cd^{1/4}\delta_{\mathcal{C}}^*\Big[\Big{|}\log \delta_{\mathcal{C}}^*\Big{|}\Big]^{1/2},    
\end{align}
where,
\begin{align*}
\delta_{\mathcal{C}}^*&=\mbox{min}\Bigg\{\underbrace{\Big{\|}\bm{I}_d-Var_{*}(\tilde{\bm{S}}_n^{-1/2}\hat{\bm{T}}_n^{*(PB)})\Big{\|}_{H.S}}_{\text{Term (I)}}\\
&\;\;\;\;\;\;\;\;\;\;+\Bigg[\underbrace{\frac{1}{n^2}\sum_{i=1}^{n}\mathbf{E}_*\Bigg{\|}\tilde{\bm{S}}_n^{-1/2}\hat{\bm{L}}_n^{-1}\bm{x}_i\Big\{y_i-g^{-1}(\bm{x}_i^\top\hat{\bm{\beta}}_n)\Big\}h^\prime\Big(\bm{x}_i^\top\hat{\bm{\beta}}_n\Big)\Bigg(\frac{G_i^*}{\mu_{G^*}}-1\Bigg)\Bigg{\|}^4}_{\text{Term (II)}}\Bigg]^{1/2},\dfrac{1}{e}\Bigg\}    
\end{align*}
Now due to Lemma \ref{lem:smallHSnorm},
\begin{align}\label{eqn:211}
\Big{\|}\bm{I}_d-Var_{*}(\tilde{\bm{S}}_n^{-1/2}\hat{\bm{T}}_n^{*(PB)})\Big{\|}_{H.S}= O_p\Bigg[\dfrac{d^2}{n^{1-2\alpha_1}}\Bigg]^{1/2}      
\end{align}
Now under all the assumptions, along with the fact that $\mathbf{E}_*(|G_1^*|^4)<\infty$, exact similar calculations as in Lemma \ref{lem:mainlemmafangandkoikeconvex} and \ref{lem:smallHSnorm} will lead us to the fact that, 
\begin{align}\label{eqn:212}
\frac{1}{n^2}\sum_{i=1}^{n}\mathbf{E}_*\Bigg{\|}\tilde{\bm{S}}_n^{-1/2}\hat{\bm{L}}_n^{-1}\bm{x}_i\Big\{y_i-g^{-1}(\bm{x}_i^\top\hat{\bm{\beta}}_n)\Big\}h^\prime\Big(\bm{x}_i^\top\hat{\bm{\beta}}_n\Big)\Bigg(\frac{G_i^*}{\mu_{G^*}}-1\Bigg)\Bigg{\|}^4=O_p\Bigg[\dfrac{d^2}{n^{1-2\alpha_1}}\Bigg]
\end{align}
Now combining equations (\ref{eqn:211}) and (\ref{eqn:212}), we can finally conclude from equation (\ref{eqn:210}) that,
\begin{align}\label{eqn:213}
&\mbox{sup}_{A\in\mathcal{C}}\Big{|}\mathbf{P}_*\Big(\hat{\bm{T}}_n^{*(PB)}\in A\Big)-\mathbf{P}\Big(\bm{G}_{1n}\in A\Big)\Big{|}=O_p\Bigg\{\Bigg[\dfrac{d^{5/2}}{n^{1-2\alpha_1}}\Bigg]^{1/2}\Bigg[\log \Bigg(\dfrac{d^2}{n^{1-2\alpha_1}}\Bigg)\Bigg]^{1/2}\Bigg\}=o_p(1)
\end{align}
provided $d=o[n^{2\{1-2\alpha_1\}/5}]$ for $\alpha_1<1/2$ and therefore the proof is complete. \hfill $\blacksquare$

\begin{lemma}\label{lem:PBLeadingtermball}
 Consider the assumptions and set-up as in Lemma \ref{lem:smallHSnorm}. Suppose $\bm{G}_{1n}\sim N_d(\bm{0},\Tilde{\bm{S}}_n)$. Then provided $d=o[n^{\{1-2\alpha_1\}/2}]$ for $\alpha_1<1/2$, we have;
$$\mbox{sup}_{A\in\mathcal{B}}\Big{|}\mathbf{P}_*\Big(\hat{\bm{T}}_n^{*(PB)}\in A\Big)-\mathbf{P}\Big(\bm{G}_{1n}\in A\Big)\Big{|}=o_p(1),$$
where $\mathcal{B}$ be the class of measurable Euclidean Balls.   
\end{lemma}
Proof of Lemma \ref{lem:PBLeadingtermball} We will skip this proof as we have to follow the exact same path as in Lemma \ref{lem:PBLeadingtermconvex} using Lemma \ref{lem:fangkoikeball} and \ref{lem:mainlemmafangandkoikeball}. \hfill $\blacksquare$

\begin{lemma}\label{lem:pbglmconc}
 Suppose the set-up of Lemma \ref{lem:glmconc} is true. Then 
$$\Big{\|}\sum_{i=1}^{n}\Psi_*(y_i,\hat{\bm{\beta}}_n^{*(PB)})\Big{\|}_2=o_{p_*}(1)\;\text{in probability}.$$    
\end{lemma}
Proof of Lemma \ref{lem:pbglmconc} This proof follows from the definition. \hfill $\blacksquare$

\begin{lemma}\label{lem:PBconcleading}
Suppose (A.2)-(A.4), (A.6)-(A.7) are true.  Then provided $d=o(n^{\frac{1}{2}-\alpha_1})$, we have: $$\Big{\|}\hat{\bm{W}}_n^{*(PB)}\Big{\|}_2=O_{p_*}(d^{1/2})\;\text{in probability}.$$  
\end{lemma}
Proof of Lemma \ref{lem:PBconcleading} Recall that $\hat{\bm{W}}_n^{*(PB)}=n^{-1/2}\sum_{i=1}^{n}\bm{Z}_i^*$ with $\bm{Z}_i^*=\big\{y_i-g^{-1}\big(\bm{x}_i^\top\hat{\bm{\beta}}_n\big)\big\}h^\prime(\bm{x}_i^\top\hat{\bm{\beta}}_n)\bm{x}_i\big[\mu_{G^*}^{-1}G_i^*-1\big]$ and $\{y_i-g^{-1}(\bm{x}_i^\top \bm{\beta})\}$ and $G_i^*$ being independent sequence of each other and within themselves for all $i$. Therefore, it's enough to show that $$\mathbf{E}_*\Big[\Big{\|}\hat{\bm{W}}_n^{*(PB)}\Big{\|}_2^2\Big]=O_p(d).$$
 Next for some $\tilde{z}_i$ such that for all $i$, $|\tilde{z}_i-\bm{x}_i^\top\bm{\beta}|\leq \bm{x}_i^\top(\hat{\bm{\beta}}_n-\bm{\beta})$, due to Taylor's expansion, we will find that;
\begin{align}\label{eqn:214}
&\frac{1}{n}\sum_{i=1}^{n}\Big{|}h^\prime\big(\bm{x}_i^\top\hat{\bm{\beta}}_n\big)\Big{|}^2\Big{|}y_i-g^{-1}\big(\bm{x}_i^\top\hat{\bm{\beta}}_n\big)\Big{|}^2\nonumber\\
&\leq\dfrac{4}{n}\sum_{i=1}^{n}\Bigg[\Big{|}y_i-g^{-1}\big(\bm{x}_i^\top\bm{\beta}\big)\Big{|}^2+\Big{|}\bm{x}_i^\top\big(\hat{\bm{\beta}}_n-\bm{\beta}\big)\Big{|}^2\Big{|}(g^{-1})^\prime(\tilde{z}_i)\Big{|}^2\Bigg]\Bigg[\Big{|}h^{\prime}\big(\bm{x}_i^\top\bm{\beta}\big)\Big{|}^2+\Big{|}\bm{x}_i^\top\big(\hat{\bm{\beta}}_n-\bm{\beta}\big)\Big{|}^2\Big{|}h^{\prime\prime}(\tilde{z}_i)\Big{|}^2\Bigg]\nonumber\\
&\leq 4\Bigg[\Bigg(\Big\{n^{-1}\sum_{i=1}^{n}\Big{|}y_i-g^{-1}\big(\bm{x}_i^\top\bm{\beta}\big)\Big{|}^2\Big\}+\Big\{d^{1/2}\mbox{max}_{i\in\{1,..,n\}}\|\bm{x}_i\|_\infty\Big\}^2\Big{\|}\hat{\bm{\beta}}_n-\bm{\beta}\Big{\|}_2^2\Big\{n^{-1}\sum_{i=1}^{n}\Big{|}(g^{-1})^\prime\big(\tilde{z}_i\big)\Big{|}^2\Big\}\Bigg)\nonumber\\
&\times\Big\{\mbox{max}_{i\in\{1,..,n\}}\Big{|}h^{\prime}\big(\bm{x}_i^\top\bm{\beta}\big)\Big{|}\Big\}^{2}+\Bigg(\Big\{d^{1/2}\mbox{max}_{i\in\{1,..,n\}}\|\bm{x}_i\|_\infty\Big\}^2\Big{\|}\hat{\bm{\beta}}_n-\bm{\beta}\Big{\|}_2^2\Big\{n^{-1}\sum_{i=1}^{n}\Big{|}y_i-g^{-1}\big(\bm{x}_i^\top\bm{\beta}\big)\Big{|}^4\Big\}^{1/2}\nonumber\\
&+\Big\{d^{1/2}\mbox{max}_{i\in\{1,..,n\}}\|\bm{x}_i\|_\infty\Big\}^4\Big{\|}\hat{\bm{\beta}}_n-\bm{\beta}\Big{\|}_2^4\Big\{n^{-1}\sum_{i=1}^{n}\Big{|}(g^{-1})^\prime\big(\tilde{z}_i\big)\Big{|}^4\Big\}^{1/2}\Bigg)\Big\{n^{-1}\sum_{i=1}^{n}\Big{|}h^{\prime\prime}\big(\tilde{z}_i\big)\Big{|}^{4}\Big\}^{1/2}\Bigg]\nonumber\\
&=O_p\Bigg[1+\Big(\dfrac{d}{n^{\frac{1}{2}-\alpha_1}}\Big)^2+\Big(\dfrac{d}{n^{\frac{1}{2}-\alpha_1}}\Big)^4\Bigg]
\end{align}
Therefore equation (\ref{eqn:214}) and the assumption $d=o(n^{\frac{1}{2}-\alpha_1})$ will imply that,
\begin{align*}
&\mathbf{E}_*\Big[\Big{\|}\hat{\bm{W}}_n^{*(PB)}\Big{\|}_2^2\Big]=\frac{1}{n}\mathbf{E}_*\Big[\Big{\|}\sum_{i=1}^{n}\bm{Z}_i^*\Big{\|}_2^2\Big]=\frac{1}{n}\mathbf{E}_*\Big[\sum_{i=1}^{n}\Big{\|}\bm{Z}_i^*\Big{\|}_2^2+\sum_{i\neq j}\bm{Z}_i^{*\top} \bm{Z}_j^*\Big]\\
&=\frac{1}{n}\Bigg[\sum_{i=1}^{n}\Big(|y_i-g^{-1}(\bm{x}_i^\top \hat{\bm{\beta}}_n)|^2\Big)|h^\prime (\bm{x}_i^\top \hat{\bm{\beta}}_n)|^2\big{\|}\bm{x}_i\big{\|}_2^2\underbrace{\mathbf{E}_*\Big(\mu_{G^*}^{-1}G_i^*-1\Big)^2}_{=1}\\
&+\sum_{i\neq j}\Big\{y_i-g^{-1}(\bm{x}_i^\top \hat{\bm{\beta}}_n)\Big\}\Big\{y_j-g^{-1}(\bm{x}_j^\top \hat{\bm{\beta}}_n)\Big\}\big[h^\prime (\bm{x}_i^\top \hat{\bm{\beta}}_n)h^\prime (\bm{x}_j^\top \hat{\bm{\beta}}_n)\big]\bm{x}_i^\top \bm{x}_j\\
&\qquad\qquad\qquad\qquad\qquad\qquad \qquad\qquad\qquad \times\underbrace{\mathbf{E}_*\Big[\big\{\mu_{G^*}^{-1}G_i^*-1\big\}\big\{\mu_{G^*}^{-1}G_j^*-1\big\}\Big]}_{=0}\Bigg]\\
&\leq \Big\{d^{1/2}\mbox{max}_{i\in\{1,..,n\}}\|\bm{x}_i\|_{\infty}\Big\}^2\Bigg[\frac{1}{n}\sum_{i=1}^{n}|y_i-g^{-1}(\bm{x}_i^\top \hat{\bm{\beta}}_n)|^2|h^\prime (\bm{x}_i^\top \hat{\bm{\beta}}_n)|^2\Bigg]\\
&=O_p\Bigg[d\Bigg\{1+\Big(\dfrac{d}{n^{\frac{1}{2}-\alpha_1}}\Big)^2+\Big(\dfrac{d}{n^{\frac{1}{2}-\alpha_1}}\Big)^4\Bigg\}\Bigg]=O_p(d)
\end{align*}
Therefore the proof is complete. \hfill $\blacksquare$

\begin{lemma}\label{lem:PBremainderbound}
Consider the same set-up of Lemma \ref{lem:remainderbound}. Then for any $0<B<\infty$, we have the following: 
$$\operatorname*{\sup}_{||\bm{\alpha}||=1}\operatorname*{\sup}_{\{\bm{\tau}:\|\hat{\bm{L}}_n(\bm{\tau}-\hat{\bm{\beta}}_n)\|_2\leq B(d/n)^{1/2}\}}\Bigg{|}\bm{\alpha}^\top \sum_{i=1}^{n}\mathbf{E}_*\Big[\Psi_*(y_i,\bm{\tau})-\Psi_*(y_i,\hat{\bm{\beta}}_n)\Big]-n\bm{\alpha}^\top \hat{\bm{L}}_n\big(\bm{\tau}-\hat{\bm{\beta}}_n\big)\Bigg{|}=O_p(dn^{2\alpha_1}).$$
\end{lemma}
Proof of Lemma \ref{lem:PBremainderbound} We utilize suitable Taylor's expansion with respect to $\bm{x}_i^\top\hat{\bm{\beta}}_n$ to note that, for some $z_i^*$ with $|z_i^*-\bm{x}_i^\top\hat{\bm{\beta}}_n|\leq \bm{x}_i^\top(\bm{\tau}-\hat{\bm{\beta}}_n)$, we have;

\begin{align}\label{eqn:215}
&\sum_{i=1}^n\mathbf{E}_*\Big[\Psi_*(y_i,\bm{\tau})-\Psi_*(y_i,\hat{\bm{\beta}}_n)\Big]\nonumber\\
&=-\sum_{i=1}^{n}\Big\{y_{i}h^\prime(\bm{x}_{i}^\top\bm{\tau}) -h_1^\prime(\bm{x}_{i}^\top\bm{\tau}) \Big\}\bm{x}_i+\sum_{i=1}^{n}\Big\{y_{i}h^\prime(\bm{x}_{i}^\top\hat{\bm{\beta}}_n) -h_1^\prime(\bm{x}_{i}^\top\hat{\bm{\beta}}_n) \Big\}\bm{x}_i\nonumber\\
&=\sum_{i=1}^{n}\bm{x}_i\bm{x}_i^\top\Big[(g^{-1})^\prime(\bm{x}_i^\top \hat{\bm{\beta}}_n)h^\prime (\bm{x}_{i}^\top\hat{\bm{\beta}}_n)-\Big\{y_i-g^{-1}(\bm{x}_i^\top \hat{\bm{\beta}}_n)\Big\}h^{\prime\prime}(\bm{x}_i^\top \hat{\bm{\beta}}_n)\Big]\big(\bm{\tau}-\hat{\bm{\beta}}_n\big)\nonumber\\
&+2^{-1}\sum_{i=1}^{n}\bm{x}_i
\Big[-h^{\prime\prime\prime}(z_i^*)\big\{y_i-g^{-1}(z_i^*)\big\}+(g^{-1})^{\prime\prime}(z_i^*)h^\prime (z_i^*)+2(g^{-1})^{\prime}(z_i^*)h^{\prime\prime} (z_i^*)\Big]\Big\{(\bm{\tau}-\hat{\bm{\beta}}_n)^\top \bm{x}_i\Big\}^2\nonumber\\
&=n\hat{\bm{L}}_n\big(\bm{\tau}-\hat{\bm{\beta}}_n\big)+2^{-1}\sum_{i=1}^{n}\bm{x}_i
\Big[-h^{\prime\prime\prime}(z_i^*)\big\{y_i-g^{-1}(z_i^*)\big\}+(g^{-1})^{\prime\prime}(z_i^*)h^\prime (z_i^*)+2(g^{-1})^{\prime}(z_i^*)h^{\prime\prime} (z_i^*)\Big]\nonumber\\
&\times\Big\{(\bm{\tau}-\hat{\bm{\beta}}_n)^\top \bm{x}_i\Big\}^2-\Bigg[\sum_{i=1}^{n}\bm{x}_i\bm{x}_i^\top\Big\{y_i-g^{-1}(\bm{x}_i^\top \hat{\bm{\beta}}_n)\Big\}h^{\prime\prime}(\bm{x}_i^\top \hat{\bm{\beta}}_n)\Bigg](\bm{\tau}-\hat{\bm{\beta}}_n)
\end{align}
Now note that, 
\begin{align*}
&\operatorname*{\sup}_{||\bm{\alpha}||=1}\operatorname*{\sup}_{\{\bm{\tau}:\|\hat{\bm{L}}_n(\bm{\tau}-\hat{\bm{\beta}}_n)\|_2\leq B(d/n)^{1/2}\}}\Bigg|\bm{\alpha}^\top\sum_{i=1}^{n}\bm{x}_i\bm{x}_i^\top\Big\{y_i-g^{-1}(\bm{x}_i^\top \hat{\bm{\beta}}_n)\Big\}h^{\prime\prime}(\bm{x}_i^\top \hat{\bm{\beta}}_n)\Bigg|\\
&=O_p\Big[n^{1/2}\frac{d^{1/2}}{n^{\frac12-\alpha_1}}+n\frac{d}{n^{1-2\alpha_1}}+n\frac{d^{3/2}}{n^{\frac32-3\alpha_1}}\Big]=O_p(dn^{2\alpha_1}),\quad\text{provided}\;\; d=o(n^{1-2\alpha_1}).
\end{align*}
Rest of the proof will follow exactly in the similar steps as in equation (\ref{eqn:finalerrorbound}) of Lemma \ref{lem:remainderbound} under the given assumptions. Therefore the proof is complete. \hfill $\blacksquare$

\begin{lemma}\label{lem:PBconcentration}
 Suppose the assumptions of Lemma \ref{lem:pbglmconc}-\ref{lem:PBremainderbound} hold true. Then provided $d=o\Big(n^{1-4\alpha_1}\Big)$ with $\alpha_1<1/4$, we have:
 $$\big{\|}\hat{\bm{\beta}}_n^{*(PB)}-\hat{\bm{\beta}}_n\big{\|}_2=O_{p_*}\Big[\Big(\dfrac{d}{n^{1-2\alpha_1}}\Big)^{1/2}\Big]\;\text{in probability}.$$
\end{lemma}
Proof of Lemma \ref{lem:PBconcentration} 
This proof will follow similarly as Lemma \ref{lem:concentration}. So we skip the details. \hfill $\blacksquare$

\begin{lemma}\label{lem:PBerrorbound}
Suppose the assumptions of Lemma \ref{lem:PBremainderbound} hold true. Let $\hat{\bm{r}}_n^{*(PB)}$ be the remainder term in Bahadur's representation of $n^{1/2}(\hat{\bm{\beta}}_n^{*(PB)}-\hat{\bm{\beta}}_n)$. Then following is true:
$$\|\hat{\bm{r}}_n^{*(PB)}\|=o_{p_*}(1)\;\text{in probability},$$
provided (i) $d=o(n^{\frac{2}{5}[1-6\alpha_1-\alpha_2]})$ for class $\mathcal{C}$ and (ii) $d=o(n^{\frac{1}{2}[1-6\alpha_1-\alpha_2]})$ for class $\mathcal{B}$ with $6\alpha_1+\alpha_2<1$.
\end{lemma}
Proof of Lemma \ref{lem:PBerrorbound} Suppose, $\bm{u}^*=\hat{\bm{\beta}}_n^{*(PB)}-\hat{\bm{\beta}}_n$. Hence just like in original case, the Taylor's expansion with respect to $\bm{x}_i^\top\hat{\bm{\beta}}_n$, we can obtain the Bahadur's representation of $n^{1/2}(\hat{\bm{\beta}}_n^{*(PB)}-\hat{\bm{\beta}}_n)$ as:
\begin{align}\label{eqn:004boot}
n^{1/2}(\hat{\bm{\beta}}_n^{*(PB)}-\hat{\bm{\beta}}_n)=[\hat{\mathbf{L}}_n]^{-1}\hat{\bm{W}}_n^{*(PB)}+\hat{\bm{r}}_n^{*(PB)}=\hat{\bm{T}}_n^{*(PB)}+\hat{\bm{r}}_n^{*(PB)},    
\end{align}
where we write the remainder term as:
\begin{align}\label{eqn:005boot}
\hat{\bm{r}}_n^{*(PB)}&=\dfrac{\hat{\mathbf{L}}_n^{-1}}{2n^{1/2}}\sum_{i=1}^{n}\bm{x}_i\Bigg[\Big\{y_i-g^{-1}(z_i^*)\Big\}h^{\prime\prime\prime}(z_i^*)\nonumber\\
&-2(g^{-1})^\prime(z_i^*)h^{\prime\prime}(z_i^*)-(g^{-1})^{\prime\prime}(z_i^*)h^{\prime}(z_i^*)\Bigg]\Big[\bm{x}_i^\top(\hat{\bm{\beta}}_n^{*(PB)}-\hat{\bm{\beta}}_n)\Big]^2\nonumber\\
&+\hat{L}_n^{-1}\Bigg[\frac1n\sum_{i=1}^{n}\bm{x}_i\bm{x}_i^\top\Big\{y_i-g^{-1}(\bm{x}_i^\top\hat{\bm{\beta}}_n)\Big\}h^{\prime\prime}(\bm{x}_i^\top\hat{\bm{\beta}}_n)\Bigg]\Big[n^{1/2}(\hat{\bm{\beta}}_n^{*(PB)}-\hat{\bm{\beta}}_n)\Big]
\end{align}
for some $z_i^*$ such that $|z_i^*-\bm{x}_i^\top\hat{\bm{\beta}}_n|\leq \bm{x}_i^\top\bm{u}^*$ for all $i\in\{1,..,n\}$. Then,
\begin{align}\label{eqn:418}
 n^{1/2}\hat{\bm{L}}_n\hat{\bm{r}}_n^{*(PB)}&=n\hat{\bm{L}}_n(\hat{\bm{\beta}}_n^{*(PB)}-\hat{\bm{\beta}}_n)-\sum_{i=1}^{n}\{y_i-g^{-1}(\bm{x}_i^\top\hat{\bm{\beta}}_n)\}h^\prime (\bm{x}_i^\top\hat{\bm{\beta}}_n)\bm{x}_i\Big\{\frac{G_i^*}{\mu_{G^*}}-1\Big\}\nonumber\\
&=\underbrace{\sum_{i=1}^{n}\Psi_*(y_i,\hat{\bm{\beta}}_n^{*(PB)})}_{=0}-\Bigg\{\sum_{i=1}^{n}\mathbf{E}_*\Big[\Psi_*(y_i,\hat{\bm{\beta}}_n^{*(PB)})-\Psi_*(y_i,\hat{\bm{\beta}}_n)\Big]-n\hat{\bm{L}}_n(\hat{\bm{\beta}}_n^{*(PB)}-\hat{\bm{\beta}}_n)\Bigg\}\nonumber\\
 &-\Bigg\{\underbrace{\sum_{i=1}^{n}\Big[\Psi_*(y_i,\hat{\bm{\beta}}_n^{*(PB)})-\Psi_*(y_i,\hat{\bm{\beta}}_n)\Big]-\sum_{i=1}^{n}\mathbf{E}_*\Big[\Psi_*(y_i,\hat{\bm{\beta}}_n^{*(PB)})-\Psi_*(y_i,\hat{\bm{\beta}}_n)\Big]}_{=0}\Bigg\}
\end{align}
Then Lemma \ref{lem:pbglmconc}, \ref{lem:PBconcleading}, \ref{lem:PBremainderbound} and \ref{lem:PBconcentration} along with equation (\ref{eqn:418})  will imply that,
\begin{align}\label{eqn:119419}
&\|\hat{\bm{r}}_n^{*(PB)}\|=O_{p^*}\Bigg[\frac{d}{n^{\frac{1}{2}-3\alpha_1}}\Bigg]\quad\text{in probability}.
\end{align}

Rest of the argument is similar to Lemma \ref{lem:errorbound}. Therefore, the proof is complete. \hfill $\blacksquare$

\subsection{Proofs of Main Results}\label{sec:proofmain}
In this section, we provide the proofs of our main results namely Theorem \ref{thm:convex1}, \ref{thm:ball1}, \ref{thm:pbconvexoriginal1} and \ref{thm:prbconvexoriginal1}.

\subsubsection{Proof of Theorem \ref{thm:convex1}}\label{sec:thm4.1}
Recall that $\hat{\bm{\beta}}_n$ is the GLM estimator given by (\ref{eqn:defineglmest12}). Suppose $\bm{G}_{1n}\sim N_d(\bm{0},\Tilde{\bm{S}}_n)$. For class of measurable convex sets $\mathcal{C}$, we are to show that,
\begin{align}\label{eqn:120}
\sup_{A\in\mathcal{C}}\Big{|}\mathbf{P}\Big[n^{1/2}(\hat{\bm{\beta}}_n-\bm{\beta})\in A\Big]-\mathbf{P}\Big[\bm{G}_{1n}\in A\Big]\Big{|}=o(1).    
\end{align}
Recall the Bahadur's representation as in (\ref{eqn:expansion}) given by: $$n^{1/2}[\hat{\bm{\beta}}_n-\bm{\beta}]= [\mathbf{E(L_n)}]^{-1}\bm{W}_n +\bm{r}_n=\bm{T}_n+\bm{r}_n.$$ 
Now we write,
\begin{align}\label{eqn:121}
&\sup_{A\in\mathcal{C}}\Big{|}\mathbf{P}\Big[n^{1/2}(\hat{\bm{\beta}}_n-\bm{\beta})\in A\Big]-\mathbf{P}\Big[\bm{G}_{1n}\in A\Big]\Big{|}= \sup_{A\in\mathcal{C}}\Big{|}\mathbf{P}\Big[\bm{T}_n+\bm{r}_n\in A\Big]-\mathbf{P}\Big[\bm{G}_{1n}\in A\Big]\Big{|}\nonumber\\
&\leq \sup_{A\in\mathcal{C}}\Big{|}\mathbf{P}\Big[\bm{T}_n+\bm{r}_n\in A\Big]-\mathbf{P}\Big[\bm{T}_n\in A\Big]\Big{|}+\sup_{A\in\mathcal{C}}\Big{|}\mathbf{P}\Big[\bm{T}_n\in A\Big]-\mathbf{P}\Big[\bm{G}_{1n}\in A\Big]\Big{|}\nonumber\\
&\leq \sup_{A\in\mathcal{C}}\Big{|}\mathbf{P}\Big[\bm{T}_n+\bm{r}_n\in A,\bm{T}_n\in A^c\Big]-\mathbf{P}\Big[\bm{T}_n\in A,\bm{T}_n+\bm{r}_n\in A^c\Big]\Big{|}+\sup_{A\in\mathcal{C}}\Big{|}\mathbf{P}\Big[\bm{T}_n\in A\Big]-\mathbf{P}\Big[\bm{G}_{1n}\in A\Big]\Big{|}\nonumber\\
&\leq \sup_{A\in\mathcal{C}}\Big{|}\mathbf{P}\Big[\bm{T}_n+\bm{r}_n\in A,\bm{T}_n\in A^c,\|\bm{r}_n\|>\epsilon\Big]+\mathbf{P}\Big[\bm{T}_n+\bm{r}_n\in A,\bm{T}_n\in A^c,\|\bm{r}_n\|\leq\epsilon\Big]\nonumber\\
&\;\;\;\;\;\;\;\;\;\;+\mathbf{P}\Big[\bm{T}_n\in A,\bm{T}_n+\bm{r}_n\in A^c,\|\bm{r}_n\|>\epsilon\Big]+\mathbf{P}\Big[\bm{T}_n\in A,\bm{T}_n+\bm{r}_n\in A^c,\|\bm{r}_n\|\leq\epsilon\Big]\Big{|}\nonumber\\
&\;\;\;\;\;\;\;\;\;\;\;\;\;\;\;\;\;\;\;\;\;\;\;\;\;\;\;\;\;\;\;\;\;\;\;\;\;\;\;\;\;\;\;\;\;+\sup_{A\in\mathcal{C}}\Big{|}\mathbf{P}\Big[\bm{T}_n\in A\Big]-\mathbf{P}\Big[\bm{G}_{1n}\in A\Big]\Big{|}\nonumber\\
&\leq 2\mathbf{P}\big[\|\bm{r}_n\|>\epsilon\big]+2\sup_{A\in\mathcal{C}}\big{|}\mathbf{P}\big[\bm{G}_{1n}\in (\partial{A})^{\epsilon}\big]\big{|}+2\sup_{A\in\mathcal{C}}\big{|}\mathbf{P}\big[\bm{T}_n\in (\partial{A})^{\epsilon}\big]-\mathbf{P}\big[\bm{G}_{1n}\in (\partial{A})^{\epsilon}\big]\big{|}\nonumber\\
&\;\;\;\;\;\;\;\;\;\;\;\;\;\;\;\;\;\;\;\;\;\;\;\;\;\;\;\;\;\;\;\;\;\;\;\;\;\;\;\;\;\;\;\;\;\;\;\;\;\;\;\;\;\;\;\;\;\;+\sup_{A\in\mathcal{C}}\Big{|}\mathbf{P}\Big[\bm{T}_n\in A\Big]-\mathbf{P}\Big[\bm{G}_{1n}\in A\Big]\Big{|}\nonumber\\
&=o(1),
\end{align}
The last line collectively follows from Lemma \ref{lem:errorbound}, \ref{lem:mainlemmagaussisoperimetric} and \ref{lem:mainlemmafangandkoikeconvex} under the assumption that $d=o[n^{2\{1-6\alpha_1-\alpha_2\}/5}]$ for $6\alpha_1+\alpha_2<1.$ Therefore we are done. \hfill $\blacksquare$

\subsubsection{Proof of Theorem \ref{thm:ball1}}\label{sec:thm4.2}
This proof follows from Lemma \ref{lem:errorbound}, \ref{lem:zhilovamainlemmagaussisoperimetric} and \ref{lem:mainlemmafangandkoikeball} provided $d=o[n^{\{1-6\alpha_1-\alpha_2\}/2}]$ with $6\alpha_1+\alpha_2<1$ after following the exact proof steps till equation (\ref{eqn:121}) as in the proof of Theorem \ref{thm:convex1} in section \ref{sec:thm4.1}. Hence we are done. \hfill $\blacksquare$

\subsubsection{Proof of Theorem \ref{thm:pbconvexoriginal1}}\label{sec:thm4.3}
Recall that $\hat{\bm{\beta}}_n^{*(PB)}$ is the PB-GLM estimator given by (\ref{eqn:defpb}). Suppose $\bm{G}_{1n}\sim N_d(\bm{0},\Tilde{\bm{S}}_n)$. For class of measurable convex sets $\mathcal{C}$, we are to show that,
\begin{align}\label{eqn:450}
\sup_{A\in\mathcal{C}}\Big{|}\mathbf{P}_*\Big[n^{1/2}(\hat{\bm{\beta}}_n^{*(PB)}-\hat{\bm{\beta}}_n)\in A\Big]-\mathbf{P}\Big[n^{1/2}(\hat{\bm{\beta}}_n-\bm{\beta})\in A\Big]\Big{|}=o_p(1).    
\end{align}
Now it's obvious to have that,
\begin{align}\label{eqn:451}
&\sup_{A\in\mathcal{C}}\Big{|}\mathbf{P}_*\Big[n^{1/2}(\hat{\bm{\beta}}_n^{*(PB)}-\hat{\bm{\beta}}_n)\in A\Big]-\mathbf{P}\Big[n^{1/2}(\hat{\bm{\beta}}_n-\bm{\beta})\in A\Big]\Big{|}\nonumber\\
&\leq \underbrace{\sup_{A\in\mathcal{C}}\Big{|}\mathbf{P}_*\Big[n^{1/2}(\hat{\bm{\beta}}_n^{*(PB)}-\hat{\bm{\beta}}_n)\in A\Big]-\mathbf{P}\Big[\bm{G}_{1n}\in A\Big]\Big{|}}_{\text{Term (I)}}+\underbrace{\sup_{A\in\mathcal{C}}\Big{|}\mathbf{P}\Big[n^{1/2}(\hat{\bm{\beta}}_n-\bm{\beta})\in A\Big]-\mathbf{P}\Big[\bm{G}_{1n}\in A\Big]\Big{|}}_{\text{Term (II)}}
\end{align}
Now Term (II) in equation (\ref{eqn:451}) is $o(1)$ due to equation (\ref{eqn:121}) in the proof of Theorem \ref{thm:convex1}. Next we will handle the Term (I) in equation (\ref{eqn:451}) exactly in the same footprints of equation (\ref{eqn:121}) with only change being the conditional probabilities instead of unconditional one. So finally we will have in that manner;
\begin{align}\label{eqn:452}
&\sup_{A\in\mathcal{C}}\Big{|}\mathbf{P}_*\Big[n^{1/2}(\hat{\bm{\beta}}_n^{*(PB)}-\hat{\bm{\beta}}_n)\in A\Big]-\mathbf{P}\Big[\bm{G}_{1n}\in A\Big]\Big{|}\nonumber\\
&\leq 2\sup_{A\in\mathcal{C}}\big{|}\mathbf{P}_*\big[\|\hat{\bm{r}}_n^{*(PB)}\|>\epsilon\big]\big{|}+2\sup_{A\in\mathcal{C}}\big{|}\mathbf{P}\big[\bm{G}_{1n}\in (\partial{A})^{\epsilon}\big]\big{|}\nonumber\\
&\;\;+2\sup_{A\in\mathcal{C}}\big{|}\mathbf{P}_*\big[\hat{\bm{T}}_n^{*(PB)}\in (\partial{A})^{\epsilon}\big]-\mathbf{P}\big[\bm{G}_{1n}\in (\partial{A})^{\epsilon}\big]\big{|}+\sup_{A\in\mathcal{C}}\Big{|}\mathbf{P}_*\Big[\hat{\bm{T}}_n^{*(PB)}\in A\Big]-\mathbf{P}\Big[\bm{G}_{1n}\in A\Big]\Big{|}\nonumber\\
&=o_p(1)
\end{align}
Last line collectively follows from Lemma \ref{lem:PBerrorbound}, \ref{lem:mainlemmagaussisoperimetric} and \ref{lem:PBLeadingtermconvex} under the existing assumption. Therefore we are done.\\

 The proof for class of Euclidean Balls follows from Lemma \ref{lem:PBerrorbound}, \ref{lem:zhilovamainlemmagaussisoperimetric} and \ref{lem:PBLeadingtermball} under the assumption that $d=o[n^{\{1-6\alpha_1-\alpha_2\}/2}]$ with $6\alpha_1+\alpha_2<1$ after following the exact proof steps till equation (\ref{eqn:452}). Therefore we are done.\hfill $\blacksquare$

\subsubsection{\bf Proof of Theorem \ref{thm:prbconvexoriginal1}}\label{sec:prbthm4.3} 
We will consider everything on the set $\Big\{\Big\|\hat{\bm{\beta}}_n-\bm{\beta}\Big\|_2\le B\frac{d^{1/2}}{n^{\frac12-\alpha_1}}\Big\}$. Now note that,
\begin{align}\label{eqn:666644444}
&n^{1/2}(\hat{\bm{\beta}}_n^{*(PRB)}-\hat{\bm{\beta}}_n)\nonumber\\
&=\underbrace{n^{-1/2}\sum_{i=1}^{n}\sqrt{b^{\prime\prime}\Big(h(\bm{x}_i^\top\hat{\bm{\beta}}_n)\Big)}h^\prime(\bm{x}_i^\top\hat{\bm{\beta}}_n)\Big\{(\mathbf{E}L_n)^{-1}\bm{x}_i\Big\}e_i^*}_{=\hat{\bm{T}}_n^{*(PRB)}}+\underbrace{\Bigg[\Bigg(\frac{\hat{G}_n^\top\hat{G}_n}{n}\Bigg)^{-1}-(\mathbf{E}L_n)^{-1}\Bigg]\frac{\hat{G}_n^\top\bm{e}^*}{n^{1/2}}}_{=\hat{\bm{r}}_n^{*(PRB)}}
\end{align}
Denote, $\hat{\sigma}_n^2=\frac1n\sum_{k=1}^n(e_k^\dagger-\bar{e}^\dagger)^2$, $\hat{\mu}_k:=g^{-1}(\bm{x}_k^\top\hat{\bm{\beta}}_n)$, $\hat{v}_k:=b^{\prime\prime}(h(\bm{x}_k^\top\hat{\bm{\beta}}_n))$ and $e_k^\dagger:=\frac{y_k-\hat{\mu}_k}{\sqrt{\hat{v}}_k}$ for all $1\le k\le n$. Similarly we can define the population counterpart, $e_k:=\frac{y_k-\mu_k}{\sqrt{v}_k}$ with $\mathbf{E}(e_k)=0$ and $\text{var}(e_k)=1$ for all $k$. Also let us denote,  $\tilde{S}_n=[\mathbf{E}(L_n)]^{-1}$. Then,
\[
V_n^*:=\text{var}_*\Big(\tilde{S}_n^{-1/2}\hat{T}_n^{*(PRB)}\Big)=\hat{\sigma}_n^2[\mathbf{E}(L_n)]^{-1/2}\Big(n^{-1}\hat{G}_n^\top\hat{G}_n\Big)[\mathbf{E}(L_n)]^{-1/2}
\]
Now,
\begin{align*}
 \|I_d-V_n^*\|_{HS}&\le\underbrace{\Big\|I_d-[\mathbf{E}(L_n)]^{-1/2}\Big(n^{-1}\hat{G}_n^\top\hat{G}_n\Big)[\mathbf{E}(L_n)]^{-1/2}\Big\|_{HS}}_{:=B_1\;\text{(say)}}\\
 &+\underbrace{|1-\hat{\sigma}_n^2|\Big\|[\mathbf{E}(L_n)]^{-1/2}\Big(n^{-1}\hat{G}_n^\top\hat{G}_n-\mathbf{E}(L_n)\Big)[\mathbf{E}(L_n)]^{-1/2}\Big\|_{HS}}_{:=B_2\;\text{(say)}}+\underbrace{d^{1/2}|1-\hat{\sigma}_n^2|}_{:=B_3\;\text{(say)}}   
\end{align*}
Following the exact similar treatment so far with Taylor's expansion, it's only algebra to have that,
\begin{align}\label{eqn:iiiiiiiiii}
\Big\|n^{-1}\hat{G}_n^\top\hat{G}_n-\mathbf{E}(L_n)\Big\|_{op}=O_p\Bigg[\frac{d}{n^{1-2\alpha_1}}+\frac{d^{1/2}}{n^{\frac12-\alpha_1}}\Bigg]    
\end{align}
Using (\ref{eqn:iiiiiiiiii}) and standard norm inequality, we get,
\begin{align}\label{eqn:jjjjjjjjjj}
 &B_1:= \Big\|[\mathbf{E}(L_n)]^{-1/2}\Big(n^{-1}\hat{G}_n^\top\hat{G}_n-\mathbf{E}(L_n)\Big)[\mathbf{E}(L_n)]^{-1/2}\Big\|_{HS} \nonumber\\
 &\le \Big\|[\mathbf{E}(L_n)]^{-1/2}\Big\|_{op}\Big\|n^{-1}\hat{G}_n^\top\hat{G}_n-\mathbf{E}(L_n)\Big\|_{op}\Big\|[\mathbf{E}(L_n)]^{-1/2}\Big\|_{HS}=O_p\Bigg[\frac{d^{3/2}}{n^{1-3\alpha_1}}+\frac{d}{n^{\frac12-2\alpha_1}}\Bigg]
\end{align}
Now, we decompose,
\begin{align*}
&\hat{\sigma}_n^2-1=\underbrace{\Bigg[\frac1n\sum_{i=1}^n\frac{(y_k-\hat{\mu}_k)^2}{\hat{v}_k}-1\Bigg]}_{\text{Term I}}-\underbrace{\Bigg[\frac1n\sum_{i=1}^n\frac{(y_k-\hat{\mu}_k)}{\sqrt{\hat{v}_k}}\Bigg]^2}_{\text{Term II}}
\end{align*}
Under the existing assumptions and using Taylor's expansion,
\begin{align*}
(y_k-\hat{\mu}_k)^2=(y_k-\mu_k)^2-&2(y_k-\mu_k)(g^{-1})^\prime(z_k)\Big[\bm{x}_k^\top(\hat{\bm{\beta}}_n-\bm{\beta})\Big]+\Big[(g^{-1})^\prime(z_k)\Big]^2\Big[\bm{x}_k^\top(\hat{\bm{\beta}}_n-\bm{\beta})\Big]^2\\
&\hat{v}_k^{-1}=v_k^{-1}+O\Big[\big|\bm{x}_k^\top(\hat{\bm{\beta}}_n-\bm{\beta})\big|\Big],
\end{align*}
for some $z_k$ such that, $|z_k-\bm{x}_k^\top\bm{\beta}|\le \bm{x}_k^\top(\hat{\bm{\beta}}_n-\bm{\beta})$ for all $1\le k\le n$. Careful algebraic manipulation will provide us that,
\begin{align}\label{eqn:kkkkkkkkkkk}
|1-\hat{\sigma}_n^2|=O_p\Bigg[\frac{1}{\sqrt{n}}+\frac{d^{1/2}}{n^{\frac12-\alpha_1}}+\frac{d}{n^{1-2\alpha_1}}+\frac{d^{3/2}}{n^{\frac32-3\alpha_1}}\Bigg]=O_p\Big(\frac{d^{1/2}}{n^{\frac12-\alpha_1}}\Big)  
\end{align}
That will in turn imply,
\begin{align}\label{eqn:hhhhhhhhhhhhh}
\Big\|I_d-V_n^*\Big\|_{HS}=O_p\Bigg[\frac{d}{n^{\frac12-2\alpha_1}}+\frac{d^{3/2}}{n^{1-3\alpha_1}}+\frac{d^2}{n^{\frac32-4\alpha_1}}\Bigg]=O_p\Bigg[\frac{d^2}{n^{1-4\alpha_1}}\Bigg]^{1/2}    
\end{align}
Note that, $\Big\|\tilde{S}_n^{-1/2}[\mathbf{E}(L_n)]^{-1}\Big\|^4=O_p(n^{2\alpha_1})$. This will eventually give,
\begin{align}\label{eqn:vvvvvvvv}
\Bigg[\frac{1}{n^2}\sum_{i=1}^n\mathbf{E}_*\Bigg\|\tilde{S}_n^{-1/2}[\mathbf{E}(L_n)]^{-1}\bm{x}_i\sqrt{b^{\prime\prime}\Big(h(\bm{x}_i^\top\hat{\bm{\beta}}_n)\Big)}h^\prime(\bm{x}_i^\top\hat{\bm{\beta}}_n)e_i^*\Bigg\|_2^4\Bigg]^{1/2}=O_p\Bigg[\frac{d^2}{n^{1-2\alpha_1}}\Bigg]^{1/2}        
\end{align}

Now combine (\ref{eqn:hhhhhhhhhhhhh}) and (\ref{eqn:vvvvvvvv}) with steps of Lemma \ref{lem:PBLeadingtermconvex} we will have,
\begin{align}\label{eqn:213uuuuuuuuuu}
\mbox{sup}_{A\in\mathcal{C}}\Big{|}\mathbf{P}_*\Big(\hat{\bm{T}}_n^{*(PRB)}\in A\Big)-\mathbf{P}\Big(\bm{G}_{1n}\in A\Big)\Big{|}
&=O_p\Bigg\{\Bigg[\dfrac{d^{5/2}}{n^{1-4\alpha_1}}\Bigg]^{1/2}\Bigg[\log \Bigg(\dfrac{d^2}{n^{1-4\alpha_1}}\Bigg)\Bigg]^{1/2}\Bigg\}\nonumber\\
&=o_p(1)
\end{align}
provided $d=o[n^{2\{1-4\alpha_1\}/5}]$ for $\alpha_1<1/4$. Similarly for class of Euclidean Balls,
\begin{align}\label{eqn:213kkkkkkkk}
\mbox{sup}_{A\in\mathcal{B}}\Big{|}\mathbf{P}_*\Big(\hat{\bm{T}}_n^{*(PRB)}\in A\Big)-\mathbf{P}\Big(\bm{G}_{1n}\in A\Big)\Big{|}
&=O_p\Bigg\{\Bigg[\dfrac{d^{2}}{n^{1-4\alpha_1}}\Bigg]^{1/2}\Bigg[\log \Bigg(\dfrac{d^2}{n^{1-4\alpha_1}}\Bigg)\Bigg]^{1/2}\Bigg\}\nonumber\\
&=o_p(1)
\end{align}
provided $d=o[n^{\{1-4\alpha_1\}/2}]$ for $\alpha_1<1/4$.\\

Recall that $\{e_1^*,...,e_n^*\}$ are iid from the distribution of $\{e_1^\dagger-\bar{e}^\dagger,.....,e_n^\dagger-\bar{e}^\dagger\}$. Then $\mathbf{E}_*(\bm{e}^*)=\bm{0}$ and $\mathbf{E}_*(\bm{e}^*\bm{e}^{*\top})=\hat{\sigma}_n^2I_n$. Let us denote, $\hat{\bm{W}}_n^{*(PRB)}=n^{-1/2}\hat{G}_n\bm{e}^*$. Therefore we can have,
\begin{align}\label{eqn:mkmkmkmk}
\mathbf{E}_*\Big\|\hat{\bm{W}}_n^{*(PRB)}\Big\|_2^2&=\hat{\sigma}_n^2\;\text{trace}\Bigg[\frac{\hat{G}_n^\top\hat{G}_n}{n}\Bigg]\nonumber\\
&=(\hat{\sigma}_n^2-1)\Bigg\{\text{trace}[\mathbf{E}(L_n)]+\text{trace}\Big[n^{-1}\hat{G}_n^\top\hat{G}_n-\mathbf{E}(L_n)\Big]\Bigg\}\nonumber\\
&\qquad\quad+\Bigg\{\text{trace}[\mathbf{E}(L_n)]+\text{trace}\Big[n^{-1}\hat{G}_n^\top\hat{G}_n-\mathbf{E}(L_n)\Big]\Bigg\}\nonumber\\
&\le \Big[|\hat{\sigma}_n^2-1|+1\Big]\Bigg\{d\lambda_{max}(\mathbf{E}[L_n])+d\Big\|n^{-1}\hat{G}_n^\top\hat{G}_n-\mathbf{E}(L_n)\Big\|_{op}\Bigg\}\nonumber\\
&=O_p\Bigg[d\Bigg\{\frac{d^{1/2}}{n^{\frac12-\alpha_1}}+1\Bigg\}\Bigg\{n^{\alpha_2}+\frac{d}{n^{1-2\alpha_1}}+\frac{d^{1/2}}{n^{\frac12-\alpha_1}}\Bigg\}\Bigg]=O_p(dn^{\alpha_2}),
\end{align}
provided $d=o(n^{1-2\alpha_1})$ with $\alpha_1<1/2$. This will imply that, $\Big\|\hat{\bm{W}}_n^{*(PRB)}\Big\|_2=O_{p^*}(d^{1/2}n^{\alpha_2/2})$ in probability. Now from equations (\ref{eqn:666644444}) and (\ref{eqn:iiiiiiiiii}), we can see that,

\begin{align}\label{eqn:vfvfvfvfvffvfvfvffv}
\Big\|\hat{\bm{r}}_n^{*(PRB)}\Big\|_2=O_{p^*}\Bigg[d^{1/2}n^{2\alpha_1+\frac{\alpha_2}{2}}\Bigg\{\frac{d}{n^{1-2\alpha_1}}+\frac{d^{1/2}}{n^{\frac12-\alpha_1}}\Bigg\}\Bigg]=o_{p^*}(1)\;\;\text{in probability},
\end{align}
provided $d=o[n^{2\{1-6\alpha_1-2\alpha_2\}/5}]$ when $\mathcal{A}=\mathcal{C}$, and provided $d=o[n^{\{1-6\alpha_1-2\alpha_2\}/2}]$ when $\mathcal{A}=\mathcal{B}$. Finally following the decomposition as in the proof of Theorem \ref{thm:pbconvexoriginal1}, we write,
\begin{align}\label{eqn:452vbgh}
&\sup_{A\in\mathcal{A}}\Big{|}\mathbf{P}_*\Big[n^{1/2}(\hat{\bm{\beta}}_n^{*(PRB)}-\hat{\bm{\beta}}_n)\in A\Big]-\mathbf{P}\Big[\bm{G}_{1n}\in A\Big]\Big{|}\nonumber\\
&\leq 2\sup_{A\in\mathcal{A}}\big{|}\mathbf{P}_*\big[\|\hat{\bm{r}}_n^{*(PRB)}\|>\epsilon\big]\big{|}+2\sup_{A\in\mathcal{A}}\big{|}\mathbf{P}\big[\bm{G}_{1n}\in (\partial{A})^{\epsilon}\big]\big{|}\nonumber\\
&+2\sup_{A\in\mathcal{A}}\big{|}\mathbf{P}_*\big[\hat{\bm{T}}_n^{*(PRB)}\in (\partial{A})^{\epsilon}\big]-\mathbf{P}\big[\bm{G}_{1n}\in (\partial{A})^{\epsilon}\big]\big{|}+\sup_{A\in\mathcal{A}}\Big{|}\mathbf{P}_*\Big[\hat{\bm{T}}_n^{*(PRB)}\in A\Big]-\mathbf{P}\Big[\bm{G}_{1n}\in A\Big]\Big{|}\nonumber\\
&=o_p(1),
\end{align}
which follows collectively from above arguments. Hence we are done. \hfill $\blacksquare$

\section{Regime II: Proofs of Requisite Lemmas and Main Results}\label{sec:proofreg2}

In this section, lemmas related to Proposition \ref{prop:solutionkktglm} and Theorem \ref{thm:failgauss} have been stated and proved in section \ref{sec:lemmaC}. Whereas, lemmas corresponding to Theorem \ref{thm:bootapproxglmvsc} and \ref{thm:prbbootapproxglmvsc} have been addressed in section \ref{sec:lemmaD}. Proofs of Proposition \ref{prop:solutionkktglm}, Theorem \ref{thm:failgauss}, \ref{thm:bootapproxglmvsc} and \ref{thm:prbbootapproxglmvsc} have been provided in section \ref{sec:mainresults}.

\subsection{Lemmas related to Proposition \ref{prop:solutionkktglm} and Theorem \ref{thm:failgauss}}\label{sec:lemmaC}
\begin{lemma}\label{lem:leadingpart2}
  Suppose (A.2)-(A.4), (A.6) and (B.1) hold true. Consider a Gaussian random vector  $\bm{G}_{2n}\sim N_{d_0}(\bm{0},\tilde{\bm{S}}_{n,11}).$ Then provided $d_0=o[n^{2\{1-2\gamma_1\}/5}]$  we have:
\begin{align*}
    &\sup_{A\in\mathcal{C}}\Big{|}\mathbf{P}\Big(\bm{T}_n^{(1)}\in A\Big)-\mathbf{P}\Big(\bm{G}_{2n}\in A\Big)\Big{|}=o(1).
\end{align*}   
\end{lemma}
Proof of Lemma \ref{lem:leadingpart2} This proof follows exactly in the same path as the proof of Lemma \ref{lem:mainlemmafangandkoikeconvex}. \hfill $\blacksquare$

\begin{rem}
 We say that, a random variable $Y$ with finite expectation $\mathbf{E}(Y)$, is sub-Exponential with non-negative parameters $(\nu,b)$ (hereafter refer to as SE($\nu,b$)) if $$\mathbf{E}\Big[e^{\lambda(Y-\mathbf{E}(Y))}\Big]\leq e^{\frac{\nu^2\lambda^2}{2}}\;\text{for all}\;|\lambda|<\frac{1}{b}.$$
For a symmetric random matrix $\bm{Q}^{d\times d}$, its polynomial moments assuming that they exist, are the matrices given by $\mathbf{E}[\bm{Q}^j]$, for $j=1,2,...$ The variance of the matrix $\bm{Q}$ is given by $\bm{Var}(\bm{Q})=\mathbf{E}[\bm{Q}^2]-(\mathbf{E}[\bm{Q}])^2$. The moment generating function of $\bm{Q}$ is the matrix valued mapping $\bm{\Xi}_{\bm{Q}}:\mathbf{R}\mapsto\mathbf{R}^{d\times d}$ given by, $$\bm{\Xi}_{\bm{Q}}(t):=\mathbf{E}\big[e^{t\mathbf{Q}}\big]=\sum_{k=0}^{\infty}\dfrac{t^k}{k!}\mathbf{E}[\bm{Q}^k],$$ for all $t$ in an interval containing $0$.
\end{rem}

\begin{lemma}\label{lem:univariatebernstein}
 Suppose $Y_i$'s be zero-mean independent $SE(\nu_i,b_i)$ random variables for all $i\in\{1,..,n\}$. Denote $\bm{a}=(a_1,..,a_n)^\top$. Then for every $t\geq 0$, there exists an absolute constant $c>0$ such that,
$$\mathbf{P}\Big[\big{|}\sum_{i=1}^{n}a_iY_i\big{|}>t\Big]\leq 2\exp\Bigg\{-c\;\mbox{min}\Bigg(\frac{t^2}{\sum_{i=1}^{n}a_i^2\nu_i^2},\frac{t}{\mbox{max}_{i\in\{1,..,n\}b_i|a_i|}}\Bigg)\Bigg\}$$
\end{lemma}
Proof of Lemma \ref{lem:univariatebernstein} This lemma follows from Theorem 2.8.2 of  \citet{vershynin2018high}. \hfill $\blacksquare$

\begin{lemma}\label{lem:BerntoSE}
Suppose $Y$ be a zero-mean random variable with $\sigma^2=\mathbf{E}(Y^2)<\infty$ such that for some $b>0$ it satisfies,
\begin{align}\label{eqn:7999}
\mathbf{E}[Y^j]\leq \frac{j!}{2}b^{j-2}\sigma^2\;\text{for}\;j=3,4,....     
\end{align}
Then $Y$ is a $SE(\nu=2^{1/2}\sigma,2b)$ random variable provided $|\lambda|<\frac{1}{2b}$.
\end{lemma}
Proof of Lemma \ref{lem:BerntoSE} This lemma follows from the definition of SE random variable. \hfill $\blacksquare$

\begin{lemma}\label{lem:bernstein}
Suppose a symmetric zero-mean random matrix $\bm{Q}$ satisfies:
\begin{align}\label{eqn:800}
\mathbf{E}[\bm{Q}^j]\leq \frac{j!}{2}b^{j-2}\bm{Var}(\bm{Q})\;\text{for}\;b>0\;\text{and}\; j=3,4,....  \end{align}
Then $$\bm{\Xi}_{\bm{Q}}(t)\leq \exp \Bigg(\dfrac{t^2\bm{Var}(\bm{Q})}{2[1-b|t|]}\Bigg)\;\text{for all}\; |t|<\frac{1}{b}.$$
\end{lemma}
Proof of Lemma \ref{lem:bernstein} This result is just the matrix analogue of Lemma \ref{lem:BerntoSE} and proved as Lemma 6.11 in \citet{wainwright2019high}. \hfill $\blacksquare$

\begin{lemma}\label{lem:wainwright}
Suppose $\{\bm{Q}_i\}_{i=1}^{n}$ be a sequence of independent, zero-mean, symmetric random matrices satisfying condition (\ref{eqn:800}) with parameter $b>0$. Then for $\epsilon>0$;
\begin{align}\label{eqn:801}
\mathbf{P}\Bigg[\Bigg{\|}\frac{1}{n}\sum_{i=1}^{n}\bm{Q}_i\Bigg{\|}>\epsilon\Bigg]\leq 2\Big[rank\Big(\sum_{i=1}^{n}\bm{Var}(\bm{Q}_i)\Big)\Big]\exp\Bigg\{-\frac{n\epsilon^2}{2(\sigma^2+b\epsilon)}\Bigg\}    
\end{align}
where, $\sigma^2=\Bigg{\|}\frac{1}{n}\sum_{i=1}^{n}\bm{Var}(\bm{Q}_i)\Bigg{\|}$.
\end{lemma}
Proof of Lemma \ref{lem:wainwright} This result is proved as Theorem 6.17 in \citet{wainwright2019high}. \hfill $\blacksquare$

\begin{lemma}\label{lem:closeL21EL21}
Assume (A.3), (A.4) and (A.6) hold true. Then for $\tau>0$ and provided $\log (dd_0)=o\Big(n^{2\tau/3}\Big)$ with $0<\tau<3/2$, we have,
$$max_{j\notin\mathcal{A}_n}\Big{\|}(\mathbf{L}_{n,21})_{j\cdot}-\{\mathbf{E}(\mathbf{L}_{n,21})\}_{j\cdot}\Big{\|}=o\Bigg(\frac{d_0^{1/2}}{n^{1/2}}n^{\tau/3}\Bigg)\;\;\text{with probability at least}\; 1-o(1).$$
\end{lemma}
Proof of Lemma \ref{lem:closeL21EL21} Recall that, 
$$\bm{L}_{n,21}=n^{-1}\sum_{i=1}^{n}\bm{x}_i^{(2)}\bm{x}_i^{(1)^\top}\Big\{\big[(g^{-1})^\prime(\bm{x}_i^\top\bm{\beta})\big]h^\prime(\bm{x}_i^\top\bm{\beta})-\{y_i-g^{-1}(\bm{x}_i^\top\bm{\beta})\}h^{\prime\prime}(\bm{x}_i^\top\bm{\beta})\Big\}$$. 
This simply implies that for any $j\notin\mathcal{A}_n$ and $k\in\mathcal{A}_n$,
$$(\mathbf{L}_{n,21})_{jk}-\{\mathbf{E}(\mathbf{L}_{n,21})\}_{jk}=-n^{-1}\sum_{i=1}^{n}x_{ij}^{(2)}x_{ik}^{(1)}\Big\{y_i-g^{-1}(\bm{x}_i^\top\bm{\beta})\Big\}h^{\prime\prime}(\bm{x}_i^\top\bm{\beta})$$
Hence for any $\epsilon>0$ we have,
\begin{align}\label{eqn:802}
&\mathbf{P}\Bigg[max_{j\notin\mathcal{A}_n}\Big{\|}(\mathbf{L}_{n,21})_{j\cdot}-\{\mathbf{E}(\mathbf{L}_{n,21})\}_{j\cdot}\Big{\|}>\epsilon\Bigg(\frac{d_0^{1/2}}{n^{1/2}}n^{\tau/3}\Bigg)\Bigg]\nonumber\\
&\leq \sum_{j=d_0+1}^{d}\mathbf{P}\Bigg[d_0^{1/2}\Big{\|}(\mathbf{L}_{n,21})_{j\cdot}-\{\mathbf{E}(\mathbf{L}_{n,21})\}_{j\cdot}\Big{\|}_{\infty}>\epsilon\Bigg(\frac{d_0^{1/2}}{n^{1/2}}n^{\tau/3}\Bigg)\Bigg]\nonumber\\
&\leq \sum_{j=d_0+1}^{d}\sum_{k=1}^{d_0}\mathbf{P}\Bigg[\Big{|}n^{-1/2}\sum_{i=1}^{n}x_{ij}^{(2)}x_{ik}^{(1)}\Big\{y_i-g^{-1}(\bm{x}_i^\top\bm{\beta})\Big\}h^{\prime\prime}(\bm{x}_i^\top\bm{\beta})\Big{|}>\epsilon n^{\tau/3}\Bigg]
\end{align}
Now $\{y_i-g^{-1}(\bm{x}_i^\top\bm{\beta})\}_{i=1}^{n}$ is an independent zero-mean sequence of random variables satisfying (\ref{eqn:7999}) with $|\lambda|<\frac{1}{2b}$. This implies that, $y_i-g^{-1}(\bm{x}_i^\top\bm{\beta})$ is $SE(2^{1/2}\sigma_i,2b)$ for all $i\in\{1,..,n\}$ due to Lemma \ref{lem:BerntoSE} with $\sigma_i^2=\mathbf{E}(y_i-g^{-1}(\bm{x}_i^\top\bm{\beta}))^2$.
Now consider $a_i=\frac{1}{n^{1/2}}x_{ij}^{(2)}x_{ik}^{(1)}h^{\prime\prime}(\bm{x}_i^\top\bm{\beta})$, $\nu_i=2^{1/2}\sigma_i$ and $b_i=2b$ for all $i$, in the notations of Lemma \ref{lem:univariatebernstein}. Then for $t=\epsilon n^{\frac{\tau}{3}}$ and due the assumptions, there exist constants $c,C_1,C_2>0$ such that,
from (\ref{eqn:802}) and Lemma \ref{lem:univariatebernstein} give us,
\begin{align}\label{eqn:803}
& \mathbf{P}\Bigg[max_{j\notin\mathcal{A}_n}\Big{\|}(\mathbf{L}_{n,21})_{j\cdot}-\{\mathbf{E}(\mathbf{L}_{n,21})\}_{j\cdot}\Big{\|}>\epsilon\Bigg(\frac{d_0^{1/2}}{n^{1/2}}n^{\tau/3}\Bigg)\Bigg]\nonumber\\
&\leq 2dd_0\exp\Bigg\{-\;\mbox{min}\Bigg(\frac{c\epsilon^2n^{\frac{2\tau}{3}}}{2C_1},\frac{c\epsilon n^{\frac{1}{2}+\frac{\tau}{3}}}{2bC_2}\Bigg)\Bigg\},
\end{align}
which is $o(1)$ under the assumption. Therefore, we are done.\hfill $\blacksquare$

\begin{lemma}\label{lem:closeL11EL11}
 Assume (A.3)-(A.6) are true. Then provided $d_0=o\Big(n^{1-\frac{2\tau}{3}}\Big)$ with $0<\tau<3/2$,  we have,
$$\Big{\|}\mathbf{L}_{n,11}-\mathbf{E}(\mathbf{L}_{n,11})\Big{\|}=o\Bigg(\frac{d_0^{1/2}}{n^{1/2}}n^{\tau/3}\Bigg)\;\;\text{with probability at least}\; 1-o(1).$$
\end{lemma}
Proof of Lemma \ref{lem:closeL11EL11} Recall that,
\begin{align*}
&\mathbf{L}_{n,11}-\mathbf{E}(\mathbf{L}_{n,11})=\frac{1}{n}\sum_{i=1}^{n}\underbrace{\Big\{y_i-g^{-1}(\bm{x}_i^\top\bm{\beta})\Big\}[-h^{\prime\prime}(\bm{x}_i^\top\bm{\beta})]\bm{x}_i^{(1)}\bm{x}_i^{(1)^\top}}_{\bm{Q}_i\;\text{(say)}}
\end{align*}
Note that $\bm{Q}_i$'s are zero-mean symmetric random matrices with $$Var(\bm{Q}_i)=\Big[\bm{x}_i^{(1)}\bm{x}_i^{(1)^\top}\Big]^2[-h^{\prime\prime}(\bm{x}_i^\top\bm{\beta})]^2\mathbf{E}\Big[y_i-g^{-1}(\bm{x}_i^\top\bm{\beta})\Big]^2.$$
Now since, $\{y_i-\mathbf{E}(y_i)\}_{i=1}^{n}$ be a sequence of independent random variables satisfying (\ref{eqn:7999}) with $b>0$, therefore it is easy to check that,
$\bm{Q}_i$ satisfies condition (\ref{eqn:800}) with parameters $b^*=bd_0[-h^{\prime\prime}(\bm{x}_i^\top\bm{\beta})]>0$ and $Var(\bm{Q}_i)$. Also under the assumptions, it's a simple calculation that 
$$\sigma^2=\Bigg{\|}\frac{1}{n}\sum_{i=1}^{n}\bm{Var}(\bm{Q}_i)\Bigg{\|}=O(d_0).$$
Therefore noting this, there exist two constants $K_1,K_2>0$ such that for $\epsilon>0$, Lemma \ref{lem:wainwright} gives us,
\begin{align*}
&\mathbf{P}\Bigg[\Big{\|}\mathbf{L}_{n,11}-\mathbf{E}(\mathbf{L}_{n,11})\Big{\|}>\epsilon\Bigg(\frac{d_0^{1/2}}{n^{1/2}}n^{\tau/3}\Bigg)\Bigg]\leq 2d_0\exp\Bigg[-\dfrac{d_0\epsilon^{2}n^{\frac{2\tau}{3}}}{2\Big\{K_1d_0+K_2d_0b\epsilon\frac{d_0^{1/2}}{n^{\frac{1}{2}-\frac{\tau}{3}}}\Big\}}\Bigg]=o(1).
\end{align*}
The proof is complete.\hfill$\blacksquare$

\begin{lemma}\label{lem:inversecloseL11EL11}
 Suppose the assumptions of Lemma \ref{lem:closeL11EL11} hold true. Also assume that (B.1) is true. Then provided $d_0=o\Big(n^{1-2\gamma_1-\frac{2\tau}{3}}\Big)$ with $\gamma_1+\frac{\tau}{3}<1/2$ we have,
$$\Big{\|}\big\{\mathbf{L}_{n,11}\big\}^{-1}-\big\{\mathbf{E}(\mathbf{L}_{n,11})\big\}^{-1}\Big{\|}=o\Bigg(\frac{d_0^{1/2}}{n^{\frac{1}{2}-2\gamma_1}}n^{\tau/3}\Bigg)\;\;\text{with probability at least}\; 1-o(1).$$
\end{lemma}
Proof of Lemma \ref{lem:inversecloseL11EL11} We note that,
$$\Big{\|}\big\{\mathbf{L}_{n,11}\big\}^{-1}-\big\{\mathbf{E}(\mathbf{L}_{n,11})\big\}^{-1}\Big{\|}\leq \Big{\|}\big\{\mathbf{L}_{n,11}\big\}^{-1}\Big{\|}\Big{\|}\mathbf{L}_{n,11}-\mathbf{E}(\mathbf{L}_{n,11})\Big{\|}\Big{\|}\big\{\mathbf{E}(\mathbf{L}_{n,11})\big\}^{-1}\Big{\|}$$
Then the proof will follow from Lemma \ref{lem:closeL11EL11} and Lemma \ref{lem:orderinvLnhat}. Hence we skip the details. \hfill $\blacksquare$

\begin{lemma}\label{lem:orderEL21nj}
Suppose, (A.3) and (A.6) are true. Then we have,
$$\mbox{max}_{j\notin \mathcal{A}_n}\Big{\|}\{\mathbf{E}(\mathbf{L}_{n,21})\}_{j\cdot}\Big{\|}=O(d_0^{1/2})$$
\end{lemma}
Proof of Lemma \ref{lem:orderEL21nj} We note that, 
\begin{align*}
\mbox{max}_{j\notin \mathcal{A}_n}\Big{\|}\{\mathbf{E}(\mathbf{L}_{n,21})\}_{j\cdot}\Big{\|}&\leq d_0^{1/2}\mbox{max}_{j\notin \mathcal{A}_n}\mbox{max}_{k\in \mathcal{A}_n}\Big{|}\{\mathbf{E}(\mathbf{L}_{n,21})\}_{jk}\Big{|}\\
&\leq d_0^{1/2}\mbox{max}_{j\notin \mathcal{A}_n}\mbox{max}_{k\in \mathcal{A}_n}\Bigg{|}n^{-1}\sum_{i=1}^{n}x_{ij}^{(2)}x_{ik}^{(1)}\big[(g^{-1})^\prime(\bm{x}_i^\top\bm{\beta})\big]h^\prime(\bm{x}_i^\top\bm{\beta})\Bigg{|}\\
&\leq d_0^{1/2}\Big[\mbox{max}_{j\notin \mathcal{A}_n}\mbox{max}_{i\in\{1,..,n\}}|x_{ij}^{(2)}|\Big]\Big[\mbox{max}_{k\in \mathcal{A}_n}\mbox{max}_{i\in\{1,..,n\}}|x_{ik}^{(1)}|\Big]\\
&\;\;\;\;\;\;\;\;\;\;\;\;\;\;\;\;\;\;\;\;\;\;\;\;\;\;\times\Big[\mbox{max}_{i\in\{1,..,n\}}|(g^{-1})^\prime(\bm{x}_i^\top\bm{\beta})|\Big]\Big[\mbox{max}_{i\in\{1,..,n\}}|h^\prime(\bm{x}_i^\top\bm{\beta})|\Big]\\
&=O(d_0^{1/2})
\end{align*}
The proof is complete.\hfill $\blacksquare$

\begin{lemma}\label{lem:EL21inverseEL11nWn1}
 Assume that (A.4), (A.6) and (B.2) are true. Then provided $\log(d)=o(n^{2\tau})$ with $\tau<1/2$, we have,
$$\Big{\|}\big\{\mathbf{E}(\bm{L}_{n,21})\big\}\big\{\mathbf{E}(\mathbf{L}_{n,11})\big\}^{-1}\bm{W}_n^{(1)}\Big{\|}_{\infty}=o(n^{\tau})\;\;\text{with probability at least}\; 1-o(1)$$
\end{lemma}
Proof of Lemma \ref{lem:EL21inverseEL11nWn1} For any $\epsilon>0$ we note that,
\begin{align}\label{eqn:invfer}
&\mathbf{P}\Bigg[\Big{\|}\big\{\mathbf{E}(\bm{L}_{n,21})\big\}\big\{\mathbf{E}(\mathbf{L}_{n,11})\big\}^{-1}\bm{W}_n^{(1)}\Big{\|}_{\infty}>\epsilon n^{\tau}\Bigg]\nonumber\\
&\leq \mathbf{P}\Bigg[\mbox{max}_{j\notin\mathcal{A}_n}\Bigg{|}\Bigg(\big\{\mathbf{E}(\bm{L}_{n,21})\big\}\big\{\mathbf{E}(\mathbf{L}_{n,11})\big\}^{-1}\Bigg)_{j\cdot}^\top\bm{W}_n^{(1)}\Bigg{|}>\epsilon n^{\tau}\Bigg]\nonumber\\
&\leq \sum_{j=d_0+1}^{d}\mathbf{P}\Bigg[\Bigg{|}\sum_{i=1}^{n}\frac{1}{n^{1/2}}\Bigg(\big\{\mathbf{E}(\bm{L}_{n,21})\big\}\big\{\mathbf{E}(\mathbf{L}_{n,11})\big\}^{-1}\Bigg)_{j\cdot}^\top\bm{x}_i^{(1)}h^\prime(\bm{x}_i^\top\bm{\beta})\Big\{y_i-g^{-1}(\bm{x}_i^\top\bm{\beta})\Big\}\Bigg{|}>\epsilon n^{\tau}\Bigg]
\end{align}
Now $\{y_i-g^{-1}(\bm{x}_i^\top\bm{\beta})\}_{i=1}^{n}$ is an independent zero-mean sequence of random variables satisfying (\ref{eqn:7999}) with $|\lambda|<\frac{1}{2b}$. This implies that, $y_i-g^{-1}(\bm{x}_i^\top\bm{\beta})$ is $SE(2^{1/2}\sigma_i,2b)$ for all $i\in\{1,..,n\}$ due to Lemma \ref{lem:BerntoSE} with $\sigma_i^2=\mathbf{E}(y_i-g^{-1}(\bm{x}_i^\top\bm{\beta}))^2$.\\
Now consider $a_i=\frac{1}{n^{1/2}}\Bigg(\big\{\mathbf{E}(\bm{L}_{n,21})\big\}\big\{\mathbf{E}(\mathbf{L}_{n,11})\big\}^{-1}\Bigg)_{j\cdot}^\top\bm{x}_i^{(1)}h^\prime(\bm{x}_i^\top\bm{\beta})$, $\nu_i=2^{1/2}\sigma_i$ and $b_i=2b$ for all $i$, in the notations of Lemma \ref{lem:univariatebernstein}. Then for $t=\epsilon n^{\tau}$ and due the assumptions, there exist constants $c,B_1,B_2>0$ such that,
from (\ref{eqn:invfer}) and Lemma \ref{lem:univariatebernstein} give us,
\begin{align}\label{eqn:pr}
&\mathbf{P}\Bigg[\Big{\|}\big\{\mathbf{E}(\bm{L}_{n,21})\big\}\big\{\mathbf{E}(\mathbf{L}_{n,11})\big\}^{-1}\bm{W}_n^{(1)}\Big{\|}_{\infty}>\epsilon n^{\tau}\Bigg]\nonumber\\
&\leq 2(d-d_0)\exp\Bigg\{-\min\Bigg(\frac{c\epsilon^2 n^{2\tau}}{2B_1},\frac{c\epsilon n^{\frac{1}{2}+\tau}}{2bB_2}\Bigg)\Bigg\}=o(1)   \end{align}

Therefore, we are done.\hfill $\blacksquare$

\begin{lemma}\label{lem:orderWn1}
Assume that (A.3), (A.4) and  (A.6) are true. Then provided $\log (d_0)=o(n^{2\tau/3})$ with $\tau<3/2$ we have,  
$$\Big{\|}\bm{W}_n^{(1)}\Big{\|}=o\Big(d_0^{1/2}n^{\frac{\tau}{3}}\Big)\;\text{with probability at least}\; 1-o(1).$$
\end{lemma}
Proof of Lemma \ref{lem:orderWn1} Note that we can consider $a_i=\frac{1}{n^{1/2}}x_{ik}^{(1)}h^\prime(\bm{x}_i^\top\bm{\beta})$ in the set-up of Lemma \ref{lem:univariatebernstein} to see that $\bm{W}_n^{(1)}=\sum_{i=1}^{n}a_i\big\{y_i-g^{-1}(\bm{x}_i^\top\bm{\beta})\big\}$. Then we follow the proof steps of earlier lemmas to conclude it. So we skip the details. \hfill $\blacksquare$

\begin{lemma}\label{lem:orderWn2}
 Assume that (A.3), (A.4) and  (A.6) are true. Then provided $\log(d)=o\big(n^{2\tau}\big)$ with $\tau<1/2$, we have,  
$$\Big{\|}\bm{W}_n^{(2)}\Big{\|}_{\infty}=o\big(n^{\tau}\big)\;\text{with probability at least}\; 1-o(1).$$    
\end{lemma}
Proof of Lemma \ref{lem:orderWn2} This proof follows similarly from Lemma \ref{lem:orderWn1}. \hfill $\blacksquare$

\begin{lemma}\label{lem:combined}
Suppose the assumptions of Lemma \ref{lem:closeL21EL21}, \ref{lem:inversecloseL11EL11}, \ref{lem:orderEL21nj}, \ref{lem:EL21inverseEL11nWn1} and \ref{lem:orderWn1} hold true. Then provided $d_0=o\Big(n^{\frac{1}{3}+\frac{2\tau}{9}-\frac{4\gamma_1}{3}}\Big)$ with $0<\tau<3/2$ holds, we have
$$\Big{\|}\bm{L}_{n,21}\big\{\bm{L}_{n,11}\big\}^{-1}\bm{W}_n^{(1)}\Big{\|}_{\infty}=o(n^\tau)\;\text{with probability at least}\; 1-o(1).$$
\end{lemma}
Proof of Lemma \ref{lem:combined} Note the following;
\begin{align*}
&\Big{\|}\bm{L}_{n,21}\big\{\bm{L}_{n,11}\big\}^{-1}\bm{W}_n^{(1)}\Big{\|}_{\infty}\\
&\leq\Bigg[max_{j\notin\mathcal{A}_n}\Big{\|}(\mathbf{L}_{n,21})_{j\cdot}-\{\mathbf{E}(\mathbf{L}_{n,21})\}_{j\cdot}\Big{\|}\Bigg]\Bigg[\Big{\|}\big\{\mathbf{L}_{n,11}\big\}^{-1}-\big\{\mathbf{E}(\mathbf{L}_{n,11})\big\}^{-1}\Big{\|}\Bigg]\Big{\|}\bm{W}_n^{(1)}\Big{\|}\\
&\;\;\;\;\;\;\;\;\;\;\;\;\;\;+\Bigg[\mbox{max}_{j\notin \mathcal{A}_n}\Big{\|}\{\mathbf{E}(\mathbf{L}_{n,21})\}_{j\cdot}\Big{\|}\Bigg]\Bigg[\Big{\|}\big\{\mathbf{L}_{n,11}\big\}^{-1}-\big\{\mathbf{E}(\mathbf{L}_{n,11})\big\}^{-1}\Big{\|}\Bigg]\Big{\|}\bm{W}_n^{(1)}\Big{\|}\\
&+\Bigg[max_{j\notin\mathcal{A}_n}\Big{\|}(\mathbf{L}_{n,21})_{j\cdot}-\{\mathbf{E}(\mathbf{L}_{n,21})\}_{j\cdot}\Big{\|}\Bigg]\Bigg[\Big{\|}\big\{\mathbf{E}(\mathbf{L}_{n,11})\big\}^{-1}\Big{\|}\Bigg]\Big{\|}\bm{W}_n^{(1)}\Big{\|}\\
&\qquad\qquad\qquad\qquad\qquad\qquad\qquad\qquad\qquad+\Big{\|}\big\{\mathbf{E}(\bm{L}_{n,21})\big\}\big\{\mathbf{E}(\mathbf{L}_{n,11})\big\}^{-1}\bm{W}_n^{(1)}\Big{\|}_{\infty}
\end{align*}
We have handled each of the term separately under mentioned assumptions in previous lemmas. This will give us that,
$$\Big{\|}\bm{L}_{n,21}\big\{\bm{L}_{n,11}\big\}^{-1}\bm{W}_n^{(1)}\Big{\|}_{\infty}=o\Bigg[n^{\tau}\Bigg\{1+\frac{d_0}{n^{\frac{1}{2}+\frac{\tau}{3}-\gamma_1}}+\frac{d_0^{3/2}}{n^{\frac{1}{2}+\frac{\tau}{3}-2\gamma_1}}+\frac{d_0^{3/2}}{n^{1-2\gamma_1}}\Bigg\}\Bigg]=o(n^\tau),$$ with probability at least $1-o(1)$ under the assumption $d_0=o\Big(n^{\frac{1}{3}+\frac{2\tau}{9}-\frac{4\gamma_1}{3}}\Big)$. Therefore the proof is complete. \hfill $\blacksquare$

\begin{lemma}\label{lem:InvLn11Wn1}
Suppose the set-up of Lemma \ref{lem:inversecloseL11EL11} and \ref{lem:orderWn1} are true. Then provided (B.3.i) holds, we have,
$$\Big{\|}\big\{\mathbf{E}(\bm{L}_{n,11})\big\}^{-1}\bm{W}_n^{(1)}\Big{\|}=O\Bigg(\frac{d_0^{1/2}}{n^{\frac{1}{2}-\gamma_1}}\lambda_n\Bigg)\;\text{with probability at least}\; 1-o(1).$$
\end{lemma}
Proof of Lemma \ref{lem:InvLn11Wn1} This proof follows from Lemma \ref{lem:inversecloseL11EL11} and \ref{lem:orderWn1} under the condition provided. \hfill $\blacksquare$

\begin{lemma}\label{lem:KKTconditions}
The solution to equation (\ref{eqn:deflasso}) should satisfy,
\begin{align}\label{eqn:12345}
&-\bm{W}_n+\bm{L}_{n}\bar{\bm{u}}_n+\frac{1}{2n^{3/2}}\sum_{i=1}^{n}\bm{x}_i\Bigg[h_1^{\prime\prime\prime}(z_i)-y_ih^{\prime\prime\prime}(z_i)\Bigg][\bm{x}_i^\top\bar{\bm{u}}_n]^2= -\frac{\lambda_n}{n^{1/2}}\bm{sgn}(\bar{\bm{\beta}}_n),\nonumber\\
&\;\;\;\;\;\;\;\;\;\;\;\;\;\;\;\;\;\;\;\;\;\;\;\;\;\;\;\;\;\;\;\;\;\;\;\;\;\;\;\;\;\;\;\;\;\;\;\;\;\;\;\;\;\;\;\;\;\;\;\;\;\;\;\;\;\;\;\;\;\;\;\;\;\;\;\;\;\;\;\;\;\;\;\;\;\;\;\;\;\;\;\;\;\;\;\;\;\;\;\text{provided}\;\bar{\bm{\beta}}_n\neq \bm{0}\nonumber\\
&-\frac{\lambda_n}{n^{1/2}}\mathbf{1}\leq -\bm{W}_n+\bm{L}_{n}\bar{\bm{u}}_n+\frac{1}{2n^{3/2}}\sum_{i=1}^{n}\bm{x}_i\Bigg[h_1^{\prime\prime\prime}(z_i)-y_ih^{\prime\prime\prime}(z_i)\Bigg][\bm{x}_i^\top\bar{\bm{u}}_n]^2 \leq \frac{\lambda_n}{n^{1/2}}\mathbf{1}\nonumber\\
&\;\;\;\;\;\;\;\;\;\;\;\;\;\;\;\;\;\;\;\;\;\;\;\;\;\;\;\;\;\;\;\;\;\;\;\;\;\;\;\;\;\;\;\;\;\;\;\;\;\;\;\;\;\;\;\;\;\;\;\;\;\;\;\;\;\;\;\;\;\;\;\;\;\;\;\;\;\;\;\;\;\;\;\;\;\;\;\;\;\;\;\;\;\;\;\;\;\;\;\text{provided}\;\bar{\bm{\beta}}_n=\bm{0}
\end{align}
for some $z_i$ such that $|z_i-\bm{x}_i^\top\bm{\beta}|\leq \frac{\bm{x}_i^\top\bar{\bm{u}}_n}{n^{1/2}}$ for all $i\in\{1,...,n\}$.
\end{lemma}
Proof of Lemma \ref{lem:KKTconditions} Since $Z_n(v)$, the objective function is  strictly convex, therefore (\ref{eqn:12345}) is a direct consequence of Taylor's expansion with respect to $\bm{x}_i^\top\bm{\beta}$ and KKT conditions which are necessary and sufficient.\hfill $\blacksquare$

\begin{lemma}\label{lem:concentrationbarun1}
Suppose (A.2)-(A.7), (B.1) and (B.3) hold true. Then we have,
$$\Big{\|}n^{1/2}\Big(\bar{\bm{\beta}}_n^{(1)}-\bm{\beta}^{(1)}\Big)\Big{\|}=O\Bigg(\frac{d_0^{1/2}}{n^{\frac{1}{2}-\gamma_1}}\lambda_n\Bigg)\;\text{with probability at least}\; 1-o(1),$$
where;
\begin{align}\label{eqn:solnbarun1}
& n^{1/2}\Big(\bar{\bm{\beta}}_n^{(1)}-\bm{\beta}^{(1)}\Big)=\{\mathbf{E}(\bm{L}_{n,11})\}^{-1}\bm{W}_n^{(1)}-\frac{\lambda_n}{n^{1/2}}\{\mathbf{E}(\bm{L}_{n,11})\}^{-1}\bm{sgn}(\bm{\beta}^{(1)})+\bm{Q}_{1n}(\bar{\bm{u}}_n^{(1)}),\nonumber\\
&\text{with}\;\bm{Q}_{1n}(\bar{\bm{u}}_n^{(1)})=\Big[\big\{\mathbf{L}_{n,11}\big\}^{-1}-\big\{\mathbf{E}(\mathbf{L}_{n,11})\big\}^{-1}\Big]\bm{W}_n^{(1)}-\frac{\lambda_n}{n^{1/2}}\Big[\big\{\mathbf{L}_{n,11}\big\}^{-1}-\big\{\mathbf{E}(\mathbf{L}_{n,11})\big\}^{-1}\Big]\bm{sgn}(\bm{\beta}^{(1)})\nonumber\\
&\;\;\;\;\;\;\;\;\;\;\;\;\;\;\;\;\;\;\;\;\;\;\;\;\;\;\;\;\;\;\;\;+\frac{\big\{\mathbf{E(L}_{n,11})\big\}^{-1}}{2n^{3/2}}\sum_{i=1}^{n}\bm{x}_i^{(1)}\Bigg[h_1^{\prime\prime\prime}(z_i)-y_ih^{\prime\prime\prime}(z_i)\Bigg]\Bigg[\bm{x}_i^{(1)\top}\Big\{n^{1/2}\Big(\bar{\bm{\beta}}_n^{(1)}-\bm{\beta}^{(1)}\Big)\Big\}\Bigg]^2\nonumber\\
 &+\frac{\Big[\big\{\mathbf{L}_{n,11}\big\}^{-1}-\big\{\mathbf{E}(\mathbf{L}_{n,11})\big\}^{-1}\Big]}{2n^{3/2}}\sum_{i=1}^{n}\bm{x}_i^{(1)}\Bigg[h_1^{\prime\prime\prime}(z_i)-y_ih^{\prime\prime\prime}(z_i)\Bigg]\Bigg[\bm{x}_i^{(1)\top}\Big\{n^{1/2}\Big(\bar{\bm{\beta}}_n^{(1)}-\bm{\beta}^{(1)}\Big)\Big\}\Bigg]^2
 \end{align}
\end{lemma}

Proof of Lemma \ref{lem:concentrationbarun1} Note that we can write (\ref{eqn:solnbarun1}) as, 
\begin{align}\label{eqn:brower}
n^{1/2}\Big(\bar{\bm{\beta}}_n^{(1)}-\bm{\beta}^{(1)}\Big)=\bm{f}_n\Big(n^{1/2}\Big(\bar{\bm{\beta}}_n^{(1)}-\bm{\beta}^{(1)}\Big)\Big)\end{align}
with $\bm{f}_n:\mathbb{R}^{d_0}\mapsto\mathbb{R}^{d_0}$ being some continuous function.  Now due to Lemma \ref{lem:InvLn11Wn1} we have, $$\Big{\|}\big\{\mathbf{E}(\bm{L}_{n,11})\big\}^{-1}\bm{W}_n^{(1)}\Big{\|}=O\Bigg(\frac{d_0^{1/2}}{n^{\frac{1}{2}-\gamma_1}}\lambda_n\Bigg),\;\text{with probability at least}\;1-o(1).$$
Due to assumption (B.1), $\Big{\|}\frac{\lambda_n}{n^{1/2}}\{\mathbf{E}(\bm{L}_{n,11})\}^{-1}\bm{sgn}(\bm{\beta}^{(1)})\Big{\|}=O\Bigg(\frac{d_0^{1/2}}{n^{\frac{1}{2}-\gamma_1}}\lambda_n\Bigg)$. Next for the remainder term,
\begin{align}\label{eqn:concentrationremainder}
&\Big{\|}\bm{Q}_{1n}(\bar{\bm{u}}_n^{(1)})\Big{\|}\leq \Big{\|}\big\{\mathbf{L}_{n,11}\big\}^{-1}-\big\{\mathbf{E}(\mathbf{L}_{n,11})\big\}^{-1}\Big{\|}\Big{\|}\bm{W}_n^{(1)}\Big{\|}+\frac{\lambda_n}{n^{1/2}}\Big{\|}\big\{\mathbf{L}_{n,11}\big\}^{-1}-\big\{\mathbf{E}(\mathbf{L}_{n,11})\big\}^{-1}\Big{\|}\nonumber\\
&\times \Big{\|}\bm{sgn}(\bm{\beta}^{(1)})\Big{\|}+\frac{1}{2n^{1/2}}\Bigg[\Big{\|}\{\mathbf{E}(\bm{L}_{n,11})\}^{-1}\Big{\|}+\Big{\|}\big\{\mathbf{L}_{n,11}\big\}^{-1}-\big\{\mathbf{E}(\mathbf{L}_{n,11})\big\}^{-1}\Big{\|}\Bigg]\nonumber\\
&\times\Bigg[\Big{\|}n^{1/2}\Big(\bar{\bm{\beta}}_n^{(1)}-\bm{\beta}^{(1)}\Big)\Big{\|}^2\Bigg] \Bigg[\frac{1}{n}\mbox{sup}_{\{||\bm{\alpha}||=1,||\bm{\kappa}||=1\}}\sum_{i=1}^{n}|\bm{\alpha}^\top \bm{x}_i^{(1)}||\bm{\kappa}^\top \bm{x}_i^{(1)}|^2\Bigg\{|(g^{-1})^{\prime\prime}(z_i)||h^{\prime}(z_i)|\nonumber\\
&\;\;\;\;\;\;\;\;\;\;\;\;\;\;\;\;\;\;\;\;\;\;\;\;\;\;\;\;\;\;\;\;\;\;\;\;\;\;\;\;\;\;\;\;\;\;\;\;\;\;\;+2|(g^{-1})^{\prime}(z_i)||h^{\prime\prime}(z_i)|+|h^{\prime\prime\prime}(z_i)||y_i-g^{-1}(z_i)|\Bigg\}\Bigg]\nonumber\\
&=O\Bigg[\frac{d_0}{n^{\frac{1}{2}-2\gamma_1-\frac{2\tau}{3}}}+\frac{\lambda_n}{n^{1/2}}\frac{d_0}{n^{\frac{1}{2}-2\gamma_1-\frac{\tau}{3}}}+\frac{1}{n^{\frac{1}{2}-\gamma_1}}\Bigg\{\frac{d_0^{1/2}}{n^{\frac{1}{2}-\gamma_1}}\lambda_n\Bigg\}^2+\frac{d_0^{1/2}}{n^{1-2\gamma_1-\frac{\tau}{3}}}\Bigg\{\frac{d_0^{1/2}}{n^{\frac{1}{2}-\gamma_1}}\lambda_n\Bigg\}^2\Bigg]\nonumber\\
&=O\Bigg(\frac{d_0^{1/2}}{n^{\frac{1}{2}-\gamma_1}}\lambda_n\Bigg),
\end{align}
with probability at least $1-o(1)$ whenever $\Big{\|}n^{1/2}\Big(\bar{\bm{\beta}}_n^{(1)}-\bm{\beta}^{(1)}\Big)\Big{\|}=O\Bigg(\frac{d_0^{1/2}}{n^{\frac{1}{2}-\gamma_1}}\lambda_n\Bigg)$ with probability at least $1-o(1)$ along with (B.3). Noting this fact and equation (\ref{eqn:brower}), we can write,
$$\mathbf{P}\Bigg[\Bigg{\|}\bm{f}_n\Bigg(n^{1/2}\Big(\bar{\bm{\beta}}_n^{(1)}-\bm{\beta}^{(1)}\Big)\Bigg)\Bigg{\|}\leq K_\epsilon\Bigg(\frac{d_0^{1/2}}{n^{\frac{1}{2}-\gamma_1}}\lambda_n\Bigg)\Bigg]\geq 1-\epsilon,$$
whenever $\Big{\|}n^{1/2}\Big(\bar{\bm{\beta}}_n^{(1)}-\bm{\beta}^{(1)}\Big)\Big{\|}=O\Bigg(\frac{d_0^{1/2}}{n^{\frac{1}{2}-\gamma_1}}\lambda_n\Bigg)$ with probability at least $1-o(1)$. Therefore the proof is complete due to Brouwer's fixed point theorem. \hfill $\blacksquare$

\subsection{Lemmas related to Theorem \ref{thm:bootapproxglmvsc} and \ref{thm:prbbootapproxglmvsc}}\label{sec:lemmaD}
For every $\epsilon>0$, there exists a $0<K_\epsilon<\infty$ such that the set $$\tilde{C}_{n,\epsilon}:=\Bigg\{\Big{\|}\bar{\bm{\beta}}_n^{(1)}-\bm{\beta}^{(1)}\Big{\|}\leq K_\epsilon\Big(\frac{d_0^{1/2}}{n^{\frac{1}{2}-\gamma_1}}\frac{\lambda_n}{n^{1/2}}\Big)\Bigg\}\cap\Big\{\bar{\mathcal{A}}_n=\mathcal{A}_n\Big\}$$ has probability at least $1-\epsilon$ for large enough $n$. We will consider everything on this set whenever it's needed.

\begin{lemma}\label{lem:bootleadingpart2}
  Suppose (A.3), (A.4), (A.6), (A.7) and  (B.1) hold true. Assume $\mathbf{E}|G_1^*|^4<\infty$. Consider a Gaussian random vector  $\bm{G}_{2n}\sim N_{d_0}(\bm{0},\tilde{\bm{S}}_{n,11}).$ Then provided we have that $d_0=o[n^{2\{1-2\gamma_1\}/5}]$, then:
\begin{align*}
    &\sup_{A\in\mathcal{C}}\Big{|}\mathbf{P}_*\Big(\bar{\bm{T}}_n^{*(PB)(1)}\in A\Big)-\mathbf{P}\Big(\bm{G}_{2n}\in A\Big)\Big{|}=o_p(1).
\end{align*}   
\end{lemma}
Proof of Lemma \ref{lem:bootleadingpart2} This proof follows exactly in the same path as the proof of Lemma \ref{lem:PBLeadingtermconvex}. \hfill $\blacksquare$

\begin{lemma}\label{lem:hatLn11andELn11}
 Consider the set-up of Lemma \ref{lem:closeL11EL11}. Also assume (B.3.i) and (A.6)-(A.7) hold true. Then we have,
$$\Big{\|}\bar{\mathbf{L}}_{n,11}-\mathbf{E}(\mathbf{L}_{n,11})\Big{\|}=O\Bigg(\frac{d_0^{1/2}}{n^{\frac{1}{2}-\gamma_1}}\frac{\lambda_n}{n^{1/2}}\Bigg)\;\;\text{with probability at least}\; 1-o(1).$$   
\end{lemma}
Proof of Lemma \ref{lem:hatLn11andELn11} The proof follows exactly from Lemma \ref{lem:closeLnhatELn}.\hfill$\blacksquare$

\begin{lemma}\label{lem:bootorderinvLn11hat}
Suppose (B.1) and (B.3.i) hold. Then we have;
$$\|\bar{\bm{L}}_{n,11}^{-1}\|=O(n^{\gamma_1})\;\text{with probability at least}\; 1-o(1).$$
\end{lemma}
 Proof of Lemma \ref{lem:bootorderinvLn11hat} This proof is exactly similar to Lemma \ref{lem:orderinvLnhat}.\hfill$\blacksquare$

\begin{lemma}\label{lem:bootinverseclosehatL11EL11}
 Suppose the assumptions of Lemma \ref{lem:hatLn11andELn11} and \ref{lem:bootorderinvLn11hat} hold true. Then we have,
$$\Big{\|}\big\{\bar{\mathbf{L}}_{n,11}\big\}^{-1}-\big\{\mathbf{E}(\mathbf{L}_{n,11})\big\}^{-1}\Big{\|}=O\Bigg(\frac{d_0^{1/2}}{n^{\frac{1}{2}-3\gamma_1}}\frac{\lambda_n}{n^{1/2}}\Bigg)\;\;\text{with probability at least}\; 1-o(1).$$
\end{lemma}
Proof of Lemma \ref{lem:bootinverseclosehatL11EL11} We note that,
$$\Big{\|}\big\{\bar{\mathbf{L}}_{n,11}\big\}^{-1}-\big\{\mathbf{E}(\mathbf{L}_{n,11})\big\}^{-1}\Big{\|}\leq \Big{\|}\big\{\bar{\mathbf{L}}_{n,11}\big\}^{-1}\Big{\|}\Big{\|}\bar{\mathbf{L}}_{n,11}-\mathbf{E}(\mathbf{L}_{n,11})\Big{\|}\Big{\|}\big\{\mathbf{E}(\mathbf{L}_{n,11})\big\}^{-1}\Big{\|}$$
Then the proof will follow from Lemma \ref{lem:hatLn11andELn11} and Lemma \ref{lem:bootorderinvLn11hat}. Hence we skip the details. \hfill $\blacksquare$

\begin{lemma}\label{lem:boothatLn21ELn21}
Consider the set-up of Lemma \ref{lem:concentrationbarun1}. Then we have,
$$max_{j\notin\mathcal{A}_n}\Big{\|}(\bar{\mathbf{L}}_{n,21})_{j\cdot}-\{\mathbf{E}(\mathbf{L}_{n,21})\}_{j\cdot}\Big{\|}=O\Bigg(\frac{d_0}{n^{\frac{1}{2}-\gamma_1}}\frac{\lambda_n}{n^{1/2}}\Bigg)\;\;\text{with probability at least}\; 1-o(1).$$
\end{lemma}
Proof of Lemma \ref{lem:boothatLn21ELn21} 
Note that, 
$$\bar{\mathbf{L}}_{n,21}=n^{-1}\sum_{i=1}^{n}\bm{x}_i^{(2)}\bm{x}_i^{(1)\top}\Big[\big\{(g^{-1})^\prime(\bm{x}_i^\top\bar{\bm{\beta}}_n)\big\}h^\prime(\bm{x}_i^\top\bar{\bm{\beta}}_n)\Big]$$
Then on the set $\tilde{C}_{n,\epsilon}$ it's easy to verify that,
\begin{align}\label{eqn:haabijaabi}
 &\mbox{max}_{j\notin\mathcal{A}_n}\Big{\|}(\bar{\mathbf{L}}_{n,21})_{j\cdot}-\{\mathbf{E}(\mathbf{L}_{n,21})\}_{j\cdot}\Big{\|}\nonumber\\
 &\leq d_0^{1/2}\mbox{max}_{j\notin\mathcal{A}_n}\mbox{max}_{k\in\mathcal{A}_n}\Bigg[\Bigg{|}n^{-1}\sum_{i=1}^{n}x_{ij}^{(2)}x_{ik}^{(1)}\Bigg\{(g^{-1})^\prime(\bm{x}_i^{(1)\top}\bar{\bm{\beta}}_n^{(1)})h^\prime(\bm{x}_i^{(1)\top}\bar{\bm{\beta}}_n^{(1)})\nonumber\\
 &\;\;\;\;\;\;\qquad\qquad\qquad-(g^{-1})^\prime(\bm{x}_i^{(1)\top}\bm{\beta}^{(1)})h^\prime(\bm{x}_i^{(1)\top}\bm{\beta}^{(1)})\Bigg\}\Bigg{|}\Bigg]\nonumber\\
 &\leq d_0^{1/2}\mbox{max}_{j\notin\mathcal{A}_n}\mbox{max}_{k\in\mathcal{A}_n}\Big[K_{11}\Big]\;\text{(say)},
\end{align}
Now for some $\bar{z}_i$ with $|\bar{z}_i-\bm{x}_i^{(1)\top}\bm{\beta}^{(1)}|\leq \bm{x}_i^{(1)\top}\Big(\bar{\bm{\beta}}_n^{(1)}-\bm{\beta}^{(1)}\Big)$; it's just algebraic manipulation to see that,
\begin{align*}
&d_0^{1/2}\!
\max_{j\notin\mathcal{A}_n}\!
\max_{k\in\mathcal{A}_n} K_{11}\nonumber\\
&\le 
d_0^{1/2}
\Big(\max_{j\notin\mathcal{A}_n}\max_{i}|x_{ij}^{(2)}|\Big)
\Big(\max_{k\in\mathcal{A}_n}\max_{i}|x_{ik}^{(1)}|\Big)
\|\bar{\bm\beta}_n^{(1)} - \bm\beta^{(1)}\|
\\
&\quad\times 
\Big\{\max_i \big|h'({\bm x_i^{(1)}}^\top\bm\beta^{(1)})\big|\Big\}
\Big[\sup_{\|\bm\alpha\|=1} n^{-1}\!\sum_{i}
      |\bm\alpha^\top\bm x_i^{(1)}|^2\Big]^{1/2}
\Big[n^{-1}\!\sum_i |(g^{-1})^{\prime\prime}(\bar z_i)|^2\Big]^{1/2}
\\[0.4em]
&\quad
+\, d_0^{1/2}
\Big(\max_{j\notin\mathcal{A}_n}\max_{i}|x_{ij}^{(2)}|\Big)
\Big(\max_{k\in\mathcal{A}_n}\max_{i}|x_{ik}^{(1)}|\Big)
\|\bar{\bm\beta}_n^{(1)} - \bm\beta^{(1)}\|
\\
&\quad\times
\Big\{\max_i\big|(g^{-1})'({\bm x_i^{(1)}}^\top\bm\beta^{(1)})\big|\Big\}
\Big[\sup_{\|\bm\alpha\|=1} n^{-1}\!\sum_i 
      |\bm\alpha^\top\bm x_i^{(1)}|^2\Big]^{1/2}\Big[n^{-1}\!\sum_i |h''(\bar z_i)|^2\Big]^{1/2}
\\[0.4em]
&\quad
+\, d_0^{1/2}
\Big(\max_{j\notin\mathcal{A}_n}\max_{i}|x_{ij}^{(2)}|\Big)
\Big(\max_{k\in\mathcal{A}_n}\max_{i}|x_{ik}^{(1)}|\Big)
\|\bar{\bm\beta}_n^{(1)} - \bm\beta^{(1)}\|^{2}
\\
&\quad\times
\Big[\sup_{\|\bm\alpha\|=1} n^{-1}\!\sum_i
      |\bm\alpha^\top\bm x_i^{(1)}|^{4}\Big]^{1/2}
\Big[n^{-1}\!\sum_i |h''(\bar z_i)|^{4}\Big]^{1/4}
\Big[n^{-1}\!\sum_i |(g^{-1})''(\bar z_i)|^{4}\Big]^{1/4}
\\[0.5em]
&=
O\!\left(
\frac{d_0}{n^{1/2 - \gamma_1}} \frac{\lambda_n}{n^{1/2}}
+
\frac{d_0^{3/2}}{n^{1 - 2\gamma_1}} \frac{\lambda_n^2}{n}
\right)
\end{align*}

on the set $\tilde{C}_{n,\epsilon}$ with probability at least $1-\epsilon$ under the existing assumptions. The proof is complete due to assumption (B.3.i).\hfill$\blacksquare$

\begin{lemma}\label{lem:bootleadconcen}
Suppose $\{\mu_{G^*}^{-1}G_i^*-1\}_{i=1}^{n}$ be sequence of iid random variables satisfying equation (\ref{eqn:7999}) with $b>0$ and $|\lambda|<1/2b$. Then under the assumptions (A.3), (A.4), (A.6), (A.7) and provided $\log(d_0)=o(n^{2\tau/3})$ with $0<\tau<3/2$, we have,
$$\Big{\|}\bar{\bm{W}}_n^{*(PB)(1)}\Big{\|}=o_{p^*}\Big(d_0^{1/2}n^{\frac{\tau}{3}}\Big)\;\text{with probability at least}\; 1-o(1).$$
\end{lemma}
Proof of Lemma \ref{lem:bootleadconcen} Note that for $\epsilon>0$,
\begin{align}\label{eqn:987654}
&\mathbf{P}_*\Big[\Big{\|}\bar{\bm{W}}_n^{*(PB)(1)}\Big{\|}>\epsilon d_0^{1/2}n^{\frac{\tau}{3}}\Big]\nonumber\\
&\leq \mathbf{P}_*\Bigg[\mbox{max}_{j\in\mathcal{A}_n}\Bigg{|}n^{-1/2}\sum_{i=1}^{n}\big\{y_i-g^{-1}\big(\bm{x}_i^\top\bar{\bm{\beta}}_n\big)\big\}h^\prime(\bm{x}_i^\top\bar{\bm{\beta}}_n)x_{ij}^{(1)}\big[\mu_{G^*}^{-1}G_i^*-1\big]\Bigg{|}>\epsilon n^{\tau/3}\Bigg]\nonumber\\
&\leq \sum_{j=1}^{d_0}\mathbf{P}_*\Bigg[\Bigg{|}n^{-1/2}\sum_{i=1}^{n}\big\{y_i-g^{-1}\big(\bm{x}_i^\top\bar{\bm{\beta}}_n\big)\big\}h^\prime(\bm{x}_i^\top\bar{\bm{\beta}}_n)x_{ij}^{(1)}\big[\mu_{G^*}^{-1}G_i^*-1\big]\Bigg{|}>\epsilon n^{\tau/3}\Bigg]
\end{align}
Since $\{\mu_{G^*}^{-1}G_i^*-1\}_{i=1}^{n}$ be sequence of iid random variables satisfying equation (\ref{eqn:7999}) with $b>0$ and $|\lambda|<1/2b$, therefore for all $i\in\{1,...,n\}$, we have $\mu_{G^*}^{-1}G_i^*-1$ is $SE(2^{1/2},2b)$ due to Lemma \ref{lem:BerntoSE}. Consider $a_i=n^{-1/2}\big\{y_i-g^{-1}\big(\bm{x}_i^\top\bar{\bm{\beta}}_n\big)\big\}h^\prime(\bm{x}_i^\top\bar{\bm{\beta}}_n)x_{ij}^{(1)}$, $\nu_i=2^{1/2}$ and $b_i=2b$ for all $i$ in the notation of Lemma \ref{lem:univariatebernstein}. Then choosing $t=\epsilon n^{\tau/3}$ and on the set $\tilde{C}_{n,\epsilon}$ under the given assumptions, this proof is complete following the earlier proof steps of Lemma \ref{lem:orderWn1}.\hfill$\blacksquare$

\begin{lemma}\label{lem:bootlead2concen}
Assume the set-up of Lemma \ref{lem:bootleadconcen}. Then provided $\log d=o\Big(n^{2\tau}\Big)$ with $0<\tau<1/2$, we have,
$$\Big{\|}\bar{\bm{W}}_n^{*(PB)(2)}\Big{\|}_\infty=o_{p^*}\Big(n^{\tau}\Big)\;\text{with probability at least}\; 1-o(1).$$
\end{lemma}
Proof of Lemma \ref{lem:bootlead2concen} This proof follows similarly as that of Lemma \ref{lem:bootleadconcen}.\hfill$\blacksquare$

\begin{lemma}\label{lem:bootEL21inverseEL11nhatWn1}
 Suppose $\{\mu_{G^*}^{-1}G_i^*-1\}_{i=1}^{n}$ be sequence of iid random variables satisfying equation (\ref{eqn:7999}) with $b>0$ and $|\lambda|<1/2b$. Assume that (A.4), (A.6) and (B.2) are true. Then provided $\log d=o\Big(n^{2\tau}\Big)$ with $0<\tau<1/2$ we have,
$$\Big{\|}\big\{\mathbf{E}(\bm{L}_{n,21})\big\}\big\{\mathbf{E}(\mathbf{L}_{n,11})\big\}^{-1}\bar{\bm{W}}_n^{*(PB)(1)}\Big{\|}_{\infty}=o_{p*}(n^{\tau})\;\;\text{with probability at least}\; 1-o(1)$$
\end{lemma}
Proof of Lemma \ref{lem:bootEL21inverseEL11nhatWn1} This result follows similarly from Lemma \ref{lem:EL21inverseEL11nWn1}.\hfill$\blacksquare$

\begin{lemma}\label{lem:bootcombined}
Suppose the assumptions of Lemma \ref{lem:bootinverseclosehatL11EL11}, \ref{lem:boothatLn21ELn21}, \ref{lem:bootleadconcen}, \ref{lem:bootEL21inverseEL11nhatWn1} and \ref{lem:orderEL21nj} are true. Then provided (B.3.i) holds true, we have,
$$\Big{\|}\bar{\bm{L}}_{n,21}\big\{\bar{\bm{L}}_{n,11}\big\}^{-1}\bar{\bm{W}}_n^{*(PB)(1)}\Big{\|}_{\infty}=O_{p*}(n^\tau)\;\text{with probability at least}\; 1-o(1).$$
\end{lemma}
Proof of Lemma \ref{lem:bootcombined} Note the following;
\begin{align*}
&\Big{\|}\bar{\bm{L}}_{n,21}\big\{\bar{\bm{L}}_{n,11}\big\}^{-1}\bar{\bm{W}}_n^{*(PB)(1)}\Big{\|}_{\infty}\\
&\leq\Bigg[max_{j\notin\mathcal{A}_n}\Big{\|}(\bar{\mathbf{L}}_{n,21})_{j\cdot}-\{\mathbf{E}(\mathbf{L}_{n,21})\}_{j\cdot}\Big{\|}\Bigg]\Bigg[\Big{\|}\big\{\bar{\mathbf{L}}_{n,11}\big\}^{-1}-\big\{\mathbf{E}(\mathbf{L}_{n,11})\big\}^{-1}\Big{\|}\Bigg]\Big{\|}\bar{\bm{W}}_n^{*(PB)(1)}\Big{\|}\\
&\;\;\;\;\;\;\;\;\;\;\;\;\;\;+\Bigg[\mbox{max}_{j\notin \mathcal{A}_n}\Big{\|}\{\mathbf{E}(\mathbf{L}_{n,21})\}_{j\cdot}\Big{\|}\Bigg]\Bigg[\Big{\|}\big\{\bar{\mathbf{L}}_{n,11}\big\}^{-1}-\big\{\mathbf{E}(\mathbf{L}_{n,11})\big\}^{-1}\Big{\|}\Bigg]\Big{\|}\bar{\bm{W}}_n^{*(PB)(1)}\Big{\|}\\
&+\Bigg[max_{j\notin\mathcal{A}_n}\Big{\|}(\bar{\mathbf{L}}_{n,21})_{j\cdot}-\{\mathbf{E}(\mathbf{L}_{n,21})\}_{j\cdot}\Big{\|}\Bigg]\Bigg[\Big{\|}\big\{\mathbf{E}(\mathbf{L}_{n,11})\big\}^{-1}\Big{\|}\Bigg]\Big{\|}\bar{\bm{W}}_n^{*(PB)(1)}\Big{\|}\\
&\qquad\qquad\qquad\qquad\qquad\qquad\qquad\qquad\qquad+\Big{\|}\big\{\mathbf{E}(\bm{L}_{n,21})\big\}\big\{\mathbf{E}(\mathbf{L}_{n,11})\big\}^{-1}\bar{\bm{W}}_n^{*(PB)(1)}\Big{\|}_{\infty}
\end{align*}
We have handled each of the term separately under mentioned assumptions in previous lemmas. This will give us that,
\begin{align*}
&\Big{\|}\bar{\bm{L}}_{n,21}\big\{\bar{\bm{L}}_{n,11}\big\}^{-1}\bar{\bm{W}}_n^{*(PB)(1)}\Big{\|}_{\infty}\\
&=O_{p*}\Bigg[\frac{d_0^{2}}{n^{1-4\gamma_1-\frac{\tau}{3}}}\frac{\lambda_n^2}{n}+\frac{d_0^{3/2}}{n^{\frac{1}{2}-3\gamma_1-\frac{\tau}{3}}}\frac{\lambda_n}{n^{1/2}}+\frac{d_0^{3/2}}{n^{\frac{1}{2}-2\gamma_1-\frac{\tau}{3}}}\frac{\lambda_n}{n^{1/2}}+n^{\tau}\Bigg],
\end{align*}
which is $O_{p*}(n^{\tau})$ in probability under (B.3.i). Hence the proof is complete.\hfill$\blacksquare$

\begin{lemma}\label{lem:bootkktsoln}
The solution to equation (\ref{def:pblassoglm}) should satisfy,
\begin{align}\label{eqn:123456}
&-\bar{\bm W}_n^{*(PB)}
+\check{\bm L}_{n}\bar{\bm u}_n^{*(PB)}+Q_{1n}^*(\bar{\bm{u}}_n^{*(PB)})
= -\frac{\lambda_n}{n^{1/2}}
   \bm{sgn}(\bar{\bm\beta}_n^{*(PB)}),\quad\text{provided }\bar{\bm\beta}_n^{*(PB)}\neq\bm 0.
\nonumber\\[0.8em]
&-\frac{\lambda_n}{n^{1/2}}\mathbf 1
\;\le\;
-\bar{\bm W}_n^{*(PB)}
+\check{\bm L}_{n}\bar{\bm u}_n^{*(PB)}+Q_{1n}^*(\bar{\bm{u}}_n^{*(PB)})
\;\le\;
\frac{\lambda_n}{n^{1/2}}\mathbf 1,\quad
\text{provided }\bar{\bm\beta}_n^{*(PB)}=\bm 0.
\end{align}

where, 
\begin{align*}
Q_{1n}^*(\bar{\bm{u}}_n^{*(PB)})&=\frac{1}{2n^{3/2}}
\sum_{i=1}^{n}
\bm x_i
\Big[h_1^{\prime\prime\prime}(z_i^*)
      - y_i h^{\prime\prime\prime}(z_i^*)\Big]
\,[\bm x_i^\top \bar{\bm u}_n^{*(PB)}]^2
\end{align*}

for some $z_i^*$ such that $|z_i^*-\bm{x}_i^\top\bar{\bm{\beta}}_n|\leq \frac{\bm{x}_i^\top\bar{\bm{u}}_n^{*(PB)}}{n^{1/2}}$ for all $i\in\{1,...,n\}$.  
\end{lemma}
Proof of Lemma \ref{lem:bootkktsoln} This proof follows similarly as Lemma \ref{lem:KKTconditions}. \hfill$\blacksquare$

\begin{lemma}\label{lem:bootconcentrationbarun1}
Suppose (A.2), (A.4)-(A.7), (B.1) and  (B.3) hold true. we have,
$$\Big{\|}n^{1/2}\Big(\bar{\bm{\beta}}_n^{*(PB)(1)}-\bar{\bm{\beta}}_n^{(1)}\Big)\Big{\|}=O_{p*}\Bigg(\frac{d_0^{1/2}}{n^{\frac{1}{2}-\gamma_1}}\lambda_n\Bigg)\;\text{with probability at least}\; 1-o(1),$$
where $n^{1/2}\Big(\bar{\bm{\beta}}_n^{*(1)}-\bar{\bm{\beta}}_n^{(1)}\Big)$ is given by equation (\ref{eqn:bootsolnbarun1}).
\end{lemma}
Proof of Lemma \ref{lem:bootconcentrationbarun1} This result follows from Lemma \ref{lem:concentrationbarun1}.\hfill$\blacksquare$

\begin{lemma}\label{lem:bootkktsolution}
Suppose  (B.1)-(B.4), (A.2)-(A.7) hold true. Then provided $\log d=o\big(n^{\frac{2\tau}{3}}\big)$ with $\tau<3/2$, we have the unique solution of the equation (\ref{def:pblassoglm}) as:  $\Big(\bar{\bm{u}}_n^{*(PB)(1)\top},\bar{\bm{u}}_n^{*(PB)(2)\top}\Big)^\top=\Bigg(\Big\{n^{1/2}\Big(\bar{\bm{\beta}}_n^{*(PB)(1)}-\bar{\bm{\beta}}_n^{(1)}\Big)\Big\}^{\top},\bm{0}^{1\times(d-d_0)}\Bigg)^\top$  where,
\begin{align}\label{eqn:bootsolnbarun1}
&n^{1/2}\Big(\bar{\bm{\beta}}_n^{*(PB)(1)}-\bar{\bm{\beta}}_n^{(1)}\Big)=\{\mathbf{E}(\bm{L}_{n,11})\}^{-1}\bar{\bm{W}}_n^{*(PB)(1)}-\frac{\lambda_n}{n^{1/2}}\{\mathbf{E}(\bm{L}_{n,11})\}^{-1}\bm{sgn}(\bar{\bm{\beta}}_n^{(1)})+\bm{Q}_{1n}^*(\bar{\bm{u}}_n^{*(PB)(1)}),\nonumber\\
&\text{with}\;\bm{Q}_{1n}^*(\bar{\bm{u}}_n^{*(PB)(1)})\nonumber\\
&=\Big[\big\{\check{\mathbf{L}}_{n,11}\big\}^{-1}-\big\{\mathbf{E}(\mathbf{L}_{n,11})\big\}^{-1}\Big]\bar{\bm{W}}_n^{*(PB)(1)}-\frac{\lambda_n}{n^{1/2}}\Big[\big\{\check{\mathbf{L}}_{n,11}\big\}^{-1}-\big\{\mathbf{E}(\mathbf{L}_{n,11})\big\}^{-1}\Big]\bm{sgn}(\bar{\bm{\beta}}_n^{(1)})\nonumber\\
&\;\;\;\;\;\;\;\;\;\;\;\;\;\;\;\;\;\;\;\;\;\;\;\;\;\;\;+\frac{\big\{\mathbf{E(L}_{n,11})\big\}^{-1}}{2n^{3/2}}\sum_{i=1}^{n}\bm{x}_i^{(1)}\Bigg[h_1^{\prime\prime\prime}(z_i^*)-y_ih^{\prime\prime\prime}(z_i^*)\Bigg]\Bigg[\bm{x}_i^{(1)\top}\Big\{n^{1/2}\Big(\bar{\bm{\beta}}_n^{*(PB)(1)}-\bar{\bm{\beta}}_n^{(1)}\Big)\Big\}\Bigg]^2\nonumber\\
&+\frac{\Big[\big\{\check{\mathbf{L}}_{n,11}\big\}^{-1}-\big\{\mathbf{E}(\mathbf{L}_{n,11})\big\}^{-1}\Big]}{2n^{3/2}}\sum_{i=1}^{n}\bm{x}_i^{(1)}\Bigg[h_1^{\prime\prime\prime}(z_i^*)-y_ih^{\prime\prime\prime}(z_i^*)\Bigg]\Bigg[\bm{x}_i^{(1)\top}\Big\{n^{1/2}\Big(\bar{\bm{\beta}}_n^{*(PB)(1)}-\bar{\bm{\beta}}_n^{(1)}\Big)\Big\}\Bigg]^2
\end{align}
\end{lemma}
Proof of Lemma \ref{lem:bootkktsolution} This result follows exactly similar steps of Proposition \ref{prop:solutionkktglm} as in section \ref{sec:prop5.1} when we consider everything on the set $\tilde{C}_{n,\epsilon}$.\hfill$\blacksquare$

\subsection{Proofs of Main Results}\label{sec:mainresults}
In this section, we provide the proofs of Proposition \ref{prop:solutionkktglm}, Theorem \ref{thm:failgauss}, \ref{thm:bootapproxglmvsc} and \ref{thm:prbbootapproxglmvsc} respectively.
\subsubsection{Proof of Proposition \ref{prop:solutionkktglm}}\label{sec:prop5.1}

Due to Lemma \ref{lem:KKTconditions}, the solution to equation (\ref{eqn:deflasso}) must satisfy equation (\ref{eqn:12345}). Due to strict convexity of $Z_n(\bm{v})$, these KKT conditions are necessary and sufficient. Now consider the choice of solution:  $$\bar{\bm{u}}_n^{(2)}=\bm{0}^{(d-d_0)\times 1}$$ and 
\begin{align}\label{eqn:654321}
\bar{\bm{u}}_n^{(1)}=\{\mathbf{E}(\bm{L}_{n,11})\}^{-1}\bm{W}_n^{(1)}-\frac{\lambda_n}{n^{1/2}}\{\mathbf{E}(\bm{L}_{n,11})\}^{-1}\bm{sgn}(\bm{\beta}^{(1)})+\bm{Q}_{1n}(\bar{\bm{u}}_n^{(1)}),   
\end{align}    
where we have already established in Lemma \ref{lem:concentrationbarun1} that $\Big{\|}\bm{Q}_{1n}(\bar{\bm{u}}_n^{(1)})\Big{\|}=O\Bigg(\frac{d_0^{1/2}}{n^{\frac{1}{2}-\gamma_1}}\lambda_n\Bigg)$ with probability at least $1-o(1)$. These conditions along with (B.3.ii) and Lemma \ref{lem:InvLn11Wn1}, will give us that,
with probability at least $1-o(1)$, 
\begin{align}\label{eqn:signconsistency}
&\mbox{max}_{j\in\mathcal{A}_n}\Bigg{|}\Big(\big\{\mathbf{E}(\bm{L}_{n,11})\big\}^{-1}\Big)_{j\cdot}^\top\bm{W}_n^{(1)}-\frac{\lambda_n}{n^{1/2}}\Big(\big\{\mathbf{E}(\bm{L}_{n,11})\big\}^{-1}\Big)_{j\cdot}^\top\bm{sgn}(\bm{\beta}^{(1)})+\bm{Q}_{1n,j}(\bar{\bm{u}}_n^{(1)})\Bigg{|}\nonumber\\
&\;\;\;\;\;\;\;\;\;\;\;\;\;\;\;\;\;\;\;\;\;\;\;\;\;\;\;\;\;\;\;\;\;\;\;\;\;\;\;\;\;\;\;\;\;\;\;\;\;\;\;\;\;\;\;\;\;\;\;\;\;\;\;\;\;\;\;\;\;\;<n^{1/2}\mbox{min}_{j\in\mathcal{A}_n}\big{|}\beta_j^{(1)}\big{|}\nonumber\\
&\implies \big{|}\bar{\bm{u}}_{n,j}^{(1)}\big{|}<n^{1/2}\big{|}\beta_j^{(1)}\big{|}\;\;\text{for all}\;\;j\in\mathcal{A}_n\nonumber\\
&\implies \frac{\bar{\bm{u}}_{n,j}^{(1)}}{n^{1/2}}sgn(\beta_j^{(1)})>-\big{|}\beta_j^{(1)}\big{|}\;\;\text{for all}\;\;j\in\mathcal{A}_n\nonumber\\
&\implies \Big(\bar{\beta}_{n,j}^{(1)}-\beta_j^{(1)}\Big)sgn(\beta_j^{(1)})\big{|}\beta_j^{(1)}\big{|}+\big{|}\beta_j^{(1)}\big{|}^2>0\;\;\text{for all}\;\;j\in\mathcal{A}_n\nonumber\\
&\implies sgn(\bar{\beta}_{n,j}^{(1)}) =sgn(\beta_j^{(1)})\;\;\text{for all}\;\;j\in\mathcal{A}_n
\end{align}
Therefore we can conclude that the choice $\Bigg(\Big\{n^{1/2}\Big(\bar{\bm{\beta}}_n^{(1)}-\bm{\beta}^{(1)}\Big)\Big\}^{\top},\bm{0}^{1\times(d-d_0)}\Bigg)^\top$ satisfies:
\begin{align*}
&-\bm{W}_n+\bm{L}_{n}\bar{\bm{u}}_n+\frac{1}{2n^{3/2}}\sum_{i=1}^{n}\bm{x}_i\Bigg[h_1^{\prime\prime\prime}(z_i)-y_ih^{\prime\prime\prime}(z_i)\Bigg][\bm{x}_i^\top\bar{\bm{u}}_n]^2= -\frac{\lambda_n}{n^{1/2}}\bm{sgn}(\bar{\bm{\beta}}_n),
\end{align*}
since $\mathcal{A}_n=\{j\in\{1,..,d\}:\beta_j\neq0\}=\{1,...,d_0\}$ and $sgn(\bar{\beta}_{n,j}^{(1)}) =sgn(\beta_j^{(1)})\;\;\text{for all}\;\;j\in\mathcal{A}_n$.
Now we denote,
\begin{align}\label{eqn:toomuchcalculation}
&\bm{Q}_{2n}(\bar{\bm{u}}_n^{(1)})\nonumber\\
&=\frac{1}{2n^{3/2}}\sum_{i=1}^{n}\Bigg[\bm{L}_{n,21}(\bm{L}_{n,11})^{-1}\bm{x}_i^{(1)}-\bm{x}_i^{(2)}\Bigg]\Bigg[h_1^{\prime\prime\prime}(z_i)-y_ih^{\prime\prime\prime}(z_i)\Bigg]\Bigg[\bm{x}_i^{(1)\top}\Big\{n^{1/2}\Big(\bar{\bm{\beta}}_n^{(1)}-\bm{\beta}^{(1)}\Big)\Big\}\Bigg]^2
\end{align}
Therefore due to Lemma \ref{lem:concentrationbarun1} and assumptions (A.3),(A.4),(A.6) and (B.2), it is an easy exercise to verify that with probability at least $1-o(1)$,
\begin{align}\label{eqn:secondremainder}
\Big{\|}\bm{Q}_{2n}(\bar{\bm{u}}_n^{(1)})\Big{\|}_{\infty}=O\Bigg[\frac{\lambda_n^2}{n^{1/2}}\frac{d_0}{n^{1-2\gamma_1}}\Bigg].    
\end{align}
 Now under conditions (B.3.i) and (B.4), following equation (\ref{eqn:secondremainder}), it can be verified that, with probability at least $1-o(1)$,
\begin{align}\label{eqn:secondkkt}
&\Big\|
\bm L_{n,21}\{\bm L_{n,11}\}^{-1}\bm W_n^{(1)}
   - \bm W_n^{(2)}
\Big\|_{\infty}\notag
\\
&\;\le
\min_{j\notin\mathcal A_n}
\Bigg[
    \frac{\lambda_n}{n^{1/2}}
    -\frac{\lambda_n}{n^{1/2}}
       \Big|
       \{\bm L_{n,21}(\bm L_{n,11})^{-1}\}_{j\cdot}^{\!\top}
       \bm{sgn}(\bm\beta^{(1)})
       \Big|
    -\big|\,\bm Q_{2n,j}(\bar{\bm u}_n^{(1)})\,\big|
\Bigg]
\nonumber\\[1em]
&\implies
-\frac{\lambda_n}{n^{1/2}}\mathbf 1
\;\le\;
-\bm W_n^{(2)}
+ \bm L_{n,21}\{\bm L_{n,11}\}^{-1}\bm W_n^{(1)}
\\
&\qquad\qquad\qquad
-\frac{\lambda_n}{n^{1/2}}
   \bm L_{n,21}(\bm L_{n,11})^{-1}
   \bm{sgn}(\bm\beta^{(1)})
\;-\;
\bm Q_{2n}(\bar{\bm u}_n^{(1)})
\;\le\;
\frac{\lambda_n}{n^{1/2}}\mathbf 1,
\nonumber\\[0.5em]
&\qquad\qquad\qquad\qquad\qquad\qquad\qquad\qquad\qquad\qquad\qquad\qquad
\text{componentwise for all } j\notin\mathcal A_n.
\nonumber
\end{align}

Therefore we can conclude that the choice $\Bigg(\Big\{n^{1/2}\Big(\bar{\bm{\beta}}_n^{(1)}-\bm{\beta}^{(1)}\Big)\Big\}^{\top},\bm{0}^{1\times(d-d_0)}\Bigg)^\top$ satisfies:
\begin{align*}
&-\frac{\lambda_n}{n^{1/2}}\mathbf{1}\leq -\bm{W}_n+\bm{L}_{n}\bar{\bm{u}}_n+\frac{1}{2n^{3/2}}\sum_{i=1}^{n}\bm{x}_i\Bigg[h_1^{\prime\prime\prime}(z_i)-y_ih^{\prime\prime\prime}(z_i)\Bigg][\bm{x}_i^\top\bar{\bm{u}}_n]^2 \leq \frac{\lambda_n}{n^{1/2}}\mathbf{1},    
\end{align*}
since $\bar{\bm{u}}_n^{(2)}=n^{1/2}\big(\bar{\bm{\beta}}_n^{(2)}-\bm{0}\big)=0\Longleftrightarrow \bar{\bm{\beta}}_n^{(2)}=\bm{0}$ for all $j\notin\mathcal{A}_n$. The uniqueness of this solution follows due to Lemma 2 of \citet{tibshirani2013lasso}.\hfill $\blacksquare$

\subsubsection{Proof of Theorem \ref{thm:failgauss}}\label{sec:thm5.1}

We will show that,
\begin{align}\label{eqn:failgaussball}
\Delta_n=\mbox{sup}_{B\in\mathcal{B}}\Bigg{|}\mathbf{P}\Big[n^{1/2}\Big(\bar{\bm{\beta}}_n^{(1)}-\bm{\beta}^{(1)}\Big)\in B\Big]-\mathbf{P}\Big[\bm{G}_{2n}\in B\Big]\Bigg{|}\to 1\;\text{as}\;n\to\infty, \end{align}
where $\mathcal{B}$ is class of measurable Euclidean Balls and $\bm{G}_{2n}\sim N_{d_0}(\bm{0},\tilde{\bm{S}}_{n,11})$. Denote $\bm{Z}_n\sim  N_{d_0}(\bm{0},I_{d_0})$.
Without loss of generality, we will prove this negative result when $h(u)=u$ and $\gamma_1=\gamma_2=0$. Note that for this choice, $\bm{L}_{n,11}=\mathbf{E}(\bm{L}_{n,11})$. Denote the following:
\begin{align}\label{eqn:enmihball}
&\bm{b}_n^{(1)}=-\frac{\lambda_n}{n^{1/2}}\{\mathbf{E}(\bm{L}_{n,11})\}^{-1}\bm{sgn}(\bm{\beta}^{(1)})\nonumber\\
&\bm{Q}_{1n}(\bar{\bm{u}}_n^{(1)})=\frac{\big\{\mathbf{E(L}_{n,11})\big\}^{-1}}{2n^{3/2}}\sum_{i=1}^{n}\bm{x}_i^{(1)}\Big[(g^{-1})^{\prime\prime}(z_i)\Big]\Bigg[\bm{x}_i^{(1)\top}\Big\{n^{1/2}\Big(\bar{\bm{\beta}}_n^{(1)}-\bm{\beta}^{(1)}\Big)\Big\}\Bigg]^2
\end{align}

For some $0<B<\infty$, denote the set $C_n=\Big\{\Big{\|}\bm{W}_n^{(1)}\Big{\|}_{\infty}\le B n^{\tau/3}\Big\}$ such that $\mathbf{P}(C_n)\geq 1-o(1)$.  Now for some $\tilde{B}_n\in\mathcal{B}$ we denote $B_n:=\big[\tilde{S}_{n,11}\big]^{-1/2}\tilde{B}_n$ so that $B_n$ remains a measurable Euclidean ball. Since we have assumed that
\[
\mbox{max}_{j\in\mathcal{A}_n}\Big{|}\Big(\tilde{\bm{S}}_{n,11}^{-1/2}\Big)_{j\cdot}^\top\Big\{\mathbf{E}(\bm{L}_{n,11})\bm{sgn}(\bm{\beta}^{(1)})\Big\}\Big{|}>k,
\]
This will imply that,
\begin{align}\label{eqn:ballfail}
\Big\|\big[\tilde{S}_{n,11}\big]^{-1/2}\bm{b}_n^{(1)}\Big\|_2
\geq k\frac{\lambda_n}{n^{1/2}}.
\end{align}

Define the ball
\[
B_n:=\Big\{\bm{z}\in\mathbf{R}^{d_0}:\|\bm{z}\|_2\le \frac{k}{4}\frac{\lambda_n}{n^{1/2}}\Big\}.
\]

Therefore, we have,
\begin{align}\label{eqn:ball1}
\Delta_n
&\geq \Bigg{|}\mathbf{P}\Big[n^{1/2}\Big(\bar{\bm{\beta}}_n^{(1)}-\bm{\beta}^{(1)}\Big)\in \tilde{B}_n\Big]-\mathbf{P}\Big[\bm{G}_{2n}\in \tilde{B}_n\Big]\Bigg| \nonumber\\
&\geq \Bigg{|}\mathbf{P}\Big[\Big\{\bm{T}_n^{(1)}+\bm{b}_n^{(1)}+\bm{Q}_{1n}(\bar{\bm{u}}_n^{(1)})\in \tilde{B}_n\Big\}\cap C_n\Big]-\mathbf{P}\Big[\bm{G}_{2n}\in \tilde{B}_n\Big]\Bigg|-2\mathbf{P}(C_n^c)\nonumber\\
&\geq \Bigg{|}\mathbf{P}\Big[\bm{G}_{2n}\in \tilde{B}_n\Big]-\mathbf{P}\Big[\bm{G}_{2n}\in \tilde{B}_n-\bm{b}_n^{(1)}-\bm{Q}_{1n}(\bar{\bm{u}}_n^{(1)})\Big]\Bigg|\nonumber\\
&\;\;\;\;\;\;\;\;-\Bigg{|}\mathbf{P}\Big[\bm{T}_n^{(1)}\in \tilde{B}_n-\bm{b}_n^{(1)}-\bm{Q}_{1n}(\bar{\bm{u}}_n^{(1)})\Big]-\mathbf{P}\Big[\bm{G}_{2n}\in \tilde{B}_n-\bm{b}_n^{(1)}-\bm{Q}_{1n}(\bar{\bm{u}}_n^{(1)})\Big]\Bigg|-o(1)\nonumber\\
&\geq \Bigg{|}\mathbf{P}\Big[\bm{Z}_n\in B_n\Big]-\mathbf{P}\Big[\bm{Z}_n\in B_n-\big[\tilde{S}_{n,11}\big]^{-1/2}\bm{b}_n^{(1)}-\big[\tilde{S}_{n,11}\big]^{-1/2}\bm{Q}_{1n}(\bar{\bm{u}}_n^{(1)})\Big]\Bigg|-o(1)\nonumber\\
&\ge \mathbf{P}\Big(\|\bm{Z}_n\|_2\le \frac{k}{4}\frac{\lambda_n}{n^{1/2}}\Big)
-\mathbf{P}\Bigg(\Big\|\bm{Z}_n+\big[\tilde{S}_{n,11}\big]^{-1/2}\bm{b}_n^{(1)}+\big[\tilde{S}_{n,11}\big]^{-1/2}\bm{Q}_{1n}(\bar{\bm{u}}_n^{(1)})\Big\|_2\le \frac{k}{4}\frac{\lambda_n}{n^{1/2}}\Bigg)-o(1),
\end{align}
where the third inequality follows from translation invariance of Euclidean balls together with the triangle inequality, and the penultimate inequality follows from Lemma \ref{lem:leadingpart2}. Now under the assumption (B.3.i) and equation (\ref{eqn:ballfail}), it's obvious to have that for large enough $n$,
\begin{align}\label{eqn:wheretogo}
&\Big\|\big[\tilde{S}_{n,11}\big]^{-1/2}\bm{b}_n^{(1)}+\big[\tilde{S}_{n,11}\big]^{-1/2}\bm{Q}_{1n}(\bar{\bm{u}}_n^{(1)})\Big\|_2> \frac{k}{2}\frac{\lambda_n}{n^{1/2}}    
\end{align}
Now using assumption (B.3.i), equation (\ref{eqn:wheretogo}) and reverse triangle inequality we get for large enough $n$;
\begin{align}\label{eqn:997655}
&\Bigg\{\Big\|\bm{Z}_n+\big[\tilde{S}_{n,11}\big]^{-1/2}\bm{b}_n^{(1)}+\big[\tilde{S}_{n,11}\big]^{-1/2}\bm{Q}_{1n}(\bar{\bm{u}}_n^{(1)})\Big\|_2\le \frac{k}{4}\frac{\lambda_n}{n^{1/2}}\Bigg\}\nonumber\\
&\subseteq\Bigg\{\Big\|\big[\tilde{S}_{n,11}\big]^{-1/2}\bm{b}_n^{(1)}+\big[\tilde{S}_{n,11}\big]^{-1/2}\bm{Q}_{1n}(\bar{\bm{u}}_n^{(1)})\Big\|_2-\Big\|\bm{Z}_n\Big\|_2\le \frac{k}{4}\frac{\lambda_n}{n^{1/2}}\Bigg\}\nonumber\\
&\subseteq\Big\{\Big\|\bm{Z}_n\Big\|_2\ge \frac{k}{4}\frac{\lambda_n}{n^{1/2}}\Big\}\subseteq\Big\{\Big\|\bm{Z}_n\Big\|_2> \frac{kB_1}{4}n^\tau\Big\}
\end{align}

Therefore from equation (\ref{eqn:ball1}) and (\ref{eqn:997655}), we finally write 
\begin{align}\label{eqn:222222222}
\Delta_n\ge&\;\mathbf{P}\Big[\Big\|\bm{Z}_n\Big\|_2\le \frac{kB_1}{4}n^\tau\Big]-\mathbf{P}\Big[\Big\|\bm{Z}_n\Big\|_2> \frac{kB_1}{4}n^\tau\Big]-o(1)  \end{align}

\noindent
Since $\bm Z_n \sim N_{d_0}(\bm 0,I_{d_0})$, we have
\[
\|\bm Z_n\|_2^2=\sum_{j=1}^{d_0} Z_{n,j}^2 \sim \chi^2_{d_0}.
\]
Then due to Laurent--Massart inequality (cf. Lemma 1 of \citet{laurent2000adaptive}), there exist constants $c_1,c_2>0$ such that
\[
\mathbf P\Bigg(\|\bm Z_n\|_2 \ge \frac{kB_1}{4}n^\tau\Bigg)
\le c_1 \exp{(-c_2 n^{2\tau})}\to 0,
\]
as $n\to\infty$. Therefore, the proof is complete.
\hfill $\blacksquare$

\subsubsection{Proof of Theorem \ref{thm:bootapproxglmvsc}}\label{sec:thm5.2}

We consider the proof for class of convex sets. Analysis of Euclidean Balls will be analogous. Recall that, from equation (\ref{eqn:solnbarun1}) we have;
\begin{align}\label{eqn:originalsolution}
n^{1/2}\Big(\bar{\bm{\beta}}_n^{(1)}-\bm{\beta}^{(1)}\Big)&=\{\mathbf{E}(\bm{L}_{n,11})\}^{-1}\bm{W}_n^{(1)}-\frac{\lambda_n}{n^{1/2}}\{\mathbf{E}(\bm{L}_{n,11})\}^{-1}\bm{sgn}(\bm{\beta}^{(1)})+\bm{Q}_{1n}(\bar{\bm{u}}_n^{(1)})\nonumber\\
&=\bm{T}_n^{(1)}+\bm{b}_n^{(1)}+\bm{r}_n^{(1)},\;\;\text{(say)}
\end{align}
Similarly from equation (\ref{eqn:bootsolnbarun1}) we have seen that,
\begin{align}\label{eqn:bootsolution}
 n^{1/2}\Big(\bar{\bm{\beta}}_n^{*(PB)(1)}-\bar{\bm{\beta}}_n^{(1)}\Big)&=\{\mathbf{E}(\bm{L}_{n,11})\}^{-1}\bar{\bm{W}}_n^{*(PB)(1)}-\frac{\lambda_n}{n^{1/2}}\{\mathbf{E}(\bm{L}_{n,11})\}^{-1}\bm{sgn}(\bar{\bm{\beta}}_n^{(1)})\nonumber\\
 &\qquad\qquad\qquad\qquad+\bm{Q}_{1n}^*(\bar{\bm{u}}_n^{*(PB)(1)})\nonumber\\
 &=\bar{\bm{T}}_n^{*(PB)(1)}+\bar{\bm{b}}_n^{*(PB)(1)}+\bar{\bm{r}}_n^{*(PB)(1)}\;\;\text{(say)}.
\end{align}
For any convex set $B\in\mathcal{C}$ and a vector $\bm{\xi}$, it is well-known that $B-\xi\in\mathcal{C}$. Also $\bm{Z}+\xi\in B\implies \bm{Z}\in B-\xi$. Suppose $\bm{G}_{2n}\sim N_{d_0}(\bm{0},\tilde{\bm{S}}_{n,11})$ and $,\tilde{\bm{S}}_{n,11}=var(\tilde{\bm{T}}_n^{(1)})$. Now on the set $\tilde{C}_{n,\epsilon}$ with $\mathbf{P}(\tilde{C}_{n,\epsilon})\geq 1-\epsilon$, we have seen that $\bar{b}_{n,j}^{*(PB)(1)}=b_{n,j}^{(1)}$ for all $j\in\mathcal{A}_n$. Now for any $\epsilon>0$ and sufficiently large $n$, we note that,
\begin{align}\label{eqn:final}
&\Delta_{n,2}^{*(PB)}(\mathcal{C})=\mathbf{P}\Bigg[\mbox{sup}_{B\in\mathcal{C}}\Bigg{|}\mathbf{P}_*\Big[n^{1/2}\Big(\bar{\bm{\beta}}_n^{*(PB)(1)}-\bar{\bm{\beta}}_n^{(1)}\Big)\in B\Big]-\mathbf{P}\Big[n^{1/2}\Big(\bar{\bm{\beta}}_n^{(1)}-\bm{\beta}^{(1)}\Big)\in B\Big]\Bigg{|}>\epsilon\Bigg]\nonumber\\
&\leq \mathbf{P}\Bigg[\Bigg\{\mbox{sup}_{B\in\mathcal{C}}\Bigg{|}\mathbf{P}_*\Big[n^{1/2}\Big(\bar{\bm{\beta}}_n^{*(PB)(1)}-\bar{\bm{\beta}}_n^{(1)}\Big)\in B\Big]-\mathbf{P}\Big[n^{1/2}\Big(\bar{\bm{\beta}}_n^{(1)}-\bm{\beta}^{(1)}\Big)\in B\Big]\Bigg{|}>\epsilon\Bigg\}\cap \tilde{C}_{n,\epsilon}\Bigg]\nonumber\\
&\qquad\qquad\qquad\qquad\qquad\qquad\qquad\qquad\qquad+\mathbf{P}\Big(\tilde{C}_{n,\epsilon}^c\Big)\nonumber\\
&\leq \mathbf{P}\Bigg[\Bigg\{\mbox{sup}_{B\in\mathcal{C}}\Bigg{|}\mathbf{P}_*\Big[\bar{\bm{T}}_n^{*(PB)(1)}+\bm{b}_n^{(1)}+\bar{\bm{r}}_n^{*(PB)(1)}\in B\Big]-\mathbf{P}\Big[\bm{T}_n^{(1)}+\bm{b}_n^{(1)}+\bm{r}_n^{(1)}\in B\Big]\Bigg{|}>\epsilon\Bigg\}\Bigg]+\epsilon/2\nonumber\\
&\leq \mathbf{P}\Bigg[\Bigg\{\mbox{sup}_{B\in\mathcal{C}}\Bigg{|}\mathbf{P}_*\Big[\bar{\bm{T}}_n^{*(PB)(1)}+\bar{\bm{r}}_n^{*(PB)(1)}\in B-\bm{b}_n^{(1)}\Big]-\mathbf{P}\Big[\bm{T}_n^{(1)}+\bm{r}_n^{(1)}\in \underbrace{B-\bm{b}_n^{(1)}}_{D\;,\text{say;}}\Big]\Bigg{|}>\epsilon\Bigg\}\Bigg]+\epsilon/2\nonumber\\
&\leq \mathbf{P}\Bigg[\Bigg\{\mbox{sup}_{D\in\mathcal{C}}\Bigg{|}\mathbf{P}_*\Big[\bar{\bm{T}}_n^{*(PB)(1)}+\bar{\bm{r}}_n^{*(PB)(1)}\in D\Big]-\mathbf{P}\Big[\bm{T}_n^{(1)}+\bm{r}_n^{(1)}\in D\Big]\Bigg{|}>\epsilon\Bigg\}\Bigg]+\epsilon/2
\end{align}
This will give us that,
\begin{align}
\Delta_{n,2}^{*(PB)}(\mathcal{C})&\leq \mathbf{P}\Bigg[\underbrace{\mbox{sup}_{D\in\mathcal{C}}\Bigg{|}\mathbf{P}_*\Big[\bar{\bm{T}}_n^{*(PB)(1)}+\bar{\bm{r}}_n^{*(PB)(1)}\in D\Big]-\mathbf{P}_*\Big[\bar{\bm{T}}_n^{*(PB)(1)}\in D\Big]\Bigg{|}}_{\text{Term I}}\nonumber\\
&+\underbrace{\mbox{sup}_{D\in\mathcal{C}}\Bigg{|}\mathbf{P}_*\Big[\bar{\bm{T}}_n^{*(PB)(1)}\in D\Big]-\mathbf{P}\Big[\bm{G}_{2n}\in D\Big]\Bigg{|}}_{\text{Term II}}+\underbrace{\mbox{sup}_{D\in\mathcal{C}}\Bigg{|}\mathbf{P}\Big[\bm{T}_n^{(1)}+\bm{r}_n^{(1)}\in D\Big]-\mathbf{P}\Big[\bm{T}_n^{(1)}\in D\Big]\Bigg{|}}_{\text{Term III}}\nonumber\\
&+\underbrace{\mbox{sup}_{D\in\mathcal{C}}\Bigg{|}\mathbf{P}\Big[\bm{T}_n^{(1)}\in D\Big]-\mathbf{P}\Big[\bm{G}_{2n}\in D\Big]\Bigg{|}}_{\text{Term IV}}>\epsilon\Bigg]+\epsilon/2
\end{align}
Now for large enough $n$ and under the assumption, Lemma \ref{lem:leadingpart2} gives us that, $\text{Term IV}=o(1)$. From Lemma \ref{lem:concentrationbarun1}, we have already seen that under given assumptions,
$$\|\bm{r}_n^{(1)}\|=O_p\Bigg[\frac{d_0}{n^{\frac{1}{2}-2\gamma_1-\frac{2\tau}{3}}}+\frac{\lambda_n}{n^{1/2}}\frac{d_0}{n^{\frac{1}{2}-2\gamma_1-\frac{\tau}{3}}}+\frac{1}{n^{\frac{1}{2}-\gamma_1}}\Bigg\{\frac{d_0^{1/2}}{n^{\frac{1}{2}-\gamma_1}}\lambda_n\Bigg\}^2+\frac{d_0^{1/2}}{n^{1-2\gamma_1-\frac{\tau}{3}}}\Bigg\{\frac{d_0^{1/2}}{n^{\frac{1}{2}-\gamma_1}}\lambda_n\Bigg\}^2\Bigg]=o_p(1),$$
provided assumption (B.3.i) and $d_0=o(n^{\frac13-2\gamma_1})$ hold. Then following exactly similar lines of equation (\ref{eqn:121}) from the proof of Theorem \ref{thm:convex1} will give us that, for large enough $n$, $\text{Term III}=o(1)$. Hence equation (\ref{eqn:final}) reduces to,
\begin{align}\label{eqn:final1}
\Delta_{n,2}^{*(PB)}(\mathcal{C})&\leq \mathbf{P}\Bigg[\underbrace{\mbox{sup}_{D\in\mathcal{C}}\Bigg{|}\mathbf{P}_*\Big[\bar{\bm{T}}_n^{*(PB)(1)}+\bar{\bm{r}}_n^{*(PB)(1)}\in D\Big]-\mathbf{P}_*\Big[\bar{\bm{T}}_n^{*(PB)(1)}\in D\Big]\Bigg{|}}_{\text{Term I}}\nonumber\\
&\;\;\;\;\;\;\;\;\;\;\;\;\;\;\;\;\;\;\;\;\;\;\;\;\;\;\;\;\;+\underbrace{\mbox{sup}_{D\in\mathcal{C}}\Bigg{|}\mathbf{P}_*\Big[\bar{\bm{T}}_n^{*(PB)(1)}\in D\Big]-\mathbf{P}\Big[\bm{G}_{2n}\in D\Big]\Bigg{|}}_{\text{Term II}}>\epsilon/2\Bigg]+\epsilon/2
\end{align}
This will in turn imply that,
\begin{align}
 \Delta_{n,2}^{*(PB)}(\mathcal{C})&\leq \underbrace{\mathbf{P}\Bigg[\mbox{sup}_{D\in\mathcal{C}}\Bigg{|}\mathbf{P}_*\Big[\bar{\bm{T}}_n^{*(PB)(1)}+\bar{\bm{r}}_n^{*(PB)(1)}\in D\Big]-\mathbf{P}_*\Big[\bar{\bm{T}}_n^{*(PB)(1)}\in D\Big]\Bigg{|}>\epsilon/4\Bigg]}_{\text{Term I}}\nonumber\\
&\;\;\;\;\;\;\;\;\;\;\;\;\;\;\;\;\;\;\;\;\;\;\;\;\;\;\;\;\;+\underbrace{\mathbf{P}\Bigg[\mbox{sup}_{D\in\mathcal{C}}\Bigg{|}\mathbf{P}_*\Big[\bar{\bm{T}}_n^{*(PB)(1)}\in D\Big]-\mathbf{P}\Big[\bm{G}_{2n}\in D\Big]\Bigg{|}>\epsilon/4\Bigg]}_{\text{Term II}}+\epsilon/2   
\end{align}
Now following Lemma \ref{lem:PBLeadingtermconvex} under the assumptions, it can be concluded that $\text{Term II}<\epsilon/4$ for large enough $n$. Similar to proof of Theorem \ref{thm:pbconvexoriginal1}, following equation (\ref{eqn:452}) and the fact that $\|\bar{\bm{r}}_n^{*(PB)(1)}\|=o_p(1)$, we can infer that $\text{Term I}<\epsilon/4$ for large enough $n$ as well. That concludes the proof.\hfill $\blacksquare$

\subsubsection{\bf Proof of Theorem \ref{thm:prbbootapproxglmvsc}}\label{sec:thm5.3}
As earlier, we will consider everything on this set $\tilde{C}_{n,\epsilon}$ whenever it's needed. Let us partition $\bar{L}_n^{(PRB)}=n^{-1}\bar{G}_n\bar{G}_n$ as per first $d_0$ many components as,
\[
\bar{L}_n^{(PRB)} = \begin{bmatrix}
\bar{L}_{n,11}^{(PRB)} & \bar{L}_{n,12}^{(PRB)} \\
\bar{L}_{n,21}^{(PRB)} & \bar{L}_{n,22}^{(PRB)}
\end{bmatrix}.
\]
Similarly we partition $\bar{W}_n^{*(PRB)}=n^{-1/2}\bar{G}_n^\top\bm{e}^{**}$ as,

\[
\bar{W}_n^{*(PRB)} = \begin{bmatrix}
n^{-1/2}\sum_{i=1}^{n}\sqrt{b^{\prime\prime}(h(\bm{x}_i^\top\bar{\bm{\beta}}_n))}h^\prime(\bm{x}_i^\top\bar{\bm{\beta}}_n)\bm{x}_i^{(1)}e_i^{**} \\
n^{-1/2}\sum_{i=1}^{n}\sqrt{b^{\prime\prime}(h(\bm{x}_i^\top\bar{\bm{\beta}}_n))}h^\prime(\bm{x}_i^\top\bar{\bm{\beta}}_n)\bm{x}_i^{(2)}e_i^{**}  
\end{bmatrix}=
\begin{bmatrix}
\bar{W}_{n,1}^{*(PRB)}\\
\bar{W}_{n,2}^{*(PRB)}
\end{bmatrix}.
\]
We denote $\bar{\bm{u}}_n^{*(PRB)}=n^{1/2}(\bar{\bm{\beta}}_n^{*(PRB)}-\bar{\bm{\beta}}_n)$ and we can partition this as above. Following this definition and KKT conditions, it is possible to have following,

\begin{align}\label{eqn:1234567}
&-\bar{\bm W}_n^{*(PRB)}
+\bar{\bm L}_{n}^{(PRB)}\bar{\bm u}_n^{*(PRB)}
= -\frac{\lambda_n}{n^{1/2}}
   \bm{sgn}(\bar{\bm\beta}_n^{*(PRB)}),\;\;\text{provided }\bar{\bm\beta}_n^{*(PRB)}\neq\bm 0.\nonumber\\[0.8em]
&-\frac{\lambda_n}{n^{1/2}}\mathbf 1
\;\le\;
-\bar{\bm W}_n^{*(PRB)}
+\bar{\bm L}_{n}^{(PRB)}\bar{\bm u}_n^{*(PRB)}
\;\le\;
\frac{\lambda_n}{n^{1/2}}\mathbf 1,\quad
\text{provided }\bar{\bm\beta}_n^{*(PRB)}=\bm 0.
\end{align}
Next we define;
\begin{align*}
 & \bar{\bm{u}}_n^{*(PRB)(1)}:=\bar{\bm{T}}_n^{*(PRB)(1)}+\bar{\bm{b}}_n^{*(PRB)(1)}+ \bar{\bm{r}}_n^{*(PRB)(1)}\\
 &\bar{\bm{T}}_n^{*(PRB)(1)}:=(\mathbf{E}L_{n,11})^{-1}\bar{\bm{W}}_{n,1}^{*(PRB)},\;\;\bar{\bm{b}}_n^{*(PRB)(1)}:=-(\mathbf{E}L_{n,11})^{-1}\frac{\lambda_n}{n^{1/2}}\bm{sgn}(\bar{\bm{\beta}}_n^{(1)})\\
 &\;\bar{\bm{r}}_n^{*(PRB)(1)}:=\Big[\bar{L}_{n,11}^{(PRB)-1}-(\mathbf{E}L_{n,11})^{-1}\Big]\Big\{\bar{\bm{W}}_{n,1}^{*(PRB)}-\frac{\lambda_n}{n^{1/2}}\bm{sgn}(\bar{\bm{\beta}}_n^{(1)})\Big\}
\end{align*}
Denote, $\bar{\sigma}_n^2=\frac1n\sum_{k=1}^n(e_k^\#-\bar{e}^\#)^2$, $\bar{\mu}_k:=g^{-1}(\bm{x}_k^{(1)\top}\bar{\bm{\beta}}_n^{(1)})$, $\bar{v}_k:=b^{\prime\prime}(h(\bm{x}_k^{(1)\top}\bar{\bm{\beta}}_n^{(1)}))$ and $e_k^\#:=\frac{y_k-\bar{\mu}_k}{\sqrt{\bar{v}}_k}$ for all $1\le k\le n$. Similarly we can define the population counterpart, $e_k:=\frac{y_k-\mu_k}{\sqrt{v}_k}$ with $\mathbf{E}(e_k)=0$ and $\text{var}(e_k)=1$ for all $k$. Also let us denote,  $\tilde{S}_{n,11}=[\mathbf{E}(L_{n,11})]^{-1}$. Then,
\[
\bar V_n^*:=\text{var}_*\Big(\tilde{S}_{n,11}^{-1/2}\bar{T}_n^{*(PRB)(1)}\Big)=\bar{\sigma}_n^2[\mathbf{E}(L_{n,11})]^{-1/2}\bar{L}_{n,11}^{(PRB)}[\mathbf{E}(L_{n,11})]^{-1/2}
\]
Now,
\begin{align*}
 \|I_{d_0}-\bar V_n^*\|_{HS}&\le\bar{\sigma}_n^2\underbrace{\Big\|[\mathbf{E}(L_{n,11})]^{-1/2}\bar{L}_{n,11}^{(PRB)}[\mathbf{E}(L_{n,11})]^{-1/2}\Big\|_{HS}}_{:=B_1\;\text{(say)}}\\
 &+\underbrace{\Big\|[\mathbf{E}(L_{n,11})]^{-1/2}[S_{n,11}][\mathbf{E}(L_{n,11})]^{-1/2}\Big\|_{HS}}_{:=B_2\;\text{(say)}}   
\end{align*}
Now following exact similar argument of equation (\ref{eqn:kkkkkkkkkkk}) will give us that,

\begin{align}\label{eqn:kkkkkllllllllmmmmmmm}
|\bar{\sigma}_n^2|=O_p\Bigg[\frac{1}{\sqrt{n}}+\frac{d_0^{1/2}}{n^{\frac{1}{2}-\gamma_1}}\frac{\lambda_n}{n^{1/2}}+\frac{d_0}{n^{1-2\gamma_1}}\frac{\lambda_n^2}{n}+\frac{d_0^{3/2}}{n^{\frac32-3\gamma_1}}\frac{\lambda_n^3}{n^{3/2}}\Bigg]=O_p\big(1\big),  
\end{align}
provided $\lambda_n=O[d_0^{-1/2}n^{1-\gamma_1}]$. Next we observe that from Lemma \ref{lem:frobHSnorm}, 
\begin{align}\label{eqn:mmmmmmnnnbbb}
&B_2=\Bigg{\|}\frac{1}{n}\sum_{i=1}^{n}\Big\{[\mathbf{E}(L_{n,11})]^{-1/2}\bm{x}_i^{(1)}\Big\}\Big\{[\mathbf{E}(L_{n,11})]^{-1/2}\bm{x}_i^{(1)}\Big\}^\top\Big[h^\prime\big(\bm{x}_i^\top\bm{\beta}\big)\Big]^2\mathbf{E}\Big[y_i-g^{-1}\big(\bm{x}_i^\top\bm{\beta}\big)\Big]^2\Bigg{\|}_{H.S} \nonumber\\
& \leq\Bigg[\frac{2}{n^2}\sum_{i=1}^{n}\Bigg{\|}\Big\{[\mathbf{E}(L_{n,11})]^{-1/2}\bm{x}_i^{(1)}\Big\}\Big\{[\mathbf{E}(L_{n,11})]^{-1/2}\bm{x}_i^{(1)}\Big\}^\top\Big[h^\prime\big(\bm{x}_i^\top\bm{\beta}\big)\Big]^2\mathbf{E}\Big[y_i-g^{-1}\big(\bm{x}_i^\top\bm{\beta}\big)\Big]^2\Bigg{\|}_{F}^2\Bigg]^{1/2}\nonumber\\
& \leq \Bigg[\dfrac{2}{n}\Big\{d_0^{1/2}\max_{i\in\{1,..,n\}}\|\bm{x}_i^{(1)}\|_{\infty}\Big\}^4\Big{\|}\big\{\mathbf{E}(\bm{L}_{n,11})\big\}^{-1/2}\Big{\|}^4\Big\{\max_{i\in\{1,..,n\}}\big{|}h^\prime (\bm{x}_i^\top \bm{\beta})\big{|}\Big\}^{4}\nonumber\\
&\;\;\;\;\;\;\;\;\;\;\;\;\;\;\;\;\;\;\;\;\;\;\;\;\;\;\;\;\;\;\;\;\;\;\;\;\;\;\;\;\;\;\;\;\;\;\;\;\;\;\;\;\;\;\;\;\;\;\;\;\;\;\;\;\;\;\;\;\;\;\times\Big(n^{-1}\sum_{i=1}^{n}\mathbf{E}\Big{|}\big\{y_i-g^{-1}(\bm{x}_i^\top \bm{\beta})\big\}\Big{|}^4\Big)\Bigg]^{1/2}\nonumber\\
 &=O\Bigg[\dfrac{d_0^2}{n^{1-2\gamma_1}}\Bigg]^{1/2}
\end{align}
Next we note that,
\begin{align}\label{eqn:hhhhhkkkkkkkhhhhhhhh}
&B_1\le \Bigg[\frac{2}{n^2}\sum_{i=1}^{n}\Bigg{\|}\Big\{[\mathbf{E}(L_{n,11})]^{-1/2}\bm{x}_i^{(1)}\Big\}\Big\{[\mathbf{E}(L_{n,11})]^{-1/2}\bm{x}_i^{(1)}\Big\}^\top\Big[h^\prime\big(\bm{x}_i^{(1)\top}\bar{\bm{\beta}}_n^{(1)}\big)\Big]\Big[(g^{-1})^\prime\big(\bm{x}_i^{(1)\top}\bar{\bm{\beta}}_n^{(1)}\big)\Big]\Bigg{\|}_{F}^2\Bigg]^{1/2}\nonumber\\
&=O_p\Bigg[\dfrac{d_0^2}{n^{1-2\gamma_1}}\Bigg\{1+\frac{d_0}{n^{1-2\gamma_1}}\frac{\lambda_n^2}{n}+\frac{d_0^2}{n^{2-4\gamma_1}}\frac{\lambda_n^4}{n^2}\Bigg\}\Bigg]^{1/2}=O_p\Bigg[\dfrac{d_0^2}{n^{1-2\gamma_1}}\Bigg]^{1/2},
\end{align}
provided $\lambda_n=O[d_0^{-1/2}n^{\frac12-\gamma_1}]$. Next we follow equation (\ref{eqn:vvvvvvvv}) to get,
\begin{align}\label{eqn:vvvvvbbbb}
\Bigg[\frac{1}{n^2}\sum_{i=1}^n\mathbf{E}_*\Bigg\|\tilde{S}_{n,11}^{-1/2}[\mathbf{E}(L_{n,11})]^{-1}\bm{x}_i^{(1)}\sqrt{b^{\prime\prime}\Big(h(\bm{x}_i^{(1)\top}\bar{\bm{\beta}}_n^{(1)})\Big)}h^\prime(\bm{x}_i^{(1)\top}\bar{\bm{\beta}}_n^{(1)})e_i^{**}\Bigg\|_2^4\Bigg]^{1/2}=O_p\Bigg[\frac{d_0^2}{n^{1-2\gamma_1}}\Bigg]^{1/2} \end{align}

This will eventually imply that,
\begin{align}\label{eqn:213uuuuuuhhhh}
\mbox{sup}_{A\in\mathcal{C}}\Big{|}\mathbf{P}_*\Big(\bar{\bm{T}}_n^{*(PRB)(1)}\in A\Big)-\mathbf{P}\Big(\bm{G}_{2n}\in A\Big)\Big{|}
&=O_p\Bigg\{\Bigg[\dfrac{d_0^{5/2}}{n^{1-2\alpha_1}}\Bigg]^{1/2}\Bigg[\log \Bigg(\dfrac{d_0^2}{n^{1-2\gamma_1}}\Bigg)\Bigg]^{1/2}\Bigg\}\nonumber\\
&=o_p(1)
\end{align}
provided $d_0=o[n^{2\{1-2\gamma_1\}/5}]$ for $\gamma_1<1/2$. Similarly for class of Euclidean Balls,
\begin{align}\label{eqn:213kkkkkkjjj}
\mbox{sup}_{A\in\mathcal{B}}\Big{|}\mathbf{P}_*\Big(\bar{\bm{T}}_n^{*(PRB)(1)}\in A\Big)-\mathbf{P}\Big(\bm{G}_{2n}\in A\Big)\Big{|}
&=O_p\Bigg\{\Bigg[\dfrac{d_0^{2}}{n^{1-2\gamma_1}}\Bigg]^{1/2}\Bigg[\log \Bigg(\dfrac{d_0^2}{n^{1-2\gamma_1}}\Bigg)\Bigg]^{1/2}\Bigg\}\nonumber\\
&=o_p(1)
\end{align}
provided $d_0=o[n^{\{1-2\gamma_1\}/2}]$ for $\gamma_1<1/2$. It's simple calculation to check the following:
\begin{itemize}
    \item[$(\nu.1)$] $\Bigg\|\bar{L}_{n,11}^{(PRB)}-\mathbf{E}(L_{n,11})\Bigg\|_{op}=O_p\Bigg[\frac{d_0^{1/2}}{n^{\frac12-\gamma_1}}\frac{\lambda_n}{n^{1/2}}\Bigg]$, provided $\lambda_n=O(d_0^{-1/2}n^{1-\gamma_1})$.
    \item[$(\nu.2)$] $\Big\|\bar{L}_{n,11}^{(PRB)-1}\Big\|_{op}=O_p(n^{\gamma_1})$, provided $\lambda_n=O(d_0^{-1/2}n^{1-2\gamma_1})$.
    \item[$(\nu.3)$] $\Bigg\|\bar{L}_{n,11}^{(PRB)-1}-\{\mathbf{E}(L_{n,11})\}^{-1}\Bigg\|_{op}=O_p\Bigg[\frac{d_0^{1/2}}{n^{\frac12-3\gamma_1}}\frac{\lambda_n}{n^{1/2}}\Bigg]$.
    \item[$(\nu.4)$] $max_{j\in\{d_0+1,...,d\}}\Big{\|}(\bar{L}_{n,21}^{(PRB)})_{j\cdot}-\{\mathbf{E}(L_{n,21})\}_{j\cdot}\Big{\|}_{op}=O_p\Bigg[\frac{d_0}{n^{\frac12-\gamma_1}}\frac{\lambda_n}{n^{1/2}}\Bigg]$, provided $\lambda_n=O(d_0^{-1/2}n^{1-\gamma_1})$.
\end{itemize}

Next we observe that for all $j=3,4,...$,
\begin{align}\label{eqn:spl}
\mathbf{E}_*(e_i^{**j})&\le \frac{j!}{2}\Bigg[\frac1n\sum_{k=1}^{n}(e_k^\#-\bar{e}^\#)^j\Bigg] \nonumber\\
&\le \frac{j!}{2}\Bigg[\underbrace{\mbox{max}_{1\le k\le n}\Big|e_k^\#-\bar{e}^\#\Big|}_{:=b_*\;(say)}\Bigg]^{j-2}\bar{\sigma}_n^2=\frac{j!}{2} b_*\text{var}_*(e_i^{**})
\end{align}
Since for all $i=1,..,n$, the iid centered sequence $e_i^{**}$ satisfies Bernstein condition (\ref{eqn:spl}), from Lemma \ref{lem:BerntoSE} we can claim that, 
$$e_i^{**}\sim \text{SE}(\sqrt{2}\bar{\sigma}_n,2b_*)\;\text{for all}\;\; i=1,..,n.$$
Now since we have assumed $y_k-g^{-1}(\bm{x}_k^{\top}\bm{\beta})\sim \text{SE}(\sqrt{2}\sigma_k,2b)$ for some $b>0$ and independently for all $i=1,..,n$, it can be shown that (cf. Lemma 2.2.10 of \citet{van1996weak}),
$$\mbox{max}_{1\le k\le n}\Big|y_k-g^{-1}(\bm{x}_k^{\top}\bm{\beta})\Big|=O_p(\log n).$$
Note that, $\mbox{max}_{1\le k\le n}\Big|\bm{x}_k^{(1)\top}\Big(\bar{\bm{\beta}}_n^{(1)}-\bm{\beta}^{(1)}\Big)\Big|=O_p\Bigg[\frac{d_0}{n^{\frac12-\gamma_1}}\frac{\lambda_n}{n^{1/2}}\Bigg].$ Then simple algebra and existing assumptions will guide us to,
\begin{align}\label{eqn:kjkjkjkjkjkj}
b_*= \mbox{max}_{1\le k\le n}\Big|e_k^\#-\bar{e}^\#\Big|&=O_p\Bigg[\log n+\frac{d_0}{n^{\frac12-\gamma_1}}\frac{\lambda_n}{n^{1/2}}+\log n\Bigg( \frac{d_0}{n^{\frac12-\gamma_1}}\frac{\lambda_n}{n^{1/2}}\Bigg)+\Bigg(\frac{d_0}{n^{\frac12-\gamma_1}}\frac{\lambda_n}{n^{1/2}}\Bigg)^2\Bigg]\nonumber\\
&=O_p(\log n),\quad\text{provided}\;\; \lambda_n=O(d_0^{-1}n^{1-\gamma_1}).
\end{align}
Now choose $a_i:=n^{-1/2}\sum_{i=1}^{n}\sqrt{b^{\prime\prime}(h(\bm{x}_i^{(1)\top}\bar{\bm{\beta}}_n^{(1)}))}h^\prime(\bm{x}_i^{(1)\top}\bar{\bm{\beta}}_n^{(1)})x_{ij}^{(1)}$ for all $j=1,..,d_0$. Then under the existing assumptions, it's straight forward to have that, $\bar{\sigma}_n^2\sum_{i=1}^na_i^2=O_p(1)$ and $n^{1/2}\mbox{max}_{1\le i\le n}|a_i|=O_p(1)$. Combining everything and using Lemma \ref{lem:univariatebernstein} we get,
$$\Big\|\bar{W}_{n,1}^{*(PRB)}\Big\|_2=o_{p*}(d_0^{1/2}n^{\tau/3})\quad\text{in probability provided}\; \log d_0=o(n^{2\tau/3}).$$
Similar to Lemma \ref{lem:bootlead2concen}-\ref{lem:bootcombined} we get the following:
\begin{itemize}
    \item[$(\kappa.1)$] $\Big\|\bar{W}_{n,2}^{*(PRB)}\Big\|_\infty=o_{p*}(n^{\tau})\quad\text{in probability provided}\; \log d=o(n^{2\tau}),\;\text{with}\; 0<\tau<1/2.$  
\item[$(\kappa.2)$] $\Big{\|}\big\{\mathbf{E}(\bm{L}_{n,21})\big\}\big\{\mathbf{E}(\mathbf{L}_{n,11})\big\}^{-1}\bar{\bm{W}}_{n,1}^{*(PRB)}\Big{\|}_{\infty}=o_{p*}(n^{\tau})\;\;\text{in probability provided}\; \log d=o(n^{2\tau}),\;\text{with}\; 0<\tau<1/2.$
\item[$(\kappa.3)$] $\Big{\|}\bar{\bm{L}}_{n,21}^{(PRB)}\big\{\bar{\bm{L}}_{n,11}^{(PRB)}\big\}^{-1}\bar{\bm{W}}_{n,1}^{*(PRB)}\Big{\|}_{\infty}=O_{p*}(n^\tau)\;\text{in probability}.$
\end{itemize}
Having everything on board, finally we can conclude,
$$\Big{\|}n^{1/2}\Big(\bar{\bm{\beta}}_n^{*(PRB)(1)}-\bar{\bm{\beta}}_n^{(1)}\Big)\Big{\|}_2=O_{p*}\Bigg(\frac{d_0^{1/2}}{n^{\frac{1}{2}-\gamma_1}}\lambda_n\Bigg)\;\text{with probability at least}\; 1-o(1),$$
provided $d_0=o(n^{1-4\gamma_1-2\tau/3})$ with $0<2\gamma_1+\tau/3<1/2$. Rest of the proof is routine following the proof of Theorem \ref{thm:bootapproxglmvsc} and writing the KKT conditions carefully. Hence we skip the details. \hfill $\blacksquare$

\bibliographystyle{amsplain}

\end{document}